%% file: paper.tex
\documentclass[aps,prc,groupedaddress,superscriptaddress,amsmath,floatfix,twocolumn]{revtex4}

\usepackage{nicefrac}
\usepackage{pstricks}
\usepackage{pst-node}
\usepackage{graphicx}

\begin{document}
\title{Hybrid model calculations of direct photons in high-energy nuclear collisions}
\author{Bj\o{}rn B\"auchle}
\email{baeuchle@th.physik.uni-frankfurt.de}
\affiliation{Frankfurt Institute for Advanced Studies, Frankfurt am Main, Germany}
\affiliation{Institut f\"ur Theoretische Physik, Goethe-Universit\"at,
Frankfurt am Main, Germany}
\author{Marcus Bleicher}
\affiliation{Institut f\"ur Theoretische Physik, Goethe-Universit\"at,
Frankfurt am Main, Germany}

\begin{abstract}

Direct photon emission in heavy-ion collisions is calculated within a
relativistic micro+macro hybrid model and compared to the microscopic
transport model UrQMD.  In the hybrid approach, the high-density part of the
evolution is replaced by an ideal 3-dimensional hydrodynamic calculation.
This allows to examine the effects of viscosity and full local
thermalization, in comparison of the transport model to the ideal
fluid-dynamics. We study the origin of high-$p_\bot$ photons as well as the
impact of elementary high-$\sqrt{s}$ collisions. We further explore the
contribution of different production channels and non-thermal radiation to
the spectrum of direct photons. Detailed comparison to the measurements by
the WA98-collaboration are also undertaken.

\end{abstract}

\maketitle

\section{Introduction}\label{sec:intro}

Creating and studying high-density and -temperature nuclear matter is the
major goal of heavy-ion experiments. A state of quasi-free partonic degrees
of freedom, the Quark-Gluon-Plasma (QGP)~\cite{Harris:1996zx,Bass:1998vz}
may be formed, if the energy density in the reaction is high enough. Strong
jet quenching, high elliptic flow and other observations made at the
Relativistic Heavy Ion Collider (BNL-RHIC) suggest the successful creation
of a strongly coupled QGP
(sQGP) at these energies~\cite{Adams:2005dq,Back:2004je,Arsene:2004fa,Adcox:2004mh}. Possible
evidence for the creation of this new state of matter has also been put
forward by collaborations at the Super Proton Synchrotron (CERN-SPS), as for
instance the step in the mean transverse mass excitation function of
protons, kaons and pions and the enhanced $K^+/\pi^+$-ratio \cite{:2007fe}.

Out of the many possible observables, electromagnetic probes have the
advantage of leaving the hot and dense region undisturbed: once they are
created, they escape freely from
the reaction zone, due to their negligible rescattering cross-sections.
Besides dileptons, direct photon emission is therefore of greatest interest
to gain insight into the early, hot and therefore possibly partonic stages
of the reaction.  Direct photons are distinguished from the bulk of photons as
those coming from collisions and not decays.

Unfortunately, the overwhelming amount of photons in heavy-ion collisions
comes from hadronic decays in the late stages, mostly $\pi^0 \rightarrow
\gamma\gamma$.  These decay-photons impose a serious challenge for the
experimental extraction of direct photon data. Up
to now, several experiments have gone through the challenge to obtain the spectra of direct
photons: Helios, WA80 and CERES (all at CERN-SPS) could publish upper
limits, while WA98 (CERN-SPS)~\cite{Aggarwal:2000ps} and PHENIX
(BNL-RHIC)~\cite{Adler:2005ig,:2008fqa} have published explicit data
points for direct photons.

On the theoretical side, calculations for the elementary photon production
processes are known since long, see e.g.\ Kapusta {\it et
al.}~\cite{Kapusta:1991qp} and Xiong {\it et al.}~\cite{Xiong:1992ui}. The
major problem here is the difficulty to describe the time evolution of the produced
matter, which is, up to now, not possible from first principle Quantum
Chromodynamics (QCD). One has to rely on well-developed dynamical models
 to describe the space-time evolution of the nuclear interactions
in the hot and dense stage of the reaction. A well-established approach to
explore the dynamics of heavy-ion reactions is
relativistic transport
theory~\cite{Geiger:1997pf,Bass:1998ca,Bleicher:1999xi,Ehehalt:1995is,Molnar:2004yh,Xu:2004mz,Lin:2004en,Burau:2004ev,Bass:2007hy}.
In this kind of microscopic description, the hadronic and/or partonic stage
of the collision is described under certain approximations. Most transport
models, for instance, cannot describe collisions with more than two incoming
particles, which restricts the applicability to low particle densities,
where multi-particle interactions are less important. Some attempts to
include multi-particle interactions do exist
\cite{Barz:2000zz,Cassing:2001ds,Xu:2004mz,Larionov:2007hy,Bleibel:2006xx,Bleibel:2007se},
but this field of study is still rather new. The coupling of a partonic
phase with a hadronic phase poses another challenge on transport models,
because the microscopic details of that transition are not well known.
Another complication in the transport approach is that all microscopic scatterings are
explicitly treated in the model and therefore the cross-sections for all processes
must be known or extrapolated. However, for many processes high quality experimental
data are not available, and therefore a large fraction of the cross-sections have to be
calculated or parametrized by additional models.

Relativistic, (non-)viscous fluid- or hydrodynamics is a different approach
to explore the space-time evolution of a heavy-ion
collision~\cite{Cleymans:1985wp,McLerran:1986nc,PHRVA.D34.794,Kataja:1988iq,Srivastava:1991ju,Srivastava:1991nc,Srivastava:1991dm,Srivastava:1992gh,Cleymans:1992zc,Rischke:1995mt,Hirano:2001eu,Huovinen:2001wx,Huovinen:2002im,Kolb:2003dz,Nonaka:2006yn,Frodermann:2007ab}.
It constitutes a macroscopic description of the matter that is created,
assuming that at every time and in every place the matter is in local
thermal equilibrium.  This assumption can only be true if the matter is
sufficiently dense, so in the late stages of a heavy-ion collision
fluid-dynamics looses applicability. In addition, the requirement of local
thermal equilibrium restricts the starting time of the hydrodynamic model.
An advantage is that in the dense stages, hydrodynamics can propagate any
kind of matter, and also allows for transitions between two types of matter,
e.g.\ QGP and hadron gas, if an appropriate Equation of State (EoS) is
provided. Fluid-dynamics can therefore be used to study hadronic and
partonic matter in one common framework.

The restrictions of these kinds of models can be loosened as well. By
introducing viscosity and heat conductivity, perfect thermal equilibrium
does not have to be present at any point. However, even with second order
corrections the matter has to be close to
equilibrium~\cite{Dusling:2007gi,Baier:2006um,Song:2008si}.

Input to solve the hydrodynamic differential equations are the boundaries,
i.e.\ the initial state (the distributions of all relevant densities and
currents at the time the evolution starts), the Equation of State providing
the pressure as function of the energy and baryon densities, that describes
the behaviour of the matter that is considered, and the freeze-out
hypersurface.

Finally, approaches to the theoretical description of direct photon spectra
may include calculations in perturbative Quantum-Chromo-Dynamics (pQCD).
Calculations based on pQCD describe the high-$p_\bot$ photon data in
proton-proton collisions very well and, if scaled by the number of binary
nucleon-nucleon
collisions, also those in heavy-ion reactions. The range of applicability of
these calculations is, however, limited to high transverse momenta
$p^{\gamma}_\bot \gg 1~{\rm GeV}$.

The different approaches have different applicability limits. E.g., thermal
rates can only be applied if the assumption of local thermal equilibrium is
fulfilled. Photon emission rates can then be calculated by folding the
particle distribution functions of the participating particle species with
the respective cross-sections.  This framework can be applied to either
static models, simplified hydrodynamics-inspired models such as the blast
wave model and to full fluid-dynamic calculations. The space-time evolution
of a reaction as predicted by microscopic theories can be averaged over in
order to apply thermal rates to the coarse-grained
distributions~\cite{Huovinen:2002im}.  The application of microscopic
cross-sections can only be undertaken in a model where all microscopic
collisions are known. That limits the field of use to transport models.
For previous calculations of photon spectra from transport
models see e.g.\ ~\cite{Dumitru:1998sd,Srivastava:1998tf,Bratkovskaya:2008iq}.

In this paper, we investigate the spectra of direct photons coming from
microscopic hadronic scatterings, thermal hadronic and partonic emission and
hard initial pQCD scatterings. We compare results from a purely microscopic
model to those from an integrated micro+macro hydrodynamic approach that
embeds a hydrodynamic phase into the UrQMD approach. The paper is organized
as follows.  First, we explain the hybrid model (Section~\ref{sec:hybrid}).
In the following Section~\ref{sec:photons}, we elaborate on the photon
sources considered in our model. In Sections~\ref{sec:rates},
\ref{sec:spectra} and \ref{sec:times}, we apply our model to compare thermal
rates from microscopic theory to those from the literature, compare
different physics assumptions with experimental data from the WA98
collaboration~\cite{Aggarwal:2000ps} and analyse the sources of photon
emission. 

\section{The hybrid model}\label{sec:hybrid}

\subsection{Transport model}

UrQMD v2.3 (Ultra-relativistic Quantum Molecular Dynamics) is a microscopic
transport model~\cite{Bass:1998ca,Bleicher:1999xi,Petersen:2008kb}. It includes all
hadrons and resonances up to masses $m \approx 2.2~{\rm GeV}$ and at high
energies can excite and fragment strings. The cross-sections are either
parametrized, calculated via detailed balance or taken from the additive
quark model (AQM), if no experimental values are available.  At high parton
momentum transfers, PYTHIA~\cite{Sjostrand:2006za} is employed for pQCD
scatterings. 

\begin{figure}
 \input{pythia.tex}
 \caption{(Color Online.) Charged particle spectra from
 proton-proton-collisions at $E_{\rm lab} = 158$~GeV calculated with UrQMD
 with (solid black line) and without (dashed red line) PYTHIA.}
 \label{fig:urqmd:pythia}
\end{figure}
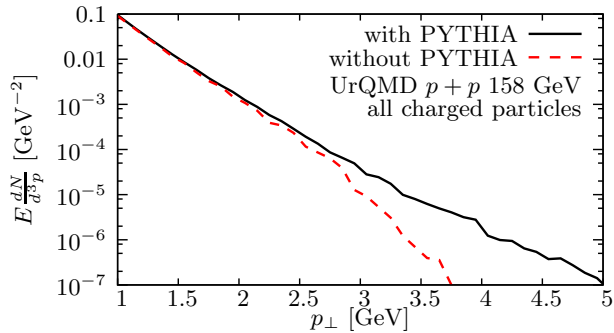

UrQMD differentiates between two regimes for the excitation and
fragmentation of strings. Below a momentum transfer of $Q < 1.5$~GeV a
maximum of two longitudinal strings are excited according to the LUND
picture, at momentum transfers above $Q > 1.5$~GeV hard interactions are
modelled via PYTHIA.  Figure~\ref{fig:urqmd:totxsect} shows the total
cross-sections of resonant hadronic interactions, string excitation and
(hard) PYTHIA-scatterings as a function of the center-of-mass energy of the
collision $\sqrt{s_{\rm coll}}$. The contribution of hard scatterings to the
total $\pi^+\pi^-$ cross-section at the highest SPS-energies ($\sqrt{s}
\approx 17.3$~GeV) is about 4~\%.  Figure~\ref{fig:urqmd:pythia} shows a
comparison between charged particle spectra from proton-proton collisions
calculated in UrQMD with and without the PYTHIA contribution.  For detailed
information on the inclusion of PYTHIA, the reader is referred to Section~II
of~\cite{Petersen:2008kb}.

In the UrQMD framework, all particle properties (mass, width, spectral
shape) are taken at their vacuum values, the propagation is performed
without potentials (cascade mode). UrQMD has been used by Dumitru {\it et al.} to
study direct photon emission earlier~\cite{Dumitru:1998sd}; a brief
comparison between their results and the results obtained with this approach
can be found in Appendix~\ref{app:dumitru}.

\begin{figure}
 \input{sigmatot.tex}
 \caption{(Color Online.) UrQMD-cross-sections for $\pi^+\pi^-$-collisions
 as function of center-of-mass energy. We show the resonant hadronic cross-section
 (red dashed line), the cross-section for the formation of strings (blue
 dotted line) and for hard scatterings via PYTHIA (green dash-dotted line).
 The peak at the $\rho$-meson pole mass has been cut out for better
 visibility.}
 \label{fig:urqmd:totxsect}
\end{figure}
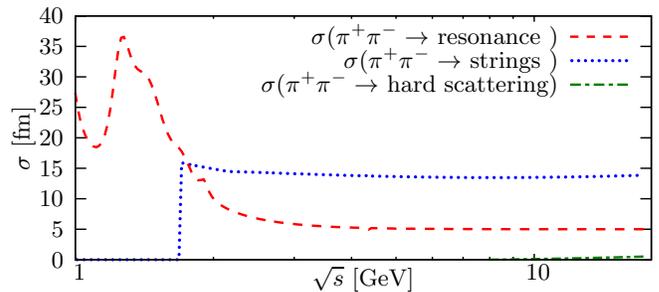

\subsection{Hybrid model}

In the following, we compare results from this microscopic model to results
obtained with a hybrid model description~\cite{Petersen:2008dd}. Here, the
high-density part of the reaction is modelled using ideal 3+1-dimensional
fluid-dynamics.  The unequilibrated initial state and the low-density final
state are described by UrQMD. Thus, those stages are mainly governed by
string dynamics (initial state) and hadronic rescattering (final state).

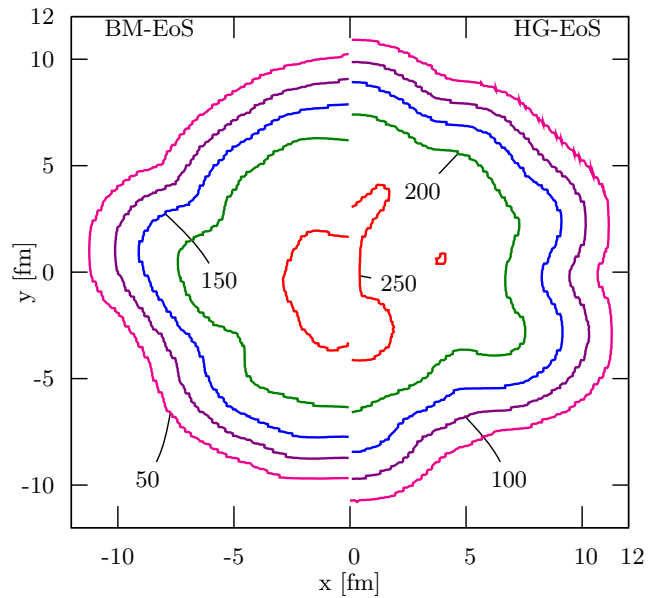
\begin{figure}
 \input{initialtemperature.tex}
 \caption{(Color Online). Temperature profiles after switching to
 fluid-dynamic description, calculated with the Bag model (left half) and
 hadron gas Equation of State (right half). Along the lines, the temperature
 is constant, going from $T = 50$~MeV at the outermost line to $T = 250$~MeV
 at the innermost. Calculations have been done for $Pb+Pb$-collisions at
 $E_{\rm Lab} = 158$~AGeV with $b = 0$~fm.} \label{fig:temperature}
\end{figure}

To connect the initial transport phase to the fluid-dynamic phase, the
baryon-number-, energy- and momentum-densities are smoothed and put into the
hydrodynamic calculation after the incoming nuclei have passed through each
other. Note that in non-central collisions, the spectators are propagated in
the cascade. Thus, the initial state for the hydrodynamic stage is subject
to both geometrical and event-by-event fluctuations. Temperature, chemical
potential, pressure and other macroscopic quantities are determined from the
densities by the Equation of State (EoS) used in the current calculation.
During this transition, the system is forced into an equilibrated state,
regardless of the actual level of equilibration before the transition. The
initial temperature profile at $z = 0$ for a sample $Pb+Pb$-event with $b =
0$~fm and $E_{\rm Lab} = 158$~AGeV for two different EoS (Bag Model and
Hadron Gas) is shown in Figure~\ref{fig:temperature}. Then the ideal (3+1)
dimensional hydrodynamic equations are solved on a grid using the
SHASTA-algorithm~\cite{Rischke:1995mt}.

After the local rest frame energy density has dropped below a threshold
value of $\epsilon_{\rm crit} = 730~{\rm MeV}/{\rm fm}^3\,(\approx 5
\epsilon_0)$, particles are created on an isochronous hyper-surface from the
densities by means of the Cooper-Frye formula and propagation is continued
in UrQMD. 

The transition scenario chosen for the
present studies is always isochronous, i.e.\ the whole system must meet the
criterion at the same calculational-frame time before the transition is
performed.  Earlier investigations within this hybrid model include an
extensive analysis of the effect of changing the transition
criterion~\cite{Petersen:2008dd}, strangeness
production~\cite{Steinheimer:2008hr,Petersen:2009zi}, HBT
correlations~\cite{Li:2008qm}, transverse mass~\cite{Petersen:2009mz} and
elliptic flow~\cite{Petersen:1900zz}.

\subsection{Equations of State}

For the investigations presented here, different Equations of State are used
for the hydrodynamic phase. The base line calculations are done with a
hadron gas Equation of State (HG-EoS), which includes the same degrees of
freedom as present in the transport phase.  This allows to explore the
effects due to the change of the dynamic description. Secondly, a MIT-Bag
model EoS (BM-EoS) with a partonic phase and a first order phase
transition~\cite{Rischke:1995mt} is employed. The BM-EoS thus allows for
investigations of photon emission from the QGP. In order to obtain
meaningful values of temperature and chemical potentials from the densities,
the BM-EoS is smoothly transferred to the HG-EoS just above the transition
energy density. 

\section{Photon emission sources}\label{sec:photons}

Photon emission is calculated perturbatively in both scenarios, hydrodynamics
and transport, because the evolution of the underlying event is not altered
by the emission of photons due to their very small emission probability. The
channels considered for photon emission may differ between the hybrid
approach and the binary scattering model.  Emission from a
Quark-Gluon-Plasma can only happen in the hydrodynamic phase, and only if
the Equation of State used has partonic degrees of freedom. Photons from
baryonic interactions are neglected in the present calculation. Emission of
hard photons from early pQCD-scatterings of nucleons is calculated
separately and incoherently added to the simulated spectra.

\subsection{Photons from microscopic collisions}\label{sec:photons:urqmd}

In the transport part of the (hybrid) model, each scattering is examined and the
cross-section for photon emission is calculated. Here, we employ the
well-established cross-sections
from Kapusta {\it et al.}~\cite{Kapusta:1991qp} and Xiong {\it et
al.}~\cite{Xiong:1992ui}. Kapusta and collaborators based their calculations
on the photon self-energy derived from a Lagrange density involving the
pion, $\rho$ and photon-fields
\begin{equation} \mathcal{L} = |D_\mu \Phi|^2 - m_\pi^2 |\Phi|^2 -
\frac{1}{4} \rho_{\mu\nu} \rho^{\mu\nu} + \frac{1}{2} m_\rho^2 \rho_\mu
\rho^\mu - \frac{1}{4} F_{\mu\nu} F^{\mu\nu} \quad.\label{eq:lagrangian:kls}
\end{equation}

Here, $\Phi$ is the pion field, $\rho_{\mu\nu} = \partial_\mu
\rho_\nu - \partial_\nu \rho_\mu$ and $F_{\mu\nu} = \partial_\mu A_\nu -
\partial_\nu A_\mu$ are the $\rho$ and photon field-strength tensors and $D_\mu
= \partial_\mu - \imath e A_\mu - \imath g_\rho \rho_\mu$ is the covariant
derivative. The $\rho$ decay constant $g_\rho$ is calculated from the total
width $\Gamma^\rho_{\rm tot}$ of the $\rho$ meson:

\begin{equation}
g_\rho^2 = 48\pi\frac{\Gamma^\rho_{\rm tot}
m_\rho^2}{\left(\sqrt{m_\rho^2-4m_\pi^2}\right)^3}\quad.
\end{equation}

The differential cross-sections used for the present
investigation~\cite{Kapusta:1991qp,Xiong:1992ui} are given
in Appendix~\ref{app:differential}.

All scatterings during the transport phase are examined in order to obtain
direct photon spectra. For every scattering that may produce photons (i.e.\
those that have initial states equal to the processes listed in
Appendix~\ref{app:differential}), the corresponding fraction of a photon,
\begin{equation} N_\gamma = \frac{\sigma_{\rm em}}{\sigma_{\rm tot}}
\label{eq:Ngamma_trans}\quad,\end{equation}
is produced. Here, $\sigma_{\rm tot}$ is the sum of the total
hadronic cross-section for a collision with these ingoing particles (as
provided by UrQMD) and the electromagnetic cross-section $\sigma_{\rm em}$ as
calculated by the aforementioned formul\ae{}. In order to
obtain the correct angular distribution of the produced photons and to
enhance statistics, for each scattering many fractional photons are created
that populate all kinematically allowed momentum transfers $t$. In this
procedure, each photon is given a weight $\Delta N^{\rm t}_\gamma$ according
to
\begin{equation} \Delta N^{\rm t}_\gamma = \frac{\frac{d\sigma_{\rm em}}{dt}
(s,t)
\Delta t}{\sigma_{\rm tot} (s)}\quad,
\label{eq:Ngamma_int} \end{equation}
and the photons are distributed evenly in the azimuthal angle $\varphi$. The integral
$\sigma_{\rm em}(s) = \int \nicefrac{d\sigma_{\rm em}}{dt} dt$ is performed analytically for each
channel. The resulting formul\ae{} are shown in
Appendix~\ref{app:integrated}.

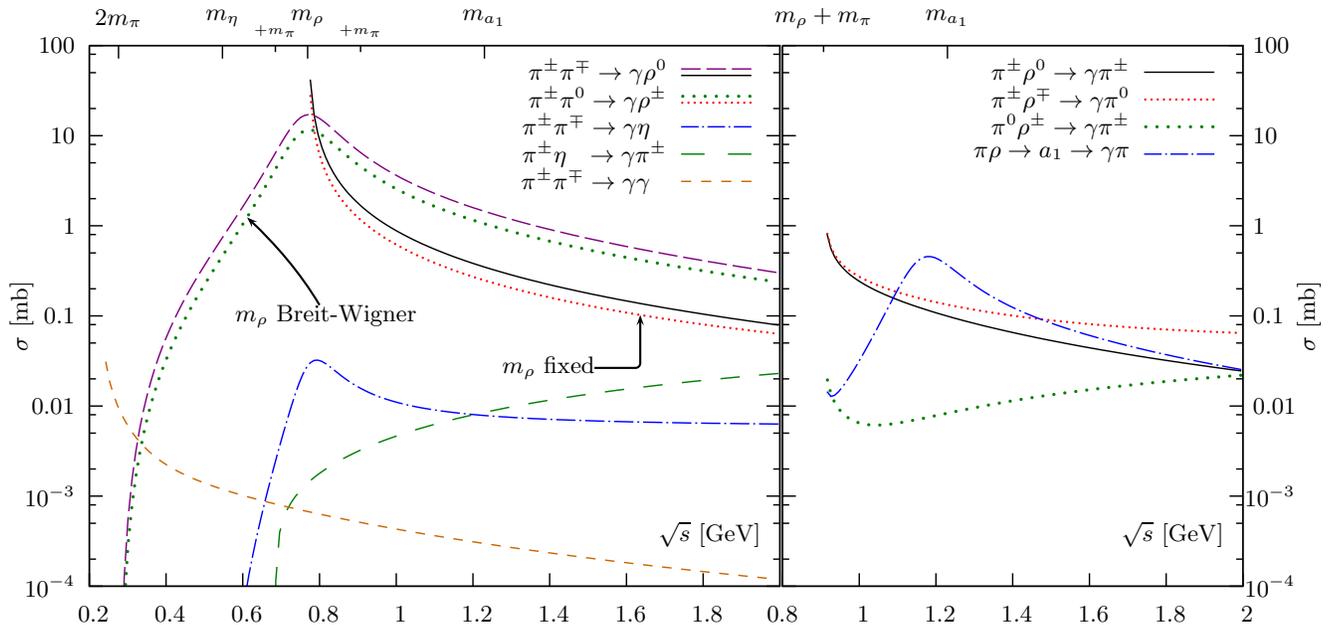
\begin{figure*}
\input{sigma.tex}
\caption{(Color Online.) Cross-sections for all included channels as a
function of $\sqrt{s}$. For visibility, the
cross-sections for all processes $\pi\rho\rightarrow\gamma\pi$ are
shown separately. They have been calculated with a $\rho$ mass
$m_\rho = 0.769$~GeV. The left plot shows the cross-sections for
$\pi\pi\rightarrow\gamma\rho$ both for fixed $\rho$ mass ($m_\rho = 0.769$~GeV,
labelled ``$m_\rho$ fixed'') and for variable $\rho$ mass (labelled ``$m_\rho$
Breit-Wigner'').}
\label{fig:crosssections}
\end{figure*}

Since the width of the $\rho$-meson is not negligible, its mass distribution
has to be taken into account. For the processes with a $\rho$-meson in the
initial state, the actual mass $m_\rho = \sqrt{p_\mu p^\mu}$ of the incoming
meson is used for the calculation of the cross-section. If there is a
$\rho$-meson in the final state, then first the mass of the $\rho$ is chosen
randomly according to a Breit-Wigner distribution with mass-dependent width.
This mass is then used for all further calculations of this process.
Figure~\ref{fig:crosssections} shows the cross-sections of the channels
listed above as a function of $\sqrt{s}$.

\subsection{Photons from hydrodynamics}\label{sec:photons:hydro}

In the hydrodynamic phase photons are produced fractionally from every cell
on the hydrodynamic grid whose energy density is above a threshold
$\varepsilon_{\rm thr} = 10^{-12}~\varepsilon_0$ using the parametrizations
by Turbide, Rapp and Gale~\cite{Turbide:2003si}.  They use an effective
non-linear $\sigma$-model Lagrange density in which the vector and axial
vector fields are implemented as massive gauge fields of the chiral $U(3)_L
\times U(3)_R$ symmetry to obtain the rates. For details on this ansatz, the
reader is referred to the original publication \cite{Turbide:2003si}.

As mentioned earlier, the processes calculated by Turbide {\it et al.}
differ from those considered by Kapusta {\it et al}. Only the processes
$\pi\pi\rightarrow\gamma\rho$ and $\pi\rho\rightarrow\gamma\pi$ are
therefore common in both models. The rate of Turbide {\it et al.} for
$\pi\rho\rightarrow\gamma\pi$ directly includes the process with an
intermediate $a_1$-meson.

To simplify the calculations, all photon rates in~\cite{Turbide:2003si} are
parametrized by the general form
\begin{equation} E\frac{dR}{d^3p} = A \exp{\left(\frac{B}{(2ET)^C} - D
\frac{E}{T}\right)}\quad,\label{eq:rate:shape}\end{equation}
where $A$, $B$, $C$ and $D$ are linear functions of some power of the
temperature $T$: $A(T) = A_1 + A_2 T^{A_3}$. 
The parameter set can be obtained from~\cite{Turbide:2003si}.
In the rates, the energy $E$ and temperature $T$ are to be given
in units of GeV, and the result will have the unit GeV$^{-2}$\ fm$^{-4}$. We
also employ the hadronic form factor introduced in~\cite{Turbide:2003si}.

In the Quark-Gluon-Plasma, the rate used is taken from Ref.~\cite{Arnold:2001ms}. They
computed the full leading-order result as

\begin{widetext}
\begin{equation}\label{eq:rate:qgp}
E\frac{dR}{d^3p} = \sum_{i=1}^{N_f}q_i^2 \frac{ \alpha_{\rm em}\alpha_{\rm
S} }{ 2\pi^2} T^2 \frac{1}{e^x+1} \left (
\ln{\left(\frac{\sqrt{3}}{g}\right)}+\frac{1}{2}\ln{\left(2x\right)}+C_{22}(x)
+C_{\rm brems}(x)+C_{\rm ann}(x)\right )\ ,
\end{equation}
and give convenient parametrizations for the contribution of
$2\leftrightarrow2$-, bremsstrahlung- and annihilation-processes ($C_{22}$,
$C_{\rm brems}$ and $C_{\rm ann}$, respectively)
\begin{subequations}\label{eq:rate:contributions}
\begin{eqnarray}\label{eq:rate:c22}
C_{22}(x) &=& 0.041 x^{-1} - 0.3615 + 1.01 \exp\left(-1.35
x\right)\ \,\\\label{eq:rate:cba}
C_{\rm brems}(x) + C_{\rm ann}(x) &=& \sqrt{1+\frac{N_f}{6}}\left[ \frac{0.548
\ln{\left(12.28 + \frac{1}{x}\right)}}{x^{\frac{3}{2}}} + \frac{0.133
x}{\sqrt{1 + \frac{x}{16.27}}}\right]\quad.
\end{eqnarray}
\end{subequations}
\end{widetext}

In Equations~\eqref{eq:rate:qgp} and \eqref{eq:rate:contributions}, $x =
\nicefrac{E}{T}$, $q_i$ is the charge of quark-flavour $i$, $\alpha_{\rm em}$
and $\alpha_{\rm S} = \nicefrac{g^2}{4\pi}$ are the electromagnetic and QCD
coupling constants, respectively. In our calculations, we use $N_f = 3$, and
therefore $\sum_i q_i^2 = \nicefrac{2}{3}$. The temperature dependence of
$\alpha_{\rm S}$ is taken from \cite{Liu:2007tw} as
\begin{equation}\label{eq:alphas}
\alpha_{\rm S}(T) = \frac{6\pi}{(33-2N_f)\ln{\left(\frac{8T}{T_C}\right)}}  ,
\end{equation}
and the critical temperature at $\mu_B = 0$ to be $T_C = 170$~MeV.

\subsection{Photons from primordial pQCD-scatterings}\label{sec:photons:pqcd}

At high transverse momenta, a major contribution to the photon
yield is the emission of photons from hard pQCD-scatterings of the partons
in the incoming protons. In the intermediate and low $p_\bot$-regions, the
contribution may be comparable to or smaller than the yield from other sources.

We apply the results extracted by Turbide {\it et al.}~\cite{Turbide:2003si}.
They first scale the photon spectrum from proton-proton-collisions by the
number of binary collisions in Pb+Pb-collisions, and then add a
Gaussian-shaped additional $k_\bot$-smearing to the result. The
width of the Gaussian is obtained by fitting this procedure to the data from
proton-nucleus collisions. The results shown here are obtained with a
$\langle \Delta k_\bot^2\rangle = 0.2$~GeV$^2$.

For comparison, we also show pQCD spectra obtained earlier by
Gale~\cite{Gale:2001yh} following Wong {\it et al.}~\cite{Wong:1998pq}. They
follow the same procedure as explained above. The authors of
\cite{Gale:2001yh} obtain a higher intrinsic transverse parton momentum of
$\langle \Delta k_\bot^2\rangle = 0.9$~GeV$^2$, yet lower spectra.

In order to compare our calculations to experimental data in
Figure~\ref{fig:results:comparepqcd}, we use the newer calculations by
Turbide {\it et al.}~\cite{Turbide:2003si}.

\section{Rates from transport and hydrodynamics}\label{sec:rates}

Before comparing photon spectra from complex nucleus-nucleus collisions
between cascade- and hybrid model, we check if both approaches give
similar results for the setup of a fully thermalized box.

I.e., we perform UrQMD calculations in a box~\cite{nucl-th/9804058},
allowing only $\pi$-, $\rho$- and a$_1$-mesons to be present and to scatter.
When the matter in the box has reached thermal and chemical equilibrium, the
rate of photon emission is extracted based on the microscopic scatterings
with the procedure described in Section~\ref{sec:photons:urqmd}.  Then, we
compare the microscopic rates to the hydrodynamic rates from
Equation~\eqref{eq:rate:shape} with the parameters
from~\cite{Turbide:2003si}. Since the available rates in the cascade and
hydrodynamic modes differ, as pointed out above, we restrict the comparison
to the common rates $\pi\pi\rightarrow\gamma\rho$ and
$\pi\rho\rightarrow\gamma\pi$. The cascade-rates are explicitly summed over
all charge combinations.

\begin{figure}
\input{ratecomp.tex}
\caption{(Color Online.) Comparisons between the rates from
\cite{Turbide:2003si} (lines) and box calculations restricted to a
$\pi,\rho,a_1$-system (points) with UrQMD at $T
= 150$~MeV.  } \label{fig:results:rates} \end{figure}
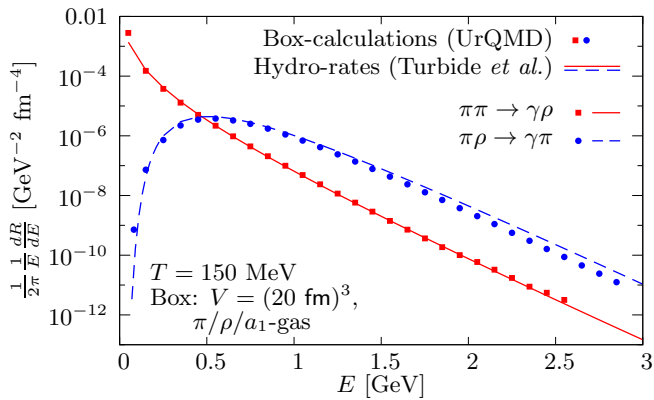

Figure~\ref{fig:results:rates} shows the comparison of the rates at a
temperature $T = 150$~MeV. It can be seen that the microscopically obtained
rates agree very well with the thermodynamic rates.

\section{Photon spectra}\label{sec:spectra}

\subsection{Emission stages}\label{sec:spectra:stages}

For the present investigation, the evolution of a heavy-ion reaction is
divided into three stages: ``initial stage'', ``intermediate stage'' and
``final stage''. The time spans of these stages are defined by the
calculations in the hybrid model as before ($t < 1.4$~fm), during
($1.4$~fm~$< t < 13.25$~fm for the hadron gas and cascade calculations and
$1.4$~fm~$< t < 30$~fm for the Bag model calculations), and after the
hydrodynamic description ($t > 13.25$~fm (HG, cascade) and $t > 30$~fm
(BM)). I.e., the initial phase denotes the initial string dynamics until the
transition to hydrodynamics. Since all differences between the models start
at the transition to hydrodynamics, this stage and its contribution is
always the same.

Secondly, the ``intermediate stage'' denotes the phase described by
hydrodynamics. It starts with the transition to hydrodynamics and ends with
the transition back to the cascade. Within this stage, the degrees of
freedom may be partonic, hadronic and partonic (in case of a Bag model EoS
calculation) or hadronic (in all cases). The ``final stage'' starts at the
transition from hydrodynamics to the cascade.  On average this happens at
$13.25$~fm for the hadron gas EoS and $30$~fm for the Bag model EoS. For
pure transport calculations, this phase is also set to start at $13.25$~fm,
so that a comparison between hadron gas EoS and cascade calculations is
possible. The degrees of freedom in the ``final stage'' are hadrons and
resonances. The calculations are done with UrQMD in cascade mode, and
proceed until the last collision has happened.

\subsection{Influence of EoS and dynamics}\label{sec:spectra:influence}

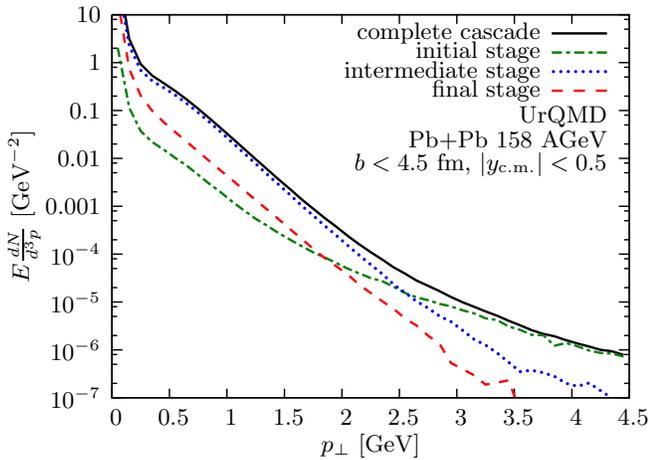
\begin{figure}
\input{stages_casc.tex}
\caption{(Color Online.) UrQMD calculation. Contributions of the initial
($t < 1.4$~fm, dash-dotted green line),
intermediate ($1.4$~fm~$ < t < 13.25$~fm, dotted blue line) and
final ($13.25$~fm~$< t$, dashed red line) stages to the spectrum from
pure cascade calculations (solid black line).}
\label{fig:results:stages:cascade}
\end{figure}

\begin{figure}
\input{stages_hg.tex}
\caption{(Color Online.) Hybrid model calculation with HG-EoS. Contributions
of the initial (green dash-dotted line), intermediate (blue dotted line) and
final (red dashed line) stages to the inclusive spectrum (solid black
line). In the intermediate stage, the matter
is described by hydrodynamics.}
\label{fig:results:stages:hg}
\end{figure}
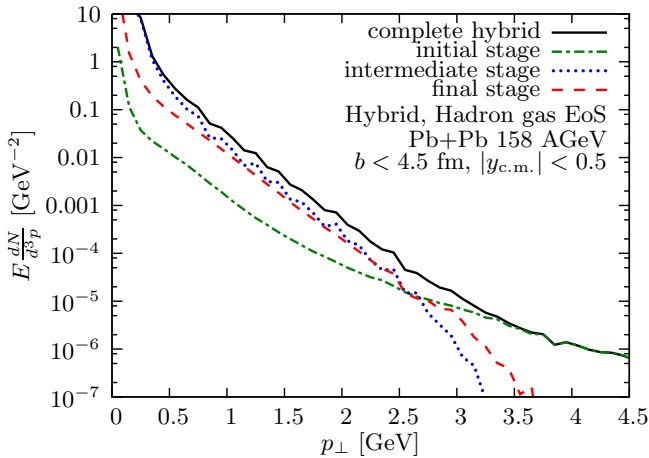

The contributions of each of these phases to the final spectra are shown in
Figure~\ref{fig:results:stages:cascade} for the pure cascade calculation and
in Figure~\ref{fig:results:stages:hg} for the hybrid calculation with the
hadron gas EoS. In both figures, the black solid lines show the complete
direct photon spectra, the green dash-dotted lines show the initial
stage-contribution, the red dashed lines show the photon spectrum from the
final phase and the blue dotted lines show the contribution from the
intermediate stage.  In the hybrid model, the evolution during the
intermediate stage is calculated within the hydrodynamic framework. We find
that the contribution from this intermediate phase ($1.4$~fm~$< t <
13.25$~fm) is similar in both models. However, at high transverse momenta
$p_\bot > 3$~GeV, the cascade calculation yields more photons from this
intermediate time span. This can be related to imperfect thermalization of
the system at the transition from the initial non-equilibrium state, which
is forced to thermalization at the transition to the hydrodynamic phase, but
preserved when doing cascade-only calculations.  However,
Figure~\ref{fig:results:stages:cascade} suggests photon emission towards
high transverse momenta from the intermediate stage is in any case strongly
suppressed with respect to photon emission from the initial stage. It is
therefore justified to neglect non-equilibrium effects from the intermediate
phase in the hybrid model.  The major difference between cascade and hybrid
model calculations is the magnitude of the contributions from the final
stage (red dashed lines). Here, both models describe the system in the same
way. The contribution from this phase to the hybrid model direct photon
spectrum is very similar to the contribution from the intermediate stage.
In the pure cascade calculation, on the contrary, the final phase
contributes roughly a factor of 5 less to the spectrum than the intermediate
stage does.  A possible explanation for this is that the transition
procedure from hydrodynamic to transport description enhances the number of
meson-meson collisions in the late phase.

Interactions at early times ($t < 1.4$~fm) are a significant source
of high transverse momentum $p_\bot > 3$~GeV photons. Here, the thermalized
system of the later phase provides only minor contributions to the inclusive
spectrum.

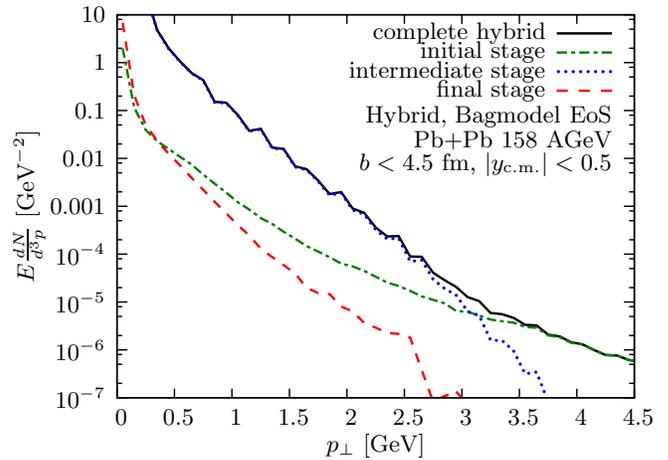
\begin{figure}
\input{stages_bm.tex}
\caption{(Color Online.) Hybrid model calculation with BM-EoS.
Contributions to the initial (green dash-dotted line), intermediate (blue
dotted line) and final (red dashed line) stages to the inclusive spectrum
(solid black line). In the intermediate stage, the matter is described by
hydrodynamics.}
\label{fig:results:stages:bm}
\end{figure}

In the same analysis done for the hybrid model with the Bag model EoS (see
Fig.~\ref{fig:results:stages:bm}), the picture is different.  The
contribution from the hydrodynamic intermediate stage is strongly enhanced,
and the contribution of the final stage after the transition from the
hydrodynamic phase is, in turn, reduced with respect to the cases presented
before. The total photon spectrum is completely dominated by emission from
the hydrodynamic phase at low and intermediate transverse photon momenta
$p_\bot < 3$~GeV.  This coincides with the observation that the length of
the hydrodynamic phase in this model ($\approx 29$~fm) is much longer than
in the calculations with hadron gas EoS ($\approx 12$~fm), due to the large
latent heat in the setup with the Bag model EoS. At high transverse photon
momenta, the initial stage non-equilibrium cascade phase dominates the
spectrum, as in the other cases discussed above.

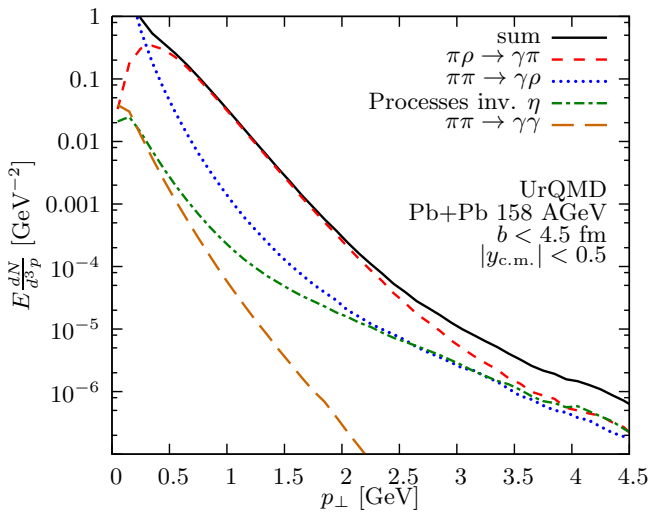
\begin{figure}
\input{cascade.tex}
\caption{(Color Online.) UrQMD calculation. Contributions of the different
channels.}
\label{fig:results:chan:cascade}
\end{figure}

\begin{figure}
\input{hadrongas.tex}
\caption{(Color Online.) Hybrid model calculation with hadron gas EoS.
Contributions of the different channels.}
\label{fig:results:chan:hadrongas}
\end{figure}
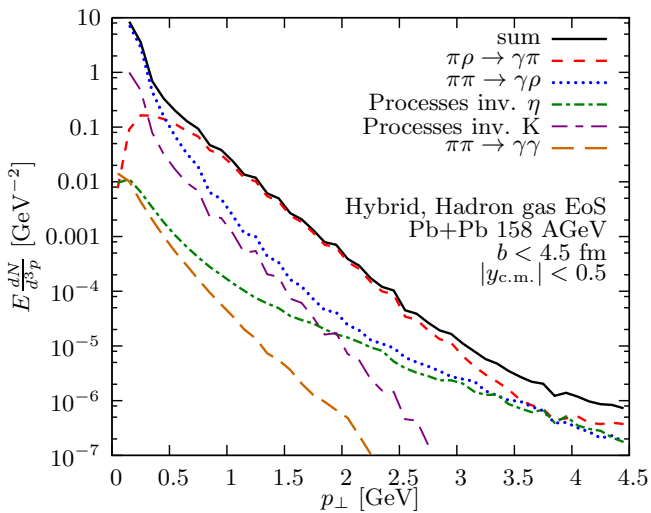

\subsection{Channel decomposition}\label{sec:spectra:channel}

The contributions of the different channels to the hadronic calculations,
both in the pure cascade mode and the hybrid approach with a hadron gas EoS,
is shown in Figures~\ref{fig:results:chan:cascade} and
\ref{fig:results:chan:hadrongas}.  The dominant contributions in both models
are very similar; at low transverse momenta $p_\bot < 0.5$~GeV, the process
with two initial pions $\pi\pi\rightarrow\gamma\rho$ is dominant, while in
the broad range of $0.5$~GeV $< p_\bot < 3.5$~GeV,
$\pi\rho\rightarrow\gamma\pi$ is the major source of photons. Processes with
an eta-meson ($\pi\pi\rightarrow\gamma\eta$ and
$\pi\eta\rightarrow\gamma\pi$) become important at high transverse momenta,
where they contribute in similar magnitude as the two aforementioned
channels. Processes with kaons contribute less to the spectrum than the
corresponding non-strange channels, and the process
$\pi\pi\rightarrow\gamma\gamma$ is the least significant in all
calculations, as expected.

In the photon spectrum extracted from the cascade calculation, one can
observe a flattening of the spectrum at high transverse momenta $p_\bot
\approx 3$~GeV. At this point, $\pi\pi\rightarrow\gamma\rho$ and
$\eta$-processes start to provide significant contributions to the photon
spectrum and lead to a flatter slope already at $p_\bot\approx 2$~GeV. In
Section~\ref{sec:times}, we come back to this slope change and show it to be
consistent with the average emission times of the photons at these
transverse momenta.

\begin{figure}
\input{bagmodel.tex}
\caption{(Color Online.) Hybrid model calculation with Bag model EoS.
Contributions of the different channels in hybrid calculation with Bag model
EoS.}
\label{fig:results:chan:bagmodel}
\end{figure}
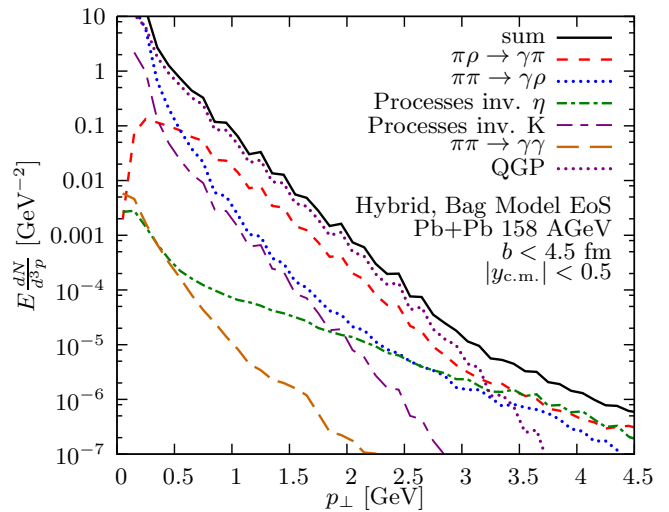

The calculations with the Bag model EoS yield a different picture (see
Figure~\ref{fig:results:chan:bagmodel}). Here, the dominant contribution
comes from the Quark Gluon Plasma, whose emission magnitude is about two
times higher than the combined contribution from all hadronic processes.
Again, initial stage (pre-equilibrium) processes are dominant at high
transverse momenta $p_\bot > 3$~GeV. Apart from that, we observe a similar
distribution among the reaction channels as we did above (see discussion of
Figure~\ref{fig:results:chan:hadrongas}); the process
$\pi\rho\rightarrow\gamma\pi$ is dominant in the intermediate
$p_\bot$-region, $\pi\pi\rightarrow\gamma\rho$ dominates the hadronic
contribution at low $p_\bot$, the early scatterings in those channels and
the channels involving $\eta$-mesons contribute to the high $p_\bot$-region
in approximately equal amounts.

\begin{figure}
\input{directcomp.tex}
\caption{(Color Online.) Comparison of direct photon spectra from the hybrid
model with hadron gas EoS and transport model, using only common channels.}
\label{fig:results:directcompare}
\end{figure}
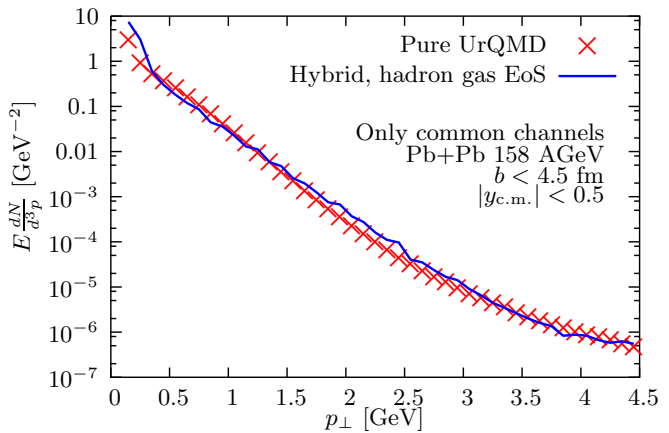

Figure~\ref{fig:results:directcompare} shows a comparison of hybrid model
calculations with hadron gas EoS and transport-calculations, using only
common channels applied to a dynamic system (see
Figure~\ref{fig:results:rates} for a static comparison). One observes that
the direct photon spectrum is not sensitive to the change in the underlying
dynamics (e.g.\ finite viscosities vs.\ ideal fluid) and indicates that even
the two-body collision dynamic in UrQMD drives the system into equilibrium
in the $\pi-\rho-a_1$ channel.

\subsection{Comparison to data}\label{sec:spectra:data}

\begin{figure}
\input{comparison.tex}
\caption{(Color Online.) Comparison of the direct photon spectra from all
variations of the model to the experimental data
by the WA98-collaboration~\cite{Aggarwal:2000ps}. Calculations
without intermediate hydrodynamic stage (pure cascade mode) are shown as red
crosses, hybrid calculations with hadron gas EoS as red solid line and Bag
model calculations are depicted by the dark-green dotted line. The data
contain no photons from initial proton-proton collisions.}
\label{fig:results:compare}
\end{figure}
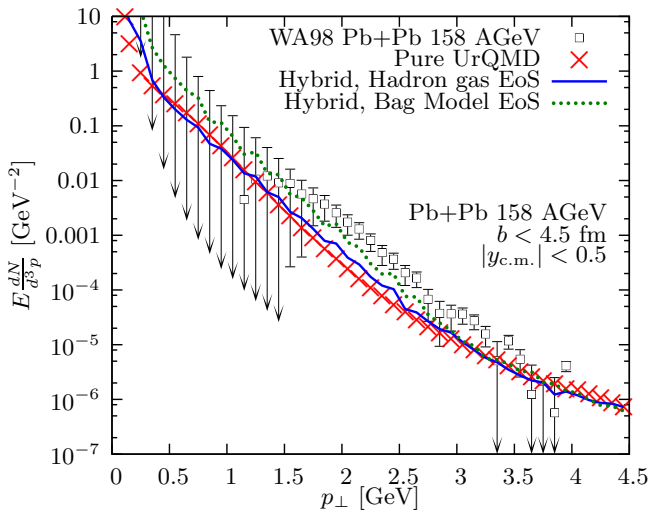

A comprehensive comparison between the models presented above and
experimental data is shown in Figure~\ref{fig:results:compare}. Here, we
show the sum over all channels and all stages in the cases of pure
UrQMD-calculations and hybrid model with hadron gas and Bag model EoS. The Bag
model Equation of State yields the highest photon spectra and provides a
reasonable description of the data from the
WA98-collaboration~\cite{Aggarwal:2000ps}. The purely hadronic calculations,
with and without intermediate hydrodynamic calculation, give smaller yields
than the scenario with a QGP.

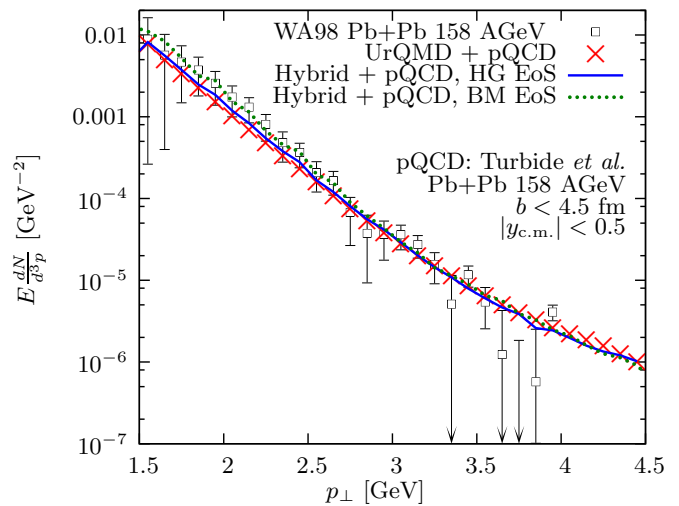
\begin{figure}
\input{comparison_pqcd.tex}
\caption{(Color Online.) Comparison of the direct photon spectra from the
model, with added pQCD-photons from~\cite{Turbide:2003si}, to the
experimental data by the WA98-collaboration~\cite{Aggarwal:2000ps}.
Calculations without intermediate hydrodynamic stage (pure cascade mode) are
shown as red crosses, hybrid calculations with hadron gas EoS as red solid
line and Bag model calculations are depicted by the dark-green dotted line.}
\label{fig:results:comparepqcd}
\end{figure}

After adding the pQCD-spectra as extracted by~\cite{Turbide:2003si}, we
obtain Figure~\ref{fig:results:comparepqcd}. Due to the rather large
pQCD-contribution the difference between the spectra with varying EoS is
reduced, and all calculations agree very well with the data.

\section{Sensitivity to different emission times and processes}\label{sec:times}

\begin{figure}
\input{filterp.tex}
\caption{(Color Online.) UrQMD calculation. The $p_\bot$-spectra for all
photons (solid black line), photons from boosted collisions (dotted blue
line) and collisions with high center-of-mass energy (dashed red line). The
grey-shaded area indicates the range of the $p_\bot$-region used for the
solid black lines in Figures~\ref{fig:results:fil:qt} and
\ref{fig:results:fil:ss}.}
\label{fig:results:fil:pt}
\end{figure}
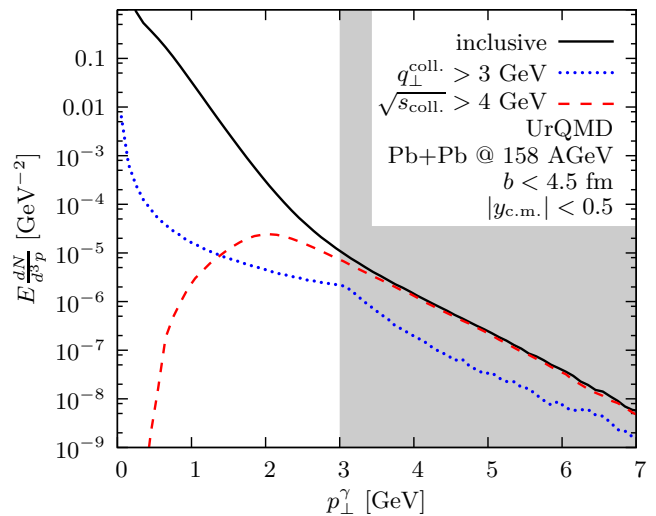

\begin{figure}
\input{filterq.tex}
\caption{(Color Online.) UrQMD calculation. The number of photons as a
function of the transverse center-of-mass momentum of the elementary
collision $q^{\rm coll.}$. We show all photons (blue dotted line), photons
that have high transverse momentum (black solid line) and photons from
collisions with high center-of-mass energy (red dashed line). The
light-blue-shaded area indicates the range of the $q_\bot$-region used for
the dotted blue lines in Figures~\ref{fig:results:fil:pt} and
\ref{fig:results:fil:ss}.}
\label{fig:results:fil:qt}
\end{figure}
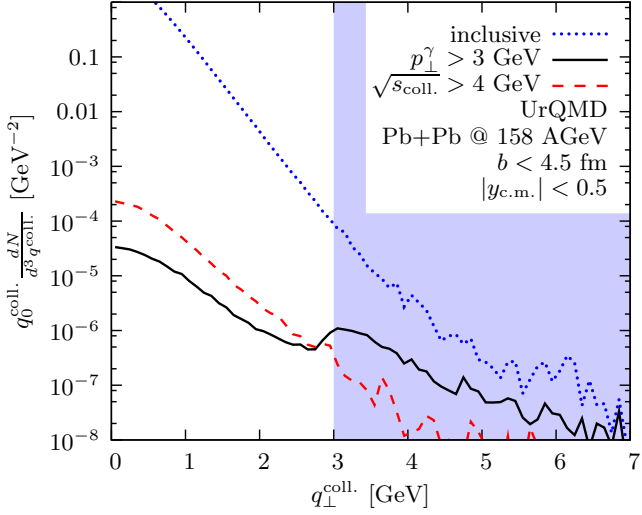

\begin{figure}
\input{filters.tex}
\caption{(Color Online.) UrQMD calculation. The number of photons as a function of
the center-of-mass energy of the elementary collision. We show
all photons (red dashed line), photons that have high transverse momentum
(black solid line) and photons from collisions of with high transverse
center-of-mass momentum. The light-red-shaded area shows the range of the
$\sqrt{s}$-region used for the red dashed lines in
Figures~\ref{fig:results:fil:pt} and \ref{fig:results:fil:qt}.}
\label{fig:results:fil:ss}
\end{figure}
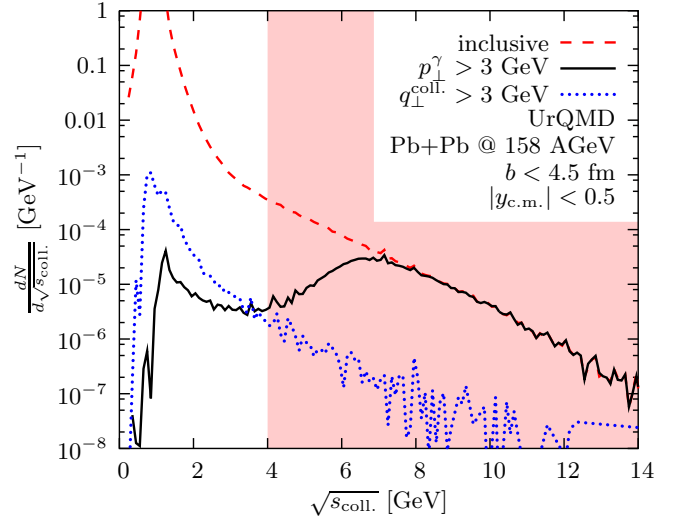

To investigate the major sources for photons to the transverse momentum
spectrum and to explore the sensitivity to the different reaction stages, we
investigate the origin of the change of slope of the spectra at high
transverse momenta $p_\bot \approx 3$~GeV in
Figures~\ref{fig:results:fil:pt}, \ref{fig:results:fil:qt} and
\ref{fig:results:fil:ss} within UrQMD. Two processes may contribute to high
$p_\bot$-photons: (I) collisions with a large $\sqrt{s}$ in elementary
reactions (i.e. early stage collisions) and (II) collisions of particles
with large transverse flow $q_\bot$ but rather small $\sqrt{s}$ (i.e., late
stage collisions).  To disentangle these effects, we determine the
contributions of scatterings with high center-of-mass energy $\sqrt{s} >
4$~GeV and high center-of-mass transverse momentum $q_\bot > 3$~GeV.
Figure~\ref{fig:results:fil:pt} shows the transverse momentum spectrum of
photons split up into the two contributions. Nearly all photons at high
transverse momenta (grey-shaded area) come from collisions with high
center-of-mass energies, whereas the contribution of high center-of-mass
transverse momenta only shows a trivial structure at $p_\bot \approx
q_\bot^{\rm threshold} = 3$~GeV.

One cross-check for Figure~\ref{fig:results:fil:pt} is shown in
Figure~\ref{fig:results:fil:qt}. Here, we observe that the collision
spectrum is exponential. Only at high transverse boosts, a
deviation from an exponential spectrum can be seen. This indicates that most
photons with high transverse momentum come from unboosted collisions with
$q_\bot < 1$~GeV. The center-of-mass energy and transverse momentum of a
collision show no correlation.

The latter can be seen in Figure~\ref{fig:results:fil:ss}, where the
photon production rate is shown as a function of the center-of-mass energy
of the individual collisions. The
figure confirms the notion that most photons with high transverse momenta
come from collisions with high center-of-mass energies. It is worthwhile to
observe, however, that starting at $\sqrt{s_{\rm coll.}} = 7$~GeV, each
elementary collision produces essentially only photons with transverse
momenta $p_\bot > 3$~GeV.

The distribution of center-of-mass energies also shows that the vast
majority of collisions happen around the $\rho$- and $a_1$-pole masses.

\begin{figure}
\input{timeavg.tex}
\caption{(Color Online.) UrQMD calculation. Average photon
emission times as a function of the transverse momentum for all photons,
$\pi\rho\rightarrow\gamma\pi$-, $\pi\pi\rightarrow\gamma\rho$- and other
processes (black solid, red dashed, blue dotted and green dash-dotted lines,
respectively). The shaded areas
correspond to the $p_\bot$-regions used for the curves shown in
Figure~\ref{fig:results:timeslice}.}
\label{fig:results:timeavg}
\end{figure}
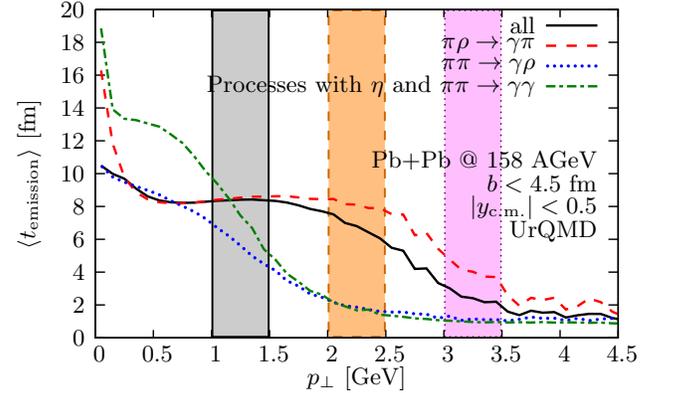

Figure~\ref{fig:results:timeavg} shows the average emission times of photons
as a function of the transverse momentum for the various channels. It is
noticeable that over a very broad momentum range, $0.3~{\rm GeV} < p_\bot <
2.1~{\rm GeV}$, the average emission time stays at a constant level of about
$\langle t_{\rm emission}\rangle \approx 8$~fm. This coincides with the
region where the process $\pi\rho\rightarrow\gamma\pi$ dominates. Only at
high transverse momenta, the early times dominate. This is consistent
with the findings in Section~\ref{sec:spectra}, that the spectrum
clearly shows two different temperatures, one in the region below $p_\bot =
2.5$~GeV and a different temperature above $p_\bot > 3$~GeV. It also
explains why the spectral contributions from $\pi\pi\rightarrow\gamma\rho$
and $\eta$-processes show a much flatter slope already at $p_\bot \approx
2$~GeV.  At late times, when the average center-of-mass energy has
decreased, photons are predominantly produced at low transverse
momenta.

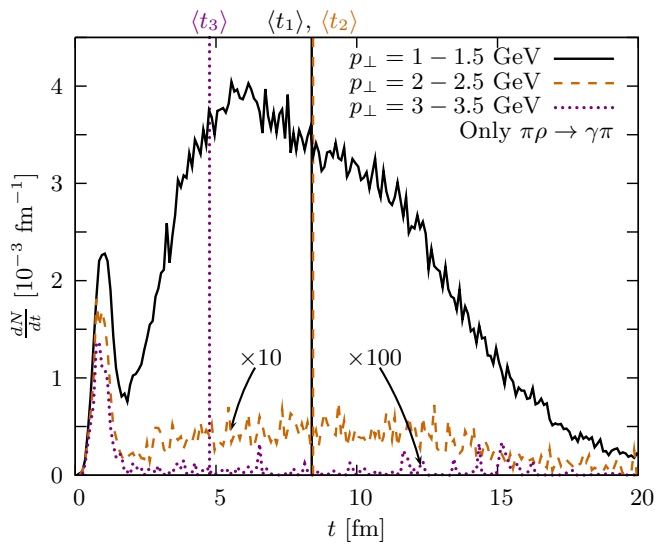
\begin{figure}
\input{timefilter.tex}
\caption{(Color Online.) UrQMD calculation. Emission time distribution of
photons from $\pi\rho$-scatterings for the different photon transverse
momentum regions indicated in Figure~\ref{fig:results:timeavg}. The vertical
lines indicate the average emission time in the corresponding $p_\bot$-bin.}
\label{fig:results:timeslice}
\end{figure}

In Figure~\ref{fig:results:timeslice}, we show the emission time
distribution of photons in various $p_\bot$-bins for the
$\pi\rho\rightarrow\gamma\pi$-processes. It is interesting to see that at
all transverse momenta, there is an initial flash of photons emitted at very
early times $t \approx 1$~fm. In the low $p_\bot$-bin $1~{\rm GeV} < p_\bot
< 1.5~{\rm GeV}$, a very strong contribution from the bulk emission in the
hot and dense stage between $t = 4$~fm and $t = 12$~fm raises the average
emission time. In the intermediate $p_\bot$-region $2~{\rm GeV} < p_\bot <
2.5~{\rm GeV}$, the bulk contribution is greatly reduced and shines less
bright than the initial flash. In the highest $p_\bot$-region $3~{\rm GeV} <
p_\bot < 3.5~{\rm GeV}$, the late bulk contribution is small and the initial
stage dominates. However, one should note that due to the long lifetime of
the intermediate stage, the average emission times are shifted to higher
values.

\section{Summary}\label{sec:summary}

In this work, we have studied direct photon emission from hadronic and
partonic sources within three different dynamical models. In
Section~\ref{sec:photons}, we presented our model for photon emission from
microscopic collisions (Subsection~\ref{sec:photons:urqmd}) and from thermal
rates (Subsection~\ref{sec:photons:hydro}). Then, we introduced the
cross-sections and thermal rates used for the present calculations.

In Section~\ref{sec:rates}, we showed that the emission rates from a
thermalized microscopic cascade calculation agree very well with the thermal
rates used in the hydrodynamic part of the present model. We discussed the
contributions from the different stages before, during and after the
high-density part of the evolution to the direct photon spectra for cascade
calculations as well as hadron gas and Bag model hybrid calculations. It was
found that the relative contributions of photons in the hybrid calculation
with hadron gas EoS and cascade simulations are similar. In contrast, in the
Bag model calculations the intermediate high-density hydrodynamic phase
takes a much longer time and contributes substantially more to the photon
spectra.

Investigations that differentiate between the different channels showed that
the process $\pi\rho\rightarrow\gamma\pi$ is the dominant
hadronic source for photon production at intermediate photon transverse
momenta $0.5$~GeV $< p_\bot < 3$~GeV, while
$\pi\pi\rightarrow\gamma\rho$ is dominant at low photon transverse
momenta $p_\bot < 0.5$~GeV, only.

When comparing the different variations of the model (see
Fig.~\ref{fig:results:compare}), we found that both the hybrid model and the
cascade model can explain the spectra measured by the
WA98-collaboration~\cite{Aggarwal:2000ps}, if pQCD-photons are included. By
comparison of Figures~\ref{fig:results:directcompare} and
\ref{fig:results:compare}, we can also conclude that the photon yields from
the hybrid-model calculation with hadron gas Equation of State and pure
transport calculation are equal within uncertainties, if the same sets of
channels is used in the calculations.

We also found that photons at high transverse momenta $p_\bot > 3$~GeV show
a significantly flatter slope (and therefore higher effective temperature)
than photons with lower $p_\bot$. This effect was attributed to higher center
of mass-energies that produce these photons, and we find that it is not
significantly influenced by elementary collisions that have a high center of
mass-boost in transverse direction. The analysis of average photon emission
times showed that photons at $p_\bot > 3$~GeV are emitted significantly
earlier than at lower transverse momenta.

We also discussed different model assumptions (see Appendix~\ref{app:string}
and~\ref{app:rho}), namely
photon emission from colliding string ends and compared
those to pQCD-spectra. The result indicated that direct photon emission from string ends
is restricted to the early phase of the collision where the medium is hot.
Neglecting those collisions lowers the effective temperature at high
transverse momenta $p_\bot > 2.5$~GeV. In any case, string end contributions
can be neglected in comparison to the pQCD contribution.

We found that the inclusion of a finite width of the produced $\rho$-mesons
is relevant only at low transverse momenta. I.e., only in the lowest
$p_\bot$-bin, the effect of assigning the pole mass or a mass chosen
randomly according to the spectral function is important.

\section{Outlook}

The good agreement between the calculations presented here and the
experimental data shows that the cascade+hydrodynamic hybrid model provides an
excellent tool to explore the properties of QCD-matter at energies where the
(onset of) deconfinement is expected. We plan to extend the investigation
to a more realistic equation of state with chiral restoration, critical
endpoint and rapid cross-over~\cite{Steinheimer:2009nn}. Also, the decays of short-lived mesons and
baryons which cannot be subtracted by the experiment may play a major role
in enhancing the direct photon spectra and will therefore be investigated.

Furthermore, direct photon emission will be investigated for the systems studied
or planned to be studied at RHIC (PHENIX), LHC (ALICE), FAIR (CBM) and
SIS-100 (HADES).

More differential observables like direct photon elliptic flow can also be
explored in the present model.

\section{Acknowledgements}

This work has been supported by the Frankfurt Center for Scientific
Computing (CSC), the GSI and the BMBF. The authors thank Hannah Petersen for
providing the hybrid- and Dirk Rischke for the hydrodynamic code, as well as Jan
Steinheimer-Froschauer for the equation of state. B.\ B\"auchle gratefully
acknowledges support from the Deutsche Telekom Stiftung and the Helmholtz
Research School on Quark Matter Studies. This work was supported by the
Hessian LOEWE initiative through the Helmholtz International Center for FAIR.

The authors thank Elvira Santini and Pasi Huovinnen for valuable discussions
and Klaus Reygers for experimental clarifications.

\appendix

\section{Discussion: String ends}\label{app:string}

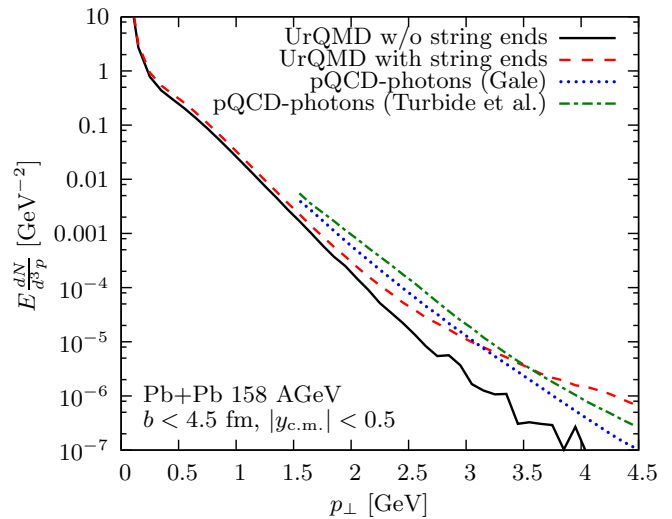
\begin{figure}
\input{strings.tex}
\caption{(Color Online.) UrQMD calculation. Total yields of photons 
with (dashed line) and without (solid line) the contributions
from colliding string ends. For comparison, the spectrum obtained from
PYTHIA is shown as well.}
\label{fig:results:strings}
\end{figure}

In UrQMD, the leading particles from a string have a reduced cross-section
during their formation time. For all other purposes, they are treated as
hadrons. (Hadronic) scatterings of the string ends happen typically at high
center-of-mass energies $\sqrt{s}$. Collisions from string ends can only
produce photons if the collision of fully formed hadrons of the same type
would produce photons. Thus, their contribution is treated as an
addition to the prompt contribution from primordial nucleus-nucleus
interactions. Photons from those collisions contribute significantly to the
spectra at high transverse momenta. Since these particles are not fully
formed hadrons, but effectively represent quarks or di-quarks, a hadronic
treatment of those processes is questionable.

The effects of including the photons from colliding string ends, i.e.\
interaction of leading (di-)quarks, in the calculation can be seen in
Figure~\ref{fig:results:strings}. The spectrum obtained by neglecting the
collision of string ends is exponential and does not exhibit the flattening
at high transverse photon momenta. The inclusion of (di-)quark
scatterings, however, leads to a strong increase of the photon yield at high
$p_\bot$. The contribution of pQCD-photons to the inclusive spectrum starts
to be significant already at relatively low transverse photon momenta
$p_\bot \approx 1$~GeV, although the magnitude of the contribution differs
between the different parametrizations in \cite{Turbide:2003si} and
\cite{Gale:2001yh}.

\section{Discussion: $\rho$-meson width}\label{app:rho}

\begin{figure} \input{onoffpole.tex} \caption{(Color Online.) UrQMD
calculation. Photon spectra from collisions that produce a $\rho$ meson in the
final state for the production of $\rho$ mesons at its pole-mass (solid and
dashed lines) and for the production of $\rho$ mesons according to a
Breit-Wigner-mass distribution (dotted and dash-dotted lines).}
\label{fig:results:pole} \end{figure}
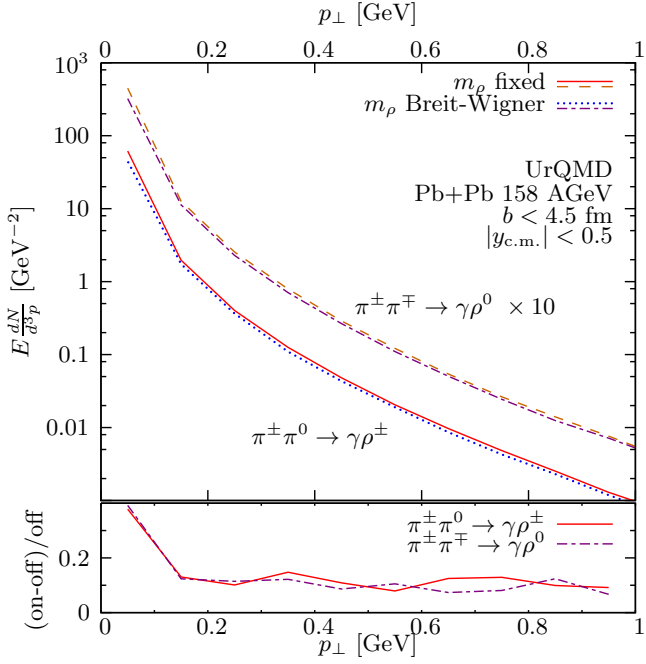

Earlier, we discussed the handling of the $\rho$-meson's finite width.
Figure~\ref{fig:results:pole} shows the effects of following the calculation
outlined there. In both channels, $\pi^\pm\pi^\mp\rightarrow\gamma\rho^0$
and $\pi^\pm\pi^0\rightarrow\gamma\rho^\pm$, the yield is about $10~\%$
higher for $\rho$s produced at their pole mass, and only at very low momenta
this excess becomes as large as $40~\%$.

This behaviour can be explained by kinematic arguments: The by far highest
scattering rate in $\pi+\pi$-collisions is at $\sqrt{s} \approx m_\rho^0$.
Here, the photon cross-section with fixed pole mass is much higher than the
extended calculation with variable $\rho$ mass. At all other center-of-mass
energies, the extended model gives a higher cross-section, but these
comparatively rare processes provide only a minor contribution to the
spectrum. The processes at low $\sqrt{s}$ will contribute primarily to the
low-$p_\bot$-region, because the production of the $\rho$-meson consumes
most of the available energy.  Therefore, the enhancement in the model is
most pronounced at low $p_\bot$.

\section{Comparison to older works with UrQMD}\label{app:dumitru}

\begin{figure}[t]
 \input{dumitru.tex}
 \caption{(Color Online.) Comparison of the current model (red dashed line)
 to calculations from Dumitru {\it et al.}~(\cite{Dumitru:1998sd}, black
 solid line). For further comparision, we apply our model to an earlier
 UrQMD version (blue dotted line).}
 \label{fig:dumitru}
\end{figure}
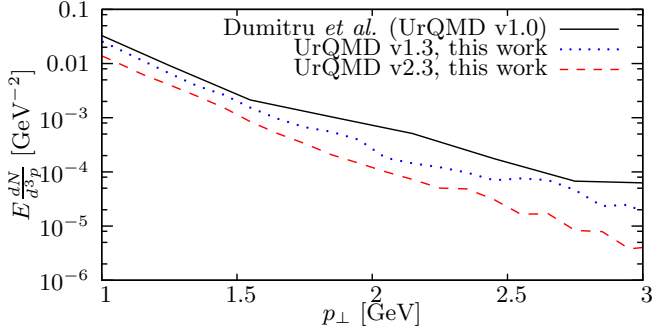

As mentioned in Section~\ref{sec:hybrid}, UrQMD has been used previously in
order to obtain direct photon spectra~\cite{Dumitru:1998sd}. The authors
calculated transverse momentum spectra for central ($b = 0$~fm)
Pb+Pb-collisions at $E_{\rm Lab} = 160$~AGeV. Limiting ourselves to the same
conditions, we can compare our work to that of Dumitru {\it et.\ al}.
In Figure~\ref{fig:dumitru}, we compare results from the current
UrQMD-version 2.3 to those obtained with the earlier versions 1.3 (using
our own photon-analysis) and 1.0 (taken from \cite{Dumitru:1998sd}). The
older UrQMD-versions yield significantly higher photon spectra at
intermediate $p_\bot$. 

We can identify two changes in UrQMD that lead to this behaviour. Between
versions 1.0 (used by Dumitru {\it et al.}) and 1.3, the angular
distributions of various processes have been altered. Since the collisions
of pions with a high difference in rapidity provide significant
contributions to the spectra from \cite{Dumitru:1998sd}, the improved
angular distributions reduce the photon production cross-section.
Furthermore, in versions prior to 2.3, the number of pions has been
unphysically high. The correction of this leads to fewer collisions
involving pions and hence to a further reduction of the spectra.

\section{Differential cross-sections}\label{app:differential}

For completeness, we list the differential cross-sections for the processes
used in \cite{Kapusta:1991qp,Xiong:1992ui}.

\begin{subequations}
\begin{widetext}
\label{eq:dsigmadt}
\begin{eqnarray}
\frac{d\sigma}{dt}\left(\pi^\pm\pi^\mp\rightarrow\gamma\rho^0\right)&=&
\frac{\alpha g_\rho^2}{4 s p_{\rm c.m.}^2} \left \{ 2 - ( m_\rho^2 - 4 m_\pi^2 )
\left [ \frac{s - 2m_\pi^2}{s-m_\rho^2} \frac{1}{t-m_\pi^2} +
\frac{m_\pi^2}{(t-m_\pi^2)^2} + (t \leftrightarrow u) \right ] \right
\}\quad,\label{eq:dsigmadt:picpicrho}\\
\frac{d\sigma}{dt}\left ( \pi^\pm\pi^0\rightarrow\gamma\rho^\pm \right ) &=&
-\frac{\alpha g_\rho^2}{16 s p_{\rm c.m.}^2} \left [ \frac{(s-2m_\rho^2)
(t-m_\pi^2)^2} {m_\rho^2(s-m_\rho^2)^2}+ \frac{(s-6m_\rho^2)(t-m_\pi^2)}
{m_\rho^2(s-m_\rho^2)}+ \frac{4s(m_\rho^2-4m_\pi^2)} {(s-m_\rho^2)^2}
\right.\nonumber\\
&&+\left.  \frac{4(m_\rho^2-4m_\pi^2)}{t-m_\pi^2}\left(
\frac{s}{s-m_\rho^2}+\frac{m_\pi^2}{t-m_\pi^2}
\right)+\frac{m_\pi^2}{m_\rho^2} -\frac{9}{2} + (t\leftrightarrow u) \right
]\quad,\label{eq:dsigmadt:picpi0}
\end{eqnarray}
\begin{eqnarray}
\frac{d\sigma}{dt}\left(\pi^\pm\rho^0\rightarrow\gamma\pi^\pm\right) &=&
\frac{\alpha g_\rho^2}{12 s p_{\rm c.m.}^2} \left [
2-\frac{s(m_\rho^2-4m_\pi^2)}{(s-m_\pi^2)^2} -
\frac{(m_\rho^2-4m_\pi^2)}{t-m_\pi^2}
\left(\frac{s-m_\rho^2+m_\pi^2}{(s-m_\pi^2)(t-m_\pi^2)}+
\frac{m_\pi^2}{(t-m_\pi^2)}\right)\right]\quad,\label{eq:dsigmadt:picrh0}\\
\frac{d\sigma}{dt}(\pi^\pm\rho^\mp\rightarrow\gamma\pi^0) &=&
-\frac{\alpha g_\rho^2}{48 s p_{\rm c.m.}^2} \left\{
4(m_\rho^2-4m_\pi^2)
\left[ \frac{t}{(t-m_\pi^2)^2} + \frac{u}{(u-m_\rho^2)^2} -
\frac{m_\rho^2}{s-m_\pi^2}
\left(\frac{1}{t-m_\pi^2}+\frac{1}{u-m_\rho^2}\right)
\right]\right.\nonumber \\\label{eq:dsigmadt:picrhc}
&&\left.+\left(3+\frac{s-m_\pi^2}{m_\rho^2}\right) \frac{s-m_\pi^2}{u-m_\rho^2}
-\frac{1}{2}+\frac{s}{m_\rho^2}-\left(\frac{s-m_\pi^2}{u-m_\rho^2}\right)^2\right\}\quad,
\\
\frac{d\sigma}{dt}\left(\pi^0\rho^\pm\rightarrow\gamma\pi^\pm\right) &=&
\frac{\alpha g_\rho^2}{48 s p_{\rm c.m.}^2} \left[ \frac{9}{2} -
\frac{s}{m_\rho^2} - \frac{4s(m_\rho^2-4m_\pi^2)}{(s-m_\pi^2)^2}
+\frac{(s-m_\pi^2)^2-4m_\rho^2(m_\rho^2-4m_\pi^2)}{(u-m_\rho^2)^2}
\right.\nonumber\\\label{eq:dsigmadt:pi0rhc}
&&\left.+\frac{1}{u-m_\rho^2}\left(5(s-m_\pi^2)- \frac{(s-m_\pi^2)^2}{m_\rho^2}-
\frac{4(m_\rho^2-4m_\pi^2)(s-m_\pi^2+m_\rho^2)}{s-m_\pi^2}\right)\right]\quad,\\
\frac{d\sigma}{dt}\left(\pi^\pm\pi^\mp\rightarrow\gamma\eta\right)&=&
\frac{\pi\alpha A\left|F_\pi(s)\right|^2}{16 m_\eta^2m_\rho^4 s p_{\rm
c.m.}^2}
\left[ s(u-m_\pi^2)(t-m_\pi^2) -
m_\pi^2(s-m_\eta^2)^2\right]\quad,\label{eq:dsigmadt:picpiceta}\\
\frac{d\sigma}{dt}\left(\pi^\pm\eta\rightarrow\gamma\pi^\pm\right)&=&
\frac{\pi\alpha A\left|F_\pi(u)\right|^2}{16 m_\eta^2m_\rho^4 s p_{\rm
c.m.}^2}
\left[ u(s-m_\pi^2)(t-m_\pi^2) -
m_\pi^2(u-m_\eta^2)^2\right]\quad,\label{eq:dsigmadt:piceta}\\
\frac{d\sigma}{dt}\left(\pi^\pm\pi^\mp\rightarrow\gamma\gamma\right)&=&
\frac{2\pi\alpha^2}{s p_{\rm c.m.}^2} \left[ 1
+ 2m_\pi^2 \left( \frac{1}{t-m_\pi^2}+\frac{1}{u-m_\pi^2} \right)
+ 2m_\pi^4 \left( \frac{1}{t-m_\pi^2}+\frac{1}{u-m_\pi^2} \right)^2 \right ]
\quad.\label{eq:dsigmadt:picpicgam}
\end{eqnarray}
\end{widetext}

In these equations, $t = (p_\pi - p_\gamma)^2$ is always the momentum
transfer from the pion to the photon~\footnote{All cases with two pions in
the initial state are symmetric in $t \leftrightarrow u$. Therefore, it is
not necessary to specify which pion is the reference for $t$.} (unlike the
convention used in \cite{Kapusta:1991qp}) and $p_{\rm c.m.} =
(2\sqrt{s})^{-1} \sqrt{s^2-2s(m_1^2+m_2^2) + (m_1^2-m_2^2)^2}$ is the
three-momentum of the incoming particles in the center-of-mass frame ($m_1$
and $m_2$ being their masses). The value of $A$ is, consistent with
\cite{Kapusta:1991qp}, $A = g_{\eta\rho\rho}^2g_\rho^2/4\pi\gamma_\rho^2 =
4.7$, and the pion electromagnetic form factor is $$F_\pi(s) =
\frac{m_\rho^4}{ (s-m_\rho^2)+\Gamma_\rho^2m_\rho^2}\quad.$$

In their 1992 paper, Xiong {\it et al.}~\cite{Xiong:1992ui} calculate the
cross-section for the formation of an intermediate $a_1$-meson during
$\pi\rho$-scattering, averaged over all possible charge combinations:
\begin{widetext}
\begin{eqnarray} \frac{d\sigma}{dt}\left(\pi\rho\rightarrow
a_1\rightarrow\gamma\pi\right) & =
&\frac{\pi^2\sqrt{s}}{2 p_{\rm c.m.}^3 (s-m_\pi)^2}
\frac{\Gamma_{a_1\rightarrow\pi\rho}\Gamma_{a_1\rightarrow\gamma\pi}}
{(\sqrt{s}-m_{a_1})^2 + (\Gamma_{a_1 \rightarrow\pi\rho} +
\Gamma_{a_1\rightarrow\gamma\pi})^2/4}\quad.
\label{eq:dsigmadt:a1}
\end{eqnarray}
\end{widetext}
\end{subequations}
This channel is not included in Kapusta {\it et al}. Xiong {\it et al.}
obtain this from a Lagrange-density involving only the pion-, photon-,
$\rho$- and $a_1$-fields
\begin{widetext}
\begin{equation}\label{eq:lagrangian:xiong}
\mathcal{L} = G_\rho a_1^\mu ( g_{\mu\nu} (p_\pi \cdot p_\rho) - {p_\pi}_\mu
{p_\rho}_\nu ) \rho^\nu \Phi + G_\rho \frac{e}{g_\rho} a_1^\mu ( g_{\mu\nu}
(p_\pi \cdot p_\gamma) - {p_\pi}_\mu
{p_\gamma}_\nu ) A^\nu \Phi\quad,
\end{equation}
where $G_\rho = 14.8$~GeV$^{-1}$. The partial widths of the $a_1$,
$\Gamma_{a_1\rightarrow\pi\rho}$ and $\Gamma_{a_1\rightarrow\gamma\pi}$, are
estimated to be
\begin{subequations}\begin{eqnarray}
\Gamma_{a_1\rightarrow\pi\rho} &=&
\frac{G_\rho^2 p_{\rm c.m.}}{24\pi m_{a_1}^2}\left\{
\frac{m_\rho^2}{4s}\left[ s-(m_\rho^2-m_\pi^2) \right] +
\frac{1}{2}(s-m_\rho^2-m_\pi^2)
\right\}\label{eq:width:a1pirho}\quad\text{and}\\
\Gamma_{a_1\rightarrow\gamma\pi} &=&
\frac{G_\rho^2 \alpha p_{\rm c.m.}}{12 g_\rho^2 m_{a_1}^2} ( s-m_\pi^2 )
\label{eq:width:a1gammapi}\quad.
\end{eqnarray}\end{subequations}
\end{widetext}

\section{Integrated Cross-sections}\label{app:integrated}

The integration over the cross-sections listed in
appendix~\ref{app:differential} yield the following results:

\begin{subequations}
\begin{widetext}
\begin{eqnarray}
\sigma\left ( \pi^\pm\pi^\mp\rightarrow\gamma\rho^0 \right ) &=&
\frac{\alpha g_\rho^2}{4 s p_{\rm c.m.}^2}
\left \{ 2\Delta t - ( m_\rho^2 - 4 m_\pi^2 )
\left [ \frac{s - 2m_\pi^2}{s-m_\rho^2}
\ln{\frac{(t_--m_\pi^2)}{(t_+-m_\pi^2)}}
\right.\right.
\nonumber\\&&\left.\left.
+ \frac{m_\pi^2\Delta t}{(m_\pi^2 - t_+)(m_\pi^2 - t_-)}
+ (t_\pm \leftrightarrow u_\mp)
\right ] \right \}
\quad,\label{eq:sigma:picpicrho}\\
\sigma\left ( \pi^\pm\pi^0\rightarrow\gamma\rho^\pm \right ) &=&
-\frac{\alpha g_\rho^2}{16 s p_{\rm c.m.}^2} \left [ 
\frac{s-2m_\rho^2}{  m_\rho^2(s-m_\rho^2)^2} \frac{1}{3}(t_-^3-t_+^3)
+ \frac{s-6m_\rho^2}{m_\rho^2(s-m_\rho^2)}   \frac{1}{2}(t_-^2-t_+^2)
\right.\nonumber\\
&&\left.
+ \left( \frac{4s(m_\rho^2-4m_\pi^2)}{(s-m_\rho^2)^2} +
\frac{m_\pi^2}{m_\rho^2} - \frac{9}{2} \right ) \Delta t
+\frac{4s(m_\rho^2-4m_\pi^2)}{(s-m_\rho^2)}
\ln{\frac{t_--m_\pi^2}{t_+-m_\pi^2}}
\right.\nonumber\\
&&\left.
+\frac{4 m_\pi^2 (m_\rho^2-4m_\pi^2) \Delta t}{(m_\pi^2 - t_+)(m_\pi^2 - t_-)}
+ (t_\pm \leftrightarrow u_\mp)
\right]
\quad,\label{eq:sigma:picpi0}\\
\sigma\left(\pi^\pm\rho^0\rightarrow\gamma\pi^\pm\right) &=&
\frac{\alpha g_\rho^2}{12 s p_{\rm c.m.}^2} \left [
2\Delta t-\frac{s(m_\rho^2-4m_\pi^2)}{(s-m_\pi^2)^2}\Delta t
- (m_\rho^2-4m_\pi^2) \left(
\frac{s-m_\rho^2+m_\pi^2}{s-m_\pi^2} \ln{\frac{t_--m_\pi^2}{t_+-m_\pi^2}}
\right.\right.
\nonumber\\&&\left.\left.
+ \frac{m_\pi^2\Delta t}{(m_\pi^2 - t_+)(m_\pi^2 - t_-)}
\right)\right]
\quad,\label{eq:sigma:picrh0}\\
\sigma(\pi^\pm\rho^\mp\rightarrow\gamma\pi^0) &=&
-\frac{\alpha g_\rho^2}{48 s p_{\rm c.m.}^2} \left\{
4(m_\rho^2-4m_\pi^2) \left[
 \frac{m_\pi^2\Delta t}{(u_+-m_\rho^2)(u_--m_\rho^2)}
+\frac{m_\pi^2\Delta t}{(t_+-m_\pi^2 )(t_--m_\pi^2 )}
\right.\right.
\nonumber\\&&\left.\left.
+\ln{\frac{u_+-m_\rho^2}{u_--m_\rho^2}}
+\ln{\frac{t_--m_\pi^2 }{t_+-m_\pi^2 }}
-\frac{m_\rho^2}{s-m_\pi^2}\left(
 \ln{\frac{t_--m_\pi^2 }{t_+-m_\pi^2}}
+\ln{\frac{u_+-m_\rho^2}{u_--m_\rho^2}}
\right)
\right]
\right.
\nonumber\\&&\left.
+(s-m_\pi^2)\left(3+\frac{s-m_\pi^2}{m_\rho^2}\right)
\ln{\frac{u_+-m_\rho^2}{u_--m_\rho^2}}
+ \Delta t \left(
\frac{s}{m_\rho^2} - \frac{1}{2}
-\frac{(s-m_\pi^2)^2}{(u_+-m_\rho^2)(u_--m_\rho^2)}
\right)
\right\}
\quad,\label{eq:sigma:picrhc}\\
\sigma\left(\pi^0\rho^\pm\rightarrow\gamma\pi^\pm\right) &=&
\frac{\alpha g_\rho^2}{48 s p_{\rm c.m.}^2} \left[
\Delta t
 \left( \frac{9}{2} -\frac{s}{m_\rho^2} 
       -\frac{4s(m_\rho^2-4m_\pi^2)}{(s-m_\pi^2)^2}
       +\frac{(s-m_\pi^2)^2-4m_\rho^2(m_\rho^2-4m_\pi^2)}{(u_+-m_\rho^2)(u_--m_\rho^2)}
 \right)
\right.\nonumber\\
&&\left.
+ \left(
 5(s-m_\pi^2)
 -\frac{(s-m_\pi^2)^2}{m_\rho^2}
 -\frac{4(m_\rho^2-4m_\pi^2)(s-m_\pi^2+m_\rho^2)}{s-m_\pi^2}
 \right)
 \ln{\frac{u_+-m_\rho^2}{u_--m_\rho^2}}
\right]
\quad,\label{eq:sigma:pi0rhc}\\
\sigma\left(\pi^\pm\pi^\mp\rightarrow\gamma\eta\right)&=&
\frac{\pi\alpha A\left|F_\pi(s)\right|^2}{16 m_\eta^2m_\rho^4 s p_{\rm
c.m.}^2}
\left[
 (2m_\pi^2+m_\eta^2-s)\frac{s}{2}(t_+^2-t_-^2)
-\frac{s}{3}(t_+^3-t_-^3)
-m_\pi^2\left(m_\eta^4+s(m_\pi^2-m_\eta^2)\right)
\right]
\quad,\label{eq:sigma:picpiceta}\\
\sigma\left(\pi^\pm\eta\rightarrow\gamma\pi^\pm\right)&=&
\frac{\pi\alpha A}{16 m_\eta^2 s p_{\rm c.m.}^2} \left\{
 -m_\pi^2\left[(t_-+u_-)(s-m_\pi^2)+(2m_\pi^2-s)^2\right] I_0 \phantom{\frac{1}{1}}
\right.
\nonumber\\&&\left.
+\left[(s-m_\pi^2)(m_\pi^2+t_-+u_-)-2m_\pi^2(s-2m_\pi^2)\right]
 \left[(t_-+u_--m_\rho^2) I_0 + \frac{1}{2} I_1 \right] 
\right.
\nonumber\\&&\left.
\phantom{\frac{1}{2}}-s\left[ \Delta t + (t_-+u_--m_\rho^2) I_1 +
(t_-+u_--m_\rho^2)^2 I_0 - m_\rho^2\Gamma_\rho^2 I_0 
\right]
\right\}
\quad,\label{eq:sigma:piceta}\\
\sigma\left(\pi^\pm\pi^\mp\rightarrow\gamma\gamma\right)&=&
\frac{2\pi\alpha^2}{s p_{\rm c.m.}^2} \left\{ \Delta t
+2m_\pi^2 \left[
  \left( 1 - \frac{2m_\pi^2}{s} \right )
  \ln\frac{t_--m_\pi^2}{t_+-m_\pi^2}
 +\frac{m_\pi^2\Delta t}{(t_--m_\pi^2)(t_+-m_\pi^2)}
 +(t_\pm \leftrightarrow u_\mp)
\right]\right\}
\quad,\label{eq:sigma:picpicgam}\\
\sigma\left(\pi\rho\rightarrow{}a_1\rightarrow\gamma\pi\right) &=&
4 p_1 \omega \frac{d\sigma}{dt}\left(\pi\rho\rightarrow
a_1\rightarrow\gamma\pi\right)
\quad.\label{eq:sigma:a1}
\end{eqnarray}
\end{widetext}
\end{subequations}

In those equations, $t_\pm = m_1^2-2\omega(E_1\pm p_1)$ are the minimal and
maximal allowed momentum transfers, and correspondingly $u_\pm =
m_1^2+m_2^2+m_4^2-s-t_\pm$, and $\Delta t$ is a shorthand for $t_--t_+$. The
indices 1, 2 and 4 denote the particles in the order given on the left hand
side of the equations (3 is the photon). $E_1$, $p_1$ and $\omega = E_3$
denote momentum and energy of the respective particles in the
center-of-mass frame. In the cross-section
$\pi^\pm\eta\rightarrow\gamma\pi^\pm$ (Eq.~\eqref{eq:sigma:piceta}), the
following notations have been used for simplicity:

\begin{subequations}\label{eq:piceta}
\begin{widetext}
\begin{eqnarray}
I_0 &=& \frac{1}{m_\rho\Gamma_\rho}
        \left[
         {\rm atan}{\left(\frac{u_+-m_\rho^2}{m_\rho\Gamma_\rho}\right)}
        -{\rm atan}{\left(\frac{u_--m_\rho^2}{m_\rho\Gamma_\rho}\right)}
        \right]\quad\text{and}\label{eq:i0}\\
I_1 &=& \ln{ \left(
  \frac{(u_--m_\rho^2)^2+m_\rho^2\Gamma_\rho^2}
       {(u_+-m_\rho^2)^2+m_\rho^2\Gamma_\rho^2}
             \right)}\quad.\label{eq:i1}
\end{eqnarray}
\end{widetext}
\end{subequations}

\end{document}

%% file: sigmatot.tex
\ifx\PSTloaded\undefined
\def\PSTloaded{t}
\psset{arrowsize=.01 3.2 1.4 .3}
\psset{dotsize=.01}
\catcode`@=11

\definecolor{darkgreen}{rgb}{0,.5,0}
\newpsobject{PST@Border}{psline}{linewidth=.0015,linestyle=solid}
\newpsobject{PST@total}{psline}{linecolor=black,    linewidth=.0025,linestyle=solid}
\newpsobject{PST@hadron}{psline}{linecolor=red,      linewidth=.0025,linestyle=dashed,dash=.01 .01}
\newpsobject{PST@string}{psline}{linecolor=blue,     linewidth=.0035,linestyle=dotted,dotsep=.004}
\newpsobject{PST@pythia}{psline}{linecolor=darkgreen,linewidth=.0025,linestyle=dashed,dash=.01 .004 .004 .004}
\catcode`@=12

\fi
\psset{unit=5.0in,xunit=1.124\columnwidth,yunit=.45\columnwidth}
\pspicture(0.09,0.0400)(0.98,0.988)
\ifx\nofigs\undefined
\catcode`@=11

\PST@Border(0.1750,0.1344)
(0.1900,0.1344)

\PST@Border(0.9630,0.1344)
(0.9480,0.1344)

\rput[r](0.1590,0.1344){ 0}
\PST@Border(0.1750,0.2386)
(0.1900,0.2386)

\PST@Border(0.9630,0.2386)
(0.9480,0.2386)

\rput[r](0.1590,0.2386){ 5}
\PST@Border(0.1750,0.3428)
(0.1900,0.3428)

\PST@Border(0.9630,0.3428)
(0.9480,0.3428)

\rput[r](0.1590,0.3428){ 10}
\PST@Border(0.1750,0.4470)
(0.1900,0.4470)

\PST@Border(0.9630,0.4470)
(0.9480,0.4470)

\rput[r](0.1590,0.4470){ 15}
\PST@Border(0.1750,0.5512)
(0.1900,0.5512)

\PST@Border(0.9630,0.5512)
(0.9480,0.5512)

\rput[r](0.1590,0.5512){ 20}
\PST@Border(0.1750,0.6554)
(0.1900,0.6554)

\PST@Border(0.9630,0.6554)
(0.9480,0.6554)

\rput[r](0.1590,0.6554){ 25}
\PST@Border(0.1750,0.7596)
(0.1900,0.7596)

\PST@Border(0.9630,0.7596)
(0.9480,0.7596)

\rput[r](0.1590,0.7596){ 30}
\PST@Border(0.1750,0.8638)
(0.1900,0.8638)

\PST@Border(0.9630,0.8638)
(0.9480,0.8638)

\rput[r](0.1590,0.8638){ 35}
\PST@Border(0.1750,0.9680)
(0.1900,0.9680)

\PST@Border(0.9630,0.9680)
(0.9480,0.9680)

\rput[r](0.1590,0.9680){ 40}
\PST@Border(0.1750,0.1344)
(0.1750,0.1544)

\PST@Border(0.1750,0.9680)
(0.1750,0.9480)

\rput(0.1750,0.0924){ 1}
\PST@Border(0.3640,0.1344)
(0.3640,0.1444)

\PST@Border(0.3640,0.9680)
(0.3640,0.9580)

\PST@Border(0.4745,0.1344)
(0.4745,0.1444)

\PST@Border(0.4745,0.9680)
(0.4745,0.9580)

\PST@Border(0.5529,0.1344)
(0.5529,0.1444)

\PST@Border(0.5529,0.9680)
(0.5529,0.9580)

\PST@Border(0.6138,0.1344)
(0.6138,0.1444)

\PST@Border(0.6138,0.9680)
(0.6138,0.9580)

\PST@Border(0.6635,0.1344)
(0.6635,0.1444)

\PST@Border(0.6635,0.9680)
(0.6635,0.9580)

\PST@Border(0.7055,0.1344)
(0.7055,0.1444)

\PST@Border(0.7055,0.9680)
(0.7055,0.9580)

\PST@Border(0.7419,0.1344)
(0.7419,0.1444)

\PST@Border(0.7419,0.9680)
(0.7419,0.9580)

\PST@Border(0.7740,0.1344)
(0.7740,0.1444)

\PST@Border(0.7740,0.9680)
(0.7740,0.9580)

\PST@Border(0.8028,0.1344)
(0.8028,0.1544)

\PST@Border(0.8028,0.9680)
(0.8028,0.9480)

\rput(0.8028,0.0924){ 10}
\PST@Border(0.1750,0.9680)
(0.1750,0.1344)
(0.9630,0.1344)
(0.9630,0.9680)
(0.1750,0.9680)

\rput{L}(0.1040,0.5512){$\sigma$~[fm]}
\rput(0.5690,0.0694){$\sqrt{s}$~[GeV]}
\rput[r](0.8360,0.8150){$\sigma ( \pi^+\pi^- \rightarrow $ strings )}
\PST@string(0.8520,0.8150) (0.9310,0.8150)
\rput[r](0.8360,0.8970){$\sigma ( \pi^+\pi^- \rightarrow $ resonance )}
\PST@hadron(0.8520,0.8970) (0.9310,0.8970)
\rput[r](0.8360,0.7330){$\sigma( \pi^+\pi^- \rightarrow $ hard scattering)}
\PST@pythia(0.8520,0.7330) (0.9310,0.7330)

\PST@string(0.1750,0.1344) (0.1777,0.1344) (0.1815,0.1344) (0.1852,0.1344)
(0.1888,0.1344) (0.1927,0.1344) (0.1965,0.1344) (0.2002,0.1344) (0.2039,0.1344)
(0.2076,0.1344) (0.2114,0.1344) (0.2152,0.1344) (0.2190,0.1344) (0.2227,0.1344)
(0.2265,0.1344) (0.2303,0.1344) (0.2341,0.1344) (0.2378,0.1344) (0.2414,0.1344)
(0.2453,0.1344) (0.2490,0.1344) (0.2527,0.1344) (0.2566,0.1344) (0.2604,0.1344)
(0.2642,0.1344) (0.2679,0.1344) (0.2716,0.1344) (0.2754,0.1344) (0.2791,0.1344)
(0.2830,0.1344) (0.2866,0.1344) (0.2904,0.1344) (0.2941,0.1344) (0.2980,0.1344)
(0.3018,0.1344) (0.3055,0.1344) (0.3092,0.1344) (0.3130,0.1344) (0.3168,0.1344)
(0.3205,0.4655) (0.3243,0.4637) (0.3280,0.4618) (0.3317,0.4597) (0.3356,0.4636)
(0.3393,0.4562) (0.3430,0.4544) (0.3468,0.4526) (0.3506,0.4509) (0.3544,0.4492)
(0.3580,0.4474) (0.3619,0.4459) (0.3656,0.4442) (0.3694,0.4424) (0.3731,0.4409)
(0.3769,0.4393) (0.3806,0.4376) (0.3844,0.4362) (0.3882,0.4356) (0.3919,0.4354)
(0.3957,0.4350) (0.3995,0.4348) (0.4033,0.4347) (0.4070,0.4343) (0.4107,0.4341)
(0.4145,0.4337) (0.4183,0.4335) (0.4221,0.4331) (0.4258,0.4330) (0.4296,0.4324)
(0.4333,0.4322) (0.4371,0.4318) (0.4408,0.4370) (0.4446,0.4312) (0.4483,0.4308)
(0.4521,0.4305) (0.4559,0.4301) (0.4597,0.4299) (0.4634,0.4295) (0.4671,0.4291)
(0.4709,0.4289) (0.4747,0.4283) (0.4784,0.4282) (0.5010,0.4264) (0.5160,0.4250)
(0.5198,0.4247) (0.5311,0.4240) (0.5348,0.4236) (0.5386,0.4233) (0.5424,0.4230)
(0.5461,0.4227) (0.5499,0.4225) (0.5536,0.4222) (0.5574,0.4219) (0.5611,0.4217)
(0.5649,0.4214) (0.5686,0.4212) (0.5724,0.4210) (0.5762,0.4259) (0.5799,0.4205)
(0.5837,0.4202) (0.5875,0.4201) (0.5912,0.4198) (0.5950,0.4196) (0.5988,0.4194)
(0.6025,0.4192) (0.6063,0.4190) (0.6100,0.4188) (0.6138,0.4191) (0.6176,0.4184)
(0.6213,0.4183) (0.6250,0.4182) (0.6288,0.4180) (0.6326,0.4176) (0.6364,0.4175)
(0.6401,0.4175) (0.6438,0.4173) (0.6476,0.4172) (0.6514,0.4169) (0.6551,0.4169)
(0.6589,0.4168) (0.6627,0.4166) (0.6664,0.4164) (0.6702,0.4164) (0.6740,0.4162)
(0.6777,0.4161) (0.6815,0.4159) (0.6852,0.4159) (0.6890,0.4160) (0.6928,0.4158)
(0.6965,0.4157) (0.7003,0.4157) (0.7040,0.4155) (0.7078,0.4155) (0.7116,0.4155)
(0.7153,0.4153) (0.7191,0.4153) (0.7228,0.4153) (0.7266,0.4153) (0.7304,0.4153)
(0.7341,0.4153) (0.7379,0.4153) (0.7416,0.4153) (0.7454,0.4153) (0.7491,0.4153)
(0.7529,0.4153) (0.7567,0.4153) (0.7604,0.4153) (0.7642,0.4153) (0.7679,0.4153)
(0.7717,0.4153) (0.7755,0.4153) (0.7792,0.4155) (0.7830,0.4155) (0.7868,0.4155)
(0.7905,0.4155) (0.7943,0.4157) (0.7980,0.4157) (0.8018,0.4157) (0.8055,0.4160)
(0.8092,0.4160) (0.8132,0.4161) (0.8168,0.4161) (0.8207,0.4164) (0.8242,0.4164)
(0.8280,0.4166) (0.8319,0.4166) (0.8356,0.4168) (0.8394,0.4170) (0.8432,0.4170)
(0.8470,0.4172) (0.8506,0.4174) (0.8545,0.4174) (0.8581,0.4176) (0.8618,0.4178)
(0.8658,0.4180) (0.8694,0.4183) (0.8732,0.4182) (0.8770,0.4184) (0.8807,0.4187)
(0.8846,0.4189) (0.8882,0.4191) (0.8919,0.4193) (0.8958,0.4195) (0.8995,0.4197)
(0.9033,0.4199) (0.9070,0.4201) (0.9107,0.4203) (0.9146,0.4207) (0.9183,0.4209)
(0.9221,0.4212) (0.9259,0.4214) (0.9297,0.4218) (0.9334,0.4220) (0.9371,0.4222)
(0.9409,0.4224) (0.9447,0.4228) (0.9484,0.4230) (0.9522,0.4234)

\PST@hadron(0.1750,0.7055) (0.1777,0.6604) (0.1815,0.6150) (0.1852,0.5827)
(0.1888,0.5589) (0.1927,0.5410) (0.1965,0.5285) (0.2002,0.5210) (0.2039,0.5191)
(0.2076,0.5239) (0.2114,0.5366) (0.2152,0.5595) (0.2190,0.5956) (0.2227,0.6469)
(0.2265,0.7135) (0.2303,0.7886) (0.2341,0.8550) (0.2378,0.8934) (0.2414,0.8961)
(0.2453,0.8740) (0.2490,0.8434) (0.2527,0.8163) (0.2566,0.7971) (0.2604,0.7857)
(0.2642,0.7792) (0.2679,0.7742) (0.2716,0.7667) (0.2754,0.7536) (0.2791,0.7336)
(0.2830,0.7079) (0.2866,0.6789) (0.2904,0.6477) (0.2941,0.6156) (0.2980,0.5870)
(0.3018,0.5639) (0.3055,0.5464) (0.3092,0.5341) (0.3130,0.5252) (0.3168,0.5164)
(0.3205,0.5046) (0.3243,0.4878) (0.3280,0.4679) (0.3317,0.4474) (0.3356,0.4233)
(0.3393,0.4145) (0.3430,0.4061) (0.3468,0.4066) (0.3506,0.4093) (0.3544,0.3908)
(0.3580,0.3668) (0.3619,0.3500) (0.3656,0.3385) (0.3694,0.3299) (0.3731,0.3226)
(0.3769,0.3168) (0.3806,0.3116) (0.3844,0.3067) (0.3882,0.3025) (0.3919,0.2986)
(0.3957,0.2952) (0.3995,0.2919) (0.4033,0.2886) (0.4070,0.2859) (0.4107,0.2832)
(0.4145,0.2807) (0.4183,0.2784) (0.4221,0.2763) (0.4258,0.2741) (0.4296,0.2723)
(0.4333,0.2705) (0.4371,0.2688) (0.4408,0.2617) (0.4446,0.2657) (0.4483,0.2642)
(0.4521,0.2628) (0.4559,0.2615) (0.4597,0.2603) (0.4634,0.2592) (0.4671,0.2582)
(0.4709,0.2571) (0.4747,0.2563) (0.4784,0.2551) (0.5010,0.2505) (0.5160,0.2483)
(0.5198,0.2478) (0.5311,0.2462) (0.5348,0.2459) (0.5386,0.2456) (0.5424,0.2453)
(0.5461,0.2449) (0.5499,0.2446) (0.5536,0.2442) (0.5574,0.2439) (0.5611,0.2437)
(0.5649,0.2434) (0.5686,0.2431) (0.5724,0.2429) (0.5762,0.2374) (0.5799,0.2424)
(0.5837,0.2422) (0.5875,0.2420) (0.5912,0.2418) (0.5950,0.2416) (0.5988,0.2416)
(0.6025,0.2414) (0.6063,0.2412) (0.6100,0.2411) (0.6138,0.2405) (0.6176,0.2409)
(0.6213,0.2407) (0.6250,0.2405) (0.6288,0.2403) (0.6326,0.2404) (0.6364,0.2404)
(0.6401,0.2402) (0.6438,0.2402) (0.6476,0.2400) (0.6514,0.2402) (0.6551,0.2399)
(0.6589,0.2399) (0.6627,0.2398) (0.6664,0.2399) (0.6702,0.2396) (0.6740,0.2396)
(0.6777,0.2395) (0.6815,0.2396) (0.6852,0.2394) (0.6890,0.2392) (0.6928,0.2394)
(0.6965,0.2392) (0.7003,0.2392) (0.7040,0.2392) (0.7078,0.2392) (0.7116,0.2390)
(0.7153,0.2392) (0.7191,0.2392) (0.7228,0.2390) (0.7266,0.2390) (0.7304,0.2390)
(0.7341,0.2390) (0.7379,0.2388) (0.7416,0.2388) (0.7454,0.2388) (0.7491,0.2388)
(0.7529,0.2388) (0.7567,0.2388) (0.7604,0.2388) (0.7642,0.2388) (0.7679,0.2388)
(0.7717,0.2388) (0.7755,0.2388) (0.7792,0.2388) (0.7830,0.2388) (0.7868,0.2388)
(0.7905,0.2388) (0.7943,0.2386) (0.7980,0.2388) (0.8018,0.2388) (0.8055,0.2386)
(0.8092,0.2388) (0.8132,0.2386) (0.8168,0.2388) (0.8207,0.2386) (0.8242,0.2388)
(0.8280,0.2386) (0.8319,0.2388) (0.8356,0.2386) (0.8394,0.2386) (0.8432,0.2388)
(0.8470,0.2386) (0.8506,0.2386) (0.8545,0.2388) (0.8581,0.2386) (0.8618,0.2386)
(0.8658,0.2386) (0.8694,0.2386) (0.8732,0.2388) (0.8770,0.2388) (0.8807,0.2386)
(0.8846,0.2386) (0.8882,0.2386) (0.8919,0.2386) (0.8958,0.2386) (0.8995,0.2386)
(0.9033,0.2386) (0.9070,0.2388) (0.9107,0.2388) (0.9146,0.2386) (0.9183,0.2386)
(0.9221,0.2386) (0.9259,0.2386) (0.9297,0.2386) (0.9334,0.2386) (0.9371,0.2386)
(0.9409,0.2388) (0.9447,0.2386) (0.9484,0.2386) (0.9522,0.2386)

\PST@pythia(0.7418,0.1344) (0.7454,0.1345) (0.7491,0.1347) (0.7529,0.1348)
(0.7567,0.1349) (0.7604,0.1350) (0.7642,0.1352) (0.7679,0.1353) (0.7717,0.1355)
(0.7755,0.1356) (0.7792,0.1357) (0.7830,0.1359) (0.7868,0.1360) (0.7905,0.1362)
(0.7943,0.1363) (0.7980,0.1365) (0.8018,0.1366) (0.8055,0.1368) (0.8092,0.1369)
(0.8132,0.1371) (0.8168,0.1373) (0.8207,0.1375) (0.8242,0.1376) (0.8280,0.1378)
(0.8319,0.1380) (0.8356,0.1381) (0.8394,0.1383) (0.8432,0.1385) (0.8470,0.1387)
(0.8506,0.1389) (0.8545,0.1391) (0.8581,0.1392) (0.8618,0.1394) (0.8658,0.1396)
(0.8694,0.1398) (0.8732,0.1400) (0.8770,0.1402) (0.8807,0.1405) (0.8846,0.1407)
(0.8882,0.1409) (0.8919,0.1411) (0.8958,0.1413) (0.8995,0.1416) (0.9033,0.1418)
(0.9070,0.1420) (0.9107,0.1422) (0.9146,0.1425) (0.9183,0.1427) (0.9221,0.1430)
(0.9259,0.1432) (0.9297,0.1435) (0.9334,0.1437) (0.9371,0.1440) (0.9409,0.1443)
(0.9447,0.1445) (0.9484,0.1448) (0.9522,0.1451)

\PST@Border(0.1750,0.9680)
(0.1750,0.1344)
(0.9630,0.1344)
(0.9630,0.9680)
(0.1750,0.9680)

\catcode`@=12
\fi
\endpspicture

%% file: pythia.tex
\ifx\PSTloaded\undefined
\def\PSTloaded{t}
\psset{arrowsize=.01 3.2 1.4 .3}
\psset{dotsize=.01}
\catcode`@=11

\newpsobject{PST@Border}{psline}{linewidth=.0015,linestyle=solid}
\newpsobject{PST@Axes}{psline}{linewidth=.0015,linestyle=dotted,dotsep=.004}
\newpsobject{PST@Solid}{psline}{linecolor=black,    linewidth=.0025,linestyle=solid}
\newpsobject{PST@Dashed}{psline}{linecolor=red,      linewidth=.0025,linestyle=dashed,dash=.01 .01}
\catcode`@=12

\fi
\psset{unit=5.0in,xunit=1.031\columnwidth,yunit=.5\columnwidth}
\pspicture(0.1,0.002)(0.98,0.98)
\ifx\nofigs\undefined
\catcode`@=11

\PST@Border(0.2390,0.1344)
(0.2540,0.1344)

\PST@Border(0.9630,0.1344)
(0.9480,0.1344)

\rput[r](0.2230,0.1344){$10^{-7}$}
\PST@Border(0.2390,0.1762)
(0.2465,0.1762)

\PST@Border(0.9630,0.1762)
(0.9555,0.1762)

\PST@Border(0.2390,0.2315)
(0.2465,0.2315)

\PST@Border(0.9630,0.2315)
(0.9555,0.2315)

\PST@Border(0.2390,0.2599)
(0.2465,0.2599)

\PST@Border(0.9630,0.2599)
(0.9555,0.2599)

\PST@Border(0.2390,0.2733)
(0.2540,0.2733)

\PST@Border(0.9630,0.2733)
(0.9480,0.2733)

\rput[r](0.2230,0.2733){$10^{-6}$}
\PST@Border(0.2390,0.3152)
(0.2465,0.3152)

\PST@Border(0.9630,0.3152)
(0.9555,0.3152)

\PST@Border(0.2390,0.3704)
(0.2465,0.3704)

\PST@Border(0.9630,0.3704)
(0.9555,0.3704)

\PST@Border(0.2390,0.3988)
(0.2465,0.3988)

\PST@Border(0.9630,0.3988)
(0.9555,0.3988)

\PST@Border(0.2390,0.4123)
(0.2540,0.4123)

\PST@Border(0.9630,0.4123)
(0.9480,0.4123)

\rput[r](0.2230,0.4123){$10^{-5}$}
\PST@Border(0.2390,0.4541)
(0.2465,0.4541)

\PST@Border(0.9630,0.4541)
(0.9555,0.4541)

\PST@Border(0.2390,0.5094)
(0.2465,0.5094)

\PST@Border(0.9630,0.5094)
(0.9555,0.5094)

\PST@Border(0.2390,0.5377)
(0.2465,0.5377)

\PST@Border(0.9630,0.5377)
(0.9555,0.5377)

\PST@Border(0.2390,0.5512)
(0.2540,0.5512)

\PST@Border(0.9630,0.5512)
(0.9480,0.5512)

\rput[r](0.2230,0.5512){$10^{-4}$}
\PST@Border(0.2390,0.5930)
(0.2465,0.5930)

\PST@Border(0.9630,0.5930)
(0.9555,0.5930)

\PST@Border(0.2390,0.6483)
(0.2465,0.6483)

\PST@Border(0.9630,0.6483)
(0.9555,0.6483)

\PST@Border(0.2390,0.6767)
(0.2465,0.6767)

\PST@Border(0.9630,0.6767)
(0.9555,0.6767)

\PST@Border(0.2390,0.6901)
(0.2540,0.6901)

\PST@Border(0.9630,0.6901)
(0.9480,0.6901)

\rput[r](0.2230,0.6901){$10^{-3}$}
\PST@Border(0.2390,0.7320)
(0.2465,0.7320)

\PST@Border(0.9630,0.7320)
(0.9555,0.7320)

\PST@Border(0.2390,0.7872)
(0.2465,0.7872)

\PST@Border(0.9630,0.7872)
(0.9555,0.7872)

\PST@Border(0.2390,0.8156)
(0.2465,0.8156)

\PST@Border(0.9630,0.8156)
(0.9555,0.8156)

\PST@Border(0.2390,0.8291)
(0.2540,0.8291)

\PST@Border(0.9630,0.8291)
(0.9480,0.8291)

\rput[r](0.2230,0.8291){ 0.01}
\PST@Border(0.2390,0.8709)
(0.2465,0.8709)

\PST@Border(0.9630,0.8709)
(0.9555,0.8709)

\PST@Border(0.2390,0.9262)
(0.2465,0.9262)

\PST@Border(0.9630,0.9262)
(0.9555,0.9262)

\PST@Border(0.2390,0.9545)
(0.2465,0.9545)

\PST@Border(0.9630,0.9545)
(0.9555,0.9545)

\PST@Border(0.2390,0.9680)
(0.2540,0.9680)

\PST@Border(0.9630,0.9680)
(0.9480,0.9680)

\rput[r](0.2230,0.9680){ 0.1}
\PST@Border(0.2390,0.1344)
(0.2390,0.1544)

\PST@Border(0.2390,0.9680)
(0.2390,0.9480)

\rput(0.2390,0.0924){ 1}
\PST@Border(0.3295,0.1344)
(0.3295,0.1544)

\PST@Border(0.3295,0.9680)
(0.3295,0.9480)

\rput(0.3295,0.0924){ 1.5}
\PST@Border(0.4200,0.1344)
(0.4200,0.1544)

\PST@Border(0.4200,0.9680)
(0.4200,0.9480)

\rput(0.4200,0.0924){ 2}
\PST@Border(0.5105,0.1344)
(0.5105,0.1544)

\PST@Border(0.5105,0.9680)
(0.5105,0.9480)

\rput(0.5105,0.0924){ 2.5}
\PST@Border(0.6010,0.1344)
(0.6010,0.1544)

\PST@Border(0.6010,0.9680)
(0.6010,0.9480)

\rput(0.6010,0.0924){ 3}
\PST@Border(0.6915,0.1344)
(0.6915,0.1544)

\PST@Border(0.6915,0.9680)
(0.6915,0.9480)

\rput(0.6915,0.0924){ 3.5}
\PST@Border(0.7820,0.1344)
(0.7820,0.1544)

\PST@Border(0.7820,0.9680)
(0.7820,0.9480)

\rput(0.7820,0.0924){ 4}
\PST@Border(0.8725,0.1344)
(0.8725,0.1544)

\PST@Border(0.8725,0.9680)
(0.8725,0.9480)

\rput(0.8725,0.0924){ 4.5}
\PST@Border(0.9630,0.1344)
(0.9630,0.1544)

\PST@Border(0.9630,0.9680)
(0.9630,0.9480)

\rput(0.9630,0.0924){ 5}
\PST@Border(0.2390,0.9680)
(0.2390,0.1344)
(0.9630,0.1344)
(0.9630,0.9680)
(0.2390,0.9680)

\rput{L}(0.1040,0.5512){$E\frac{dN}{d^3p}~[{\rm GeV}^{-2}]$}
\rput(0.6010,0.0294){$p_\bot$~[GeV]}

\rput[r](0.8360,0.9070){with PYTHIA}
\PST@Solid(0.8520,0.9070)(0.9310,0.9070)
\rput[r](0.8360,0.8350){without PYTHIA}
\PST@Dashed(0.8520,0.8350)(0.9310,0.8350)
\rput[r](0.9310,0.7430){UrQMD $p+p$ 158~GeV}
\rput[r](0.9310,0.6710){all charged particles}

\PST@Solid(0.2390,0.9651) (0.2481,0.9508) (0.2662,0.9228) (0.2843,0.8961)
(0.3024,0.8698) (0.3205,0.8439) (0.3386,0.8196) (0.3567,0.7963) (0.3748,0.7710)
(0.3929,0.7487) (0.4110,0.7282) (0.4291,0.7031) (0.4472,0.6823) (0.4653,0.6572)
(0.4834,0.6385) (0.5015,0.6147) (0.5196,0.5904) (0.5377,0.5691) (0.5558,0.5424)
(0.5739,0.5258) (0.5920,0.5086) (0.6101,0.4747) (0.6282,0.4660) (0.6463,0.4458)
(0.6644,0.4117) (0.6825,0.3988) (0.7006,0.3837) (0.7187,0.3699) (0.7368,0.3578)
(0.7549,0.3429) (0.7730,0.3351) (0.7911,0.2860) (0.8092,0.2726) (0.8273,0.2697)
(0.8454,0.2468) (0.8635,0.2361) (0.8816,0.2139) (0.8997,0.2161) (0.9178,0.1951)
(0.9359,0.1717) (0.9540,0.1553) (0.9630,0.1378)

\PST@Dashed(0.2390,0.9626) (0.2481,0.9482) (0.2662,0.9202) (0.2843,0.8929)
(0.3024,0.8662) (0.3205,0.8376) (0.3386,0.8170) (0.3567,0.7883) (0.3748,0.7619)
(0.3929,0.7452) (0.4110,0.7136) (0.4291,0.6944) (0.4472,0.6707) (0.4653,0.6334)
(0.4834,0.6252) (0.5015,0.6023) (0.5196,0.5600) (0.5377,0.5419) (0.5558,0.5242)
(0.5739,0.4944) (0.5920,0.4269) (0.6101,0.4071) (0.6463,0.3411) (0.6644,0.2855)
(0.7006,0.2172) (0.7187,0.2110) (0.7368,0.1344)

\PST@Border(0.2390,0.9680)
(0.2390,0.1344)
(0.9630,0.1344)
(0.9630,0.9680)
(0.2390,0.9680)

\catcode`@=12
\fi
\endpspicture

%% file: initialtemperature.tex
\ifx\PSTloaded\undefined
\def\PSTloaded{t}
\psset{arrowsize=.01 3.2 1.4 .3}
\psset{dotsize=.01}
\catcode`@=11

\definecolor{violett}{rgb}{.5,0,.5}
\definecolor{darkgreen}{rgb}{0,.5,0}
\newpsobject{PST@Border}{psline}{linewidth=.0015,linestyle=solid}
\newpsobject{PST@Axes}{psline}{linewidth=.0015,linestyle=dotted,dotsep=.004}
\newpsobject{PST@lineone}{psline}{linewidth=.0025,linestyle=solid,linecolor=red}
\newpsobject{PST@linetwo}{psline}{linewidth=.0025,linestyle=solid,linecolor=darkgreen}
\newpsobject{PST@linethr}{psline}{linewidth=.0025,linestyle=solid,linecolor=blue}
\newpsobject{PST@linefou}{psline}{linewidth=.0025,linestyle=solid,linecolor=violett}
\newpsobject{PST@linefiv}{psline}{linewidth=.0025,linestyle=solid,linecolor=magenta}
\catcode`@=12

\fi
\psset{unit=5.0in,xunit=1.273\columnwidth,yunit=1.273\columnwidth}
\pspicture(0.08,0.12)(0.865,0.845)
\ifx\nofigs\undefined
\catcode`@=11

\rput[l](0.2034,0.8165){BM-EoS}
\rput[r](0.8035,0.8165){HG-EoS}


\psset{linewidth=.0015,nodesepB=.01}
\pnode(0.5118,0.5166){A250}
\rput[tl](0.5370,0.5167){\rnode{B250}{250}}
\ncline{A250}{B250}

\pnode(0.6302,0.6640){A200}
\rput[br](0.6100,0.6100){\rnode{B200}{200}}
\ncline{A200}{B200}

\pnode(0.2771,0.5913){A150}
\rput[tl](0.3200,0.5200){\rnode{B150}{150}}
\ncarc{A150}{B150}

\pnode(0.6390,0.3465){A100}
\rput[tl](0.6700,0.2800){\rnode{B100}{100}}
\ncarc{A100}{B100}

\pnode(0.2831,0.3511){A050}
\rput[tr](0.2700,0.2800){\rnode{B050}{50}}
\ncarc{A050}{B050}

\PST@lineone(0.5021,0.5995) (0.5052,0.6013) (0.5077,0.6021) (0.5079,0.6050)
(0.5084,0.6053) (0.5116,0.6063) (0.5136,0.6080) (0.5143,0.6109)
(0.5148,0.6116) (0.5180,0.6117) (0.5198,0.6139) (0.5212,0.6155)
(0.5244,0.6155) (0.5259,0.6168) (0.5260,0.6198) (0.5275,0.6214)
(0.5307,0.6222) (0.5317,0.6227) (0.5319,0.6257) (0.5339,0.6267)
(0.5371,0.6262) (0.5403,0.6262) (0.5417,0.6257) (0.5417,0.6227)
(0.5435,0.6217) (0.5466,0.6217) (0.5474,0.6198) (0.5474,0.6168)
(0.5473,0.6139) (0.5473,0.6139) (0.5469,0.6109) (0.5469,0.6080)
(0.5466,0.6074) (0.5435,0.6074) (0.5422,0.6050) (0.5422,0.6021)
(0.5405,0.5991) (0.5405,0.5962) (0.5403,0.5958) (0.5371,0.5958)
(0.5353,0.5932) (0.5339,0.5912) (0.5330,0.5903) (0.5326,0.5873)
(0.5307,0.5860) (0.5275,0.5846) (0.5274,0.5844) (0.5274,0.5814)
(0.5252,0.5785) (0.5252,0.5755) (0.5244,0.5741) (0.5212,0.5741)
(0.5204,0.5726) (0.5191,0.5696) (0.5190,0.5667) (0.5180,0.5640)
(0.5148,0.5640) (0.5147,0.5637) (0.5147,0.5608) (0.5137,0.5578)
(0.5138,0.5549) (0.5131,0.5519) (0.5125,0.5490) (0.5124,0.5460)
(0.5124,0.5431) (0.5124,0.5401) (0.5121,0.5372) (0.5121,0.5342)
(0.5119,0.5313) (0.5117,0.5283) (0.5117,0.5254) (0.5116,0.5224)
(0.5116,0.5196) (0.5118,0.5166) (0.5118,0.5137) (0.5119,0.5107)
(0.5119,0.5078) (0.5119,0.5048) (0.5125,0.5019) (0.5125,0.4989)
(0.5138,0.4960) (0.5140,0.4930) (0.5148,0.4921) (0.5180,0.4920)
(0.5209,0.4901) (0.5212,0.4900) (0.5244,0.4900) (0.5275,0.4889)
(0.5307,0.4882) (0.5339,0.4876) (0.5346,0.4871) (0.5347,0.4842)
(0.5371,0.4823) (0.5403,0.4823) (0.5412,0.4812) (0.5412,0.4783)
(0.5435,0.4753) (0.5435,0.4751) (0.5466,0.4750) (0.5467,0.4724)
(0.5480,0.4694) (0.5492,0.4665) (0.5492,0.4635) (0.5498,0.4615)
(0.5513,0.4606) (0.5513,0.4576) (0.5530,0.4551) (0.5531,0.4547)
(0.5530,0.4544) (0.5517,0.4517) (0.5530,0.4488) (0.5530,0.4488)
(0.5530,0.4488) (0.5530,0.4488) (0.5504,0.4458) (0.5505,0.4429)
(0.5498,0.4425) (0.5490,0.4399) (0.5490,0.4370) (0.5476,0.4340)
(0.5471,0.4311) (0.5466,0.4301) (0.5435,0.4300) (0.5424,0.4281)
(0.5403,0.4257) (0.5371,0.4257) (0.5365,0.4252) (0.5365,0.4222)
(0.5339,0.4206) (0.5307,0.4194) (0.5300,0.4193) (0.5296,0.4163)
(0.5275,0.4160) (0.5244,0.4150) (0.5212,0.4150) (0.5180,0.4144)
(0.5148,0.4144) (0.5116,0.4142) (0.5084,0.4143) (0.5052,0.4144)
(0.5021,0.4152)
\PST@lineone(0.6137,0.5313) (0.6135,0.5311) (0.6103,0.5311) (0.6072,0.5311)
(0.6044,0.5313) (0.6052,0.5342) (0.6040,0.5372) (0.6040,0.5372)
(0.6072,0.5379) (0.6091,0.5401) (0.6091,0.5431) (0.6103,0.5435)
(0.6135,0.5435) (0.6151,0.5431) (0.6151,0.5401) (0.6152,0.5372)
(0.6151,0.5342) (0.6137,0.5313)

\PST@linetwo(0.5021,0.7114) (0.5052,0.7116) (0.5084,0.7115) (0.5116,0.7114)
(0.5127,0.7112) (0.5148,0.7109) (0.5180,0.7109) (0.5212,0.7101)
(0.5244,0.7101) (0.5275,0.7091) (0.5307,0.7085) (0.5321,0.7083)
(0.5326,0.7053) (0.5339,0.7050) (0.5371,0.7037) (0.5403,0.7037)
(0.5434,0.7024) (0.5434,0.6994) (0.5435,0.6994) (0.5466,0.6994)
(0.5498,0.6978) (0.5530,0.6976) (0.5554,0.6965) (0.5562,0.6955)
(0.5576,0.6935) (0.5594,0.6920) (0.5626,0.6918) (0.5638,0.6906)
(0.5658,0.6892) (0.5689,0.6892) (0.5711,0.6876) (0.5712,0.6847)
(0.5721,0.6840) (0.5753,0.6838) (0.5785,0.6818) (0.5786,0.6817)
(0.5786,0.6788) (0.5817,0.6767) (0.5849,0.6766) (0.5863,0.6758)
(0.5881,0.6738) (0.5912,0.6737) (0.5921,0.6729) (0.5944,0.6710)
(0.5976,0.6709) (0.5994,0.6699) (0.6008,0.6695) (0.6040,0.6691)
(0.6072,0.6691) (0.6103,0.6686) (0.6135,0.6686) (0.6167,0.6682)
(0.6167,0.6682) (0.6199,0.6682) (0.6231,0.6677) (0.6263,0.6672)
(0.6295,0.6671) (0.6301,0.6670) (0.6302,0.6640) (0.6326,0.6633)
(0.6358,0.6633) (0.6390,0.6621) (0.6422,0.6621) (0.6444,0.6611)
(0.6444,0.6581) (0.6454,0.6576) (0.6486,0.6566) (0.6517,0.6554)
(0.6520,0.6552) (0.6540,0.6522) (0.6549,0.6500) (0.6581,0.6499)
(0.6584,0.6493) (0.6607,0.6463) (0.6607,0.6434) (0.6613,0.6426)
(0.6645,0.6426) (0.6661,0.6404) (0.6661,0.6375) (0.6677,0.6351)
(0.6696,0.6345) (0.6709,0.6335) (0.6717,0.6316) (0.6730,0.6286)
(0.6740,0.6273) (0.6752,0.6257) (0.6752,0.6227) (0.6772,0.6200)
(0.6804,0.6200) (0.6806,0.6198) (0.6806,0.6168) (0.6824,0.6139)
(0.6836,0.6113) (0.6868,0.6112) (0.6869,0.6109) (0.6873,0.6080)
(0.6892,0.6050) (0.6892,0.6021) (0.6900,0.6010) (0.6931,0.6008)
(0.6950,0.5991) (0.6950,0.5962) (0.6963,0.5945) (0.6973,0.5932)
(0.6991,0.5903) (0.6993,0.5873) (0.6995,0.5869) (0.7027,0.5868)
(0.7043,0.5844) (0.7043,0.5814) (0.7045,0.5785) (0.7045,0.5755)
(0.7042,0.5726) (0.7034,0.5696) (0.7033,0.5667) (0.7027,0.5653)
(0.6995,0.5652) (0.6989,0.5637) (0.6989,0.5608) (0.6975,0.5578)
(0.6975,0.5549) (0.6963,0.5520) (0.6963,0.5519) (0.6951,0.5490)
(0.6952,0.5460) (0.6941,0.5431) (0.6941,0.5401) (0.6931,0.5372)
(0.6931,0.5372) (0.6914,0.5342) (0.6900,0.5334) (0.6894,0.5313)
(0.6888,0.5283) (0.6887,0.5254) (0.6881,0.5224) (0.6881,0.5196)
(0.6876,0.5166) (0.6876,0.5137) (0.6875,0.5107) (0.6876,0.5078)
(0.6876,0.5048) (0.6880,0.5019) (0.6880,0.4989) (0.6889,0.4960)
(0.6890,0.4930) (0.6900,0.4905) (0.6921,0.4901) (0.6931,0.4899)
(0.6946,0.4871) (0.6948,0.4842) (0.6963,0.4819) (0.6963,0.4819)
(0.6968,0.4812) (0.6968,0.4783) (0.6990,0.4753) (0.6995,0.4724)
(0.6995,0.4723) (0.7027,0.4718) (0.7042,0.4694) (0.7059,0.4667)
(0.7091,0.4667) (0.7092,0.4665) (0.7092,0.4635) (0.7106,0.4606)
(0.7106,0.4576) (0.7116,0.4547) (0.7116,0.4517) (0.7117,0.4488)
(0.7115,0.4458) (0.7115,0.4429) (0.7105,0.4399) (0.7105,0.4370)
(0.7105,0.4370) (0.7091,0.4344) (0.7059,0.4344) (0.7056,0.4340)
(0.7050,0.4311) (0.7028,0.4281) (0.7027,0.4281) (0.6995,0.4280)
(0.6963,0.4260) (0.6946,0.4252) (0.6945,0.4222) (0.6931,0.4217)
(0.6900,0.4216) (0.6868,0.4208) (0.6836,0.4208) (0.6804,0.4203)
(0.6772,0.4203) (0.6740,0.4201) (0.6709,0.4203) (0.6677,0.4203)
(0.6645,0.4210) (0.6613,0.4210) (0.6581,0.4218) (0.6549,0.4218)
(0.6534,0.4222) (0.6534,0.4252) (0.6517,0.4257) (0.6486,0.4263)
(0.6454,0.4269) (0.6422,0.4272) (0.6390,0.4272) (0.6358,0.4273)
(0.6326,0.4273) (0.6295,0.4265) (0.6263,0.4260) (0.6244,0.4252)
(0.6242,0.4222) (0.6231,0.4218) (0.6199,0.4199) (0.6167,0.4199)
(0.6159,0.4193) (0.6159,0.4163) (0.6135,0.4146) (0.6103,0.4146)
(0.6089,0.4134) (0.6078,0.4104) (0.6072,0.4097) (0.6040,0.4091)
(0.6024,0.4075) (0.6008,0.4062) (0.5996,0.4045) (0.5996,0.4016)
(0.5976,0.3988) (0.5944,0.3987) (0.5943,0.3986) (0.5943,0.3957)
(0.5912,0.3929) (0.5881,0.3929) (0.5878,0.3927) (0.5861,0.3898)
(0.5849,0.3884) (0.5817,0.3881) (0.5804,0.3868) (0.5785,0.3858)
(0.5753,0.3841) (0.5721,0.3841) (0.5717,0.3839) (0.5717,0.3809)
(0.5689,0.3797) (0.5658,0.3797) (0.5626,0.3784) (0.5594,0.3784)
(0.5584,0.3780) (0.5584,0.3750) (0.5562,0.3741) (0.5530,0.3731)
(0.5498,0.3729) (0.5480,0.3721) (0.5466,0.3711) (0.5435,0.3710)
(0.5413,0.3691) (0.5403,0.3682) (0.5371,0.3680) (0.5352,0.3662)
(0.5339,0.3655) (0.5307,0.3644) (0.5275,0.3639) (0.5263,0.3632)
(0.5263,0.3603) (0.5244,0.3592) (0.5212,0.3592) (0.5180,0.3577)
(0.5148,0.3577) (0.5139,0.3573) (0.5139,0.3544) (0.5116,0.3534)
(0.5084,0.3531) (0.5052,0.3523) (0.5021,0.3516) (0.5021,0.3516)

\PST@linethr(0.5021,0.7507) (0.5052,0.7507) (0.5084,0.7504) (0.5116,0.7505)
(0.5148,0.7500) (0.5180,0.7500) (0.5201,0.7496) (0.5202,0.7466)
(0.5212,0.7464) (0.5244,0.7464) (0.5275,0.7455) (0.5307,0.7451)
(0.5339,0.7444) (0.5355,0.7437) (0.5355,0.7407) (0.5371,0.7400)
(0.5403,0.7400) (0.5435,0.7384) (0.5466,0.7384) (0.5479,0.7378)
(0.5488,0.7348) (0.5498,0.7341) (0.5530,0.7339) (0.5562,0.7319)
(0.5563,0.7319) (0.5594,0.7300) (0.5626,0.7300) (0.5642,0.7289)
(0.5643,0.7260) (0.5658,0.7250) (0.5689,0.7250) (0.5720,0.7230)
(0.5720,0.7201) (0.5721,0.7200) (0.5753,0.7199) (0.5785,0.7177)
(0.5794,0.7171) (0.5809,0.7142) (0.5817,0.7134) (0.5849,0.7129)
(0.5867,0.7112) (0.5881,0.7103) (0.5912,0.7103) (0.5944,0.7084)
(0.5976,0.7084) (0.5978,0.7083) (0.5979,0.7053) (0.6008,0.7041)
(0.6040,0.7031) (0.6072,0.7030) (0.6103,0.7026) (0.6135,0.7026)
(0.6165,0.7024) (0.6165,0.6994) (0.6167,0.6994) (0.6199,0.6994)
(0.6231,0.6990) (0.6263,0.6989) (0.6295,0.6987) (0.6326,0.6983)
(0.6358,0.6983) (0.6390,0.6976) (0.6422,0.6976) (0.6454,0.6968)
(0.6478,0.6965) (0.6486,0.6962) (0.6517,0.6945) (0.6527,0.6935)
(0.6549,0.6921) (0.6581,0.6920) (0.6600,0.6906) (0.6613,0.6899)
(0.6645,0.6899) (0.6677,0.6879) (0.6692,0.6876) (0.6698,0.6847)
(0.6709,0.6845) (0.6740,0.6826) (0.6752,0.6817) (0.6752,0.6788)
(0.6772,0.6769) (0.6804,0.6769) (0.6815,0.6758) (0.6829,0.6729)
(0.6836,0.6715) (0.6868,0.6714) (0.6875,0.6699) (0.6900,0.6671)
(0.6913,0.6670) (0.6920,0.6640) (0.6931,0.6639) (0.6957,0.6611)
(0.6957,0.6581) (0.6963,0.6574) (0.6982,0.6552) (0.6995,0.6531)
(0.7027,0.6525) (0.7029,0.6522) (0.7040,0.6493) (0.7059,0.6470)
(0.7091,0.6470) (0.7095,0.6463) (0.7095,0.6434) (0.7116,0.6404)
(0.7116,0.6375) (0.7123,0.6368) (0.7154,0.6367) (0.7178,0.6345)
(0.7186,0.6337) (0.7212,0.6316) (0.7218,0.6302) (0.7250,0.6295)
(0.7253,0.6286) (0.7282,0.6260) (0.7314,0.6260) (0.7318,0.6257)
(0.7318,0.6227) (0.7345,0.6208) (0.7377,0.6208) (0.7388,0.6198)
(0.7388,0.6168) (0.7409,0.6147) (0.7418,0.6139) (0.7441,0.6109)
(0.7441,0.6109) (0.7460,0.6080) (0.7473,0.6076) (0.7489,0.6050)
(0.7489,0.6021) (0.7503,0.5991) (0.7503,0.5962) (0.7505,0.5958)
(0.7537,0.5958) (0.7545,0.5932) (0.7551,0.5903) (0.7552,0.5873)
(0.7554,0.5844) (0.7554,0.5814) (0.7552,0.5785) (0.7552,0.5755)
(0.7549,0.5726) (0.7543,0.5696) (0.7543,0.5667) (0.7537,0.5644)
(0.7505,0.5644) (0.7503,0.5637) (0.7503,0.5608) (0.7491,0.5578)
(0.7490,0.5549) (0.7477,0.5519) (0.7473,0.5511) (0.7441,0.5492)
(0.7440,0.5490) (0.7433,0.5460) (0.7412,0.5431) (0.7412,0.5401)
(0.7409,0.5397) (0.7394,0.5372) (0.7393,0.5342) (0.7377,0.5314)
(0.7345,0.5313) (0.7345,0.5313) (0.7331,0.5283) (0.7331,0.5254)
(0.7319,0.5224) (0.7319,0.5196) (0.7316,0.5166) (0.7316,0.5137)
(0.7316,0.5107) (0.7323,0.5078) (0.7323,0.5048) (0.7338,0.5019)
(0.7338,0.4989) (0.7345,0.4977) (0.7377,0.4977) (0.7387,0.4960)
(0.7388,0.4930) (0.7407,0.4901) (0.7409,0.4898) (0.7430,0.4871)
(0.7431,0.4842) (0.7441,0.4828) (0.7473,0.4820) (0.7478,0.4812)
(0.7478,0.4783) (0.7494,0.4753) (0.7496,0.4724) (0.7505,0.4704)
(0.7537,0.4704) (0.7541,0.4694) (0.7552,0.4665) (0.7552,0.4635)
(0.7560,0.4606) (0.7560,0.4576) (0.7566,0.4547) (0.7566,0.4517)
(0.7566,0.4488) (0.7566,0.4458) (0.7566,0.4429) (0.7562,0.4399)
(0.7562,0.4370) (0.7555,0.4340) (0.7553,0.4311) (0.7544,0.4281)
(0.7544,0.4281) (0.7537,0.4264) (0.7505,0.4264) (0.7499,0.4252)
(0.7499,0.4222) (0.7481,0.4193) (0.7481,0.4163) (0.7473,0.4152)
(0.7441,0.4144) (0.7431,0.4134) (0.7423,0.4104) (0.7409,0.4087)
(0.7401,0.4075) (0.7377,0.4053) (0.7345,0.4053) (0.7336,0.4045)
(0.7336,0.4016) (0.7314,0.3999) (0.7282,0.3999) (0.7264,0.3986)
(0.7264,0.3957) (0.7250,0.3948) (0.7218,0.3947) (0.7186,0.3929)
(0.7183,0.3927) (0.7154,0.3900) (0.7142,0.3898) (0.7123,0.3895)
(0.7091,0.3872) (0.7059,0.3872) (0.7053,0.3868) (0.7027,0.3860)
(0.6995,0.3859) (0.6963,0.3851) (0.6931,0.3844) (0.6900,0.3843)
(0.6873,0.3839) (0.6873,0.3809) (0.6868,0.3809) (0.6836,0.3809)
(0.6804,0.3807) (0.6772,0.3807) (0.6740,0.3807) (0.6709,0.3806)
(0.6677,0.3806) (0.6645,0.3806) (0.6613,0.3806) (0.6581,0.3804)
(0.6549,0.3804) (0.6517,0.3804) (0.6486,0.3804) (0.6454,0.3803)
(0.6422,0.3803) (0.6390,0.3803) (0.6358,0.3798) (0.6326,0.3798)
(0.6295,0.3788) (0.6263,0.3783) (0.6254,0.3780) (0.6253,0.3750)
(0.6231,0.3740) (0.6204,0.3721) (0.6199,0.3715) (0.6167,0.3714)
(0.6151,0.3691) (0.6135,0.3667) (0.6103,0.3666) (0.6100,0.3662)
(0.6072,0.3644) (0.6040,0.3642) (0.6019,0.3632) (0.6018,0.3603)
(0.6008,0.3598) (0.5976,0.3583) (0.5944,0.3582) (0.5926,0.3573)
(0.5926,0.3544) (0.5912,0.3537) (0.5881,0.3537) (0.5849,0.3522)
(0.5817,0.3522) (0.5802,0.3514) (0.5785,0.3502) (0.5761,0.3485)
(0.5753,0.3472) (0.5721,0.3467) (0.5714,0.3455) (0.5689,0.3438)
(0.5658,0.3438) (0.5643,0.3426) (0.5643,0.3396) (0.5626,0.3380)
(0.5594,0.3380) (0.5581,0.3367) (0.5581,0.3337) (0.5562,0.3316)
(0.5554,0.3308) (0.5530,0.3280) (0.5527,0.3278) (0.5498,0.3251)
(0.5498,0.3249) (0.5469,0.3219) (0.5469,0.3190) (0.5466,0.3187)
(0.5435,0.3187) (0.5403,0.3161) (0.5371,0.3160) (0.5371,0.3160)
(0.5371,0.3131) (0.5339,0.3110) (0.5313,0.3101) (0.5307,0.3099)
(0.5275,0.3093) (0.5244,0.3078) (0.5212,0.3078) (0.5192,0.3072)
(0.5180,0.3053) (0.5148,0.3052) (0.5138,0.3042) (0.5116,0.3039)
(0.5084,0.3038) (0.5052,0.3038) (0.5021,0.3038)

\PST@linefou(0.5021,0.7750) (0.5052,0.7749) (0.5084,0.7751) (0.5116,0.7747)
(0.5148,0.7744) (0.5148,0.7744) (0.5180,0.7743) (0.5244,0.7738)
(0.5266,0.7732) (0.5275,0.7729) (0.5307,0.7725) (0.5339,0.7721)
(0.5371,0.7710) (0.5403,0.7710) (0.5419,0.7702) (0.5420,0.7673)
(0.5435,0.7666) (0.5466,0.7666) (0.5498,0.7652) (0.5530,0.7650)
(0.5545,0.7643) (0.5545,0.7614) (0.5562,0.7606) (0.5594,0.7591)
(0.5626,0.7591) (0.5637,0.7584) (0.5643,0.7555) (0.5658,0.7544)
(0.5689,0.7544) (0.5716,0.7525) (0.5721,0.7522) (0.5753,0.7521)
(0.5785,0.7502) (0.5795,0.7496) (0.5795,0.7466) (0.5817,0.7451)
(0.5849,0.7451) (0.5867,0.7437) (0.5867,0.7407) (0.5881,0.7397)
(0.5912,0.7397) (0.5941,0.7378) (0.5944,0.7367) (0.5976,0.7365)
(0.5981,0.7348) (0.6008,0.7329) (0.6025,0.7319) (0.6040,0.7311)
(0.6072,0.7310) (0.6103,0.7299) (0.6135,0.7299) (0.6167,0.7292)
(0.6199,0.7291) (0.6214,0.7289) (0.6215,0.7260) (0.6231,0.7257)
(0.6263,0.7255) (0.6295,0.7254) (0.6326,0.7251) (0.6358,0.7251)
(0.6390,0.7247) (0.6422,0.7247) (0.6454,0.7241) (0.6486,0.7237)
(0.6517,0.7234) (0.6530,0.7230) (0.6530,0.7201) (0.6549,0.7195)
(0.6581,0.7195) (0.6613,0.7184) (0.6645,0.7184) (0.6675,0.7171)
(0.6677,0.7167) (0.6709,0.7157) (0.6720,0.7142) (0.6740,0.7130)
(0.6763,0.7112) (0.6772,0.7106) (0.6804,0.7106) (0.6836,0.7086)
(0.6868,0.7086) (0.6873,0.7083) (0.6873,0.7053) (0.6900,0.7033)
(0.6931,0.7031) (0.6941,0.7024) (0.6942,0.6994) (0.6963,0.6977)
(0.6977,0.6965) (0.6993,0.6935) (0.6995,0.6932) (0.7027,0.6925)
(0.7040,0.6906) (0.7059,0.6886) (0.7091,0.6886) (0.7100,0.6876)
(0.7100,0.6847) (0.7123,0.6823) (0.7154,0.6822) (0.7160,0.6817)
(0.7160,0.6788) (0.7186,0.6759) (0.7187,0.6758) (0.7204,0.6729)
(0.7215,0.6699) (0.7218,0.6696) (0.7250,0.6695) (0.7275,0.6670)
(0.7275,0.6640) (0.7282,0.6633) (0.7314,0.6633) (0.7333,0.6611)
(0.7333,0.6581) (0.7345,0.6568) (0.7377,0.6567) (0.7395,0.6552)
(0.7409,0.6533) (0.7418,0.6522) (0.7432,0.6493) (0.7441,0.6485)
(0.7473,0.6480) (0.7494,0.6463) (0.7494,0.6434) (0.7505,0.6426)
(0.7537,0.6426) (0.7560,0.6404) (0.7560,0.6375) (0.7568,0.6367)
(0.7600,0.6367) (0.7622,0.6345) (0.7632,0.6330) (0.7646,0.6316)
(0.7661,0.6286) (0.7664,0.6284) (0.7696,0.6267) (0.7703,0.6257)
(0.7703,0.6227) (0.7723,0.6198) (0.7723,0.6168) (0.7728,0.6160)
(0.7759,0.6160) (0.7772,0.6139) (0.7783,0.6109) (0.7785,0.6080)
(0.7791,0.6063) (0.7823,0.6063) (0.7828,0.6050) (0.7828,0.6021)
(0.7836,0.5991) (0.7836,0.5962) (0.7841,0.5932) (0.7844,0.5903)
(0.7845,0.5873) (0.7846,0.5844) (0.7846,0.5814) (0.7847,0.5785)
(0.7847,0.5755) (0.7844,0.5726) (0.7842,0.5696) (0.7842,0.5667)
(0.7836,0.5637) (0.7836,0.5608) (0.7827,0.5578) (0.7826,0.5549)
(0.7823,0.5539) (0.7791,0.5538) (0.7785,0.5519) (0.7775,0.5490)
(0.7774,0.5460) (0.7761,0.5431) (0.7761,0.5401) (0.7759,0.5399)
(0.7728,0.5399) (0.7715,0.5372) (0.7715,0.5342) (0.7702,0.5313)
(0.7696,0.5300) (0.7680,0.5283) (0.7675,0.5254) (0.7664,0.5240)
(0.7654,0.5224) (0.7654,0.5196) (0.7650,0.5166) (0.7651,0.5137)
(0.7651,0.5107) (0.7663,0.5078) (0.7664,0.5048) (0.7664,0.5048)
(0.7696,0.5021) (0.7697,0.5019) (0.7697,0.4989) (0.7713,0.4960)
(0.7714,0.4930) (0.7728,0.4904) (0.7759,0.4904) (0.7761,0.4901)
(0.7778,0.4871) (0.7778,0.4842) (0.7791,0.4814) (0.7823,0.4814)
(0.7824,0.4812) (0.7824,0.4783) (0.7837,0.4753) (0.7838,0.4724)
(0.7847,0.4694) (0.7855,0.4665) (0.7855,0.4635) (0.7855,0.4633)
(0.7876,0.4606) (0.7876,0.4576) (0.7887,0.4551) (0.7888,0.4547)
(0.7887,0.4540) (0.7882,0.4517) (0.7887,0.4497) (0.7888,0.4488)
(0.7887,0.4458) (0.7888,0.4429) (0.7888,0.4429) (0.7887,0.4416)
(0.7883,0.4399) (0.7883,0.4370) (0.7866,0.4340) (0.7859,0.4311)
(0.7855,0.4302) (0.7851,0.4281) (0.7843,0.4252) (0.7843,0.4222)
(0.7832,0.4193) (0.7832,0.4163) (0.7823,0.4146) (0.7791,0.4146)
(0.7785,0.4134) (0.7781,0.4104) (0.7770,0.4075) (0.7759,0.4060)
(0.7728,0.4060) (0.7717,0.4045) (0.7717,0.4016) (0.7696,0.3989)
(0.7691,0.3986) (0.7691,0.3957) (0.7664,0.3944) (0.7639,0.3927)
(0.7632,0.3918) (0.7624,0.3898) (0.7605,0.3868) (0.7600,0.3864)
(0.7568,0.3864) (0.7537,0.3839) (0.7537,0.3809) (0.7537,0.3809)
(0.7505,0.3809) (0.7473,0.3786) (0.7441,0.3782) (0.7437,0.3780)
(0.7436,0.3750) (0.7409,0.3735) (0.7383,0.3721) (0.7377,0.3716)
(0.7345,0.3715) (0.7322,0.3691) (0.7314,0.3682) (0.7282,0.3681)
(0.7262,0.3662) (0.7250,0.3657) (0.7218,0.3656) (0.7186,0.3643)
(0.7155,0.3632) (0.7154,0.3618) (0.7137,0.3603) (0.7123,0.3603)
(0.7091,0.3591) (0.7059,0.3591) (0.7027,0.3583) (0.6995,0.3583)
(0.6963,0.3576) (0.6945,0.3573) (0.6941,0.3544) (0.6931,0.3543)
(0.6900,0.3541) (0.6868,0.3536) (0.6836,0.3536) (0.6804,0.3535)
(0.6772,0.3535) (0.6740,0.3532) (0.6709,0.3530) (0.6677,0.3528)
(0.6645,0.3524) (0.6613,0.3524) (0.6581,0.3517) (0.6549,0.3517)
(0.6540,0.3514) (0.6517,0.3506) (0.6486,0.3497) (0.6454,0.3492)
(0.6441,0.3485) (0.6422,0.3465) (0.6390,0.3465) (0.6382,0.3455)
(0.6358,0.3444) (0.6326,0.3444) (0.6295,0.3429) (0.6274,0.3426)
(0.6267,0.3396) (0.6263,0.3396) (0.6231,0.3384) (0.6199,0.3369)
(0.6167,0.3369) (0.6163,0.3367) (0.6163,0.3337) (0.6135,0.3324)
(0.6103,0.3324) (0.6072,0.3309) (0.6051,0.3308) (0.6040,0.3307)
(0.6008,0.3286) (0.5997,0.3278) (0.5985,0.3249) (0.5976,0.3244)
(0.5944,0.3243) (0.5912,0.3223) (0.5881,0.3223) (0.5876,0.3219)
(0.5876,0.3190) (0.5849,0.3165) (0.5817,0.3164) (0.5813,0.3160)
(0.5813,0.3131) (0.5787,0.3101) (0.5785,0.3099) (0.5765,0.3072)
(0.5762,0.3042) (0.5753,0.3033) (0.5721,0.3032) (0.5704,0.3013)
(0.5704,0.2983) (0.5689,0.2969) (0.5658,0.2969) (0.5641,0.2954)
(0.5641,0.2924) (0.5626,0.2911) (0.5594,0.2911) (0.5572,0.2895)
(0.5562,0.2887) (0.5530,0.2866) (0.5510,0.2865) (0.5498,0.2862)
(0.5493,0.2836) (0.5466,0.2821) (0.5435,0.2821) (0.5403,0.2808)
(0.5371,0.2808) (0.5368,0.2806) (0.5368,0.2777) (0.5339,0.2765)
(0.5307,0.2760) (0.5275,0.2757) (0.5244,0.2750) (0.5212,0.2750)
(0.5193,0.2747) (0.5191,0.2718) (0.5180,0.2716) (0.5148,0.2716)
(0.5116,0.2712) (0.5084,0.2715) (0.5052,0.2712) (0.5021,0.2712)

\PST@linefiv(0.5021,0.8020) (0.5052,0.8018) (0.5084,0.8018) (0.5116,0.8018)
(0.5148,0.8015) (0.5180,0.8015) (0.5212,0.8011) (0.5244,0.8011)
(0.5275,0.8005) (0.5307,0.8003) (0.5339,0.7998) (0.5342,0.7997)
(0.5348,0.7967) (0.5371,0.7960) (0.5403,0.7960) (0.5435,0.7947)
(0.5466,0.7947) (0.5486,0.7938) (0.5498,0.7933) (0.5530,0.7931)
(0.5562,0.7921) (0.5591,0.7908) (0.5593,0.7879) (0.5594,0.7879)
(0.5626,0.7879) (0.5658,0.7863) (0.5689,0.7863) (0.5715,0.7850)
(0.5715,0.7820) (0.5721,0.7817) (0.5753,0.7817) (0.5785,0.7801)
(0.5806,0.7791) (0.5815,0.7761) (0.5817,0.7760) (0.5849,0.7756)
(0.5881,0.7734) (0.5912,0.7734) (0.5916,0.7732) (0.5944,0.7712)
(0.5976,0.7711) (0.5988,0.7702) (0.5989,0.7673) (0.6008,0.7660)
(0.6037,0.7643) (0.6037,0.7614) (0.6040,0.7612) (0.6072,0.7610)
(0.6103,0.7592) (0.6135,0.7592) (0.6151,0.7584) (0.6158,0.7555)
(0.6167,0.7550) (0.6199,0.7550) (0.6231,0.7537) (0.6263,0.7530)
(0.6295,0.7526) (0.6300,0.7525) (0.6326,0.7520) (0.6358,0.7520)
(0.6390,0.7515) (0.6422,0.7515) (0.6454,0.7510) (0.6486,0.7506)
(0.6517,0.7504) (0.6549,0.7497) (0.6581,0.7497) (0.6587,0.7466)
(0.6587,0.7496) (0.6613,0.7461) (0.6645,0.7461) (0.6677,0.7453)
(0.6709,0.7451) (0.6740,0.7444) (0.6761,0.7407) (0.6761,0.7437)
(0.6772,0.7403) (0.6804,0.7403) (0.6836,0.7389) (0.6868,0.7389)
(0.6890,0.7378) (0.6900,0.7349) (0.6909,0.7348) (0.6931,0.7347)
(0.6963,0.7325) (0.6972,0.7319) (0.6995,0.7304) (0.7027,0.7304)
(0.7047,0.7260) (0.7047,0.7289) (0.7059,0.7251) (0.7091,0.7251)
(0.7117,0.7201) (0.7117,0.7230) (0.7123,0.7196) (0.7154,0.7195)
(0.7182,0.7171) (0.7186,0.7158) (0.7193,0.7142) (0.7216,0.7112)
(0.7218,0.7110) (0.7250,0.7108) (0.7275,0.7053) (0.7275,0.7083)
(0.7282,0.7045) (0.7314,0.7045) (0.7333,0.6994) (0.7333,0.7024)
(0.7345,0.6983) (0.7377,0.6983) (0.7400,0.6965) (0.7409,0.6939)
(0.7411,0.6935) (0.7432,0.6906) (0.7441,0.6898) (0.7473,0.6892)
(0.7490,0.6847) (0.7490,0.6876) (0.7505,0.6833) (0.7537,0.6833)
(0.7553,0.6788) (0.7553,0.6817) (0.7568,0.6770) (0.7600,0.6769)
(0.7612,0.6758) (0.7628,0.6729) (0.7632,0.6723) (0.7655,0.6699)
(0.7664,0.6694) (0.7696,0.6681) (0.7708,0.6640) (0.7708,0.6670)
(0.7728,0.6621) (0.7759,0.6621) (0.7770,0.6581) (0.7770,0.6611)
(0.7791,0.6559) (0.7823,0.6559) (0.7830,0.6552) (0.7846,0.6522)
(0.7855,0.6498) (0.7861,0.6493) (0.7887,0.6484) (0.7911,0.6434)
(0.7911,0.6463) (0.7919,0.6427) (0.7937,0.6375) (0.7937,0.6404)
(0.7951,0.6356) (0.7982,0.6355) (0.7989,0.6345) (0.8003,0.6316)
(0.8008,0.6286) (0.8014,0.6274) (0.8046,0.6274) (0.8055,0.6227)
(0.8055,0.6257) (0.8068,0.6168) (0.8068,0.6198) (0.8078,0.6141)
(0.8078,0.6141) (0.8081,0.6139) (0.8092,0.6109) (0.8110,0.6085)
(0.8111,0.6080) (0.8115,0.6050) (0.8115,0.6021) (0.8119,0.5991)
(0.8118,0.5962) (0.8122,0.5932) (0.8120,0.5903) (0.8124,0.5873)
(0.8125,0.5844) (0.8125,0.5814) (0.8126,0.5785) (0.8125,0.5755)
(0.8125,0.5726) (0.8122,0.5696) (0.8124,0.5667) (0.8122,0.5637)
(0.8122,0.5608) (0.8118,0.5578) (0.8117,0.5549) (0.8114,0.5519)
(0.8110,0.5504) (0.8090,0.5490) (0.8094,0.5460) (0.8078,0.5452)
(0.8070,0.5431) (0.8070,0.5401) (0.8058,0.5372) (0.8058,0.5342)
(0.8046,0.5316) (0.8014,0.5316) (0.8013,0.5313) (0.8001,0.5283)
(0.8001,0.5254) (0.7991,0.5224) (0.7991,0.5196) (0.7992,0.5166)
(0.7992,0.5137) (0.7995,0.5107) (0.8005,0.5078) (0.8006,0.5048)
(0.8014,0.5030) (0.8046,0.5030) (0.8052,0.5019) (0.8052,0.4989)
(0.8066,0.4960) (0.8067,0.4930) (0.8078,0.4904) (0.8092,0.4901)
(0.8110,0.4896) (0.8120,0.4871) (0.8122,0.4842) (0.8135,0.4812)
(0.8135,0.4783) (0.8142,0.4765) (0.8144,0.4753) (0.8145,0.4724)
(0.8150,0.4694) (0.8154,0.4665) (0.8154,0.4635) (0.8157,0.4606)
(0.8157,0.4576) (0.8160,0.4547) (0.8161,0.4517) (0.8161,0.4488)
(0.8161,0.4458) (0.8161,0.4429) (0.8160,0.4399) (0.8160,0.4370)
(0.8158,0.4340) (0.8159,0.4311) (0.8155,0.4281) (0.8150,0.4252)
(0.8150,0.4222) (0.8143,0.4193) (0.8143,0.4163) (0.8142,0.4157)
(0.8133,0.4134) (0.8127,0.4104) (0.8120,0.4075) (0.8110,0.4055)
(0.8078,0.4048) (0.8077,0.4045) (0.8077,0.4016) (0.8061,0.3986)
(0.8061,0.3957) (0.8046,0.3932) (0.8014,0.3932) (0.8011,0.3927)
(0.8004,0.3898) (0.7993,0.3868) (0.7982,0.3854) (0.7951,0.3854)
(0.7939,0.3839) (0.7939,0.3809) (0.7919,0.3787) (0.7910,0.3780)
(0.7909,0.3750) (0.7887,0.3734) (0.7855,0.3724) (0.7851,0.3721)
(0.7835,0.3691) (0.7823,0.3669) (0.7791,0.3668) (0.7788,0.3662)
(0.7759,0.3639) (0.7728,0.3639) (0.7719,0.3632) (0.7719,0.3603)
(0.7696,0.3586) (0.7664,0.3577) (0.7654,0.3573) (0.7652,0.3544)
(0.7632,0.3535) (0.7600,0.3516) (0.7568,0.3516) (0.7566,0.3514)
(0.7537,0.3493) (0.7505,0.3493) (0.7493,0.3485) (0.7473,0.3461)
(0.7441,0.3455) (0.7441,0.3455) (0.7409,0.3445) (0.7377,0.3434)
(0.7345,0.3434) (0.7320,0.3426) (0.7320,0.3396) (0.7314,0.3393)
(0.7282,0.3393) (0.7250,0.3378) (0.7218,0.3377) (0.7191,0.3367)
(0.7191,0.3337) (0.7186,0.3336) (0.7154,0.3328) (0.7123,0.3327)
(0.7091,0.3320) (0.7059,0.3320) (0.7027,0.3314) (0.6995,0.3314)
(0.6963,0.3309) (0.6952,0.3308) (0.6931,0.3306) (0.6900,0.3306)
(0.6868,0.3300) (0.6836,0.3300) (0.6804,0.3291) (0.6772,0.3290)
(0.6741,0.3278) (0.6740,0.3278) (0.6715,0.3249) (0.6709,0.3246)
(0.6677,0.3243) (0.6645,0.3227) (0.6613,0.3227) (0.6594,0.3219)
(0.6594,0.3190) (0.6581,0.3184) (0.6549,0.3184) (0.6517,0.3170)
(0.6486,0.3162) (0.6479,0.3160) (0.6473,0.3131) (0.6454,0.3125)
(0.6422,0.3109) (0.6390,0.3109) (0.6372,0.3101) (0.6358,0.3094)
(0.6326,0.3094) (0.6295,0.3075) (0.6284,0.3072) (0.6263,0.3055)
(0.6249,0.3042) (0.6231,0.3036) (0.6199,0.3021) (0.6167,0.3021)
(0.6153,0.3013) (0.6153,0.2983) (0.6135,0.2971) (0.6103,0.2971)
(0.6082,0.2954) (0.6082,0.2924) (0.6072,0.2915) (0.6040,0.2913)
(0.6020,0.2895) (0.6008,0.2883) (0.5993,0.2865) (0.5990,0.2836)
(0.5976,0.2822) (0.5944,0.2821) (0.5930,0.2806) (0.5930,0.2777)
(0.5912,0.2761) (0.5881,0.2761) (0.5866,0.2747) (0.5865,0.2718)
(0.5849,0.2703) (0.5817,0.2702) (0.5799,0.2688) (0.5785,0.2677)
(0.5761,0.2659) (0.5757,0.2629) (0.5753,0.2627) (0.5721,0.2626)
(0.5689,0.2606) (0.5658,0.2606) (0.5647,0.2600) (0.5647,0.2570)
(0.5626,0.2559) (0.5594,0.2559) (0.5562,0.2545) (0.5550,0.2541)
(0.5543,0.2512) (0.5530,0.2507) (0.5498,0.2503) (0.5466,0.2491)
(0.5435,0.2491) (0.5407,0.2482) (0.5403,0.2481) (0.5371,0.2481)
(0.5339,0.2472) (0.5307,0.2465) (0.5275,0.2465) (0.5244,0.2459)
(0.5212,0.2459) (0.5180,0.2455) (0.5148,0.2455) (0.5116,0.2453)
(0.5113,0.2453) (0.5084,0.2432) (0.5053,0.2453) (0.5052,0.2453)
(0.5045,0.2453) (0.5021,0.2447) (0.5000,0.2453)

\PST@lineone(0.4984,0.4356) (0.4972,0.4349) (0.4972,0.4321) (0.4951,0.4311)
(0.4917,0.4308) (0.4884,0.4300) (0.4854,0.4293) (0.4854,0.4266)
(0.4851,0.4265) (0.4818,0.4265) (0.4784,0.4263) (0.4751,0.4264)
(0.4729,0.4266) (0.4726,0.4293) (0.4718,0.4294) (0.4685,0.4306)
(0.4651,0.4306) (0.4619,0.4321) (0.4619,0.4349) (0.4618,0.4350)
(0.4585,0.4350) (0.4552,0.4375) (0.4549,0.4377) (0.4548,0.4405)
(0.4521,0.4433) (0.4518,0.4448) (0.4495,0.4461) (0.4485,0.4462)
(0.4466,0.4489) (0.4452,0.4514) (0.4420,0.4517) (0.4419,0.4518)
(0.4413,0.4545) (0.4396,0.4573) (0.4396,0.4601) (0.4385,0.4620)
(0.4380,0.4629) (0.4380,0.4657) (0.4364,0.4685) (0.4364,0.4713)
(0.4352,0.4736) (0.4319,0.4739) (0.4318,0.4741) (0.4317,0.4769)
(0.4302,0.4797) (0.4294,0.4824) (0.4286,0.4852) (0.4285,0.4853)
(0.4252,0.4853) (0.4241,0.4880) (0.4240,0.4908) (0.4226,0.4936)
(0.4226,0.4964) (0.4219,0.4977) (0.4204,0.4992) (0.4203,0.5020)
(0.4195,0.5048) (0.4195,0.5076) (0.4193,0.5104) (0.4191,0.5132)
(0.4200,0.5160) (0.4213,0.5188) (0.4215,0.5215) (0.4219,0.5220)
(0.4230,0.5243) (0.4230,0.5271) (0.4241,0.5299) (0.4241,0.5326)
(0.4252,0.5354) (0.4285,0.5354) (0.4286,0.5354) (0.4286,0.5382)
(0.4297,0.5410) (0.4298,0.5438) (0.4310,0.5466) (0.4319,0.5486)
(0.4340,0.5494) (0.4352,0.5502) (0.4359,0.5522) (0.4380,0.5550)
(0.4380,0.5578) (0.4385,0.5584) (0.4408,0.5606) (0.4408,0.5634)
(0.4419,0.5644) (0.4452,0.5644) (0.4478,0.5662) (0.4478,0.5690)
(0.4485,0.5695) (0.4518,0.5695) (0.4552,0.5708) (0.4585,0.5712)
(0.4618,0.5712) (0.4651,0.5705) (0.4685,0.5705) (0.4718,0.5701)
(0.4745,0.5690) (0.4746,0.5662) (0.4751,0.5660) (0.4784,0.5660)
(0.4818,0.5651) (0.4851,0.5651) (0.4884,0.5643) (0.4917,0.5641)
(0.4951,0.5640) (0.4984,0.5636) 

\PST@linetwo(0.4984,0.3576) (0.4951,0.3580) (0.4917,0.3582) (0.4884,0.3583)
(0.4851,0.3585) (0.4818,0.3585) (0.4784,0.3588) (0.4751,0.3588)
(0.4718,0.3589) (0.4685,0.3590) (0.4651,0.3590) (0.4618,0.3592)
(0.4585,0.3593) (0.4559,0.3595) (0.4557,0.3623) (0.4552,0.3623)
(0.4518,0.3629) (0.4485,0.3629) (0.4452,0.3638) (0.4419,0.3638)
(0.4385,0.3646) (0.4368,0.3651) (0.4368,0.3679) (0.4352,0.3683)
(0.4319,0.3684) (0.4285,0.3693) (0.4252,0.3693) (0.4219,0.3702)
(0.4199,0.3707) (0.4196,0.3735) (0.4186,0.3737) (0.4152,0.3739)
(0.4119,0.3748) (0.4086,0.3748) (0.4053,0.3758) (0.4019,0.3762)
(0.4019,0.3763) (0.4007,0.3790) (0.3986,0.3797) (0.3953,0.3812)
(0.3920,0.3812) (0.3908,0.3818) (0.3886,0.3840) (0.3870,0.3846)
(0.3853,0.3853) (0.3834,0.3874) (0.3820,0.3885) (0.3803,0.3902)
(0.3803,0.3930) (0.3787,0.3952) (0.3753,0.3952) (0.3750,0.3958)
(0.3750,0.3986) (0.3737,0.4014) (0.3737,0.4042) (0.3729,0.4070)
(0.3729,0.4098) (0.3728,0.4126) (0.3728,0.4154) (0.3727,0.4182)
(0.3727,0.4210) (0.3727,0.4238) (0.3725,0.4266) (0.3725,0.4293)
(0.3720,0.4319) (0.3699,0.4321) (0.3697,0.4349) (0.3687,0.4350)
(0.3672,0.4377) (0.3672,0.4405) (0.3654,0.4424) (0.3645,0.4433)
(0.3641,0.4461) (0.3620,0.4477) (0.3587,0.4478) (0.3570,0.4489)
(0.3554,0.4499) (0.3521,0.4499) (0.3488,0.4517) (0.3487,0.4517)
(0.3477,0.4545) (0.3454,0.4554) (0.3421,0.4555) (0.3388,0.4569)
(0.3354,0.4570) (0.3348,0.4573) (0.3348,0.4601) (0.3321,0.4616)
(0.3298,0.4629) (0.3298,0.4657) (0.3288,0.4662) (0.3255,0.4666)
(0.3231,0.4685) (0.3231,0.4713) (0.3221,0.4721) (0.3188,0.4721)
(0.3167,0.4741) (0.3167,0.4769) (0.3155,0.4782) (0.3131,0.4797)
(0.3122,0.4809) (0.3108,0.4824) (0.3088,0.4851) (0.3088,0.4852)
(0.3070,0.4880) (0.3069,0.4908) (0.3055,0.4934) (0.3022,0.4934)
(0.3020,0.4936) (0.3020,0.4964) (0.3007,0.4992) (0.3007,0.5020)
(0.2994,0.5048) (0.2994,0.5076) (0.2989,0.5089) (0.2955,0.5103)
(0.2955,0.5104) (0.2955,0.5125) (0.2956,0.5132) (0.2955,0.5132)
(0.2942,0.5160) (0.2933,0.5188) (0.2931,0.5215) (0.2923,0.5243)
(0.2923,0.5271) (0.2922,0.5284) (0.2921,0.5299) (0.2921,0.5326)
(0.2921,0.5354) (0.2921,0.5382) (0.2922,0.5390) (0.2927,0.5410)
(0.2929,0.5438) (0.2940,0.5466) (0.2955,0.5487) (0.2973,0.5494)
(0.2989,0.5519) (0.2989,0.5522) (0.3020,0.5550) (0.3020,0.5578)
(0.3022,0.5579) (0.3055,0.5579) (0.3088,0.5599) (0.3112,0.5606)
(0.3114,0.5634) (0.3122,0.5636) (0.3155,0.5645) (0.3188,0.5662)
(0.3221,0.5662) (0.3222,0.5662) (0.3222,0.5690) (0.3255,0.5703)
(0.3288,0.5704) (0.3321,0.5717) (0.3323,0.5718) (0.3324,0.5746)
(0.3354,0.5768) (0.3388,0.5768) (0.3393,0.5774) (0.3404,0.5802)
(0.3414,0.5829) (0.3421,0.5841) (0.3454,0.5845) (0.3460,0.5857)
(0.3462,0.5885) (0.3473,0.5913) (0.3473,0.5941) (0.3483,0.5969)
(0.3483,0.5997) (0.3487,0.6007) (0.3497,0.6025) (0.3497,0.6053)
(0.3511,0.6081) (0.3516,0.6109) (0.3521,0.6121) (0.3554,0.6124)
(0.3560,0.6137) (0.3577,0.6165) (0.3578,0.6193) (0.3587,0.6206)
(0.3620,0.6206) (0.3631,0.6221) (0.3631,0.6249) (0.3654,0.6276)
(0.3654,0.6277) (0.3654,0.6305) (0.3683,0.6332) (0.3684,0.6360)
(0.3687,0.6363) (0.3720,0.6364) (0.3746,0.6388) (0.3748,0.6416)
(0.3753,0.6421) (0.3787,0.6421) (0.3817,0.6444) (0.3820,0.6447)
(0.3852,0.6472) (0.3853,0.6475) (0.3886,0.6487) (0.3892,0.6500)
(0.3920,0.6519) (0.3953,0.6519) (0.3964,0.6528) (0.3964,0.6556)
(0.3986,0.6573) (0.4008,0.6584) (0.4009,0.6612) (0.4019,0.6616)
(0.4053,0.6626) (0.4076,0.6640) (0.4076,0.6668) (0.4086,0.6673)
(0.4119,0.6673) (0.4152,0.6690) (0.4186,0.6694) (0.4189,0.6696)
(0.4193,0.6724) (0.4219,0.6736) (0.4248,0.6752) (0.4252,0.6757)
(0.4285,0.6757) (0.4306,0.6780) (0.4319,0.6790) (0.4352,0.6792)
(0.4380,0.6808) (0.4385,0.6809) (0.4419,0.6817) (0.4452,0.6817)
(0.4485,0.6823) (0.4518,0.6824) (0.4552,0.6828) (0.4585,0.6830)
(0.4618,0.6829) (0.4651,0.6828) (0.4685,0.6828) (0.4718,0.6826)
(0.4751,0.6823) (0.4784,0.6822) (0.4818,0.6817) (0.4851,0.6817)
(0.4884,0.6812) (0.4917,0.6809) (0.4945,0.6808) (0.4951,0.6807)
(0.4984,0.6802) 

\PST@linethr(0.4984,0.3219) (0.4951,0.3220) (0.4917,0.3222) (0.4884,0.3220)
(0.4851,0.3220) (0.4818,0.3220) (0.4784,0.3218) (0.4751,0.3219)
(0.4718,0.3216) (0.4685,0.3214) (0.4651,0.3213) (0.4618,0.3212)
(0.4585,0.3213) (0.4552,0.3213) (0.4518,0.3214) (0.4485,0.3214)
(0.4452,0.3217) (0.4419,0.3217) (0.4385,0.3222) (0.4352,0.3230)
(0.4319,0.3229) (0.4307,0.3232) (0.4302,0.3260) (0.4285,0.3264)
(0.4252,0.3264) (0.4219,0.3274) (0.4186,0.3283) (0.4152,0.3285)
(0.4145,0.3287) (0.4145,0.3315) (0.4119,0.3324) (0.4086,0.3324)
(0.4053,0.3337) (0.4019,0.3342) (0.4011,0.3343) (0.4009,0.3371)
(0.3986,0.3377) (0.3953,0.3389) (0.3920,0.3389) (0.3892,0.3399)
(0.3892,0.3427) (0.3886,0.3429) (0.3853,0.3430) (0.3820,0.3443)
(0.3790,0.3455) (0.3787,0.3467) (0.3753,0.3470) (0.3751,0.3483)
(0.3720,0.3499) (0.3687,0.3499) (0.3666,0.3511) (0.3654,0.3519)
(0.3623,0.3539) (0.3620,0.3549) (0.3587,0.3551) (0.3583,0.3567)
(0.3554,0.3587) (0.3521,0.3588) (0.3510,0.3595) (0.3510,0.3623)
(0.3487,0.3641) (0.3476,0.3651) (0.3475,0.3679) (0.3454,0.3698)
(0.3421,0.3699) (0.3413,0.3707) (0.3413,0.3735) (0.3388,0.3763)
(0.3388,0.3785) (0.3375,0.3790) (0.3354,0.3791) (0.3334,0.3818)
(0.3323,0.3846) (0.3321,0.3853) (0.3315,0.3874) (0.3297,0.3902)
(0.3298,0.3930) (0.3288,0.3950) (0.3261,0.3958) (0.3261,0.3986)
(0.3255,0.3989) (0.3246,0.4014) (0.3246,0.4042) (0.3239,0.4070)
(0.3239,0.4098) (0.3231,0.4126) (0.3229,0.4154) (0.3224,0.4182)
(0.3221,0.4189) (0.3188,0.4189) (0.3181,0.4210) (0.3180,0.4238)
(0.3165,0.4266) (0.3165,0.4293) (0.3155,0.4307) (0.3135,0.4321)
(0.3134,0.4349) (0.3122,0.4357) (0.3088,0.4376) (0.3087,0.4377)
(0.3087,0.4405) (0.3060,0.4433) (0.3057,0.4461) (0.3055,0.4463)
(0.3022,0.4464) (0.2998,0.4489) (0.2989,0.4501) (0.2957,0.4517)
(0.2955,0.4519) (0.2942,0.4545) (0.2922,0.4564) (0.2914,0.4573)
(0.2914,0.4601) (0.2892,0.4629) (0.2892,0.4657) (0.2889,0.4661)
(0.2855,0.4661) (0.2837,0.4685) (0.2837,0.4713) (0.2822,0.4732)
(0.2789,0.4733) (0.2783,0.4741) (0.2783,0.4769) (0.2761,0.4797)
(0.2756,0.4809) (0.2750,0.4824) (0.2740,0.4852) (0.2722,0.4877)
(0.2689,0.4877) (0.2687,0.4880) (0.2686,0.4908) (0.2666,0.4936)
(0.2666,0.4964) (0.2656,0.4977) (0.2623,0.4978) (0.2612,0.4992)
(0.2612,0.5020) (0.2593,0.5048) (0.2593,0.5076) (0.2589,0.5082)
(0.2576,0.5104) (0.2570,0.5132) (0.2560,0.5160) (0.2556,0.5167)
(0.2523,0.5169) (0.2514,0.5188) (0.2513,0.5215) (0.2502,0.5243)
(0.2502,0.5271) (0.2492,0.5299) (0.2492,0.5326) (0.2490,0.5335)
(0.2456,0.5335) (0.2452,0.5354) (0.2452,0.5382) (0.2448,0.5410)
(0.2449,0.5438) (0.2448,0.5466) (0.2450,0.5494) (0.2449,0.5522)
(0.2455,0.5550) (0.2455,0.5578) (0.2456,0.5583) (0.2490,0.5583)
(0.2497,0.5606) (0.2497,0.5634) (0.2510,0.5662) (0.2510,0.5690)
(0.2523,0.5711) (0.2556,0.5712) (0.2560,0.5718) (0.2560,0.5746)
(0.2584,0.5774) (0.2589,0.5786) (0.2598,0.5802) (0.2614,0.5829)
(0.2623,0.5836) (0.2656,0.5837) (0.2685,0.5857) (0.2686,0.5885)
(0.2689,0.5887) (0.2722,0.5887) (0.2756,0.5906) (0.2771,0.5913)
(0.2771,0.5941) (0.2789,0.5949) (0.2822,0.5950) (0.2855,0.5963)
(0.2889,0.5963) (0.2915,0.5969) (0.2915,0.5997) (0.2922,0.5999)
(0.2955,0.6007) (0.2989,0.6009) (0.3022,0.6024) (0.3055,0.6024)
(0.3056,0.6025) (0.3056,0.6053) (0.3088,0.6075) (0.3102,0.6081)
(0.3117,0.6109) (0.3122,0.6111) (0.3149,0.6137) (0.3155,0.6141)
(0.3168,0.6165) (0.3169,0.6193) (0.3184,0.6221) (0.3184,0.6249)
(0.3188,0.6256) (0.3221,0.6256) (0.3233,0.6277) (0.3233,0.6305)
(0.3252,0.6332) (0.3252,0.6360) (0.3255,0.6364) (0.3288,0.6370)
(0.3304,0.6388) (0.3305,0.6416) (0.3321,0.6435) (0.3329,0.6444)
(0.3347,0.6472) (0.3352,0.6500) (0.3354,0.6502) (0.3388,0.6502)
(0.3412,0.6528) (0.3412,0.6556) (0.3421,0.6564) (0.3454,0.6565)
(0.3476,0.6584) (0.3476,0.6612) (0.3487,0.6621) (0.3511,0.6640)
(0.3511,0.6668) (0.3521,0.6675) (0.3554,0.6675) (0.3581,0.6696)
(0.3581,0.6724) (0.3587,0.6728) (0.3620,0.6728) (0.3654,0.6749)
(0.3659,0.6752) (0.3683,0.6780) (0.3687,0.6784) (0.3720,0.6789)
(0.3737,0.6808) (0.3753,0.6819) (0.3787,0.6819) (0.3812,0.6835)
(0.3813,0.6863) (0.3820,0.6868) (0.3853,0.6886) (0.3886,0.6888)
(0.3892,0.6891) (0.3892,0.6919) (0.3920,0.6935) (0.3953,0.6935)
(0.3973,0.6947) (0.3973,0.6975) (0.3986,0.6983) (0.4019,0.6994)
(0.4053,0.7002) (0.4054,0.7003) (0.4054,0.7031) (0.4086,0.7048)
(0.4119,0.7048) (0.4142,0.7059) (0.4151,0.7087) (0.4152,0.7088)
(0.4186,0.7089) (0.4219,0.7105) (0.4241,0.7115) (0.4252,0.7119)
(0.4285,0.7119) (0.4319,0.7131) (0.4352,0.7134) (0.4385,0.7141)
(0.4391,0.7143) (0.4403,0.7171) (0.4419,0.7174) (0.4452,0.7174)
(0.4485,0.7181) (0.4518,0.7181) (0.4552,0.7186) (0.4585,0.7191)
(0.4618,0.7190) (0.4651,0.7194) (0.4685,0.7194) (0.4718,0.7197)
(0.4732,0.7199) (0.4737,0.7227) (0.4751,0.7228) (0.4784,0.7228)
(0.4818,0.7231) (0.4851,0.7231) (0.4884,0.7233) (0.4917,0.7235)
(0.4951,0.7236) (0.4984,0.7240)

\PST@linefou(0.4984,0.2965) (0.4951,0.2965) (0.4917,0.2966) (0.4884,0.2965)
(0.4851,0.2964) (0.4818,0.2964) (0.4784,0.2962) (0.4751,0.2962)
(0.4718,0.2960) (0.4685,0.2958) (0.4651,0.2958) (0.4618,0.2956)
(0.4585,0.2956) (0.4552,0.2956) (0.4518,0.2956) (0.4485,0.2956)
(0.4452,0.2958) (0.4419,0.2958) (0.4385,0.2961) (0.4352,0.2965)
(0.4319,0.2965) (0.4285,0.2972) (0.4252,0.2972) (0.4225,0.2980)
(0.4225,0.3008) (0.4219,0.3010) (0.4186,0.3017) (0.4152,0.3019)
(0.4119,0.3029) (0.4086,0.3029) (0.4070,0.3036) (0.4070,0.3064)
(0.4053,0.3071) (0.4019,0.3075) (0.3986,0.3083) (0.3966,0.3092)
(0.3966,0.3120) (0.3953,0.3125) (0.3920,0.3125) (0.3886,0.3137)
(0.3853,0.3139) (0.3829,0.3148) (0.3820,0.3158) (0.3802,0.3176)
(0.3787,0.3183) (0.3753,0.3184) (0.3720,0.3203) (0.3687,0.3203)
(0.3686,0.3204) (0.3654,0.3216) (0.3620,0.3231) (0.3587,0.3231)
(0.3586,0.3232) (0.3584,0.3260) (0.3554,0.3273) (0.3521,0.3273)
(0.3494,0.3287) (0.3494,0.3315) (0.3487,0.3319) (0.3454,0.3336)
(0.3421,0.3337) (0.3410,0.3343) (0.3410,0.3371) (0.3388,0.3386)
(0.3354,0.3386) (0.3335,0.3399) (0.3335,0.3427) (0.3321,0.3437)
(0.3294,0.3455) (0.3288,0.3468) (0.3272,0.3483) (0.3255,0.3489)
(0.3230,0.3511) (0.3221,0.3520) (0.3188,0.3520) (0.3171,0.3539)
(0.3167,0.3567) (0.3155,0.3578) (0.3126,0.3595) (0.3124,0.3623)
(0.3122,0.3624) (0.3088,0.3643) (0.3082,0.3651) (0.3082,0.3679)
(0.3060,0.3707) (0.3060,0.3735) (0.3055,0.3743) (0.3022,0.3743)
(0.3008,0.3763) (0.3008,0.3790) (0.2992,0.3818) (0.2989,0.3833)
(0.2978,0.3846) (0.2955,0.3867) (0.2952,0.3874) (0.2940,0.3902)
(0.2938,0.3930) (0.2928,0.3958) (0.2928,0.3986) (0.2922,0.4003)
(0.2919,0.4014) (0.2919,0.4042) (0.2908,0.4070) (0.2908,0.4098)
(0.2897,0.4126) (0.2893,0.4154) (0.2889,0.4170) (0.2855,0.4172)
(0.2852,0.4182) (0.2836,0.4210) (0.2835,0.4238) (0.2822,0.4262)
(0.2789,0.4266) (0.2788,0.4293) (0.2769,0.4321) (0.2769,0.4349)
(0.2756,0.4367) (0.2748,0.4377) (0.2748,0.4405) (0.2728,0.4433)
(0.2725,0.4461) (0.2722,0.4465) (0.2689,0.4467) (0.2672,0.4489)
(0.2656,0.4513) (0.2627,0.4517) (0.2623,0.4519) (0.2616,0.4545)
(0.2595,0.4573) (0.2595,0.4601) (0.2589,0.4608) (0.2572,0.4629)
(0.2572,0.4657) (0.2556,0.4676) (0.2523,0.4677) (0.2516,0.4685)
(0.2516,0.4713) (0.2495,0.4741) (0.2495,0.4769) (0.2490,0.4775)
(0.2456,0.4775) (0.2439,0.4797) (0.2426,0.4824) (0.2423,0.4832)
(0.2415,0.4852) (0.2392,0.4880) (0.2392,0.4908) (0.2390,0.4911)
(0.2357,0.4916) (0.2340,0.4936) (0.2340,0.4964) (0.2323,0.4987)
(0.2290,0.4987) (0.2287,0.4992) (0.2287,0.5020) (0.2271,0.5048)
(0.2271,0.5076) (0.2257,0.5102) (0.2254,0.5104) (0.2247,0.5132)
(0.2229,0.5160) (0.2224,0.5168) (0.2199,0.5188) (0.2197,0.5215)
(0.2190,0.5220) (0.2180,0.5243) (0.2180,0.5271) (0.2173,0.5299)
(0.2173,0.5326) (0.2167,0.5354) (0.2167,0.5382) (0.2163,0.5410)
(0.2163,0.5438) (0.2163,0.5466) (0.2162,0.5494) (0.2163,0.5522)
(0.2166,0.5550) (0.2166,0.5578) (0.2171,0.5606) (0.2171,0.5634)
(0.2178,0.5662) (0.2178,0.5690) (0.2189,0.5718) (0.2189,0.5746)
(0.2190,0.5748) (0.2222,0.5774) (0.2224,0.5778) (0.2231,0.5802)
(0.2247,0.5829) (0.2257,0.5840) (0.2269,0.5857) (0.2269,0.5885)
(0.2290,0.5911) (0.2323,0.5911) (0.2325,0.5913) (0.2325,0.5941)
(0.2352,0.5969) (0.2352,0.5997) (0.2357,0.6002) (0.2390,0.6005)
(0.2416,0.6025) (0.2416,0.6053) (0.2423,0.6058) (0.2455,0.6081)
(0.2456,0.6085) (0.2490,0.6087) (0.2495,0.6109) (0.2523,0.6129)
(0.2556,0.6130) (0.2567,0.6137) (0.2589,0.6150) (0.2621,0.6165)
(0.2623,0.6171) (0.2656,0.6170) (0.2660,0.6193) (0.2689,0.6203)
(0.2722,0.6203) (0.2756,0.6215) (0.2771,0.6221) (0.2772,0.6249)
(0.2789,0.6256) (0.2822,0.6256) (0.2855,0.6273) (0.2889,0.6273)
(0.2894,0.6277) (0.2894,0.6305) (0.2922,0.6327) (0.2928,0.6332)
(0.2928,0.6360) (0.2955,0.6388) (0.2955,0.6389) (0.2974,0.6416)
(0.2989,0.6423) (0.3002,0.6444) (0.3016,0.6472) (0.3021,0.6500)
(0.3022,0.6501) (0.3055,0.6501) (0.3075,0.6528) (0.3075,0.6556)
(0.3088,0.6571) (0.3110,0.6584) (0.3111,0.6612) (0.3122,0.6618)
(0.3155,0.6638) (0.3156,0.6640) (0.3156,0.6668) (0.3180,0.6696)
(0.3181,0.6724) (0.3188,0.6731) (0.3221,0.6731) (0.3243,0.6752)
(0.3255,0.6779) (0.3255,0.6779) (0.3256,0.6780) (0.3288,0.6790)
(0.3304,0.6808) (0.3321,0.6820) (0.3339,0.6835) (0.3340,0.6863)
(0.3354,0.6874) (0.3388,0.6874) (0.3413,0.6891) (0.3413,0.6919)
(0.3421,0.6925) (0.3454,0.6925) (0.3487,0.6946) (0.3489,0.6947)
(0.3489,0.6975) (0.3521,0.6997) (0.3554,0.6997) (0.3563,0.7003)
(0.3563,0.7031) (0.3587,0.7047) (0.3620,0.7047) (0.3641,0.7059)
(0.3647,0.7087) (0.3654,0.7091) (0.3687,0.7112) (0.3720,0.7113)
(0.3723,0.7115) (0.3753,0.7136) (0.3787,0.7137) (0.3795,0.7143)
(0.3800,0.7171) (0.3820,0.7184) (0.3849,0.7199) (0.3850,0.7227)
(0.3853,0.7228) (0.3886,0.7230) (0.3920,0.7249) (0.3953,0.7249)
(0.3963,0.7255) (0.3963,0.7283) (0.3986,0.7296) (0.4019,0.7304)
(0.4053,0.7310) (0.4054,0.7311) (0.4054,0.7338) (0.4086,0.7353)
(0.4119,0.7353) (0.4152,0.7365) (0.4173,0.7366) (0.4186,0.7375)
(0.4192,0.7394) (0.4219,0.7405) (0.4252,0.7415) (0.4285,0.7415)
(0.4314,0.7422) (0.4319,0.7424) (0.4352,0.7425) (0.4385,0.7437)
(0.4419,0.7448) (0.4452,0.7448) (0.4460,0.7450) (0.4485,0.7465)
(0.4518,0.7467) (0.4543,0.7478) (0.4552,0.7480) (0.4585,0.7486)
(0.4618,0.7486) (0.4651,0.7492) (0.4685,0.7492) (0.4718,0.7496)
(0.4751,0.7501) (0.4784,0.7500) (0.4818,0.7504) (0.4851,0.7504)
(0.4865,0.7506) (0.4866,0.7534) (0.4884,0.7537) (0.4917,0.7541)
(0.4951,0.7542) (0.4984,0.7547)

\PST@linefiv (0.4984,0.2720) (0.4951,0.2720) (0.4917,0.2720) (0.4884,0.2720)
(0.4851,0.2719) (0.4818,0.2719) (0.4784,0.2718) (0.4751,0.2717)
(0.4718,0.2716) (0.4685,0.2715) (0.4651,0.2715) (0.4618,0.2715)
(0.4585,0.2715) (0.4552,0.2716) (0.4518,0.2716) (0.4485,0.2716)
(0.4452,0.2716) (0.4419,0.2716) (0.4385,0.2717) (0.4352,0.2718)
(0.4319,0.2718) (0.4285,0.2721) (0.4252,0.2721) (0.4219,0.2724)
(0.4186,0.2727) (0.4152,0.2728) (0.4148,0.2729) (0.4148,0.2757)
(0.4119,0.2765) (0.4086,0.2765) (0.4053,0.2776) (0.4019,0.2781)
(0.4006,0.2784) (0.3997,0.2812) (0.3986,0.2816) (0.3953,0.2830)
(0.3920,0.2830) (0.3897,0.2840) (0.3886,0.2847) (0.3853,0.2852)
(0.3824,0.2868) (0.3820,0.2872) (0.3801,0.2896) (0.3787,0.2902)
(0.3753,0.2902) (0.3720,0.2913) (0.3687,0.2914) (0.3655,0.2924)
(0.3654,0.2952) (0.3654,0.2952) (0.3620,0.2963) (0.3587,0.2963)
(0.3554,0.2974) (0.3521,0.2975) (0.3505,0.2980) (0.3505,0.3008)
(0.3487,0.3014) (0.3454,0.3028) (0.3421,0.3028) (0.3403,0.3036)
(0.3403,0.3064) (0.3388,0.3071) (0.3354,0.3071) (0.3321,0.3089)
(0.3316,0.3092) (0.3314,0.3120) (0.3288,0.3131) (0.3255,0.3135)
(0.3230,0.3148) (0.3221,0.3164) (0.3188,0.3165) (0.3183,0.3176)
(0.3155,0.3200) (0.3147,0.3204) (0.3122,0.3214) (0.3088,0.3225)
(0.3079,0.3232) (0.3078,0.3260) (0.3055,0.3277) (0.3022,0.3277)
(0.3008,0.3287) (0.3008,0.3315) (0.2989,0.3332) (0.2955,0.3337)
(0.2948,0.3343) (0.2947,0.3371) (0.2922,0.3392) (0.2915,0.3399)
(0.2915,0.3427) (0.2889,0.3455) (0.2889,0.3459) (0.2855,0.3461)
(0.2854,0.3483) (0.2831,0.3511) (0.2822,0.3523) (0.2789,0.3531)
(0.2785,0.3539) (0.2779,0.3567) (0.2763,0.3595) (0.2763,0.3623)
(0.2756,0.3637) (0.2748,0.3651) (0.2748,0.3679) (0.2733,0.3707)
(0.2733,0.3735) (0.2722,0.3757) (0.2689,0.3757) (0.2687,0.3763)
(0.2686,0.3790) (0.2674,0.3818) (0.2668,0.3846) (0.2664,0.3874)
(0.2656,0.3897) (0.2623,0.3900) (0.2622,0.3902) (0.2621,0.3930)
(0.2612,0.3958) (0.2612,0.3986) (0.2601,0.4014) (0.2601,0.4042)
(0.2589,0.4069) (0.2589,0.4070) (0.2589,0.4098) (0.2572,0.4126)
(0.2564,0.4154) (0.2556,0.4178) (0.2523,0.4180) (0.2522,0.4182)
(0.2511,0.4210) (0.2510,0.4238) (0.2496,0.4266) (0.2496,0.4293)
(0.2490,0.4304) (0.2456,0.4304) (0.2442,0.4321) (0.2442,0.4349)
(0.2423,0.4372) (0.2419,0.4377) (0.2419,0.4405) (0.2397,0.4433)
(0.2390,0.4461) (0.2389,0.4461) (0.2357,0.4473) (0.2347,0.4489)
(0.2332,0.4517) (0.2329,0.4545) (0.2323,0.4553) (0.2290,0.4553)
(0.2278,0.4573) (0.2278,0.4601) (0.2260,0.4629) (0.2260,0.4657)
(0.2257,0.4662) (0.2224,0.4680) (0.2210,0.4685) (0.2208,0.4713)
(0.2190,0.4719) (0.2164,0.4741) (0.2163,0.4769) (0.2157,0.4774)
(0.2124,0.4774) (0.2101,0.4797) (0.2091,0.4815) (0.2078,0.4824)
(0.2057,0.4838) (0.2046,0.4852) (0.2024,0.4873) (0.2018,0.4880)
(0.2018,0.4908) (0.1998,0.4936) (0.1998,0.4964) (0.1991,0.4976)
(0.1958,0.4976) (0.1948,0.4992) (0.1948,0.5020) (0.1933,0.5048)
(0.1933,0.5076) (0.1924,0.5095) (0.1891,0.5099) (0.1889,0.5104)
(0.1890,0.5132) (0.1877,0.5160) (0.1870,0.5188) (0.1868,0.5215)
(0.1861,0.5243) (0.1860,0.5271) (0.1858,0.5284) (0.1856,0.5299)
(0.1856,0.5326) (0.1853,0.5354) (0.1853,0.5382) (0.1851,0.5410)
(0.1851,0.5438) (0.1851,0.5466) (0.1850,0.5494) (0.1851,0.5522)
(0.1853,0.5550) (0.1853,0.5578) (0.1856,0.5606) (0.1856,0.5634)
(0.1858,0.5649) (0.1860,0.5662) (0.1860,0.5690) (0.1868,0.5718)
(0.1869,0.5746) (0.1877,0.5774) (0.1888,0.5802) (0.1890,0.5829)
(0.1891,0.5832) (0.1924,0.5837) (0.1933,0.5857) (0.1933,0.5885)
(0.1948,0.5913) (0.1948,0.5941) (0.1958,0.5957) (0.1991,0.5957)
(0.1997,0.5969) (0.1997,0.5997) (0.2015,0.6025) (0.2015,0.6053)
(0.2024,0.6065) (0.2041,0.6081) (0.2053,0.6109) (0.2057,0.6113)
(0.2091,0.6131) (0.2095,0.6137) (0.2121,0.6165) (0.2123,0.6193)
(0.2124,0.6193) (0.2157,0.6193) (0.2189,0.6221) (0.2189,0.6249)
(0.2190,0.6250) (0.2224,0.6259) (0.2257,0.6274) (0.2260,0.6277)
(0.2260,0.6305) (0.2290,0.6324) (0.2323,0.6324) (0.2337,0.6332)
(0.2337,0.6360) (0.2357,0.6372) (0.2390,0.6373) (0.2423,0.6388)
(0.2423,0.6390) (0.2426,0.6416) (0.2456,0.6430) (0.2490,0.6430)
(0.2523,0.6443) (0.2556,0.6443) (0.2560,0.6444) (0.2589,0.6458)
(0.2618,0.6472) (0.2623,0.6479) (0.2656,0.6480) (0.2669,0.6500)
(0.2689,0.6509) (0.2722,0.6509) (0.2745,0.6528) (0.2746,0.6556)
(0.2756,0.6566) (0.2770,0.6584) (0.2770,0.6612) (0.2788,0.6640)
(0.2788,0.6668) (0.2789,0.6669) (0.2822,0.6672) (0.2837,0.6696)
(0.2838,0.6724) (0.2855,0.6748) (0.2889,0.6748) (0.2891,0.6752)
(0.2899,0.6780) (0.2913,0.6808) (0.2922,0.6818) (0.2940,0.6835)
(0.2942,0.6863) (0.2955,0.6875) (0.2989,0.6881) (0.2998,0.6891)
(0.2998,0.6919) (0.3022,0.6946) (0.3055,0.6946) (0.3056,0.6947)
(0.3056,0.6975) (0.3088,0.7000) (0.3097,0.7003) (0.3098,0.7031)
(0.3122,0.7042) (0.3155,0.7058) (0.3155,0.7059) (0.3160,0.7087)
(0.3188,0.7113) (0.3221,0.7113) (0.3224,0.7115) (0.3255,0.7140)
(0.3263,0.7143) (0.3288,0.7166) (0.3290,0.7171) (0.3321,0.7190)
(0.3335,0.7199) (0.3335,0.7227) (0.3354,0.7238) (0.3388,0.7238)
(0.3414,0.7255) (0.3414,0.7283) (0.3421,0.7287) (0.3454,0.7287)
(0.3487,0.7307) (0.3494,0.7311) (0.3494,0.7338) (0.3521,0.7353)
(0.3554,0.7353) (0.3579,0.7366) (0.3581,0.7394) (0.3587,0.7398)
(0.3620,0.7399) (0.3654,0.7421) (0.3656,0.7422) (0.3687,0.7448)
(0.3720,0.7450) (0.3721,0.7450) (0.3735,0.7478) (0.3753,0.7488)
(0.3787,0.7488) (0.3820,0.7505) (0.3822,0.7506) (0.3823,0.7534)
(0.3853,0.7548) (0.3886,0.7550) (0.3913,0.7562) (0.3913,0.7590)
(0.3920,0.7593) (0.3953,0.7593) (0.3986,0.7607) (0.4019,0.7614)
(0.4045,0.7618) (0.4046,0.7646) (0.4053,0.7647) (0.4086,0.7658)
(0.4119,0.7658) (0.4152,0.7668) (0.4186,0.7670) (0.4204,0.7674)
(0.4208,0.7702) (0.4219,0.7705) (0.4252,0.7713) (0.4285,0.7713)
(0.4319,0.7721) (0.4352,0.7721) (0.4385,0.7728) (0.4396,0.7730)
(0.4419,0.7743) (0.4452,0.7744) (0.4475,0.7758) (0.4485,0.7761)
(0.4518,0.7761) (0.4552,0.7771) (0.4585,0.7780) (0.4618,0.7780)
(0.4639,0.7786) (0.4651,0.7788) (0.4685,0.7788) (0.4718,0.7793)
(0.4751,0.7799) (0.4784,0.7799) (0.4818,0.7804) (0.4851,0.7805)
(0.4884,0.7810) (0.4902,0.7814) (0.4917,0.7835) (0.4936,0.7841)
(0.4951,0.7842) (0.4984,0.7848)

\PST@Border(0.1634,0.8299) (0.1634,0.2120) (0.8365,0.2120) (0.8365,0.8299)
(0.1634,0.8299)

\PST@Border(0.2196,0.2121) (0.2196,0.2381) \rput(0.2196,0.1775){-10}
\PST@Border(0.3598,0.2121) (0.3598,0.2381) \rput(0.3598,0.1775){-5}
\PST@Border(0.3598,0.8299) (0.3598,0.8039) \PST@Border(0.5000,0.2121)
(0.5000,0.2381) \rput(0.5000,0.1775){ 0} \PST@Border(0.5000,0.8299)
(0.5000,0.8039) \PST@Border(0.6402,0.2121) (0.6402,0.2381)
\rput(0.6402,0.1775){ 5} \PST@Border(0.6402,0.8299) (0.6402,0.8039)
\PST@Border(0.7804,0.2121) (0.7804,0.2381) \rput(0.7804,0.1775){ 10}
\rput(0.8365,0.1775){ 12} \rput(0.5000,0.1445){x [fm]}
\PST@Border(0.1635,0.2636) (0.1830,0.2636) \rput[r](0.1427,0.2636){-10}
\PST@Border(0.8365,0.2636) (0.8170,0.2636) \PST@Border(0.1635,0.3923)
(0.1830,0.3923) \rput[r](0.1427,0.3923){-5} \PST@Border(0.8365,0.3923)
(0.8170,0.3923) \PST@Border(0.1635,0.5210) (0.1830,0.5210)
\rput[r](0.1427,0.5210){ 0} \PST@Border(0.8365,0.5210) (0.8170,0.5210)
\PST@Border(0.1635,0.6497) (0.1830,0.6497) \rput[r](0.1427,0.6497){ 5}
\PST@Border(0.8365,0.6497) (0.8170,0.6497) \PST@Border(0.1635,0.7784)
(0.1830,0.7784) \rput[r](0.1427,0.7784){ 10} \PST@Border(0.8365,0.7784)
(0.8170,0.7784) \rput[r](0.1427,0.8299){ 12} \rput{L}(0.1067,0.5210){y [fm]}
\PST@Border(0.1634,0.8299) (0.1634,0.2120) (0.8365,0.2120) (0.8365,0.8299)
(0.1634,0.8299)

\catcode`@=12
\fi
\endpspicture

%% file: sigma.tex
\ifx\PSTloaded\undefined
\def\PSTloaded{t}
\psset{arrowsize=.01 3.2 1.4 .3}
\psset{dotsize=.01}
\catcode`@=11

\definecolor{darkgreen}{rgb}{0,0.5,0}
\definecolor{violett}{rgb}{.5,0,.5}
\definecolor{orange}{rgb}{.8,.4,0}
\newpsobject{PST@Border}{psline}{linewidth=.0015,linestyle=solid}
\newpsobject{PST@Axes}{psline}{linewidth=.0015,linestyle=dotted,dotsep=.004}
\newpsobject{PST@lineone}{psline}{linewidth=.0015,linestyle=solid,linecolor=black}
\newpsobject{PST@linetwo}{psline}{linewidth=.0025,linestyle=dotted,dotsep=.004,linecolor=red}
\newpsobject{PST@linethr}{psline}{linewidth=.0015,linestyle=dashed,dash=.02 .006,linecolor=violett}
\newpsobject{PST@linefou}{psline}{linewidth=.0035,linestyle=dotted,dotsep=.008,linecolor=darkgreen}
\newpsobject{PST@linefiv}{psline}{linewidth=.0015,linestyle=dashed,dash=.02 .004 .002 .004,linecolor=blue}
\newpsobject{PST@linesix}{psline}{linewidth=.0015,linestyle=dashed,dash=.02 .02,linecolor=darkgreen}
\newpsobject{PST@linesev}{psline}{linewidth=.0015,linestyle=dashed,dash=.01 .01,linecolor=orange}
\catcode`@=12

\fi
\psset{unit=5.0in,xunit=.66\textwidth,yunit=\columnwidth}
\pspicture(0.088000,0.000000)(0.960000,1.000000)
\ifx\nofigs\undefined
\catcode`@=11

\PST@Border(0.1840,0.0840) (0.1990,0.0840)
\PST@Border(0.9600,0.0840) (0.9450,0.0840)
\rput[r](0.1680,0.0840){$10^{-4}$}

\PST@Border(0.1840,0.1257) (0.1915,0.1257)
\PST@Border(0.9600,0.1257) (0.9525,0.1257)

\PST@Border(0.1840,0.1809) (0.1915,0.1809)
\PST@Border(0.9600,0.1809) (0.9525,0.1809)

\PST@Border(0.1840,0.2092) (0.1915,0.2092)
\PST@Border(0.9600,0.2092) (0.9525,0.2092)

\PST@Border(0.1840,0.2227) (0.1990,0.2227)
\PST@Border(0.9600,0.2227) (0.9450,0.2227)
\rput[r](0.1680,0.2227){$10^{-3}$}

\PST@Border(0.1840,0.2644) (0.1915,0.2644)
\PST@Border(0.9600,0.2644) (0.9525,0.2644)

\PST@Border(0.1840,0.3196) (0.1915,0.3196)
\PST@Border(0.9600,0.3196) (0.9525,0.3196)

\PST@Border(0.1840,0.3479) (0.1915,0.3479)
\PST@Border(0.9600,0.3479) (0.9525,0.3479)

\PST@Border(0.1840,0.3613) (0.1990,0.3613)
\PST@Border(0.9600,0.3613) (0.9450,0.3613)
\rput[r](0.1680,0.3613){ 0.01}

\PST@Border(0.1840,0.4031) (0.1915,0.4031)
\PST@Border(0.9600,0.4031) (0.9525,0.4031)

\PST@Border(0.1840,0.4583) (0.1915,0.4583)
\PST@Border(0.9600,0.4583) (0.9525,0.4583)

\PST@Border(0.1840,0.4866) (0.1915,0.4866)
\PST@Border(0.9600,0.4866) (0.9525,0.4866)

\PST@Border(0.1840,0.5000) (0.1990,0.5000)
\PST@Border(0.9600,0.5000) (0.9450,0.5000)
\rput[r](0.1680,0.5000){ 0.1}

\PST@Border(0.1840,0.5417) (0.1915,0.5417)
\PST@Border(0.9600,0.5417) (0.9525,0.5417)

\PST@Border(0.1840,0.5969) (0.1915,0.5969)
\PST@Border(0.9600,0.5969) (0.9525,0.5969)

\PST@Border(0.1840,0.6252) (0.1915,0.6252)
\PST@Border(0.9600,0.6252) (0.9525,0.6252)

\PST@Border(0.1840,0.6387) (0.1990,0.6387)
\PST@Border(0.9600,0.6387) (0.9450,0.6387)
\rput[r](0.1680,0.6387){ 1}

\PST@Border(0.1840,0.6804) (0.1915,0.6804)
\PST@Border(0.9600,0.6804) (0.9525,0.6804)

\PST@Border(0.1840,0.7356) (0.1915,0.7356)
\PST@Border(0.9600,0.7356) (0.9525,0.7356)

\PST@Border(0.1840,0.7639) (0.1915,0.7639)
\PST@Border(0.9600,0.7639) (0.9525,0.7639)

\PST@Border(0.1840,0.7773) (0.1990,0.7773)
\PST@Border(0.9600,0.7773) (0.9450,0.7773)
\rput[r](0.1680,0.7773){ 10}

\PST@Border(0.1840,0.8191) (0.1915,0.8191)
\PST@Border(0.9600,0.8191) (0.9525,0.8191)

\PST@Border(0.1840,0.8743) (0.1915,0.8743)
\PST@Border(0.9600,0.8743) (0.9525,0.8743)

\PST@Border(0.1840,0.9026) (0.1915,0.9026)
\PST@Border(0.9600,0.9026) (0.9525,0.9026)

\PST@Border(0.1840,0.9160) (0.1990,0.9160)
\PST@Border(0.9600,0.9160) (0.9450,0.9160)
\rput[r](0.1680,0.9160){ 100}

\PST@Border(0.1840,0.0840) (0.1840,0.1040)
\rput(0.1840,0.0420){ 0.2}

\PST@Border(0.2702,0.0840) (0.2702,0.1040)
\rput(0.2702,0.0420){ 0.4}

\PST@Border(0.3564,0.0840) (0.3564,0.1040)
\rput(0.3564,0.0420){ 0.6}

\PST@Border(0.4427,0.0840) (0.4427,0.1040)
\rput(0.4427,0.0420){ 0.8}

\PST@Border(0.5289,0.0840) (0.5289,0.1040)
\rput(0.5289,0.0420){ 1}

\PST@Border(0.6151,0.0840) (0.6151,0.1040)
\rput(0.6151,0.0420){ 1.2}

\PST@Border(0.7013,0.0840) (0.7013,0.1040)
\rput(0.7013,0.0420){ 1.4}

\PST@Border(0.7876,0.0840) (0.7876,0.1040)
\rput(0.7876,0.0420){ 1.6}

\PST@Border(0.8738,0.0840) (0.8738,0.1040)
\rput(0.8738,0.0420){ 1.8}


\PST@Border(0.2168,0.9160) (0.2168,0.8960)
\rput(0.2168,0.9580){$2m_\pi$}

\PST@Border(0.3336,0.9160) (0.3336,0.8960)
\rput(0.3336,0.9580){$m_\eta$}

\PST@Border(0.3931,0.9160) (0.3931,0.9060)
\rput(0.3931,0.9380){\tiny $+m_\pi$}

\PST@Border(0.4293,0.9160) (0.4293,0.8960)
\rput(0.4293,0.9580){$m_\rho$}

\PST@Border(0.4888,0.9160) (0.4888,0.9060)
\rput(0.4888,0.9380){\tiny $+m_\pi$}

\PST@Border(0.6280,0.9160) (0.6280,0.8960)
\rput(0.6280,0.9580){$m_{a_1}$}

\PST@Border(0.1840,0.9160) (0.1840,0.0840) (0.9600,0.0840) (0.9600,0.9160)
(0.1840,0.9160)

\rput{L}(0.1080,.5){$\sigma~[{\rm mb}]$}
\rput(.8813,.1640){$\sqrt{s}~[{\rm GeV}]$}

\rput[r](0.8330,0.8750){$\pi^\pm\pi^\mp\rightarrow\gamma\rho^0$} \PST@lineone(0.8490,.8700)(0.9280,0.8700)

\PST@lineone(0.4323,0.8637) (0.4366,0.8089) (0.4409,0.7798)
(0.4453,0.7596) (0.4496,0.7440) (0.4539,0.7312) (0.4582,0.7204) (0.4625,0.7109)
(0.4668,0.7025) (0.4711,0.6949) (0.4754,0.6880) (0.4797,0.6817) (0.4841,0.6758)
(0.4884,0.6703) (0.4927,0.6651) (0.4970,0.6603) (0.5013,0.6557) (0.5056,0.6514)
(0.5099,0.6472) (0.5142,0.6433) (0.5185,0.6395) (0.5229,0.6359) (0.5272,0.6324)
(0.5315,0.6291) (0.5358,0.6259) (0.5401,0.6228) (0.5444,0.6198) (0.5487,0.6169)
(0.5530,0.6141) (0.5573,0.6113) (0.5617,0.6087) (0.5660,0.6061) (0.5703,0.6036)
(0.5746,0.6012) (0.5789,0.5988) (0.5832,0.5965) (0.5875,0.5942) (0.5918,0.5920)
(0.5961,0.5898) (0.6005,0.5877) (0.6048,0.5856) (0.6091,0.5836) (0.6134,0.5816)
(0.6177,0.5797) (0.6220,0.5778) (0.6263,0.5759) (0.6306,0.5741) (0.6349,0.5723)
(0.6393,0.5705) (0.6436,0.5688) (0.6479,0.5671) (0.6522,0.5654) (0.6565,0.5637)
(0.6608,0.5621) (0.6651,0.5605) (0.6694,0.5589) (0.6737,0.5574) (0.6781,0.5558)
(0.6824,0.5543) (0.6867,0.5529) (0.6910,0.5514) (0.6953,0.5499) (0.6996,0.5485)
(0.7039,0.5471) (0.7082,0.5457) (0.7125,0.5444) (0.7169,0.5430) (0.7212,0.5417)
(0.7255,0.5404) (0.7298,0.5391) (0.7341,0.5378) (0.7384,0.5365) (0.7427,0.5353)
(0.7470,0.5340) (0.7513,0.5328) (0.7557,0.5316) (0.7600,0.5304) (0.7643,0.5292)
(0.7686,0.5280) (0.7729,0.5269) (0.7772,0.5257) (0.7815,0.5246) (0.7858,0.5235)
(0.7901,0.5224) (0.7945,0.5213) (0.7988,0.5202) (0.8031,0.5191) (0.8074,0.5180)
(0.8117,0.5170) (0.8160,0.5159) (0.8203,0.5149) (0.8246,0.5139) (0.8289,0.5128)
(0.8333,0.5118) (0.8376,0.5108) (0.8419,0.5098) (0.8462,0.5089) (0.8505,0.5079)
(0.8548,0.5069) (0.8591,0.5060) (0.8634,0.5050) (0.8677,0.5041) (0.8721,0.5031)
(0.8764,0.5022) (0.8807,0.5013) (0.8850,0.5004) (0.8893,0.4995) (0.8936,0.4986)
(0.8979,0.4977) (0.9022,0.4968) (0.9065,0.4959) (0.9109,0.4951) (0.9152,0.4942)
(0.9195,0.4934) (0.9238,0.4925) (0.9281,0.4917) (0.9324,0.4908) (0.9367,0.4900)
(0.9410,0.4892) (0.9453,0.4884) (0.9497,0.4875) (0.9540,0.4867) (0.9583,0.4859)

\rput[r](0.8330,0.8330){$\pi^\pm\pi^0\rightarrow\gamma\rho^\pm$} \PST@linetwo(0.8490,.8280)(0.9280,0.8280)

\PST@linetwo(0.4323,0.8403) (0.4366,0.7857) (0.4409,0.7567)
(0.4453,0.7366) (0.4496,0.7212) (0.4539,0.7085) (0.4582,0.6978) (0.4625,0.6884)
(0.4668,0.6801) (0.4711,0.6726) (0.4754,0.6658) (0.4797,0.6595) (0.4841,0.6537)
(0.4884,0.6483) (0.4927,0.6432) (0.4970,0.6384) (0.5013,0.6339) (0.5056,0.6296)
(0.5099,0.6256) (0.5142,0.6217) (0.5185,0.6180) (0.5229,0.6144) (0.5272,0.6110)
(0.5315,0.6077) (0.5358,0.6045) (0.5401,0.6015) (0.5444,0.5985) (0.5487,0.5957)
(0.5530,0.5929) (0.5573,0.5902) (0.5617,0.5876) (0.5660,0.5850) (0.5703,0.5826)
(0.5746,0.5802) (0.5789,0.5778) (0.5832,0.5756) (0.5875,0.5733) (0.5918,0.5712)
(0.5961,0.5690) (0.6005,0.5670) (0.6048,0.5649) (0.6091,0.5630) (0.6134,0.5610)
(0.6177,0.5591) (0.6220,0.5573) (0.6263,0.5554) (0.6306,0.5536) (0.6349,0.5519)
(0.6393,0.5502) (0.6436,0.5485) (0.6479,0.5468) (0.6522,0.5452) (0.6565,0.5436)
(0.6608,0.5420) (0.6651,0.5405) (0.6694,0.5389) (0.6737,0.5375) (0.6781,0.5360)
(0.6824,0.5345) (0.6867,0.5331) (0.6910,0.5317) (0.6953,0.5303) (0.6996,0.5290)
(0.7039,0.5276) (0.7082,0.5263) (0.7125,0.5250) (0.7169,0.5237) (0.7212,0.5225)
(0.7255,0.5212) (0.7298,0.5200) (0.7341,0.5188) (0.7384,0.5176) (0.7427,0.5164)
(0.7470,0.5152) (0.7513,0.5141) (0.7557,0.5130) (0.7600,0.5118) (0.7643,0.5107)
(0.7686,0.5097) (0.7729,0.5086) (0.7772,0.5075) (0.7815,0.5065) (0.7858,0.5054)
(0.7901,0.5044) (0.7945,0.5034) (0.7988,0.5024) (0.8031,0.5014) (0.8074,0.5005)
(0.8117,0.4995) (0.8160,0.4986) (0.8203,0.4976) (0.8246,0.4967) (0.8289,0.4958)
(0.8333,0.4949) (0.8376,0.4940) (0.8419,0.4931) (0.8462,0.4922) (0.8505,0.4914)
(0.8548,0.4905) (0.8591,0.4897) (0.8634,0.4888) (0.8677,0.4880) (0.8721,0.4872)
(0.8764,0.4864) (0.8807,0.4856) (0.8850,0.4848) (0.8893,0.4840) (0.8936,0.4832)
(0.8979,0.4825) (0.9022,0.4817) (0.9065,0.4810) (0.9109,0.4802) (0.9152,0.4795)
(0.9195,0.4788) (0.9238,0.4781) (0.9281,0.4773) (0.9324,0.4766) (0.9367,0.4759)
(0.9410,0.4753) (0.9453,0.4746) (0.9497,0.4739) (0.9540,0.4732) (0.9583,0.4726)

\PST@linethr(0.8490,0.8800) (0.9280,0.8800)

\PST@linethr(0.2229,0.0840) (0.2232,0.0985) (0.2275,0.1886) (0.2319,0.2474)
(0.2362,0.2910) (0.2405,0.3257) (0.2448,0.3544) (0.2491,0.3790) (0.2534,0.4005)
(0.2577,0.4197) (0.2620,0.4371) (0.2663,0.4529) (0.2707,0.4676) (0.2750,0.4812)
(0.2793,0.4940) (0.2836,0.5061) (0.2879,0.5177) (0.2922,0.5287) (0.2965,0.5393)
(0.3008,0.5495) (0.3051,0.5594) (0.3095,0.5691) (0.3138,0.5786) (0.3181,0.5879)
(0.3224,0.5970) (0.3267,0.6061) (0.3310,0.6151) (0.3353,0.6240) (0.3396,0.6329)
(0.3439,0.6419) (0.3483,0.6508) (0.3526,0.6599) (0.3569,0.6690) (0.3612,0.6782)
(0.3655,0.6876) (0.3698,0.6971) (0.3741,0.7067) (0.3784,0.7165) (0.3827,0.7265)
(0.3871,0.7366) (0.3914,0.7467) (0.3957,0.7569) (0.4000,0.7669) (0.4043,0.7765)
(0.4086,0.7855) (0.4129,0.7936) (0.4172,0.8003) (0.4215,0.8054) (0.4259,0.8085)
(0.4302,0.8097) (0.4345,0.8091) (0.4388,0.8069) (0.4431,0.8035) (0.4474,0.7992)
(0.4517,0.7944) (0.4560,0.7893) (0.4603,0.7841) (0.4647,0.7788) (0.4690,0.7736)
(0.4733,0.7685) (0.4776,0.7636) (0.4819,0.7588) (0.4862,0.7542) (0.4905,0.7497)
(0.4948,0.7454) (0.4991,0.7412) (0.5035,0.7372) (0.5078,0.7334) (0.5121,0.7296)
(0.5164,0.7260) (0.5207,0.7226) (0.5250,0.7192) (0.5293,0.7160) (0.5336,0.7128)
(0.5379,0.7098) (0.5423,0.7068) (0.5466,0.7039) (0.5509,0.7011) (0.5552,0.6984)
(0.5595,0.6958) (0.5638,0.6932) (0.5681,0.6907) (0.5724,0.6882) (0.5767,0.6858)
(0.5811,0.6835) (0.5854,0.6812) (0.5897,0.6789) (0.5940,0.6768) (0.5983,0.6746)
(0.6026,0.6725) (0.6069,0.6704) (0.6112,0.6684) (0.6155,0.6664) (0.6199,0.6645)
(0.6242,0.6626) (0.6285,0.6607) (0.6328,0.6589) (0.6371,0.6571) (0.6414,0.6553)
(0.6457,0.6535) (0.6500,0.6518) (0.6543,0.6501) (0.6587,0.6484) (0.6630,0.6468)
(0.6673,0.6451) (0.6716,0.6435) (0.6759,0.6419) (0.6802,0.6404) (0.6845,0.6388)
(0.6888,0.6373) (0.6931,0.6358) (0.6975,0.6343) (0.7018,0.6329) (0.7061,0.6314)
(0.7104,0.6300) (0.7147,0.6286) (0.7190,0.6272) (0.7233,0.6258) (0.7276,0.6244)
(0.7319,0.6231) (0.7363,0.6218) (0.7406,0.6204) (0.7449,0.6191) (0.7492,0.6178)
(0.7535,0.6166) (0.7578,0.6153) (0.7621,0.6140) (0.7664,0.6128) (0.7707,0.6116)
(0.7751,0.6103) (0.7794,0.6091) (0.7837,0.6079) (0.7880,0.6067) (0.7923,0.6056)
(0.7966,0.6044) (0.8009,0.6032) (0.8052,0.6021) (0.8095,0.6010) (0.8139,0.5998)
(0.8182,0.5987) (0.8225,0.5976) (0.8268,0.5965) (0.8311,0.5954) (0.8354,0.5943)
(0.8397,0.5933) (0.8440,0.5922) (0.8483,0.5911) (0.8527,0.5901) (0.8570,0.5890)
(0.8613,0.5880) (0.8656,0.5870) (0.8699,0.5860) (0.8742,0.5849) (0.8785,0.5839)
(0.8828,0.5829) (0.8871,0.5820) (0.8915,0.5810) (0.8958,0.5800) (0.9001,0.5790)
(0.9044,0.5781) (0.9087,0.5771) (0.9130,0.5761) (0.9173,0.5752) (0.9216,0.5743)
(0.9259,0.5733) (0.9303,0.5724) (0.9346,0.5715) (0.9389,0.5706) (0.9432,0.5696)
(0.9475,0.5687) (0.9518,0.5678) (0.9561,0.5669)

\PST@linefou(0.8490,0.8380) (0.9280,0.8380)

\PST@linefou(0.2245,0.0840) (0.2275,0.1489) (0.2319,0.2084) (0.2362,0.2527)
(0.2405,0.2881) (0.2448,0.3174) (0.2491,0.3427) (0.2534,0.3648) (0.2577,0.3845)
(0.2620,0.4024) (0.2663,0.4187) (0.2707,0.4339) (0.2750,0.4480) (0.2793,0.4613)
(0.2836,0.4738) (0.2879,0.4857) (0.2922,0.4972) (0.2965,0.5082) (0.3008,0.5188)
(0.3051,0.5291) (0.3095,0.5391) (0.3138,0.5489) (0.3181,0.5585) (0.3224,0.5680)
(0.3267,0.5773) (0.3310,0.5866) (0.3353,0.5958) (0.3396,0.6050) (0.3439,0.6142)
(0.3483,0.6235) (0.3526,0.6328) (0.3569,0.6421) (0.3612,0.6516) (0.3655,0.6612)
(0.3698,0.6709) (0.3741,0.6808) (0.3784,0.6908) (0.3827,0.7010) (0.3871,0.7113)
(0.3914,0.7216) (0.3957,0.7320) (0.4000,0.7422) (0.4043,0.7520) (0.4086,0.7612)
(0.4129,0.7695) (0.4172,0.7764) (0.4215,0.7816) (0.4259,0.7849) (0.4302,0.7862)
(0.4345,0.7857) (0.4388,0.7837) (0.4431,0.7804) (0.4474,0.7764) (0.4517,0.7717)
(0.4560,0.7667) (0.4603,0.7616) (0.4647,0.7565) (0.4690,0.7514) (0.4733,0.7465)
(0.4776,0.7416) (0.4819,0.7369) (0.4862,0.7324) (0.4905,0.7280) (0.4948,0.7238)
(0.4991,0.7198) (0.5035,0.7159) (0.5078,0.7121) (0.5121,0.7085) (0.5164,0.7050)
(0.5207,0.7016) (0.5250,0.6984) (0.5293,0.6952) (0.5336,0.6921) (0.5379,0.6892)
(0.5423,0.6863) (0.5466,0.6835) (0.5509,0.6808) (0.5552,0.6781) (0.5595,0.6756)
(0.5638,0.6731) (0.5681,0.6706) (0.5724,0.6683) (0.5767,0.6659) (0.5811,0.6637)
(0.5854,0.6614) (0.5897,0.6593) (0.5940,0.6572) (0.5983,0.6551) (0.6026,0.6530)
(0.6069,0.6511) (0.6112,0.6491) (0.6155,0.6472) (0.6199,0.6453) (0.6242,0.6435)
(0.6285,0.6416) (0.6328,0.6399) (0.6371,0.6381) (0.6414,0.6364) (0.6457,0.6347)
(0.6500,0.6330) (0.6543,0.6314) (0.6587,0.6298) (0.6630,0.6282) (0.6673,0.6266)
(0.6716,0.6251) (0.6759,0.6235) (0.6802,0.6220) (0.6845,0.6205) (0.6888,0.6191)
(0.6931,0.6176) (0.6975,0.6162) (0.7018,0.6148) (0.7061,0.6134) (0.7104,0.6121)
(0.7147,0.6107) (0.7190,0.6094) (0.7233,0.6080) (0.7276,0.6067) (0.7319,0.6054)
(0.7363,0.6042) (0.7406,0.6029) (0.7449,0.6016) (0.7492,0.6004) (0.7535,0.5992)
(0.7578,0.5980) (0.7621,0.5968) (0.7664,0.5956) (0.7707,0.5944) (0.7751,0.5932)
(0.7794,0.5921) (0.7837,0.5910) (0.7880,0.5898) (0.7923,0.5887) (0.7966,0.5876)
(0.8009,0.5865) (0.8052,0.5854) (0.8095,0.5843) (0.8139,0.5833) (0.8182,0.5822)
(0.8225,0.5812) (0.8268,0.5801) (0.8311,0.5791) (0.8354,0.5781) (0.8397,0.5770)
(0.8440,0.5760) (0.8483,0.5750) (0.8527,0.5741) (0.8570,0.5731) (0.8613,0.5721)
(0.8656,0.5711) (0.8699,0.5702) (0.8742,0.5692) (0.8785,0.5683) (0.8828,0.5673)
(0.8871,0.5664) (0.8915,0.5655) (0.8958,0.5646) (0.9001,0.5637) (0.9044,0.5628)
(0.9087,0.5619) (0.9130,0.5610) (0.9173,0.5601) (0.9216,0.5592) (0.9259,0.5584)
(0.9303,0.5575) (0.9346,0.5567) (0.9389,0.5558) (0.9432,0.5550) (0.9475,0.5541)
(0.9518,0.5533) (0.9561,0.5525)

\rput[r](0.8330,0.7910){$\pi^\pm\pi^\mp\rightarrow\gamma\eta\phantom{^\pm}$} \PST@linefiv(0.8490,0.7910) (0.9280,0.7910)

\PST@linefiv(0.3611,0.0840) (0.3633,0.1016) (0.3677,0.1319) (0.3720,0.1598)
(0.3763,0.1857) (0.3806,0.2102) (0.3849,0.2336) (0.3892,0.2561) (0.3935,0.2780)
(0.3978,0.2994) (0.4021,0.3203) (0.4065,0.3406) (0.4108,0.3602) (0.4151,0.3786)
(0.4194,0.3953) (0.4237,0.4096) (0.4280,0.4206) (0.4323,0.4279) (0.4366,0.4315)
(0.4409,0.4319) (0.4453,0.4300) (0.4496,0.4265) (0.4539,0.4222) (0.4582,0.4176)
(0.4625,0.4129) (0.4668,0.4082) (0.4711,0.4038) (0.4754,0.3997) (0.4797,0.3958)
(0.4841,0.3922) (0.4884,0.3889) (0.4927,0.3858) (0.4970,0.3829) (0.5013,0.3803)
(0.5056,0.3778) (0.5099,0.3755) (0.5142,0.3734) (0.5185,0.3715) (0.5229,0.3696)
(0.5272,0.3679) (0.5315,0.3663) (0.5358,0.3648) (0.5401,0.3634) (0.5444,0.3620)
(0.5487,0.3608) (0.5530,0.3596) (0.5573,0.3585) (0.5617,0.3575) (0.5660,0.3565)
(0.5703,0.3556) (0.5746,0.3547) (0.5789,0.3538) (0.5832,0.3531) (0.5875,0.3523)
(0.5918,0.3516) (0.5961,0.3509) (0.6005,0.3502) (0.6048,0.3496) (0.6091,0.3490)
(0.6134,0.3485) (0.6177,0.3479) (0.6220,0.3474) (0.6263,0.3469) (0.6306,0.3464)
(0.6349,0.3460) (0.6393,0.3456) (0.6436,0.3451) (0.6479,0.3447) (0.6522,0.3444)
(0.6565,0.3440) (0.6608,0.3436) (0.6651,0.3433) (0.6694,0.3429) (0.6737,0.3426)
(0.6781,0.3423) (0.6824,0.3420) (0.6867,0.3417) (0.6910,0.3415) (0.6953,0.3412)
(0.6996,0.3409) (0.7039,0.3407) (0.7082,0.3404) (0.7125,0.3402) (0.7169,0.3400)
(0.7212,0.3398) (0.7255,0.3396) (0.7298,0.3394) (0.7341,0.3392) (0.7384,0.3390)
(0.7427,0.3388) (0.7470,0.3386) (0.7513,0.3384) (0.7557,0.3382) (0.7600,0.3381)
(0.7643,0.3379) (0.7686,0.3378) (0.7729,0.3376) (0.7772,0.3375) (0.7815,0.3373)
(0.7858,0.3372) (0.7901,0.3370) (0.7945,0.3369) (0.7988,0.3368) (0.8031,0.3366)
(0.8074,0.3365) (0.8117,0.3364) (0.8160,0.3363) (0.8203,0.3362) (0.8246,0.3361)
(0.8289,0.3359) (0.8333,0.3358) (0.8376,0.3357) (0.8419,0.3356) (0.8462,0.3355)
(0.8505,0.3354) (0.8548,0.3353) (0.8591,0.3352) (0.8634,0.3352) (0.8677,0.3351)
(0.8721,0.3350) (0.8764,0.3349) (0.8807,0.3348) (0.8850,0.3347) (0.8893,0.3347)
(0.8936,0.3346) (0.8979,0.3345) (0.9022,0.3344) (0.9065,0.3344) (0.9109,0.3343)
(0.9152,0.3342) (0.9195,0.3341) (0.9238,0.3341) (0.9281,0.3340) (0.9324,0.3340)
(0.9367,0.3339) (0.9410,0.3338) (0.9453,0.3338) (0.9497,0.3337) (0.9540,0.3336)
(0.9583,0.3336)

\rput[r](0.8330,0.7490){$\pi^\pm\eta\phantom{^\pm}\rightarrow\gamma\pi^\pm$} \PST@linesix(0.8490,0.7490) (0.9280,0.7490)

\PST@linesix(0.3935,0.0840) (0.3935,0.0920) (0.3978,0.1665) (0.4021,0.1883)
(0.4065,0.2022) (0.4108,0.2127) (0.4151,0.2213) (0.4194,0.2287) (0.4237,0.2352)
(0.4280,0.2411) (0.4323,0.2464) (0.4366,0.2514) (0.4409,0.2559) (0.4453,0.2602)
(0.4496,0.2643) (0.4539,0.2681) (0.4582,0.2717) (0.4625,0.2752) (0.4668,0.2785)
(0.4711,0.2817) (0.4754,0.2847) (0.4797,0.2876) (0.4841,0.2905) (0.4884,0.2932)
(0.4927,0.2958) (0.4970,0.2984) (0.5013,0.3008) (0.5056,0.3032) (0.5099,0.3056)
(0.5142,0.3078) (0.5185,0.3100) (0.5229,0.3121) (0.5272,0.3142) (0.5315,0.3162)
(0.5358,0.3182) (0.5401,0.3202) (0.5444,0.3220) (0.5487,0.3239) (0.5530,0.3257)
(0.5573,0.3274) (0.5617,0.3291) (0.5660,0.3308) (0.5703,0.3325) (0.5746,0.3341)
(0.5789,0.3357) (0.5832,0.3372) (0.5875,0.3387) (0.5918,0.3402) (0.5961,0.3417)
(0.6005,0.3431) (0.6048,0.3445) (0.6091,0.3459) (0.6134,0.3472) (0.6177,0.3485)
(0.6220,0.3498) (0.6263,0.3511) (0.6306,0.3524) (0.6349,0.3536) (0.6393,0.3548)
(0.6436,0.3560) (0.6479,0.3572) (0.6522,0.3584) (0.6565,0.3595) (0.6608,0.3606)
(0.6651,0.3617) (0.6694,0.3628) (0.6737,0.3639) (0.6781,0.3650) (0.6824,0.3660)
(0.6867,0.3670) (0.6910,0.3680) (0.6953,0.3690) (0.6996,0.3700) (0.7039,0.3710)
(0.7082,0.3719) (0.7125,0.3729) (0.7169,0.3738) (0.7212,0.3747) (0.7255,0.3756)
(0.7298,0.3765) (0.7341,0.3774) (0.7384,0.3782) (0.7427,0.3791) (0.7470,0.3799)
(0.7513,0.3808) (0.7557,0.3816) (0.7600,0.3824) (0.7643,0.3832) (0.7686,0.3840)
(0.7729,0.3848) (0.7772,0.3856) (0.7815,0.3863) (0.7858,0.3871) (0.7901,0.3878)
(0.7945,0.3886) (0.7988,0.3893) (0.8031,0.3900) (0.8074,0.3907) (0.8117,0.3914)
(0.8160,0.3921) (0.8203,0.3928) (0.8246,0.3935) (0.8289,0.3942) (0.8333,0.3948)
(0.8376,0.3955) (0.8419,0.3961) (0.8462,0.3968) (0.8505,0.3974) (0.8548,0.3980)
(0.8591,0.3987) (0.8634,0.3993) (0.8677,0.3999) (0.8721,0.4005) (0.8764,0.4011)
(0.8807,0.4017) (0.8850,0.4023) (0.8893,0.4028) (0.8936,0.4034) (0.8979,0.4040)
(0.9022,0.4045) (0.9065,0.4051) (0.9109,0.4057) (0.9152,0.4062) (0.9195,0.4067)
(0.9238,0.4073) (0.9281,0.4078) (0.9324,0.4083) (0.9367,0.4089) (0.9410,0.4094)
(0.9453,0.4099) (0.9497,0.4104) (0.9540,0.4109) (0.9583,0.4114)

\rput[r](0.8330,0.7070){$\pi^\pm\pi^\mp\rightarrow\gamma\gamma\phantom{^\pm}$} \PST@linesev(0.8490,0.7070) (0.9280,0.7070)

\PST@linesev(0.2024,0.4296) (0.2067,0.3994) (0.2111,0.3790) (0.2154,0.3631)
(0.2197,0.3498) (0.2241,0.3384) (0.2284,0.3285) (0.2328,0.3196) (0.2371,0.3117)
(0.2415,0.3047) (0.2458,0.2983) (0.2502,0.2925) (0.2545,0.2872) (0.2588,0.2823)
(0.2632,0.2778) (0.2675,0.2737) (0.2719,0.2699) (0.2762,0.2663) (0.2806,0.2630)
(0.2849,0.2598) (0.2893,0.2569) (0.2936,0.2540) (0.2979,0.2514) (0.3023,0.2488)
(0.3066,0.2464) (0.3110,0.2441) (0.3153,0.2418) (0.3197,0.2397) (0.3240,0.2376)
(0.3284,0.2356) (0.3327,0.2336) (0.3370,0.2317) (0.3414,0.2299) (0.3457,0.2280)
(0.3501,0.2263) (0.3544,0.2245) (0.3588,0.2229) (0.3631,0.2212) (0.3675,0.2196)
(0.3718,0.2180) (0.3761,0.2164) (0.3805,0.2149) (0.3848,0.2133) (0.3892,0.2119)
(0.3935,0.2104) (0.3979,0.2089) (0.4022,0.2075) (0.4066,0.2061) (0.4109,0.2047)
(0.4152,0.2033) (0.4196,0.2020) (0.4239,0.2006) (0.4283,0.1993) (0.4326,0.1980)
(0.4370,0.1967) (0.4413,0.1954) (0.4457,0.1941) (0.4500,0.1929) (0.4543,0.1916)
(0.4587,0.1904) (0.4630,0.1892) (0.4674,0.1880) (0.4717,0.1868) (0.4761,0.1856)
(0.4804,0.1844) (0.4848,0.1832) (0.4891,0.1821) (0.4934,0.1809) (0.4978,0.1798)
(0.5021,0.1787) (0.5065,0.1776) (0.5108,0.1765) (0.5152,0.1754) (0.5195,0.1743)
(0.5239,0.1732) (0.5282,0.1722) (0.5325,0.1711) (0.5369,0.1700) (0.5412,0.1690)
(0.5456,0.1680) (0.5499,0.1669) (0.5543,0.1659) (0.5586,0.1649) (0.5630,0.1639)
(0.5673,0.1629) (0.5716,0.1619) (0.5760,0.1609) (0.5803,0.1600) (0.5847,0.1590)
(0.5890,0.1580) (0.5934,0.1571) (0.5977,0.1561) (0.6021,0.1552) (0.6064,0.1543)
(0.6107,0.1533) (0.6151,0.1524) (0.6194,0.1515) (0.6238,0.1506) (0.6281,0.1497)
(0.6325,0.1488) (0.6368,0.1479) (0.6412,0.1470) (0.6455,0.1461) (0.6498,0.1453)
(0.6542,0.1444) (0.6585,0.1435) (0.6629,0.1427) (0.6672,0.1418) (0.6716,0.1410)
(0.6759,0.1401) (0.6803,0.1393) (0.6846,0.1385) (0.6889,0.1377) (0.6933,0.1368)
(0.6976,0.1360) (0.7020,0.1352) (0.7063,0.1344) (0.7107,0.1336) (0.7150,0.1328)
(0.7194,0.1320) (0.7237,0.1312) (0.7280,0.1305) (0.7324,0.1297) (0.7367,0.1289)
(0.7411,0.1281) (0.7454,0.1274) (0.7498,0.1266) (0.7541,0.1259) (0.7585,0.1251)
(0.7628,0.1244) (0.7671,0.1236) (0.7715,0.1229) (0.7758,0.1221) (0.7802,0.1214)
(0.7845,0.1207) (0.7889,0.1200) (0.7932,0.1192) (0.7976,0.1185) (0.8019,0.1178)
(0.8062,0.1171) (0.8106,0.1164) (0.8149,0.1157) (0.8193,0.1150) (0.8236,0.1143)
(0.8280,0.1136) (0.8323,0.1129) (0.8367,0.1123) (0.8410,0.1116) (0.8453,0.1109)
(0.8497,0.1102) (0.8540,0.1096) (0.8584,0.1089) (0.8627,0.1082) (0.8671,0.1076)
(0.8714,0.1069) (0.8758,0.1063) (0.8801,0.1056) (0.8844,0.1050) (0.8888,0.1043)
(0.8931,0.1037) (0.8975,0.1031) (0.9018,0.1024) (0.9062,0.1018) (0.9105,0.1012)
(0.9149,0.1006) (0.9192,0.0999) (0.9235,0.0993) (0.9279,0.0987) (0.9322,0.0981)
(0.9366,0.0975) (0.9409,0.0969) (0.9453,0.0963)

\rput(0.45,.50){\rnode{B}{$m_\rho$~Breit-Wigner}}
\pnode(0.3612,.6516){b}
\ncarc{<-}{b}{B}

\rput(0.7,.42){\rnode{F}{$m_\rho$~fixed}}
\pnode(0.8031,.5014){f}
\ncangle[angleA=0,angleB=-90,linearc=.005]{->}{F}{f}

\PST@Border(0.1840,0.9160) (0.1840,0.0840) (0.9600,0.0840) (0.9600,0.9160)
(0.1840,0.9160)

\catcode`@=12 \fi \endpspicture%
\hskip1pt%
\ifx\PSTloaded\undefined
\def\PSTloaded{t}
\psset{arrowsize=.01 3.2 1.4 .3}
\psset{dotsize=.01}
\catcode`@=11

\definecolor{darkgreen}{rgb}{0,0.5,0}
\definecolor{violett}{rgb}{.5,0,.5}
\newpsobject{PST@Border}{psline}{linewidth=.0015,linestyle=solid}
\newpsobject{PST@Axes}{psline}{linewidth=.0015,linestyle=dotted,dotsep=.004}
\newpsobject{PST@lineone}{psline}{linewidth=.0015,linestyle=solid,linecolor=black}
\newpsobject{PST@linetwo}{psline}{linewidth=.0025,linestyle=dotted,dotsep=.004,linecolor=red}
\newpsobject{PST@linethr}{psline}{linewidth=.0025,linestyle=dotted,dotsep=.004, linecolor=blue}
\newpsobject{PST@linefou}{psline}{linewidth=.0015,linestyle=dashed,dash=.02 .02,linecolor=darkgreen}
\newpsobject{PST@linefiv}{psline}{linewidth=.0015,linestyle=solid,linecolor=violett}
\catcode`@=12

\fi
\psset{unit=5.0in,xunit=.66\textwidth,yunit=\columnwidth}
\pspicture(0.442700,0.000000)(1.050000,1.000000)
\ifx\nofigs\undefined
\catcode`@=11

\PST@Border(0.4427,0.0840) (0.4577,0.0840)
\PST@Border(0.9600,0.0840) (0.9450,0.0840)
\rput[l](0.9620,0.0840){$10^{-4}$}

\PST@Border(0.4427,0.1257) (0.4502,0.1257)
\PST@Border(0.9600,0.1257) (0.9525,0.1257)

\PST@Border(0.4427,0.1809) (0.4502,0.1809)
\PST@Border(0.9600,0.1809) (0.9525,0.1809)

\PST@Border(0.4427,0.2092) (0.4502,0.2092)
\PST@Border(0.9600,0.2092) (0.9525,0.2092)

\PST@Border(0.4427,0.2227) (0.4577,0.2227)
\PST@Border(0.9600,0.2227) (0.9450,0.2227)
\rput[l](0.9620,0.2227){$10^{-3}$}

\PST@Border(0.4427,0.2644) (0.4502,0.2644)
\PST@Border(0.9600,0.2644) (0.9525,0.2644)

\PST@Border(0.4427,0.3196) (0.4502,0.3196)
\PST@Border(0.9600,0.3196) (0.9525,0.3196)

\PST@Border(0.4427,0.3479) (0.4502,0.3479)
\PST@Border(0.9600,0.3479) (0.9525,0.3479)

\PST@Border(0.4427,0.3613) (0.4577,0.3613)
\PST@Border(0.9600,0.3613) (0.9450,0.3613)
\rput[l](0.9620,0.3613){ 0.01}

\PST@Border(0.4427,0.4031) (0.4502,0.4031)
\PST@Border(0.9600,0.4031) (0.9525,0.4031)

\PST@Border(0.4427,0.4583) (0.4502,0.4583)
\PST@Border(0.9600,0.4583) (0.9525,0.4583)

\PST@Border(0.4427,0.4866) (0.4502,0.4866)
\PST@Border(0.9600,0.4866) (0.9525,0.4866)

\PST@Border(0.4427,0.5000) (0.4577,0.5000)
\PST@Border(0.9600,0.5000) (0.9450,0.5000)
\rput[l](0.9620,0.5000){ 0.1}

\PST@Border(0.4427,0.5417) (0.4502,0.5417)
\PST@Border(0.9600,0.5417) (0.9525,0.5417)

\PST@Border(0.4427,0.5969) (0.4502,0.5969)
\PST@Border(0.9600,0.5969) (0.9525,0.5969)

\PST@Border(0.4427,0.6252) (0.4502,0.6252)
\PST@Border(0.9600,0.6252) (0.9525,0.6252)

\PST@Border(0.4427,0.6387) (0.4577,0.6387)
\PST@Border(0.9600,0.6387) (0.9450,0.6387)
\rput[l](0.9620,0.6387){ 1}

\PST@Border(0.4427,0.6804) (0.4502,0.6804)
\PST@Border(0.9600,0.6804) (0.9525,0.6804)

\PST@Border(0.4427,0.7356) (0.4502,0.7356)
\PST@Border(0.9600,0.7356) (0.9525,0.7356)

\PST@Border(0.4427,0.7639) (0.4502,0.7639)
\PST@Border(0.9600,0.7639) (0.9525,0.7639)

\PST@Border(0.4427,0.7773) (0.4577,0.7773)
\PST@Border(0.9600,0.7773) (0.9450,0.7773)
\rput[l](0.9620,0.7773){ 10}

\PST@Border(0.4427,0.8191) (0.4502,0.8191)
\PST@Border(0.9600,0.8191) (0.9525,0.8191)

\PST@Border(0.4427,0.8743) (0.4502,0.8743)
\PST@Border(0.9600,0.8743) (0.9525,0.8743)

\PST@Border(0.4427,0.9026) (0.4502,0.9026)
\PST@Border(0.9600,0.9026) (0.9525,0.9026)

\PST@Border(0.4427,0.9160) (0.4577,0.9160)
\PST@Border(0.9600,0.9160) (0.9450,0.9160)
\rput[l](0.9620,0.9160){ 100}

\PST@Border(0.4427,0.0840) (0.4427,0.1040)
\rput(0.4427,0.0420){0.8}

\PST@Border(0.5289,0.0840) (0.5289,0.1040)
\rput(0.5289,0.0420){ 1}

\PST@Border(0.6151,0.0840) (0.6151,0.1040)
\rput(0.6151,0.0420){ 1.2}

\PST@Border(0.7013,0.0840) (0.7013,0.1040)
\rput(0.7013,0.0420){ 1.4}

\PST@Border(0.7876,0.0840) (0.7876,0.1040)
\rput(0.7876,0.0420){ 1.6}

\PST@Border(0.8738,0.0840) (0.8738,0.1040)
\rput(0.8738,0.0420){ 1.8}

\PST@Border(0.9600,0.0840) (0.9600,0.1040)
\rput(0.9600,0.0420){ 2}

\PST@Border(0.4888,0.9160) (0.4888,0.9060)
\rput(0.4888,0.9580){$m_\rho+m_\pi$}

\PST@Border(0.6280,0.9160) (0.6280,0.8960)
\rput(0.6280,0.9580){$m_{a_1}$}

\PST@Border(0.4427,0.9160) (0.4427,0.0840) (0.9600,0.0840) (0.9600,0.9160)
(0.4427,0.9160)

\rput{L}(1.0340,.5){$\sigma~[{\rm mb}]$}
\rput(.8813,.1640){$\sqrt{s}~[{\rm GeV}]$}

\rput[r](0.8330,0.8750){$\pi^\pm\rho^0\rightarrow\gamma\pi^\pm$} \PST@lineone(0.8490,0.8750) (0.9280,0.8750)

\PST@lineone(0.4927,0.6261) (0.4970,0.6036) (0.5013,0.5907)
(0.5056,0.5816) (0.5099,0.5745) (0.5142,0.5685) (0.5185,0.5634) (0.5229,0.5589)
(0.5272,0.5548) (0.5315,0.5511) (0.5358,0.5476) (0.5401,0.5444) (0.5444,0.5413)
(0.5487,0.5384) (0.5530,0.5357) (0.5573,0.5330) (0.5617,0.5305) (0.5660,0.5281)
(0.5703,0.5257) (0.5746,0.5234) (0.5789,0.5212) (0.5832,0.5191) (0.5875,0.5170)
(0.5918,0.5150) (0.5961,0.5130) (0.6005,0.5111) (0.6048,0.5092) (0.6091,0.5073)
(0.6134,0.5055) (0.6177,0.5037) (0.6220,0.5020) (0.6263,0.5003) (0.6306,0.4986)
(0.6349,0.4969) (0.6393,0.4953) (0.6436,0.4937) (0.6479,0.4921) (0.6522,0.4906)
(0.6565,0.4891) (0.6608,0.4876) (0.6651,0.4861) (0.6694,0.4846) (0.6737,0.4832)
(0.6781,0.4818) (0.6824,0.4804) (0.6867,0.4790) (0.6910,0.4777) (0.6953,0.4763)
(0.6996,0.4750) (0.7039,0.4737) (0.7082,0.4724) (0.7125,0.4711) (0.7169,0.4698)
(0.7212,0.4686) (0.7255,0.4673) (0.7298,0.4661) (0.7341,0.4649) (0.7384,0.4637)
(0.7427,0.4625) (0.7470,0.4614) (0.7513,0.4602) (0.7557,0.4591) (0.7600,0.4579)
(0.7643,0.4568) (0.7686,0.4557) (0.7729,0.4546) (0.7772,0.4535) (0.7815,0.4524)
(0.7858,0.4514) (0.7901,0.4503) (0.7945,0.4493) (0.7988,0.4482) (0.8031,0.4472)
(0.8074,0.4462) (0.8117,0.4452) (0.8160,0.4442) (0.8203,0.4432) (0.8246,0.4422)
(0.8289,0.4412) (0.8333,0.4402) (0.8376,0.4393) (0.8419,0.4383) (0.8462,0.4374)
(0.8505,0.4365) (0.8548,0.4355) (0.8591,0.4346) (0.8634,0.4337) (0.8677,0.4328)
(0.8721,0.4319) (0.8764,0.4310) (0.8807,0.4301) (0.8850,0.4293) (0.8893,0.4284)
(0.8936,0.4275) (0.8979,0.4267) (0.9022,0.4258) (0.9065,0.4250) (0.9109,0.4241)
(0.9152,0.4233) (0.9195,0.4225) (0.9238,0.4216) (0.9281,0.4208) (0.9324,0.4200)
(0.9367,0.4192) (0.9410,0.4184) (0.9453,0.4176) (0.9497,0.4168) (0.9540,0.4161)
(0.9583,0.4153)

\rput[r](0.8330,0.8330){$\pi^\pm\rho^\mp\rightarrow\gamma\pi^0$} \PST@linetwo(0.8490,0.8330) (0.9280,0.8330)

\PST@linetwo(0.4927,0.6281) (0.4970,0.6063) (0.5013,0.5942)
(0.5056,0.5857) (0.5099,0.5791) (0.5142,0.5738) (0.5185,0.5692) (0.5229,0.5652)
(0.5272,0.5616) (0.5315,0.5584) (0.5358,0.5554) (0.5401,0.5527) (0.5444,0.5501)
(0.5487,0.5477) (0.5530,0.5454) (0.5573,0.5432) (0.5617,0.5411) (0.5660,0.5392)
(0.5703,0.5373) (0.5746,0.5354) (0.5789,0.5337) (0.5832,0.5320) (0.5875,0.5304)
(0.5918,0.5288) (0.5961,0.5273) (0.6005,0.5258) (0.6048,0.5243) (0.6091,0.5229)
(0.6134,0.5216) (0.6177,0.5203) (0.6220,0.5190) (0.6263,0.5177) (0.6306,0.5165)
(0.6349,0.5153) (0.6393,0.5142) (0.6436,0.5131) (0.6479,0.5120) (0.6522,0.5109)
(0.6565,0.5099) (0.6608,0.5089) (0.6651,0.5079) (0.6694,0.5069) (0.6737,0.5059)
(0.6781,0.5050) (0.6824,0.5041) (0.6867,0.5032) (0.6910,0.5024) (0.6953,0.5015)
(0.6996,0.5007) (0.7039,0.4999) (0.7082,0.4991) (0.7125,0.4984) (0.7169,0.4976)
(0.7212,0.4969) (0.7255,0.4962) (0.7298,0.4955) (0.7341,0.4948) (0.7384,0.4941)
(0.7427,0.4934) (0.7470,0.4928) (0.7513,0.4922) (0.7557,0.4916) (0.7600,0.4910)
(0.7643,0.4904) (0.7686,0.4898) (0.7729,0.4892) (0.7772,0.4887) (0.7815,0.4881)
(0.7858,0.4876) (0.7901,0.4871) (0.7945,0.4866) (0.7988,0.4861) (0.8031,0.4856)
(0.8074,0.4851) (0.8117,0.4847) (0.8160,0.4842) (0.8203,0.4838) (0.8246,0.4834)
(0.8289,0.4829) (0.8333,0.4825) (0.8376,0.4821) (0.8419,0.4817) (0.8462,0.4813)
(0.8505,0.4810) (0.8548,0.4806) (0.8591,0.4802) (0.8634,0.4799) (0.8677,0.4795)
(0.8721,0.4792) (0.8764,0.4788) (0.8807,0.4785) (0.8850,0.4782) (0.8893,0.4779)
(0.8936,0.4776) (0.8979,0.4773) (0.9022,0.4770) (0.9065,0.4767) (0.9109,0.4764)
(0.9152,0.4761) (0.9195,0.4759) (0.9238,0.4756) (0.9281,0.4754) (0.9324,0.4751)
(0.9367,0.4749) (0.9410,0.4746) (0.9453,0.4744) (0.9497,0.4742) (0.9540,0.4739)
(0.9583,0.4737)

\rput[r](0.8330,0.7910){$\pi^0\rho^\pm\rightarrow\gamma\pi^\pm$} \PST@linefou(0.8490,0.7910) (0.9280,0.7910)

\PST@linefou(0.4927,0.4039) (0.4970,0.3802) (0.5013,0.3667)
(0.5056,0.3574) (0.5099,0.3505) (0.5142,0.3453) (0.5185,0.3413) (0.5229,0.3383)
(0.5272,0.3360) (0.5315,0.3343) (0.5358,0.3331) (0.5401,0.3324) (0.5444,0.3320)
(0.5487,0.3320) (0.5530,0.3321) (0.5573,0.3325) (0.5617,0.3331) (0.5660,0.3338)
(0.5703,0.3346) (0.5746,0.3356) (0.5789,0.3366) (0.5832,0.3376) (0.5875,0.3388)
(0.5918,0.3399) (0.5961,0.3411) (0.6005,0.3423) (0.6048,0.3436) (0.6091,0.3448)
(0.6134,0.3461) (0.6177,0.3473) (0.6220,0.3485) (0.6263,0.3498) (0.6306,0.3510)
(0.6349,0.3523) (0.6393,0.3535) (0.6436,0.3547) (0.6479,0.3559) (0.6522,0.3570)
(0.6565,0.3582) (0.6608,0.3593) (0.6651,0.3605) (0.6694,0.3616) (0.6737,0.3627)
(0.6781,0.3638) (0.6824,0.3648) (0.6867,0.3659) (0.6910,0.3669) (0.6953,0.3680)
(0.6996,0.3690) (0.7039,0.3699) (0.7082,0.3709) (0.7125,0.3719) (0.7169,0.3728)
(0.7212,0.3738) (0.7255,0.3747) (0.7298,0.3756) (0.7341,0.3765) (0.7384,0.3773)
(0.7427,0.3782) (0.7470,0.3790) (0.7513,0.3799) (0.7557,0.3807) (0.7600,0.3815)
(0.7643,0.3823) (0.7686,0.3831) (0.7729,0.3838) (0.7772,0.3846) (0.7815,0.3853)
(0.7858,0.3861) (0.7901,0.3868) (0.7945,0.3875) (0.7988,0.3882) (0.8031,0.3889)
(0.8074,0.3896) (0.8117,0.3903) (0.8160,0.3910) (0.8203,0.3916) (0.8246,0.3923)
(0.8289,0.3929) (0.8333,0.3935) (0.8376,0.3942) (0.8419,0.3948) (0.8462,0.3954)
(0.8505,0.3960) (0.8548,0.3966) (0.8591,0.3972) (0.8634,0.3977) (0.8677,0.3983)
(0.8721,0.3989) (0.8764,0.3994) (0.8807,0.4000) (0.8850,0.4005) (0.8893,0.4010)
(0.8936,0.4016) (0.8979,0.4021) (0.9022,0.4026) (0.9065,0.4031) (0.9109,0.4036)
(0.9152,0.4041) (0.9195,0.4046) (0.9238,0.4051) (0.9281,0.4055) (0.9324,0.4060)
(0.9367,0.4065) (0.9410,0.4069) (0.9453,0.4074) (0.9497,0.4079) (0.9540,0.4083)
(0.9583,0.4087)

\rput[r](0.8330,0.7490){$\pi\rho\rightarrow{}a_1\rightarrow\gamma\pi$} \PST@linefiv(0.8490,0.7490) (0.9280,0.7490)

\PST@linefiv(0.4927,0.3835) (0.4970,0.3760) (0.5013,0.3779)
(0.5056,0.3833) (0.5099,0.3904) (0.5142,0.3987) (0.5185,0.4076) (0.5229,0.4171)
(0.5272,0.4269) (0.5315,0.4371) (0.5358,0.4475) (0.5401,0.4581) (0.5444,0.4689)
(0.5487,0.4799) (0.5530,0.4911) (0.5573,0.5023) (0.5617,0.5136) (0.5660,0.5248)
(0.5703,0.5358) (0.5746,0.5465) (0.5789,0.5566) (0.5832,0.5660) (0.5875,0.5742)
(0.5918,0.5811) (0.5961,0.5863) (0.6005,0.5897) (0.6048,0.5913) (0.6091,0.5912)
(0.6134,0.5898) (0.6177,0.5872) (0.6220,0.5838) (0.6263,0.5798) (0.6306,0.5755)
(0.6349,0.5710) (0.6393,0.5664) (0.6436,0.5618) (0.6479,0.5573) (0.6522,0.5530)
(0.6565,0.5487) (0.6608,0.5447) (0.6651,0.5407) (0.6694,0.5369) (0.6737,0.5333)
(0.6781,0.5298) (0.6824,0.5264) (0.6867,0.5232) (0.6910,0.5201) (0.6953,0.5171)
(0.6996,0.5142) (0.7039,0.5114) (0.7082,0.5087) (0.7125,0.5061) (0.7169,0.5035)
(0.7212,0.5011) (0.7255,0.4987) (0.7298,0.4964) (0.7341,0.4941) (0.7384,0.4919)
(0.7427,0.4898) (0.7470,0.4877) (0.7513,0.4856) (0.7557,0.4836) (0.7600,0.4816)
(0.7643,0.4797) (0.7686,0.4779) (0.7729,0.4760) (0.7772,0.4742) (0.7815,0.4724)
(0.7858,0.4707) (0.7901,0.4690) (0.7945,0.4673) (0.7988,0.4656) (0.8031,0.4640)
(0.8074,0.4624) (0.8117,0.4608) (0.8160,0.4593) (0.8203,0.4577) (0.8246,0.4562)
(0.8289,0.4547) (0.8333,0.4533) (0.8376,0.4518) (0.8419,0.4504) (0.8462,0.4490)
(0.8505,0.4476) (0.8548,0.4462) (0.8591,0.4448) (0.8634,0.4435) (0.8677,0.4422)
(0.8721,0.4409) (0.8764,0.4396) (0.8807,0.4383) (0.8850,0.4370) (0.8893,0.4358)
(0.8936,0.4345) (0.8979,0.4333) (0.9022,0.4321) (0.9065,0.4309) (0.9109,0.4297)
(0.9152,0.4285) (0.9195,0.4274) (0.9238,0.4262) (0.9281,0.4251) (0.9324,0.4240)
(0.9367,0.4229) (0.9410,0.4218) (0.9453,0.4207) (0.9497,0.4196) (0.9540,0.4185)
(0.9583,0.4174)

\PST@Border(0.4427,0.9160) (0.4427,0.0840) (0.9600,0.0840) (0.9600,0.9160)
(0.4427,0.9160)

\catcode`@=12 \fi \endpspicture

%% file: ratecomp.tex
\ifx\PSTloaded\undefined
\def\PSTloaded{t}
\psset{arrowsize=.01 3.2 1.4 .3}
\psset{dotsize=.01}
\catcode`@=11

\definecolor{darkgreen}{rgb}{0,.5,0}
\newpsobject{PST@Border}{psline}{linewidth=.0015,linestyle=solid}
\fi
\definecolor{violett}{rgb}{.5,0,.5}
\newpsobject{lineone}{psline}{linewidth=.002,linestyle=solid,linecolor=red}
\newpsobject{linetwo}{psline}{linewidth=.002,linestyle=dashed,linecolor=blue}
\newpsobject{linethr}{psline}{linewidth=.002,linestyle=dotted,dotsep=.004,linecolor=darkgreen}
\newpsobject{pointone}{psdots}{dotsize=.01,dotstyle=square*,linewidth=.005,linecolor=red}
\newpsobject{pointtwo}{psdots}{dotsize=.01,dotstyle=*,linewidth=.005,linecolor=blue}
\newpsobject{pointthr}{psdots}{dotsize=.01,dotstyle=pentagon,linewidth=.005,linecolor=violett}
\newpsobject{pointfou}{psdots}{dotsize=.01,dotstyle=oplus,linewidth=.005,linecolor=darkgreen}
\catcode`@=12
\psset{unit=\columnwidth,xunit=1.087\columnwidth,yunit=.6\columnwidth}
\pspicture(-0.025,0.0)(0.8950,1.0)
\ifx\nofigs\undefined
\catcode`@=11

\PST@Border(0.1370,0.1260) (0.1445,0.1260) \PST@Border(0.8770,0.1260)
(0.8695,0.1260) \PST@Border(0.1370,0.2025) (0.1520,0.2025)
\PST@Border(0.8770,0.2025) (0.8620,0.2025) \rput[r](0.1210,0.2025){$10^{-12}$}
\PST@Border(0.1370,0.2791) (0.1445,0.2791) \PST@Border(0.8770,0.2791)
(0.8695,0.2791) \PST@Border(0.1370,0.3556) (0.1520,0.3556)
\PST@Border(0.8770,0.3556) (0.8620,0.3556) \rput[r](0.1210,0.3556){$10^{-10}$}
\PST@Border(0.1370,0.4322) (0.1445,0.4322) \PST@Border(0.8770,0.4322)
(0.8695,0.4322) \PST@Border(0.1370,0.5087) (0.1520,0.5087)
\PST@Border(0.8770,0.5087) (0.8620,0.5087) \rput[r](0.1210,0.5087){$10^{-8}$}
\PST@Border(0.1370,0.5853) (0.1445,0.5853) \PST@Border(0.8770,0.5853)
(0.8695,0.5853) \PST@Border(0.1370,0.6618) (0.1520,0.6618)
\PST@Border(0.8770,0.6618) (0.8620,0.6618) \rput[r](0.1210,0.6618){$10^{-6}$}
\PST@Border(0.1370,0.7384) (0.1445,0.7384) \PST@Border(0.8770,0.7384)
(0.8695,0.7384) \PST@Border(0.1370,0.8149) (0.1520,0.8149)
\PST@Border(0.8770,0.8149) (0.8620,0.8149) \rput[r](0.1210,0.8149){$10^{-4}$}
\PST@Border(0.1370,0.8915) (0.1445,0.8915) \PST@Border(0.8770,0.8915)
(0.8695,0.8915) \PST@Border(0.1370,0.9680) (0.1520,0.9680)
\PST@Border(0.8770,0.9680) (0.8620,0.9680) \rput[r](0.1210,0.9680){ 0.01}
\PST@Border(0.1370,0.1260) (0.1370,0.1460) \PST@Border(0.1370,0.9680)
(0.1370,0.9480) \rput(0.1370,0.0840){ 0} \PST@Border(0.2603,0.1260)
(0.2603,0.1460) \PST@Border(0.2603,0.9680) (0.2603,0.9480)
\rput(0.2603,0.0840){ 0.5} \PST@Border(0.3837,0.1260) (0.3837,0.1460)
\PST@Border(0.3837,0.9680) (0.3837,0.9480) \rput(0.3837,0.0840){ 1}
\PST@Border(0.5070,0.1260) (0.5070,0.1460) \PST@Border(0.5070,0.9680)
(0.5070,0.9480) \rput(0.5070,0.0840){ 1.5} \PST@Border(0.6303,0.1260)
(0.6303,0.1460) \PST@Border(0.6303,0.9680) (0.6303,0.9480)
\rput(0.6303,0.0840){ 2} \PST@Border(0.7537,0.1260) (0.7537,0.1460)
\PST@Border(0.7537,0.9680) (0.7537,0.9480) \rput(0.7537,0.0840){ 2.5}
\PST@Border(0.8770,0.1260) (0.8770,0.1460) \PST@Border(0.8770,0.9680)
(0.8770,0.9480) \rput(0.8770,0.0840){ 3} \PST@Border(0.1370,0.9680)
(0.1370,0.1260) (0.8770,0.1260) (0.8770,0.9680) (0.1370,0.9680)

\rput{L}(0.0020,0.5470){$\frac{1}{2\pi}\frac{1}{E}\frac{dR}{dE}$ [GeV$^{-2}$ fm$^{-4}$]}
\rput(0.5070,0.0210){$E$~[GeV]}

\rput[l](0.1803,0.3125){$T = 150$~MeV}
\rput[l](0.1803,0.2475){Box: $V = (20~{\sf fm})^{3}$,}
\rput[l](0.2403,0.1825){$\pi/\rho/a_1$-gas}

\rput[r](0.7500,0.9070){Box-calculations (UrQMD)}
\rput[r](0.7500,0.8350){Hydro-rates (Turbide {\it et al.})}
\rput[r](0.7500,0.7210){$\pi\pi \rightarrow \gamma\rho$}
\rput[r](0.7500,0.6490){$\pi\rho \rightarrow \gamma\pi$}

\pointone(0.7818,0.9070)\pointtwo(0.7976,0.9070)
\lineone(0.7660,0.8400)(0.8450,0.8400)
\linetwo(0.7660,0.8300)(0.8450,0.8300)
\pointone(.7857,.7210)\lineone(0.8055,0.7210)(0.8450,0.7210)
\pointtwo(.7857,.6490)\linetwo(0.8055,0.6490)(0.8450,0.6490)

\pointone(0.1493,0.9257) \pointone(0.1740,0.8286)
\pointone(0.1987,0.7822) \pointone(0.2233,0.7467)
\pointone(0.2480,0.7158) \pointone(0.2727,0.6874)
\pointone(0.2973,0.6606) \pointone(0.3220,0.6347)
\pointone(0.3467,0.6097) \pointone(0.3713,0.5853)
\pointone(0.3960,0.5612) \pointone(0.4207,0.5375)
\pointone(0.4453,0.5135) \pointone(0.4700,0.4905)
\pointone(0.4947,0.4672) \pointone(0.5193,0.4438)
\pointone(0.5440,0.4214) \pointone(0.5687,0.3989)
\pointone(0.5933,0.3763) \pointone(0.6180,0.3559)
\pointone(0.6427,0.3379) \pointone(0.6673,0.3184)
\pointone(0.6920,0.2972) \pointone(0.7167,0.2750)
\pointone(0.7413,0.2594) \pointone(0.7660,0.2406)

\pointtwo(0.1570,0.4209) \pointtwo(0.1740,0.5748)
\pointtwo(0.1987,0.6510) \pointtwo(0.2233,0.6883)
\pointtwo(0.2480,0.7035) \pointtwo(0.2727,0.7062)
\pointtwo(0.2973,0.7016) \pointtwo(0.3220,0.6923)
\pointtwo(0.3467,0.6800) \pointtwo(0.3713,0.6655)
\pointtwo(0.3960,0.6496) \pointtwo(0.4207,0.6325)
\pointtwo(0.4453,0.6146) \pointtwo(0.4700,0.5961)
\pointtwo(0.4947,0.5769) \pointtwo(0.5193,0.5574)
\pointtwo(0.5440,0.5375) \pointtwo(0.5687,0.5172)
\pointtwo(0.5933,0.4970) \pointtwo(0.6180,0.4762)
\pointtwo(0.6427,0.4555) \pointtwo(0.6673,0.4352)
\pointtwo(0.6920,0.4131) \pointtwo(0.7167,0.3928)
\pointtwo(0.7413,0.3712) \pointtwo(0.7660,0.3513)
\pointtwo(0.7907,0.3293) \pointtwo(0.8153,0.3083)
\pointtwo(0.8400,0.2863)

\lineone(0.1493,0.9019) (0.1493,0.9019) (0.1740,0.8304) (0.1987,0.7856)
(0.2233,0.7493) (0.2480,0.7172) (0.2727,0.6877) (0.2973,0.6599) (0.3220,0.6335)
(0.3467,0.6079) (0.3713,0.5832) (0.3960,0.5590) (0.4207,0.5353) (0.4453,0.5121)
(0.4700,0.4892) (0.4947,0.4666) (0.5193,0.4443) (0.5440,0.4223) (0.5687,0.4004)
(0.5933,0.3788) (0.6180,0.3573) (0.6427,0.3359) (0.6673,0.3147) (0.6920,0.2936)
(0.7167,0.2727) (0.7413,0.2518) (0.7660,0.2311) (0.7907,0.2104) (0.8153,0.1898)
(0.8400,0.1693) (0.8647,0.1489) (0.8770,0.1388)

\linetwo(0.1543,0.2427) (0.1592,0.3720) (0.1641,0.4555) (0.1691,0.5136)
(0.1740,0.5560) (0.1789,0.5882) (0.1839,0.6132) (0.1888,0.6329) (0.1937,0.6488)
(0.1987,0.6616) (0.2233,0.6982)
(0.2480,0.7104) (0.2727,0.7112) (0.2973,0.7056) (0.3220,0.6961) (0.3467,0.6841)
(0.3713,0.6702) (0.3960,0.6550) (0.4207,0.6388) (0.4453,0.6219) (0.4700,0.6044)
(0.4947,0.5864) (0.5193,0.5680) (0.5440,0.5493) (0.5687,0.5304) (0.5933,0.5112)
(0.6180,0.4918) (0.6427,0.4722) (0.6673,0.4525) (0.6920,0.4327) (0.7167,0.4128)
(0.7413,0.3928) (0.7660,0.3727) (0.7907,0.3525) (0.8153,0.3322) (0.8400,0.3119)
(0.8647,0.2915) (0.8770,0.2813)

\PST@Border(0.1370,0.9680) (0.1370,0.1260) (0.8770,0.1260) (0.8770,0.9680)
(0.1370,0.9680)

\catcode`@=12
\fi
\endpspicture

%% file: stages_casc.tex
\ifx\PSTloaded\undefined
\def\PSTloaded{t}
\psset{arrowsize=.01 3.2 1.4 .3}
\psset{dotsize=.01}
\catcode`@=11

\definecolor{darkgreen}{rgb}{0,.5,0}
\newpsobject{PST@Border}{psline}{linewidth=.0015,linestyle=solid}
\newpsobject{PST@Axes}{psline}{linewidth=.0015,linestyle=dotted,dotsep=.004}
\newpsobject{PST@inc}{psline}{linecolor=black,    linewidth=.0025,linestyle=solid}
\newpsobject{PST@aft}{psline}{linecolor=red,      linewidth=.0025,linestyle=dashed,dash=.01 .01}
\newpsobject{PST@hyd}{psline}{linecolor=blue,     linewidth=.0035,linestyle=dotted,dotsep=.004}
\newpsobject{PST@bef}{psline}{linecolor=darkgreen,linewidth=.0025,linestyle=dashed,dash=.01 .004 .004 .004}
\catcode`@=12

\fi
\psset{unit=5.0in,xunit=1.075\columnwidth,yunit=.7\columnwidth}
\pspicture(0.050000,-.030)(0.980,1.000)
\ifx\nofigs\undefined
\catcode`@=11

\PST@Border(0.2070,0.1260) (0.2220,0.1260) \PST@Border(0.9470,0.1260)
(0.9320,0.1260) \rput[r](0.1910,0.1260){$10^{-7}$} \PST@Border(0.2070,0.1577)
(0.2145,0.1577) \PST@Border(0.9470,0.1577) (0.9395,0.1577)
\PST@Border(0.2070,0.1996) (0.2145,0.1996) \PST@Border(0.9470,0.1996)
(0.9395,0.1996) \PST@Border(0.2070,0.2211) (0.2145,0.2211)
\PST@Border(0.9470,0.2211) (0.9395,0.2211) \PST@Border(0.2070,0.2313)
(0.2220,0.2313) \PST@Border(0.9470,0.2313) (0.9320,0.2313)
\rput[r](0.1910,0.2313){$10^{-6}$} \PST@Border(0.2070,0.2629) (0.2145,0.2629)
\PST@Border(0.9470,0.2629) (0.9395,0.2629) \PST@Border(0.2070,0.3048)
(0.2145,0.3048) \PST@Border(0.9470,0.3048) (0.9395,0.3048)
\PST@Border(0.2070,0.3263) (0.2145,0.3263) \PST@Border(0.9470,0.3263)
(0.9395,0.3263) \PST@Border(0.2070,0.3365) (0.2220,0.3365)
\PST@Border(0.9470,0.3365) (0.9320,0.3365) \rput[r](0.1910,0.3365){$10^{-5}$}
\PST@Border(0.2070,0.3682) (0.2145,0.3682) \PST@Border(0.9470,0.3682)
(0.9395,0.3682) \PST@Border(0.2070,0.4101) (0.2145,0.4101)
\PST@Border(0.9470,0.4101) (0.9395,0.4101) \PST@Border(0.2070,0.4316)
(0.2145,0.4316) \PST@Border(0.9470,0.4316) (0.9395,0.4316)
\PST@Border(0.2070,0.4418) (0.2220,0.4418) \PST@Border(0.9470,0.4418)
(0.9320,0.4418) \rput[r](0.1910,0.4418){$10^{-4}$} \PST@Border(0.2070,0.4734)
(0.2145,0.4734) \PST@Border(0.9470,0.4734) (0.9395,0.4734)
\PST@Border(0.2070,0.5153) (0.2145,0.5153) \PST@Border(0.9470,0.5153)
(0.9395,0.5153) \PST@Border(0.2070,0.5368) (0.2145,0.5368)
\PST@Border(0.9470,0.5368) (0.9395,0.5368) \PST@Border(0.2070,0.5470)
(0.2220,0.5470) \PST@Border(0.9470,0.5470) (0.9320,0.5470)
\rput[r](0.1910,0.5470){ 0.001} \PST@Border(0.2070,0.5787) (0.2145,0.5787)
\PST@Border(0.9470,0.5787) (0.9395,0.5787) \PST@Border(0.2070,0.6206)
(0.2145,0.6206) \PST@Border(0.9470,0.6206) (0.9395,0.6206)
\PST@Border(0.2070,0.6421) (0.2145,0.6421) \PST@Border(0.9470,0.6421)
(0.9395,0.6421) \PST@Border(0.2070,0.6523) (0.2220,0.6523)
\PST@Border(0.9470,0.6523) (0.9320,0.6523) \rput[r](0.1910,0.6523){ 0.01}
\PST@Border(0.2070,0.6839) (0.2145,0.6839) \PST@Border(0.9470,0.6839)
(0.9395,0.6839) \PST@Border(0.2070,0.7258) (0.2145,0.7258)
\PST@Border(0.9470,0.7258) (0.9395,0.7258) \PST@Border(0.2070,0.7473)
(0.2145,0.7473) \PST@Border(0.9470,0.7473) (0.9395,0.7473)
\PST@Border(0.2070,0.7575) (0.2220,0.7575) \PST@Border(0.9470,0.7575)
(0.9320,0.7575) \rput[r](0.1910,0.7575){ 0.1} \PST@Border(0.2070,0.7892)
(0.2145,0.7892) \PST@Border(0.9470,0.7892) (0.9395,0.7892)
\PST@Border(0.2070,0.8311) (0.2145,0.8311) \PST@Border(0.9470,0.8311)
(0.9395,0.8311) \PST@Border(0.2070,0.8526) (0.2145,0.8526)
\PST@Border(0.9470,0.8526) (0.9395,0.8526) \PST@Border(0.2070,0.8628)
(0.2220,0.8628) \PST@Border(0.9470,0.8628) (0.9320,0.8628)
\rput[r](0.1910,0.8628){ 1} \PST@Border(0.2070,0.8944) (0.2145,0.8944)
\PST@Border(0.9470,0.8944) (0.9395,0.8944) \PST@Border(0.2070,0.9363)
(0.2145,0.9363) \PST@Border(0.9470,0.9363) (0.9395,0.9363)
\PST@Border(0.2070,0.9578) (0.2145,0.9578) \PST@Border(0.9470,0.9578)
(0.9395,0.9578) \PST@Border(0.2070,0.9680) (0.2220,0.9680)
\PST@Border(0.9470,0.9680) (0.9320,0.9680) \rput[r](0.1910,0.9680){ 10}
\PST@Border(0.2070,0.1260) (0.2070,0.1460) \PST@Border(0.2070,0.9680)
(0.2070,0.9480) \rput(0.2070,0.0840){ 0} \PST@Border(0.2892,0.1260)
(0.2892,0.1460) \PST@Border(0.2892,0.9680) (0.2892,0.9480)
\rput(0.2892,0.0840){ 0.5} \PST@Border(0.3714,0.1260) (0.3714,0.1460)
\PST@Border(0.3714,0.9680) (0.3714,0.9480) \rput(0.3714,0.0840){ 1}
\PST@Border(0.4537,0.1260) (0.4537,0.1460) \PST@Border(0.4537,0.9680)
(0.4537,0.9480) \rput(0.4537,0.0840){ 1.5} \PST@Border(0.5359,0.1260)
(0.5359,0.1460) \PST@Border(0.5359,0.9680) (0.5359,0.9480)
\rput(0.5359,0.0840){ 2} \PST@Border(0.6181,0.1260) (0.6181,0.1460)
\PST@Border(0.6181,0.9680) (0.6181,0.9480) \rput(0.6181,0.0840){ 2.5}
\PST@Border(0.7003,0.1260) (0.7003,0.1460) \PST@Border(0.7003,0.9680)
(0.7003,0.9480) \rput(0.7003,0.0840){ 3} \PST@Border(0.7826,0.1260)
(0.7826,0.1460) \PST@Border(0.7826,0.9680) (0.7826,0.9480)
\rput(0.7826,0.0840){ 3.5} \PST@Border(0.8648,0.1260) (0.8648,0.1460)
\PST@Border(0.8648,0.9680) (0.8648,0.9480) \rput(0.8648,0.0840){ 4}
\PST@Border(0.9470,0.1260) (0.9470,0.1460) \PST@Border(0.9470,0.9680)
(0.9470,0.9480) \rput(0.9470,0.0840){ 4.5} \PST@Border(0.2070,0.9680)
(0.2070,0.1260) (0.9470,0.1260) (0.9470,0.9680) (0.2070,0.9680)

\rput{L}(0.0820,0.5470){$E\frac{dN}{d^3p}$ [GeV$^{-2}$]}
\rput(0.5770,0.0210){$p_\bot$ [GeV]}

\rput[r](0.8200,0.9270){complete cascade}
\rput[r](0.8200,0.8850){initial stage}
\rput[r](0.8200,0.8430){intermediate stage}
\rput[r](0.8200,0.8010){final stage}

\PST@inc(0.8360,0.9270)(0.9150,0.9270)
\PST@bef(0.8360,0.8850) (0.9150,0.8850)
\PST@hyd(0.8360,0.8430)(0.9150,0.8430)
\PST@aft(0.8360,0.8010)(0.9150,0.8010)

\PST@inc(0.2257,0.9680) (0.2317,0.9146) (0.2481,0.8595) (0.2646,0.8342)
(0.2810,0.8175) (0.2974,0.8006) (0.3139,0.7817) (0.3303,0.7614) (0.3468,0.7399)
(0.3632,0.7180) (0.3797,0.6957) (0.3961,0.6731) (0.4126,0.6505) (0.4290,0.6283)
(0.4454,0.6059) (0.4619,0.5845) (0.4783,0.5621) (0.4948,0.5416) (0.5112,0.5212)
(0.5277,0.5022) (0.5441,0.4820) (0.5606,0.4634) (0.5770,0.4462) (0.5934,0.4300)
(0.6099,0.4127) (0.6263,0.3989) (0.6428,0.3835) (0.6592,0.3721) (0.6757,0.3603)
(0.6921,0.3474) (0.7086,0.3359) (0.7250,0.3260) (0.7414,0.3158) (0.7579,0.3064)
(0.7743,0.2953) (0.7908,0.2853) (0.8072,0.2747) (0.8237,0.2666) (0.8401,0.2611)
(0.8566,0.2525) (0.8730,0.2453) (0.8894,0.2386) (0.9059,0.2314) (0.9223,0.2282)
(0.9388,0.2202)

\PST@bef(0.2152,0.8956) (0.2152,0.8956) (0.2317,0.7636) (0.2481,0.7122)
(0.2646,0.6882) (0.2810,0.6715) (0.2974,0.6533) (0.3139,0.6363) (0.3303,0.6163)
(0.3468,0.5957) (0.3632,0.5770) (0.3797,0.5562) (0.3961,0.5376) (0.4126,0.5194)
(0.4290,0.5042) (0.4454,0.4879) (0.4619,0.4734) (0.4783,0.4594) (0.4948,0.4457)
(0.5112,0.4346) (0.5277,0.4224) (0.5441,0.4100) (0.5606,0.3995) (0.5770,0.3892)
(0.5934,0.3825) (0.6099,0.3699) (0.6263,0.3563) (0.6428,0.3468) (0.6592,0.3404)
(0.6757,0.3324) (0.6921,0.3268) (0.7086,0.3182) (0.7250,0.3116) (0.7414,0.3006)
(0.7579,0.2968) (0.7743,0.2827) (0.7908,0.2753) (0.8072,0.2643) (0.8237,0.2638)
(0.8401,0.2406) (0.8566,0.2464) (0.8730,0.2391) (0.8894,0.2299) (0.9059,0.2245)
(0.9223,0.2231) (0.9388,0.2167)

\PST@hyd(0.2240,0.9680) (0.2317,0.9000) (0.2481,0.8458) (0.2646,0.8226)
(0.2810,0.8075) (0.2974,0.7914) (0.3139,0.7730) (0.3303,0.7529) (0.3468,0.7315)
(0.3632,0.7092) (0.3797,0.6865) (0.3961,0.6635) (0.4126,0.6404) (0.4290,0.6171)
(0.4454,0.5946) (0.4619,0.5719) (0.4783,0.5485) (0.4948,0.5264) (0.5112,0.5042)
(0.5277,0.4831) (0.5441,0.4608) (0.5606,0.4395) (0.5770,0.4188) (0.5934,0.3986)
(0.6099,0.3751) (0.6263,0.3569) (0.6428,0.3379) (0.6592,0.3231) (0.6757,0.3054)
(0.6921,0.2932) (0.7086,0.2741) (0.7250,0.2543) (0.7414,0.2418) (0.7579,0.2243)
(0.7743,0.2041) (0.7908,0.1833) (0.8072,0.1864) (0.8237,0.1813) (0.8401,0.1729)
(0.8566,0.1576) (0.8730,0.1512) (0.8894,0.1584) (0.9059,0.1417) (0.9217,0.1260)

\PST@aft(0.2188,0.9680) (0.2317,0.8488) (0.2481,0.7900) (0.2646,0.7567)
(0.2810,0.7319) (0.2974,0.7102) (0.3139,0.6888) (0.3303,0.6674) (0.3468,0.6457)
(0.3632,0.6246) (0.3797,0.6038) (0.3961,0.5822) (0.4126,0.5605) (0.4290,0.5398)
(0.4454,0.5185) (0.4619,0.4965) (0.4783,0.4754) (0.4948,0.4535) (0.5112,0.4332)
(0.5277,0.4152) (0.5441,0.3972) (0.5606,0.3744) (0.5770,0.3565) (0.5934,0.3384)
(0.6099,0.3215) (0.6263,0.3002) (0.6428,0.2847) (0.6592,0.2625) (0.6757,0.2441)
(0.6921,0.2022) (0.7086,0.1890) (0.7250,0.1753) (0.7414,0.1549) (0.7579,0.1595)
(0.7743,0.1643) (0.7836,0.1260)

\PST@Border(0.2070,0.9680) (0.2070,0.1260) (0.9470,0.1260) (0.9470,0.9680)
(0.2070,0.9680)
\rput[r](0.9150,0.7430){UrQMD}
\rput[r](0.9150,0.6910){Pb+Pb 158 AGeV}
\rput[r](0.9150,0.6390){$b < 4.5$~fm, $|y_{\rm c.m.}| < 0.5$}

\catcode`@=12
\fi
\endpspicture

%% file: stages_hg.tex
\ifx\PSTloaded\undefined
\def\PSTloaded{t}
\psset{arrowsize=.01 3.2 1.4 .3}
\psset{dotsize=.01}
\catcode`@=11

\definecolor{darkgreen}{rgb}{0,.5,0}
\newpsobject{PST@Border}{psline}{linewidth=.0015,linestyle=solid}
\newpsobject{PST@Axes}{psline}{linewidth=.0015,linestyle=dotted,dotsep=.004}
\newpsobject{PST@inc}{psline}{linecolor=black,    linewidth=.0025,linestyle=solid}
\newpsobject{PST@aft}{psline}{linecolor=red,      linewidth=.0025,linestyle=dashed,dash=.01 .01}
\newpsobject{PST@hyd}{psline}{linecolor=blue,     linewidth=.0035,linestyle=dotted,dotsep=.004}
\newpsobject{PST@bef}{psline}{linecolor=darkgreen,linewidth=.0025,linestyle=dashed,dash=.01 .004 .004 .004}
\catcode`@=12

\fi
\psset{unit=5.0in,xunit=1.075\columnwidth,yunit=.7\columnwidth}
\pspicture(0.050000,-.030)(0.980,1.000)
\ifx\nofigs\undefined
\catcode`@=11

\PST@Border(0.2070,0.1260) (0.2220,0.1260) \PST@Border(0.9470,0.1260)
(0.9320,0.1260) \rput[r](0.1910,0.1260){$10^{-7}$} \PST@Border(0.2070,0.1577)
(0.2145,0.1577) \PST@Border(0.9470,0.1577) (0.9395,0.1577)
\PST@Border(0.2070,0.1996) (0.2145,0.1996) \PST@Border(0.9470,0.1996)
(0.9395,0.1996) \PST@Border(0.2070,0.2211) (0.2145,0.2211)
\PST@Border(0.9470,0.2211) (0.9395,0.2211) \PST@Border(0.2070,0.2313)
(0.2220,0.2313) \PST@Border(0.9470,0.2313) (0.9320,0.2313)
\rput[r](0.1910,0.2313){$10^{-6}$} \PST@Border(0.2070,0.2629) (0.2145,0.2629)
\PST@Border(0.9470,0.2629) (0.9395,0.2629) \PST@Border(0.2070,0.3048)
(0.2145,0.3048) \PST@Border(0.9470,0.3048) (0.9395,0.3048)
\PST@Border(0.2070,0.3263) (0.2145,0.3263) \PST@Border(0.9470,0.3263)
(0.9395,0.3263) \PST@Border(0.2070,0.3365) (0.2220,0.3365)
\PST@Border(0.9470,0.3365) (0.9320,0.3365) \rput[r](0.1910,0.3365){$10^{-5}$}
\PST@Border(0.2070,0.3682) (0.2145,0.3682) \PST@Border(0.9470,0.3682)
(0.9395,0.3682) \PST@Border(0.2070,0.4101) (0.2145,0.4101)
\PST@Border(0.9470,0.4101) (0.9395,0.4101) \PST@Border(0.2070,0.4316)
(0.2145,0.4316) \PST@Border(0.9470,0.4316) (0.9395,0.4316)
\PST@Border(0.2070,0.4418) (0.2220,0.4418) \PST@Border(0.9470,0.4418)
(0.9320,0.4418) \rput[r](0.1910,0.4418){$10^{-4}$} \PST@Border(0.2070,0.4734)
(0.2145,0.4734) \PST@Border(0.9470,0.4734) (0.9395,0.4734)
\PST@Border(0.2070,0.5153) (0.2145,0.5153) \PST@Border(0.9470,0.5153)
(0.9395,0.5153) \PST@Border(0.2070,0.5368) (0.2145,0.5368)
\PST@Border(0.9470,0.5368) (0.9395,0.5368) \PST@Border(0.2070,0.5470)
(0.2220,0.5470) \PST@Border(0.9470,0.5470) (0.9320,0.5470)
\rput[r](0.1910,0.5470){ 0.001} \PST@Border(0.2070,0.5787) (0.2145,0.5787)
\PST@Border(0.9470,0.5787) (0.9395,0.5787) \PST@Border(0.2070,0.6206)
(0.2145,0.6206) \PST@Border(0.9470,0.6206) (0.9395,0.6206)
\PST@Border(0.2070,0.6421) (0.2145,0.6421) \PST@Border(0.9470,0.6421)
(0.9395,0.6421) \PST@Border(0.2070,0.6523) (0.2220,0.6523)
\PST@Border(0.9470,0.6523) (0.9320,0.6523) \rput[r](0.1910,0.6523){ 0.01}
\PST@Border(0.2070,0.6839) (0.2145,0.6839) \PST@Border(0.9470,0.6839)
(0.9395,0.6839) \PST@Border(0.2070,0.7258) (0.2145,0.7258)
\PST@Border(0.9470,0.7258) (0.9395,0.7258) \PST@Border(0.2070,0.7473)
(0.2145,0.7473) \PST@Border(0.9470,0.7473) (0.9395,0.7473)
\PST@Border(0.2070,0.7575) (0.2220,0.7575) \PST@Border(0.9470,0.7575)
(0.9320,0.7575) \rput[r](0.1910,0.7575){ 0.1} \PST@Border(0.2070,0.7892)
(0.2145,0.7892) \PST@Border(0.9470,0.7892) (0.9395,0.7892)
\PST@Border(0.2070,0.8311) (0.2145,0.8311) \PST@Border(0.9470,0.8311)
(0.9395,0.8311) \PST@Border(0.2070,0.8526) (0.2145,0.8526)
\PST@Border(0.9470,0.8526) (0.9395,0.8526) \PST@Border(0.2070,0.8628)
(0.2220,0.8628) \PST@Border(0.9470,0.8628) (0.9320,0.8628)
\rput[r](0.1910,0.8628){ 1} \PST@Border(0.2070,0.8944) (0.2145,0.8944)
\PST@Border(0.9470,0.8944) (0.9395,0.8944) \PST@Border(0.2070,0.9363)
(0.2145,0.9363) \PST@Border(0.9470,0.9363) (0.9395,0.9363)
\PST@Border(0.2070,0.9578) (0.2145,0.9578) \PST@Border(0.9470,0.9578)
(0.9395,0.9578) \PST@Border(0.2070,0.9680) (0.2220,0.9680)
\PST@Border(0.9470,0.9680) (0.9320,0.9680) \rput[r](0.1910,0.9680){ 10}
\PST@Border(0.2070,0.1260) (0.2070,0.1460) \PST@Border(0.2070,0.9680)
(0.2070,0.9480) \rput(0.2070,0.0840){ 0} \PST@Border(0.2892,0.1260)
(0.2892,0.1460) \PST@Border(0.2892,0.9680) (0.2892,0.9480)
\rput(0.2892,0.0840){ 0.5} \PST@Border(0.3714,0.1260) (0.3714,0.1460)
\PST@Border(0.3714,0.9680) (0.3714,0.9480) \rput(0.3714,0.0840){ 1}
\PST@Border(0.4537,0.1260) (0.4537,0.1460) \PST@Border(0.4537,0.9680)
(0.4537,0.9480) \rput(0.4537,0.0840){ 1.5} \PST@Border(0.5359,0.1260)
(0.5359,0.1460) \PST@Border(0.5359,0.9680) (0.5359,0.9480)
\rput(0.5359,0.0840){ 2} \PST@Border(0.6181,0.1260) (0.6181,0.1460)
\PST@Border(0.6181,0.9680) (0.6181,0.9480) \rput(0.6181,0.0840){ 2.5}
\PST@Border(0.7003,0.1260) (0.7003,0.1460) \PST@Border(0.7003,0.9680)
(0.7003,0.9480) \rput(0.7003,0.0840){ 3} \PST@Border(0.7826,0.1260)
(0.7826,0.1460) \PST@Border(0.7826,0.9680) (0.7826,0.9480)
\rput(0.7826,0.0840){ 3.5} \PST@Border(0.8648,0.1260) (0.8648,0.1460)
\PST@Border(0.8648,0.9680) (0.8648,0.9480) \rput(0.8648,0.0840){ 4}
\PST@Border(0.9470,0.1260) (0.9470,0.1460) \PST@Border(0.9470,0.9680)
(0.9470,0.9480) \rput(0.9470,0.0840){ 4.5} \PST@Border(0.2070,0.9680)
(0.2070,0.1260) (0.9470,0.1260) (0.9470,0.9680) (0.2070,0.9680)

\rput{L}(0.0820,0.5470){$E\frac{dN}{d^3p}$ [GeV$^{-2}$]}
\rput(0.5770,0.0210){$p_\bot$ [GeV]}
\rput[r](0.8200,0.9270){complete hybrid}
\PST@inc(0.8360,0.9270)
(0.9150,0.9270)

\PST@inc(0.2444,0.9680) (0.2481,0.9594) (0.2646,0.8752) (0.2810,0.8336)
(0.2974,0.8037) (0.3139,0.7796) (0.3303,0.7623) (0.3468,0.7272) (0.3632,0.7178)
(0.3797,0.6949) (0.3961,0.6681) (0.4126,0.6620) (0.4290,0.6312) (0.4454,0.6217)
(0.4619,0.5929) (0.4783,0.5808) (0.4948,0.5600) (0.5112,0.5367) (0.5277,0.5316)
(0.5441,0.5044) (0.5606,0.4912) (0.5770,0.4671) (0.5934,0.4504) (0.6099,0.4435)
(0.6263,0.4053) (0.6428,0.3986) (0.6592,0.3814) (0.6757,0.3666) (0.6921,0.3597)
(0.7086,0.3415) (0.7250,0.3264) (0.7414,0.3113) (0.7579,0.3026) (0.7743,0.2884)
(0.7908,0.2773) (0.8072,0.2677) (0.8237,0.2639) (0.8401,0.2406) (0.8566,0.2464)
(0.8730,0.2391) (0.8894,0.2299) (0.9059,0.2245) (0.9223,0.2231) (0.9388,0.2167)
(0.9470,0.2117)

\rput[r](0.8200,0.8850){initial stage}
\PST@bef(0.8360,0.8850)
(0.9150,0.8850)

\PST@bef(0.2152,0.8956) (0.2152,0.8956) (0.2317,0.7636) (0.2481,0.7122)
(0.2646,0.6882) (0.2810,0.6715) (0.2974,0.6533) (0.3139,0.6363) (0.3303,0.6163)
(0.3468,0.5957) (0.3632,0.5770) (0.3797,0.5562) (0.3961,0.5376) (0.4126,0.5194)
(0.4290,0.5042) (0.4454,0.4879) (0.4619,0.4734) (0.4783,0.4594) (0.4948,0.4457)
(0.5112,0.4346) (0.5277,0.4224) (0.5441,0.4100) (0.5606,0.3995) (0.5770,0.3892)
(0.5934,0.3825) (0.6099,0.3699) (0.6263,0.3563) (0.6428,0.3468) (0.6592,0.3404)
(0.6757,0.3324) (0.6921,0.3268) (0.7086,0.3182) (0.7250,0.3116) (0.7414,0.3006)
(0.7579,0.2968) (0.7743,0.2827) (0.7908,0.2753) (0.8072,0.2643) (0.8237,0.2638)
(0.8401,0.2406) (0.8566,0.2464) (0.8730,0.2391) (0.8894,0.2299) (0.9059,0.2245)
(0.9223,0.2231) (0.9388,0.2167) (0.9470,0.2117)

\rput[r](0.8200,0.8430){intermediate stage}
\PST@hyd(0.8360,0.8430)
(0.9150,0.8430)

\PST@hyd(0.2432,0.9680) (0.2481,0.9570) (0.2646,0.8671) (0.2810,0.8206)
(0.2974,0.7865) (0.3139,0.7592) (0.3303,0.7428) (0.3468,0.6956) (0.3632,0.6939)
(0.3797,0.6685) (0.3961,0.6349) (0.4126,0.6398) (0.4290,0.5996) (0.4454,0.5975)
(0.4619,0.5609) (0.4783,0.5542) (0.4948,0.5303) (0.5112,0.5008) (0.5277,0.5058)
(0.5441,0.4691) (0.5606,0.4592) (0.5770,0.4263) (0.5934,0.4054) (0.6099,0.4060)
(0.6263,0.3549) (0.6428,0.3547) (0.6592,0.3100) (0.6757,0.2849) (0.6921,0.2609)
(0.7086,0.2136) (0.7250,0.1949) (0.7399,0.1260)

\rput[r](0.8200,0.8010){final stage}
\PST@aft(0.8360,0.8010)
(0.9150,0.8010)

\PST@aft(0.2219,0.9680) (0.2317,0.8809) (0.2481,0.8212) (0.2646,0.7873)
(0.2810,0.7642) (0.2974,0.7450) (0.3139,0.7269) (0.3303,0.7084) (0.3468,0.6899)
(0.3632,0.6711) (0.3797,0.6520) (0.3961,0.6325) (0.4126,0.6129) (0.4290,0.5935)
(0.4454,0.5746) (0.4619,0.5543) (0.4783,0.5355) (0.4948,0.5176) (0.5112,0.4988)
(0.5277,0.4822) (0.5441,0.4638) (0.5606,0.4455) (0.5770,0.4263) (0.5934,0.4085)
(0.6099,0.3968) (0.6263,0.3541) (0.6428,0.3428) (0.6592,0.3375) (0.6757,0.3198)
(0.6921,0.3177) (0.7086,0.2918) (0.7250,0.2574) (0.7414,0.2365) (0.7579,0.2006)
(0.7743,0.1891) (0.7908,0.1321) (0.8072,0.1479) (0.8093,0.1260)

\rput[r](0.9150,0.7430){Hybrid, Hadron gas EoS}
\rput[r](0.9150,0.6910){Pb+Pb 158 AGeV}
\rput[r](0.9150,0.6390){$b < 4.5$~fm, $|y_{\rm c.m.}| < 0.5$}

\PST@Border(0.2070,0.9680) (0.2070,0.1260) (0.9470,0.1260) (0.9470,0.9680)
(0.2070,0.9680)

\catcode`@=12
\fi
\endpspicture

%% file: stages_bm.tex
\ifx\PSTloaded\undefined
\def\PSTloaded{t}
\psset{arrowsize=.01 3.2 1.4 .3}
\psset{dotsize=.01}
\catcode`@=11

\definecolor{darkgreen}{rgb}{0,.5,0}
\newpsobject{PST@Border}{psline}{linewidth=.0015,linestyle=solid}
\newpsobject{PST@Axes}{psline}{linewidth=.0015,linestyle=dotted,dotsep=.004}
\newpsobject{PST@inc}{psline}{linecolor=black,    linewidth=.0025,linestyle=solid}
\newpsobject{PST@aft}{psline}{linecolor=red,      linewidth=.0025,linestyle=dashed,dash=.01 .01}
\newpsobject{PST@hyd}{psline}{linecolor=blue,     linewidth=.0035,linestyle=dotted,dotsep=.004}
\newpsobject{PST@bef}{psline}{linecolor=darkgreen,linewidth=.0025,linestyle=dashed,dash=.01 .004 .004 .004}

\catcode`@=12

\fi
\psset{unit=5.0in,xunit=1.075\columnwidth,yunit=.7\columnwidth}
\pspicture(0.050000,-.030)(0.980,1.000)
\ifx\nofigs\undefined
\catcode`@=11

\PST@Border(0.2070,0.1260) (0.2220,0.1260) \PST@Border(0.9470,0.1260)
(0.9320,0.1260) \rput[r](0.1910,0.1260){$10^{-7}$} \PST@Border(0.2070,0.1577)
(0.2145,0.1577) \PST@Border(0.9470,0.1577) (0.9395,0.1577)
\PST@Border(0.2070,0.1996) (0.2145,0.1996) \PST@Border(0.9470,0.1996)
(0.9395,0.1996) \PST@Border(0.2070,0.2211) (0.2145,0.2211)
\PST@Border(0.9470,0.2211) (0.9395,0.2211) \PST@Border(0.2070,0.2313)
(0.2220,0.2313) \PST@Border(0.9470,0.2313) (0.9320,0.2313)
\rput[r](0.1910,0.2313){$10^{-6}$} \PST@Border(0.2070,0.2629) (0.2145,0.2629)
\PST@Border(0.9470,0.2629) (0.9395,0.2629) \PST@Border(0.2070,0.3048)
(0.2145,0.3048) \PST@Border(0.9470,0.3048) (0.9395,0.3048)
\PST@Border(0.2070,0.3263) (0.2145,0.3263) \PST@Border(0.9470,0.3263)
(0.9395,0.3263) \PST@Border(0.2070,0.3365) (0.2220,0.3365)
\PST@Border(0.9470,0.3365) (0.9320,0.3365) \rput[r](0.1910,0.3365){$10^{-5}$}
\PST@Border(0.2070,0.3682) (0.2145,0.3682) \PST@Border(0.9470,0.3682)
(0.9395,0.3682) \PST@Border(0.2070,0.4101) (0.2145,0.4101)
\PST@Border(0.9470,0.4101) (0.9395,0.4101) \PST@Border(0.2070,0.4316)
(0.2145,0.4316) \PST@Border(0.9470,0.4316) (0.9395,0.4316)
\PST@Border(0.2070,0.4418) (0.2220,0.4418) \PST@Border(0.9470,0.4418)
(0.9320,0.4418) \rput[r](0.1910,0.4418){$10^{-4}$} \PST@Border(0.2070,0.4734)
(0.2145,0.4734) \PST@Border(0.9470,0.4734) (0.9395,0.4734)
\PST@Border(0.2070,0.5153) (0.2145,0.5153) \PST@Border(0.9470,0.5153)
(0.9395,0.5153) \PST@Border(0.2070,0.5368) (0.2145,0.5368)
\PST@Border(0.9470,0.5368) (0.9395,0.5368) \PST@Border(0.2070,0.5470)
(0.2220,0.5470) \PST@Border(0.9470,0.5470) (0.9320,0.5470)
\rput[r](0.1910,0.5470){0.001} \PST@Border(0.2070,0.5787) (0.2145,0.5787)
\PST@Border(0.9470,0.5787) (0.9395,0.5787) \PST@Border(0.2070,0.6206)
(0.2145,0.6206) \PST@Border(0.9470,0.6206) (0.9395,0.6206)
\PST@Border(0.2070,0.6421) (0.2145,0.6421) \PST@Border(0.9470,0.6421)
(0.9395,0.6421) \PST@Border(0.2070,0.6523) (0.2220,0.6523)
\PST@Border(0.9470,0.6523) (0.9320,0.6523) \rput[r](0.1910,0.6523){ 0.01}
\PST@Border(0.2070,0.6839) (0.2145,0.6839) \PST@Border(0.9470,0.6839)
(0.9395,0.6839) \PST@Border(0.2070,0.7258) (0.2145,0.7258)
\PST@Border(0.9470,0.7258) (0.9395,0.7258) \PST@Border(0.2070,0.7473)
(0.2145,0.7473) \PST@Border(0.9470,0.7473) (0.9395,0.7473)
\PST@Border(0.2070,0.7575) (0.2220,0.7575) \PST@Border(0.9470,0.7575)
(0.9320,0.7575) \rput[r](0.1910,0.7575){ 0.1} \PST@Border(0.2070,0.7892)
(0.2145,0.7892) \PST@Border(0.9470,0.7892) (0.9395,0.7892)
\PST@Border(0.2070,0.8311) (0.2145,0.8311) \PST@Border(0.9470,0.8311)
(0.9395,0.8311) \PST@Border(0.2070,0.8526) (0.2145,0.8526)
\PST@Border(0.9470,0.8526) (0.9395,0.8526) \PST@Border(0.2070,0.8628)
(0.2220,0.8628) \PST@Border(0.9470,0.8628) (0.9320,0.8628)
\rput[r](0.1910,0.8628){ 1} \PST@Border(0.2070,0.8944) (0.2145,0.8944)
\PST@Border(0.9470,0.8944) (0.9395,0.8944) \PST@Border(0.2070,0.9363)
(0.2145,0.9363) \PST@Border(0.9470,0.9363) (0.9395,0.9363)
\PST@Border(0.2070,0.9578) (0.2145,0.9578) \PST@Border(0.9470,0.9578)
(0.9395,0.9578) \PST@Border(0.2070,0.9680) (0.2220,0.9680)
\PST@Border(0.9470,0.9680) (0.9320,0.9680) \rput[r](0.1910,0.9680){ 10}
\PST@Border(0.2070,0.1260) (0.2070,0.1460) \PST@Border(0.2070,0.9680)
(0.2070,0.9480) \rput(0.2070,0.0840){ 0} \PST@Border(0.2892,0.1260)
(0.2892,0.1460) \PST@Border(0.2892,0.9680) (0.2892,0.9480)
\rput(0.2892,0.0840){ 0.5} \PST@Border(0.3714,0.1260) (0.3714,0.1460)
\PST@Border(0.3714,0.9680) (0.3714,0.9480) \rput(0.3714,0.0840){ 1}
\PST@Border(0.4537,0.1260) (0.4537,0.1460) \PST@Border(0.4537,0.9680)
(0.4537,0.9480) \rput(0.4537,0.0840){ 1.5} \PST@Border(0.5359,0.1260)
(0.5359,0.1460) \PST@Border(0.5359,0.9680) (0.5359,0.9480)
\rput(0.5359,0.0840){ 2} \PST@Border(0.6181,0.1260) (0.6181,0.1460)
\PST@Border(0.6181,0.9680) (0.6181,0.9480) \rput(0.6181,0.0840){ 2.5}
\PST@Border(0.7003,0.1260) (0.7003,0.1460) \PST@Border(0.7003,0.9680)
(0.7003,0.9480) \rput(0.7003,0.0840){ 3} \PST@Border(0.7826,0.1260)
(0.7826,0.1460) \PST@Border(0.7826,0.9680) (0.7826,0.9480)
\rput(0.7826,0.0840){ 3.5} \PST@Border(0.8648,0.1260) (0.8648,0.1460)
\PST@Border(0.8648,0.9680) (0.8648,0.9480) \rput(0.8648,0.0840){ 4}
\PST@Border(0.9470,0.1260) (0.9470,0.1460) \PST@Border(0.9470,0.9680)
(0.9470,0.9480) \rput(0.9470,0.0840){ 4.5} \PST@Border(0.2070,0.9680)
(0.2070,0.1260) (0.9470,0.1260) (0.9470,0.9680) (0.2070,0.9680)

\rput{L}(0.0820,0.5470){$E\frac{dN}{d^3p}$ [GeV$^{-2}$]}
\rput(0.5770,0.0210){$p_\bot$ [GeV]}

\rput[r](0.8200,0.9270){complete hybrid}
\PST@inc(0.8360,0.9270)(0.9150,0.9270)

\PST@inc(0.2578,0.9680) (0.2646,0.9341) (0.2810,0.8941) (0.2974,0.8653)
(0.3139,0.8397) (0.3303,0.8244) (0.3468,0.7766) (0.3632,0.7747) (0.3797,0.7479)
(0.3961,0.7130) (0.4126,0.7172) (0.4290,0.6760) (0.4454,0.6730) (0.4619,0.6349)
(0.4783,0.6284) (0.4948,0.6035) (0.5112,0.5738) (0.5277,0.5782) (0.5441,0.5421)
(0.5606,0.5327) (0.5770,0.5006) (0.5934,0.4808) (0.6099,0.4815) (0.6263,0.4366)
(0.6428,0.4360) (0.6592,0.4010) (0.6757,0.3844) (0.6921,0.3697) (0.7086,0.3472)
(0.7250,0.3386) (0.7414,0.3124) (0.7579,0.3097) (0.7743,0.3017) (0.7908,0.2863)
(0.8072,0.2851) (0.8237,0.2648) (0.8401,0.2610) (0.8566,0.2480) (0.8730,0.2435)
(0.8894,0.2308) (0.9059,0.2200) (0.9223,0.2188) (0.9388,0.2091) (0.9470,0.2073)

\rput[r](0.8200,0.8850){initial stage}
\PST@bef(0.8360,0.8850)
(0.9150,0.8850)

\PST@bef(0.2152,0.8939) (0.2152,0.8939) (0.2317,0.7638) (0.2481,0.7149)
(0.2646,0.6899) (0.2810,0.6729) (0.2974,0.6555) (0.3139,0.6385) (0.3303,0.6166)
(0.3468,0.5966) (0.3632,0.5771) (0.3797,0.5559) (0.3961,0.5400) (0.4126,0.5211)
(0.4290,0.5082) (0.4454,0.4909) (0.4619,0.4757) (0.4783,0.4593) (0.4948,0.4474)
(0.5112,0.4349) (0.5277,0.4210) (0.5441,0.4149) (0.5606,0.4049) (0.5770,0.3921)
(0.5934,0.3794) (0.6099,0.3718) (0.6263,0.3610) (0.6428,0.3489) (0.6592,0.3413)
(0.6757,0.3335) (0.6921,0.3169) (0.7086,0.3150) (0.7250,0.3067) (0.7414,0.2980)
(0.7579,0.2941) (0.7743,0.2912) (0.7908,0.2817) (0.8072,0.2800) (0.8237,0.2630)
(0.8401,0.2591) (0.8566,0.2470) (0.8730,0.2430) (0.8894,0.2308) (0.9059,0.2200)
(0.9223,0.2188) (0.9388,0.2091) (0.9470,0.2073)

\rput[r](0.8200,0.8430){intermediate stage}
\PST@hyd(0.8360,0.8430)(0.9150,0.8430)

\PST@hyd(0.2577,0.9680) (0.2646,0.9337) (0.2810,0.8935) (0.2974,0.8645)
(0.3139,0.8389) (0.3303,0.8237) (0.3468,0.7754) (0.3632,0.7739) (0.3797,0.7470)
(0.3961,0.7117) (0.4126,0.7164) (0.4290,0.6745) (0.4454,0.6720) (0.4619,0.6332)
(0.4783,0.6271) (0.4948,0.6018) (0.5112,0.5711) (0.5277,0.5765) (0.5441,0.5388)
(0.5606,0.5296) (0.5770,0.4957) (0.5934,0.4751) (0.6099,0.4767) (0.6263,0.4257)
(0.6428,0.4285) (0.6592,0.3864) (0.6757,0.3659) (0.6921,0.3519) (0.7086,0.3156)
(0.7250,0.3068) (0.7414,0.2526) (0.7579,0.2519) (0.7743,0.2291) (0.7908,0.1787)
(0.8072,0.1825) (0.8214,0.1260)

\rput[r](0.8200,0.8010){final stage}
\PST@aft(0.8360,0.8010)(0.9150,0.8010)

\PST@aft(0.2152,0.9501) (0.2152,0.9501) (0.2317,0.7908) (0.2481,0.7253)
(0.2646,0.6910) (0.2810,0.6621) (0.2974,0.6365) (0.3139,0.6093) (0.3303,0.5832)
(0.3468,0.5568) (0.3632,0.5316) (0.3797,0.5058) (0.3961,0.4830) (0.4126,0.4542)
(0.4290,0.4353) (0.4454,0.4176) (0.4619,0.3992) (0.4783,0.3678) (0.4948,0.3562)
(0.5112,0.3538) (0.5277,0.3267) (0.5441,0.3159) (0.5606,0.2906) (0.5770,0.2802)
(0.5934,0.2688) (0.6099,0.2664) (0.6263,0.2584) (0.6428,0.1875) (0.6572,0.1260)
\PST@aft(0.6705,0.1260) (0.6757,0.1298) (0.6921,0.1396) (0.7001,0.1260)

\PST@Border(0.2070,0.9680)
(0.2070,0.1260)
(0.9470,0.1260)
(0.9470,0.9680)
(0.2070,0.9680)

\rput[r](0.9150,0.7430){Hybrid, Bagmodel EoS}
\rput[r](0.9150,0.6910){Pb+Pb 158 AGeV}
\rput[r](0.9150,0.6390){$b < 4.5$~fm, $|y_{\rm c.m.}| < 0.5$}

\catcode`@=12
\fi
\endpspicture

%% file: cascade.tex
\ifx\PSTloaded\undefined
\def\PSTloaded{t}
\psset{arrowsize=.01 3.2 1.4 .3}
\psset{dotsize=.01}
\catcode`@=11

\definecolor{darkgreen}{rgb}{0,.5,0}
\definecolor{violett}{rgb}{.5,0,.5}
\definecolor{orange}{rgb}{.8,.4,0}
\newpsobject{PST@Border}{psline}{linewidth=.0015,linestyle=solid}
\newpsobject{PST@Axes}{psline}{linewidth=.0015,linestyle=dotted,dotsep=.004}
\newpsobject{PST@inklusiv}{psline}{linecolor=black,    linewidth=.0025,linestyle=solid}
\newpsobject{PST@pirhogpi}{psline}{linecolor=red,      linewidth=.0025,linestyle=dashed,dash=.01 .01}
\newpsobject{PST@pipigrho}{psline}{linecolor=blue,     linewidth=.0035,linestyle=dotted,dotsep=.004}
\newpsobject{PST@etachann}{psline}{linecolor=darkgreen,linewidth=.0025,linestyle=dashed,dash=.01 .004 .004 .004}
\newpsobject{PST@gammagam}{psline}{linecolor=orange,   linewidth=.0025,linestyle=dashed,dash=.02 .01}
\catcode`@=12

\fi
\psset{unit=5.0in,xunit=1.075\columnwidth,yunit=.8\columnwidth}
\pspicture(0.0500,0.0220)(0.980000,0.9880)
\ifx\nofigs\undefined
\catcode`@=11

\PST@Border(0.2070,0.1260) (0.2220,0.1260) \PST@Border(0.9470,0.1260)
(0.9320,0.1260) \PST@Border(0.2070,0.1622)
(0.2145,0.1622) \PST@Border(0.9470,0.1622) (0.9395,0.1622)
\PST@Border(0.2070,0.2101) (0.2145,0.2101) \PST@Border(0.9470,0.2101)
(0.9395,0.2101) \PST@Border(0.2070,0.2346) (0.2145,0.2346)
\PST@Border(0.9470,0.2346) (0.9395,0.2346) \PST@Border(0.2070,0.2463)
(0.2220,0.2463) \PST@Border(0.9470,0.2463) (0.9320,0.2463)
\rput[r](0.1910,0.2463){$10^{-6}$} \PST@Border(0.2070,0.2825) (0.2145,0.2825)
\PST@Border(0.9470,0.2825) (0.9395,0.2825) \PST@Border(0.2070,0.3304)
(0.2145,0.3304) \PST@Border(0.9470,0.3304) (0.9395,0.3304)
\PST@Border(0.2070,0.3549) (0.2145,0.3549) \PST@Border(0.9470,0.3549)
(0.9395,0.3549) \PST@Border(0.2070,0.3666) (0.2220,0.3666)
\PST@Border(0.9470,0.3666) (0.9320,0.3666) \rput[r](0.1910,0.3666){$10^{-5}$}
\PST@Border(0.2070,0.4028) (0.2145,0.4028) \PST@Border(0.9470,0.4028)
(0.9395,0.4028) \PST@Border(0.2070,0.4506) (0.2145,0.4506)
\PST@Border(0.9470,0.4506) (0.9395,0.4506) \PST@Border(0.2070,0.4752)
(0.2145,0.4752) \PST@Border(0.9470,0.4752) (0.9395,0.4752)
\PST@Border(0.2070,0.4869) (0.2220,0.4869) \PST@Border(0.9470,0.4869)
(0.9320,0.4869) \rput[r](0.1910,0.4869){$10^{-4}$} \PST@Border(0.2070,0.5231)
(0.2145,0.5231) \PST@Border(0.9470,0.5231) (0.9395,0.5231)
\PST@Border(0.2070,0.5709) (0.2145,0.5709) \PST@Border(0.9470,0.5709)
(0.9395,0.5709) \PST@Border(0.2070,0.5955) (0.2145,0.5955)
\PST@Border(0.9470,0.5955) (0.9395,0.5955) \PST@Border(0.2070,0.6071)
(0.2220,0.6071) \PST@Border(0.9470,0.6071) (0.9320,0.6071)
\rput[r](0.1910,0.6071){ 0.001} \PST@Border(0.2070,0.6434) (0.2145,0.6434)
\PST@Border(0.9470,0.6434) (0.9395,0.6434) \PST@Border(0.2070,0.6912)
(0.2145,0.6912) \PST@Border(0.9470,0.6912) (0.9395,0.6912)
\PST@Border(0.2070,0.7158) (0.2145,0.7158) \PST@Border(0.9470,0.7158)
(0.9395,0.7158) \PST@Border(0.2070,0.7274) (0.2220,0.7274)
\PST@Border(0.9470,0.7274) (0.9320,0.7274) \rput[r](0.1910,0.7274){ 0.01}
\PST@Border(0.2070,0.7636) (0.2145,0.7636) \PST@Border(0.9470,0.7636)
(0.9395,0.7636) \PST@Border(0.2070,0.8115) (0.2145,0.8115)
\PST@Border(0.9470,0.8115) (0.9395,0.8115) \PST@Border(0.2070,0.8361)
(0.2145,0.8361) \PST@Border(0.9470,0.8361) (0.9395,0.8361)
\PST@Border(0.2070,0.8477) (0.2220,0.8477) \PST@Border(0.9470,0.8477)
(0.9320,0.8477) \rput[r](0.1910,0.8477){ 0.1} \PST@Border(0.2070,0.8839)
(0.2145,0.8839) \PST@Border(0.9470,0.8839) (0.9395,0.8839)
\PST@Border(0.2070,0.9318) (0.2145,0.9318) \PST@Border(0.9470,0.9318)
(0.9395,0.9318) \PST@Border(0.2070,0.9563) (0.2145,0.9563)
\PST@Border(0.9470,0.9563) (0.9395,0.9563) \PST@Border(0.2070,0.9680)
(0.2220,0.9680) \PST@Border(0.9470,0.9680) (0.9320,0.9680)
\rput[r](0.1910,0.9680){ 1} \PST@Border(0.2070,0.1260) (0.2070,0.1460)
\PST@Border(0.2070,0.9680) (0.2070,0.9480) \rput(0.2070,0.0840){ 0}
\PST@Border(0.2892,0.1260) (0.2892,0.1460) \PST@Border(0.2892,0.9680)
(0.2892,0.9480) \rput(0.2892,0.0840){ 0.5} \PST@Border(0.3714,0.1260)
(0.3714,0.1460) \PST@Border(0.3714,0.9680) (0.3714,0.9480)
\rput(0.3714,0.0840){ 1} \PST@Border(0.4537,0.1260) (0.4537,0.1460)
\PST@Border(0.4537,0.9680) (0.4537,0.9480) \rput(0.4537,0.0840){ 1.5}
\PST@Border(0.5359,0.1260) (0.5359,0.1460) \PST@Border(0.5359,0.9680)
(0.5359,0.9480) \rput(0.5359,0.0840){ 2} \PST@Border(0.6181,0.1260)
(0.6181,0.1460) \PST@Border(0.6181,0.9680) (0.6181,0.9480)
\rput(0.6181,0.0840){ 2.5} \PST@Border(0.7003,0.1260) (0.7003,0.1460)
\PST@Border(0.7003,0.9680) (0.7003,0.9480) \rput(0.7003,0.0840){ 3}
\PST@Border(0.7826,0.1260) (0.7826,0.1460) \PST@Border(0.7826,0.9680)
(0.7826,0.9480) \rput(0.7826,0.0840){ 3.5} \PST@Border(0.8648,0.1260)
(0.8648,0.1460) \PST@Border(0.8648,0.9680) (0.8648,0.9480)
\rput(0.8648,0.0840){ 4} \PST@Border(0.9470,0.1260) (0.9470,0.1460)
\PST@Border(0.9470,0.9680) (0.9470,0.9480) \rput(0.9470,0.0840){ 4.5}
\PST@Border(0.2070,0.9680) (0.2070,0.1260) (0.9470,0.1260) (0.9470,0.9680)
(0.2070,0.9680)

\rput{L}(0.0820,0.5470){$E\frac{dN}{d^3p}$ [GeV$^{-2}$]}
\rput(0.5770,0.0410){$p_\bot$~[GeV]}

\rput[r](0.8200,0.9270){sum}
\rput[r](0.8200,0.8850){$\pi\rho\rightarrow\gamma\pi$}
\rput[r](0.8200,0.8430){$\pi\pi\rightarrow\gamma\rho$}
\rput[r](0.8200,0.8010){Processes inv. $\eta$}
\rput[r](0.8200,0.7590){$\pi\pi\rightarrow\gamma\gamma$}
\rput[r](0.9150,0.6330){UrQMD}
\rput[r](0.9150,0.5910){Pb+Pb 158~AGeV}
\rput[r](0.9150,0.5490){$b<4.5$~fm}
\rput[r](0.9150,0.5070){$|y_{\rm c.m.}| < 0.5$}

\PST@inklusiv(0.8360,0.9270)(0.9150,0.9270)
\PST@pirhogpi(0.8360,0.8850)(0.9150,0.8850)
\PST@pipigrho(0.8360,0.8430)(0.9150,0.8430)
\PST@etachann(0.8360,0.8010)(0.9150,0.8010)
\PST@gammagam(0.8360,0.7590)(0.9150,0.7590)

\PST@inklusiv(0.2471,0.9680) (0.2481,0.9643) (0.2646,0.9353) (0.2810,0.9162)
(0.2974,0.8970) (0.3139,0.8754) (0.3303,0.8521) (0.3468,0.8276) (0.3632,0.8026)
(0.3797,0.7770) (0.3961,0.7512) (0.4126,0.7254) (0.4290,0.7001) (0.4454,0.6745)
(0.4619,0.6500) (0.4783,0.6244) (0.4948,0.6010) (0.5112,0.5776) (0.5277,0.5560)
(0.5441,0.5329) (0.5606,0.5115) (0.5770,0.4920) (0.5934,0.4735) (0.6099,0.4536)
(0.6263,0.4379) (0.6428,0.4203) (0.6592,0.4073) (0.6757,0.3938) (0.6921,0.3791)
(0.7086,0.3659) (0.7250,0.3546) (0.7414,0.3430) (0.7579,0.3321) (0.7743,0.3194)
(0.7908,0.3081) (0.8072,0.2959) (0.8237,0.2867) (0.8401,0.2804) (0.8566,0.2700)
(0.8730,0.2663) (0.8894,0.2591) (0.9059,0.2507) (0.9223,0.2396) (0.9388,0.2290)
(0.9470,0.2224)

\PST@pirhogpi(0.2152,0.7898) (0.2152,0.7898) (0.2317,0.8775) (0.2481,0.9062)
(0.2646,0.9115) (0.2810,0.9044) (0.2974,0.8898) (0.3139,0.8705) (0.3303,0.8484)
(0.3468,0.8245) (0.3632,0.7998) (0.3797,0.7744) (0.3961,0.7486) (0.4126,0.7227)
(0.4290,0.6972) (0.4454,0.6713) (0.4619,0.6464) (0.4783,0.6201) (0.4948,0.5960)
(0.5112,0.5716) (0.5277,0.5488) (0.5441,0.5245) (0.5606,0.5013) (0.5770,0.4797)
(0.5934,0.4590) (0.6099,0.4356) (0.6263,0.4169) (0.6428,0.3957) (0.6592,0.3803)
(0.6757,0.3635) (0.6921,0.3448) (0.7086,0.3292) (0.7250,0.3144) (0.7414,0.3011)
(0.7579,0.2886) (0.7743,0.2720) (0.7908,0.2616) (0.8072,0.2459) (0.8237,0.2387)
(0.8401,0.2323) (0.8566,0.2151) (0.8730,0.2099) (0.8894,0.2042) (0.9059,0.1999)
(0.9223,0.1873) (0.9388,0.1783) (0.9470,0.1680)

\PST@pipigrho(0.2427,0.9680) (0.2481,0.9412) (0.2646,0.8795) (0.2810,0.8285)
(0.2974,0.7839) (0.3139,0.7429) (0.3303,0.7051) (0.3468,0.6710) (0.3632,0.6394)
(0.3797,0.6094) (0.3961,0.5817) (0.4126,0.5569) (0.4290,0.5335) (0.4454,0.5107)
(0.4619,0.4900) (0.4783,0.4714) (0.4948,0.4528) (0.5112,0.4371) (0.5277,0.4221)
(0.5441,0.4027) (0.5606,0.3902) (0.5770,0.3768) (0.5934,0.3661) (0.6099,0.3562)
(0.6263,0.3454) (0.6428,0.3321) (0.6592,0.3217) (0.6757,0.3124) (0.6921,0.3024)
(0.7086,0.2913) (0.7250,0.2866) (0.7414,0.2751) (0.7579,0.2653) (0.7743,0.2518)
(0.7908,0.2380) (0.8072,0.2312) (0.8237,0.2195) (0.8401,0.2103) (0.8566,0.2061)
(0.8730,0.1963) (0.8894,0.1909) (0.9059,0.1788) (0.9223,0.1675) (0.9388,0.1599)
(0.9470,0.1577)

\PST@etachann(0.2152,0.7660) (0.2152,0.7660) (0.2317,0.7747) (0.2481,0.7452)
(0.2646,0.7093) (0.2810,0.6752) (0.2974,0.6438) (0.3139,0.6148) (0.3303,0.5881)
(0.3468,0.5631) (0.3632,0.5404) (0.3797,0.5195) (0.3961,0.5006) (0.4126,0.4842)
(0.4290,0.4685) (0.4454,0.4556) (0.4619,0.4424) (0.4783,0.4306) (0.4948,0.4203)
(0.5112,0.4101) (0.5277,0.3986) (0.5441,0.3899) (0.5606,0.3789) (0.5770,0.3708)
(0.5934,0.3597) (0.6099,0.3497) (0.6263,0.3421) (0.6428,0.3338) (0.6592,0.3253)
(0.6757,0.3167) (0.6921,0.3067) (0.7086,0.2966) (0.7250,0.2851) (0.7414,0.2762)
(0.7579,0.2670) (0.7743,0.2604) (0.7908,0.2498) (0.8072,0.2375) (0.8237,0.2279)
(0.8401,0.2242) (0.8566,0.2161) (0.8730,0.2182) (0.8894,0.2085) (0.9059,0.1987)
(0.9223,0.1893) (0.9388,0.1750) (0.9470,0.1684)

\PST@gammagam(0.2152,0.7976) (0.2152,0.7976) (0.2317,0.7850) (0.2481,0.7394)
(0.2646,0.6954) (0.2810,0.6536) (0.2974,0.6151) (0.3139,0.5781) (0.3303,0.5423)
(0.3468,0.5077) (0.3632,0.4741) (0.3797,0.4427) (0.3961,0.4125) (0.4126,0.3830)
(0.4290,0.3557) (0.4454,0.3272) (0.4619,0.3005) (0.4783,0.2735) (0.4948,0.2480)
(0.5112,0.2265) (0.5277,0.1977) (0.5441,0.1682) (0.5606,0.1403) (0.5690,0.1260)

\PST@Border(0.2070,0.9680) (0.2070,0.1260) (0.9470,0.1260) (0.9470,0.9680)
(0.2070,0.9680)

\catcode`@=12
\fi
\endpspicture

%% file: hadrongas.tex
\ifx\PSTloaded\undefined
\def\PSTloaded{t}
\psset{arrowsize=.01 3.2 1.4 .3}
\psset{dotsize=.01}
\catcode`@=11

\definecolor{darkgreen}{rgb}{0,.5,0}
\definecolor{violett}{rgb}{.5,0,.5}
\definecolor{orange}{rgb}{.8,.4,0}
\newpsobject{PST@Border}{psline}{linewidth=.0015,linestyle=solid}
\newpsobject{PST@Axes}{psline}{linewidth=.0015,linestyle=dotted,dotsep=.004}
\newpsobject{PST@inklusiv}{psline}{linecolor=black,    linewidth=.0025,linestyle=solid}
\newpsobject{PST@pirhogpi}{psline}{linecolor=red,      linewidth=.0025,linestyle=dashed,dash=.01 .01}
\newpsobject{PST@pipigrho}{psline}{linecolor=blue,     linewidth=.0035,linestyle=dotted,dotsep=.004}
\newpsobject{PST@etachann}{psline}{linecolor=darkgreen,linewidth=.0025,linestyle=dashed,dash=.01 .004 .004 .004}
\newpsobject{PST@kaonchan}{psline}{linecolor=violett,  linewidth=.0020,linestyle=dashed,dash=.02 .01 .01 .01}
\newpsobject{PST@gammagam}{psline}{linecolor=orange,   linewidth=.0025,linestyle=dashed,dash=.02 .01}
\catcode`@=12

\fi
\psset{unit=5.0in,xunit=1.075\columnwidth,yunit=.8\columnwidth}
\pspicture(0.0500,0.0220)(0.980000,0.9880)
\ifx\nofigs\undefined
\catcode`@=11

\PST@Border(0.2070,0.1260) (0.2220,0.1260) \PST@Border(0.9470,0.1260)
(0.9320,0.1260) \rput[r](0.1910,0.1260){$10^{-7}$} \PST@Border(0.2070,0.1577)
(0.2145,0.1577) \PST@Border(0.9470,0.1577) (0.9395,0.1577)
\PST@Border(0.2070,0.1996) (0.2145,0.1996) \PST@Border(0.9470,0.1996)
(0.9395,0.1996) \PST@Border(0.2070,0.2211) (0.2145,0.2211)
\PST@Border(0.9470,0.2211) (0.9395,0.2211) \PST@Border(0.2070,0.2313)
(0.2220,0.2313) \PST@Border(0.9470,0.2313) (0.9320,0.2313)
\rput[r](0.1910,0.2313){$10^{-6}$} \PST@Border(0.2070,0.2629) (0.2145,0.2629)
\PST@Border(0.9470,0.2629) (0.9395,0.2629) \PST@Border(0.2070,0.3048)
(0.2145,0.3048) \PST@Border(0.9470,0.3048) (0.9395,0.3048)
\PST@Border(0.2070,0.3263) (0.2145,0.3263) \PST@Border(0.9470,0.3263)
(0.9395,0.3263) \PST@Border(0.2070,0.3365) (0.2220,0.3365)
\PST@Border(0.9470,0.3365) (0.9320,0.3365) \rput[r](0.1910,0.3365){$10^{-5}$}
\PST@Border(0.2070,0.3682) (0.2145,0.3682) \PST@Border(0.9470,0.3682)
(0.9395,0.3682) \PST@Border(0.2070,0.4101) (0.2145,0.4101)
\PST@Border(0.9470,0.4101) (0.9395,0.4101) \PST@Border(0.2070,0.4316)
(0.2145,0.4316) \PST@Border(0.9470,0.4316) (0.9395,0.4316)
\PST@Border(0.2070,0.4418) (0.2220,0.4418) \PST@Border(0.9470,0.4418)
(0.9320,0.4418) \rput[r](0.1910,0.4418){$10^{-4}$} \PST@Border(0.2070,0.4734)
(0.2145,0.4734) \PST@Border(0.9470,0.4734) (0.9395,0.4734)
\PST@Border(0.2070,0.5153) (0.2145,0.5153) \PST@Border(0.9470,0.5153)
(0.9395,0.5153) \PST@Border(0.2070,0.5368) (0.2145,0.5368)
\PST@Border(0.9470,0.5368) (0.9395,0.5368) \PST@Border(0.2070,0.5470)
(0.2220,0.5470) \PST@Border(0.9470,0.5470) (0.9320,0.5470)
\rput[r](0.1910,0.5470){ 0.001} \PST@Border(0.2070,0.5787) (0.2145,0.5787)
\PST@Border(0.9470,0.5787) (0.9395,0.5787) \PST@Border(0.2070,0.6206)
(0.2145,0.6206) \PST@Border(0.9470,0.6206) (0.9395,0.6206)
\PST@Border(0.2070,0.6421) (0.2145,0.6421) \PST@Border(0.9470,0.6421)
(0.9395,0.6421) \PST@Border(0.2070,0.6523) (0.2220,0.6523)
\PST@Border(0.9470,0.6523) (0.9320,0.6523) \rput[r](0.1910,0.6523){ 0.01}
\PST@Border(0.2070,0.6839) (0.2145,0.6839) \PST@Border(0.9470,0.6839)
(0.9395,0.6839) \PST@Border(0.2070,0.7258) (0.2145,0.7258)
\PST@Border(0.9470,0.7258) (0.9395,0.7258) \PST@Border(0.2070,0.7473)
(0.2145,0.7473) \PST@Border(0.9470,0.7473) (0.9395,0.7473)
\PST@Border(0.2070,0.7575) (0.2220,0.7575) \PST@Border(0.9470,0.7575)
(0.9320,0.7575) \rput[r](0.1910,0.7575){ 0.1} \PST@Border(0.2070,0.7892)
(0.2145,0.7892) \PST@Border(0.9470,0.7892) (0.9395,0.7892)
\PST@Border(0.2070,0.8311) (0.2145,0.8311) \PST@Border(0.9470,0.8311)
(0.9395,0.8311) \PST@Border(0.2070,0.8526) (0.2145,0.8526)
\PST@Border(0.9470,0.8526) (0.9395,0.8526) \PST@Border(0.2070,0.8628)
(0.2220,0.8628) \PST@Border(0.9470,0.8628) (0.9320,0.8628)
\rput[r](0.1910,0.8628){ 1} \PST@Border(0.2070,0.8944) (0.2145,0.8944)
\PST@Border(0.9470,0.8944) (0.9395,0.8944) \PST@Border(0.2070,0.9363)
(0.2145,0.9363) \PST@Border(0.9470,0.9363) (0.9395,0.9363)
\PST@Border(0.2070,0.9578) (0.2145,0.9578) \PST@Border(0.9470,0.9578)
(0.9395,0.9578) \PST@Border(0.2070,0.9680) (0.2220,0.9680)
\PST@Border(0.9470,0.9680) (0.9320,0.9680) \rput[r](0.1910,0.9680){ 10}
\PST@Border(0.2070,0.1260) (0.2070,0.1460) \PST@Border(0.2070,0.9680)
(0.2070,0.9480) \rput(0.2070,0.0840){ 0} \PST@Border(0.2892,0.1260)
(0.2892,0.1460) \PST@Border(0.2892,0.9680) (0.2892,0.9480)
\rput(0.2892,0.0840){ 0.5} \PST@Border(0.3714,0.1260) (0.3714,0.1460)
\PST@Border(0.3714,0.9680) (0.3714,0.9480) \rput(0.3714,0.0840){ 1}
\PST@Border(0.4537,0.1260) (0.4537,0.1460) \PST@Border(0.4537,0.9680)
(0.4537,0.9480) \rput(0.4537,0.0840){ 1.5} \PST@Border(0.5359,0.1260)
(0.5359,0.1460) \PST@Border(0.5359,0.9680) (0.5359,0.9480)
\rput(0.5359,0.0840){ 2} \PST@Border(0.6181,0.1260) (0.6181,0.1460)
\PST@Border(0.6181,0.9680) (0.6181,0.9480) \rput(0.6181,0.0840){ 2.5}
\PST@Border(0.7003,0.1260) (0.7003,0.1460) \PST@Border(0.7003,0.9680)
(0.7003,0.9480) \rput(0.7003,0.0840){ 3} \PST@Border(0.7826,0.1260)
(0.7826,0.1460) \PST@Border(0.7826,0.9680) (0.7826,0.9480)
\rput(0.7826,0.0840){ 3.5} \PST@Border(0.8648,0.1260) (0.8648,0.1460)
\PST@Border(0.8648,0.9680) (0.8648,0.9480) \rput(0.8648,0.0840){ 4}
\PST@Border(0.9470,0.1260) (0.9470,0.1460) \PST@Border(0.9470,0.9680)
(0.9470,0.9480) \rput(0.9470,0.0840){ 4.5} \PST@Border(0.2070,0.9680)
(0.2070,0.1260) (0.9470,0.1260) (0.9470,0.9680) (0.2070,0.9680)

\rput{L}(0.0820,0.5470){$E\frac{dN}{d^3p}$ [GeV$^{-2}$]}
\rput(0.5770,0.0410){$p_\bot$~[GeV]}

\rput[r](0.8200,0.9270){sum}
\rput[r](0.8200,0.8850){$\pi\rho\rightarrow\gamma\pi$}
\rput[r](0.8200,0.8430){$\pi\pi\rightarrow\gamma\rho$}
\rput[r](0.8200,0.8010){Processes inv. $\eta$}
\rput[r](0.8200,0.7590){Processes inv. K}
\rput[r](0.8200,0.7170){$\pi\pi\rightarrow\gamma\gamma$}

\rput[r](0.9150,0.6010){Hybrid, Hadron gas EoS}
\rput[r](0.9150,0.5590){Pb+Pb 158 AGeV}
\rput[r](0.9150,0.5170){$b < 4.5$~fm}
\rput[r](0.9150,0.4750){$|y_{\rm c.m.}| < 0.5$}

\PST@inklusiv(0.8360,0.9270)(0.9150,0.9270)
\PST@pirhogpi(0.8360,0.8850)(0.9150,0.8850)
\PST@pipigrho(0.8360,0.8430)(0.9150,0.8430)
\PST@etachann(0.8360,0.8010)(0.9150,0.8010)
\PST@kaonchan(0.8360,0.7590)(0.9150,0.7590)
\PST@gammagam(0.8360,0.7170)(0.9150,0.7170)

\PST@inklusiv(0.2317,0.9604) (0.2481,0.9205) (0.2646,0.8458) (0.2810,0.8124)
(0.2974,0.7891) (0.3139,0.7691) (0.3303,0.7544) (0.3468,0.7230) (0.3632,0.7136)
(0.3797,0.6921) (0.3961,0.6662) (0.4126,0.6601) (0.4290,0.6300) (0.4454,0.6205)
(0.4619,0.5920) (0.4783,0.5801) (0.4948,0.5594) (0.5112,0.5362) (0.5277,0.5311)
(0.5441,0.5040) (0.5606,0.4909) (0.5770,0.4669) (0.5934,0.4502) (0.6099,0.4433)
(0.6263,0.4052) (0.6428,0.3985) (0.6592,0.3813) (0.6757,0.3665) (0.6921,0.3597)
(0.7086,0.3414) (0.7250,0.3264) (0.7414,0.3113) (0.7579,0.3026) (0.7743,0.2884)
(0.7908,0.2773) (0.8072,0.2677) (0.8237,0.2639) (0.8401,0.2406) (0.8566,0.2464)
(0.8730,0.2391) (0.8894,0.2299) (0.9059,0.2245) (0.9223,0.2231) (0.9388,0.2167)

\PST@pirhogpi(0.2152,0.6400) (0.2317,0.7531) (0.2481,0.7801) (0.2646,0.7794)
(0.2810,0.7729) (0.2974,0.7633) (0.3139,0.7503) (0.3303,0.7393) (0.3468,0.7132)
(0.3632,0.7040) (0.3797,0.6843) (0.3961,0.6598) (0.4126,0.6533) (0.4290,0.6244)
(0.4454,0.6149) (0.4619,0.5871) (0.4783,0.5750) (0.4948,0.5545) (0.5112,0.5312)
(0.5277,0.5260) (0.5441,0.4983) (0.5606,0.4847) (0.5770,0.4598) (0.5934,0.4415)
(0.6099,0.4357) (0.6263,0.3929) (0.6428,0.3868) (0.6592,0.3668) (0.6757,0.3497)
(0.6921,0.3427) (0.7086,0.3174) (0.7250,0.2977) (0.7414,0.2813) (0.7579,0.2677)
(0.7743,0.2528) (0.7908,0.2328) (0.8072,0.2225) (0.8237,0.2152) (0.8401,0.1939)
(0.8566,0.1972) (0.8730,0.2011) (0.8894,0.1888) (0.9059,0.1863) (0.9223,0.1888)
(0.9388,0.1864)

\PST@pipigrho(0.2317,0.9540) (0.2481,0.9113) (0.2646,0.8257) (0.2810,0.7782)
(0.2974,0.7398) (0.3139,0.7067) (0.3303,0.6813) (0.3468,0.6318) (0.3632,0.6193)
(0.3797,0.5876) (0.3961,0.5534) (0.4126,0.5465) (0.4290,0.5092) (0.4454,0.4972)
(0.4619,0.4643) (0.4783,0.4509) (0.4948,0.4269) (0.5112,0.4056) (0.5277,0.4010)
(0.5441,0.3770) (0.5606,0.3665) (0.5770,0.3480) (0.5934,0.3407) (0.6099,0.3327)
(0.6263,0.3144) (0.6428,0.3052) (0.6592,0.2974) (0.6757,0.2882) (0.6921,0.2775)
(0.7086,0.2728) (0.7250,0.2686) (0.7414,0.2499) (0.7579,0.2411) (0.7743,0.2322)
(0.7908,0.2301) (0.8072,0.2197) (0.8237,0.2120) (0.8401,0.1882) (0.8566,0.1895)
(0.8730,0.1790) (0.8894,0.1714) (0.9059,0.1593) (0.9223,0.1640) (0.9388,0.1521)

\PST@etachann(0.2152,0.6504) (0.2317,0.6574) (0.2481,0.6326) (0.2646,0.6030)
(0.2810,0.5744) (0.2974,0.5493) (0.3139,0.5272) (0.3303,0.5072) (0.3468,0.4886)
(0.3632,0.4732) (0.3797,0.4571) (0.3961,0.4428) (0.4126,0.4288) (0.4290,0.4183)
(0.4454,0.4093) (0.4619,0.3942) (0.4783,0.3863) (0.4948,0.3797) (0.5112,0.3687)
(0.5277,0.3590) (0.5441,0.3527) (0.5606,0.3441) (0.5770,0.3339) (0.5934,0.3272)
(0.6099,0.3067) (0.6263,0.2939) (0.6428,0.2851) (0.6592,0.2786) (0.6757,0.2704)
(0.6921,0.2705) (0.7086,0.2643) (0.7250,0.2488) (0.7414,0.2422) (0.7579,0.2433)
(0.7743,0.2248) (0.7908,0.2165) (0.8072,0.2091) (0.8237,0.2137) (0.8401,0.1888)
(0.8566,0.2010) (0.8730,0.1835) (0.8894,0.1770) (0.9059,0.1733) (0.9223,0.1603)
(0.9388,0.1518)

\PST@kaonchan(0.2317,0.8626) (0.2481,0.8284) (0.2646,0.7465) (0.2810,0.7054)
(0.2974,0.6760) (0.3139,0.6505) (0.3303,0.6347) (0.3468,0.5851) (0.3632,0.5827)
(0.3797,0.5535) (0.3961,0.5159) (0.4126,0.5192) (0.4290,0.4745) (0.4454,0.4693)
(0.4619,0.4276) (0.4783,0.4193) (0.4948,0.3914) (0.5112,0.3584) (0.5277,0.3613)
(0.5441,0.3211) (0.5606,0.3089) (0.5770,0.2727) (0.5934,0.2506) (0.6099,0.2483)
(0.6263,0.1957) (0.6428,0.1930) (0.6592,0.1465)

\PST@gammagam(0.2152,0.6683) (0.2317,0.6531) (0.2481,0.6140) (0.2646,0.5794)
(0.2810,0.5470) (0.2974,0.5178) (0.3139,0.4914) (0.3303,0.4649) (0.3468,0.4392)
(0.3632,0.4162) (0.3797,0.3936) (0.3961,0.3684) (0.4126,0.3509) (0.4290,0.3224)
(0.4454,0.3094) (0.4619,0.2906) (0.4783,0.2641) (0.4948,0.2425) (0.5112,0.2291)
(0.5277,0.2132) (0.5441,0.1981) (0.5606,0.1639) (0.5770,0.1287)

\PST@Border(0.2070,0.9680)
(0.2070,0.1260)
(0.9470,0.1260)
(0.9470,0.9680)
(0.2070,0.9680)

\catcode`@=12
\fi
\endpspicture

%% file: bagmodel.tex
\ifx\PSTloaded\undefined
\def\PSTloaded{t}
\psset{arrowsize=.01 3.2 1.4 .3}
\psset{dotsize=.01}
\catcode`@=11

\definecolor{darkgreen}{rgb}{0,.5,0}
\definecolor{violett}{rgb}{.5,0,.5}
\definecolor{orange}{rgb}{.8,.4,0}
\newpsobject{PST@Border}{psline}{linewidth=.0015,linestyle=solid}
\newpsobject{PST@Axes}{psline}{linewidth=.0015,linestyle=dotted,dotsep=.004}
\newpsobject{PST@inklusiv}{psline}{linecolor=black,    linewidth=.0025,linestyle=solid}
\newpsobject{PST@pirhogpi}{psline}{linecolor=red,      linewidth=.0025,linestyle=dashed,dash=.01 .01}
\newpsobject{PST@pipigrho}{psline}{linecolor=blue,     linewidth=.0035,linestyle=dotted,dotsep=.004}
\newpsobject{PST@etachann}{psline}{linecolor=darkgreen,linewidth=.0025,linestyle=dashed,dash=.01 .004 .004 .004}
\newpsobject{PST@kaonchan}{psline}{linecolor=violett,  linewidth=.0020,linestyle=dashed,dash=.02 .01 .01 .01}
\newpsobject{PST@gammagam}{psline}{linecolor=orange,   linewidth=.0025,linestyle=dashed,dash=.02 .01}
\newpsobject{PST@qgp}     {psline}{linecolor=violett,  linewidth=.0035,linestyle=dotted,dotsep=.004}
\catcode`@=12

\fi
\psset{unit=5.0in,xunit=1.075\columnwidth,yunit=.8\columnwidth}
\pspicture(0.0500,0.0220)(0.980000,0.9880)
\ifx\nofigs\undefined
\catcode`@=11

\PST@Border(0.2070,0.1260) (0.2220,0.1260) \PST@Border(0.9470,0.1260)
(0.9320,0.1260) \rput[r](0.1910,0.1260){$10^{-7}$} \PST@Border(0.2070,0.1577)
(0.2145,0.1577) \PST@Border(0.9470,0.1577) (0.9395,0.1577)
\PST@Border(0.2070,0.1996) (0.2145,0.1996) \PST@Border(0.9470,0.1996)
(0.9395,0.1996) \PST@Border(0.2070,0.2211) (0.2145,0.2211)
\PST@Border(0.9470,0.2211) (0.9395,0.2211) \PST@Border(0.2070,0.2313)
(0.2220,0.2313) \PST@Border(0.9470,0.2313) (0.9320,0.2313)
\rput[r](0.1910,0.2313){$10^{-6}$} \PST@Border(0.2070,0.2629) (0.2145,0.2629)
\PST@Border(0.9470,0.2629) (0.9395,0.2629) \PST@Border(0.2070,0.3048)
(0.2145,0.3048) \PST@Border(0.9470,0.3048) (0.9395,0.3048)
\PST@Border(0.2070,0.3263) (0.2145,0.3263) \PST@Border(0.9470,0.3263)
(0.9395,0.3263) \PST@Border(0.2070,0.3365) (0.2220,0.3365)
\PST@Border(0.9470,0.3365) (0.9320,0.3365) \rput[r](0.1910,0.3365){$10^{-5}$}
\PST@Border(0.2070,0.3682) (0.2145,0.3682) \PST@Border(0.9470,0.3682)
(0.9395,0.3682) \PST@Border(0.2070,0.4101) (0.2145,0.4101)
\PST@Border(0.9470,0.4101) (0.9395,0.4101) \PST@Border(0.2070,0.4316)
(0.2145,0.4316) \PST@Border(0.9470,0.4316) (0.9395,0.4316)
\PST@Border(0.2070,0.4418) (0.2220,0.4418) \PST@Border(0.9470,0.4418)
(0.9320,0.4418) \rput[r](0.1910,0.4418){$10^{-4}$} \PST@Border(0.2070,0.4734)
(0.2145,0.4734) \PST@Border(0.9470,0.4734) (0.9395,0.4734)
\PST@Border(0.2070,0.5153) (0.2145,0.5153) \PST@Border(0.9470,0.5153)
(0.9395,0.5153) \PST@Border(0.2070,0.5368) (0.2145,0.5368)
\PST@Border(0.9470,0.5368) (0.9395,0.5368) \PST@Border(0.2070,0.5470)
(0.2220,0.5470) \PST@Border(0.9470,0.5470) (0.9320,0.5470)
\rput[r](0.1910,0.5470){ 0.001} \PST@Border(0.2070,0.5787) (0.2145,0.5787)
\PST@Border(0.9470,0.5787) (0.9395,0.5787) \PST@Border(0.2070,0.6206)
(0.2145,0.6206) \PST@Border(0.9470,0.6206) (0.9395,0.6206)
\PST@Border(0.2070,0.6421) (0.2145,0.6421) \PST@Border(0.9470,0.6421)
(0.9395,0.6421) \PST@Border(0.2070,0.6523) (0.2220,0.6523)
\PST@Border(0.9470,0.6523) (0.9320,0.6523) \rput[r](0.1910,0.6523){ 0.01}
\PST@Border(0.2070,0.6839) (0.2145,0.6839) \PST@Border(0.9470,0.6839)
(0.9395,0.6839) \PST@Border(0.2070,0.7258) (0.2145,0.7258)
\PST@Border(0.9470,0.7258) (0.9395,0.7258) \PST@Border(0.2070,0.7473)
(0.2145,0.7473) \PST@Border(0.9470,0.7473) (0.9395,0.7473)
\PST@Border(0.2070,0.7575) (0.2220,0.7575) \PST@Border(0.9470,0.7575)
(0.9320,0.7575) \rput[r](0.1910,0.7575){ 0.1} \PST@Border(0.2070,0.7892)
(0.2145,0.7892) \PST@Border(0.9470,0.7892) (0.9395,0.7892)
\PST@Border(0.2070,0.8311) (0.2145,0.8311) \PST@Border(0.9470,0.8311)
(0.9395,0.8311) \PST@Border(0.2070,0.8526) (0.2145,0.8526)
\PST@Border(0.9470,0.8526) (0.9395,0.8526) \PST@Border(0.2070,0.8628)
(0.2220,0.8628) \PST@Border(0.9470,0.8628) (0.9320,0.8628)
\rput[r](0.1910,0.8628){ 1} \PST@Border(0.2070,0.8944) (0.2145,0.8944)
\PST@Border(0.9470,0.8944) (0.9395,0.8944) \PST@Border(0.2070,0.9363)
(0.2145,0.9363) \PST@Border(0.9470,0.9363) (0.9395,0.9363)
\PST@Border(0.2070,0.9578) (0.2145,0.9578) \PST@Border(0.9470,0.9578)
(0.9395,0.9578) \PST@Border(0.2070,0.9680) (0.2220,0.9680)
\PST@Border(0.9470,0.9680) (0.9320,0.9680) \rput[r](0.1910,0.9680){ 10}
\PST@Border(0.2070,0.1260) (0.2070,0.1460) \PST@Border(0.2070,0.9680)
(0.2070,0.9480) \rput(0.2070,0.0840){ 0} \PST@Border(0.2892,0.1260)
(0.2892,0.1460) \PST@Border(0.2892,0.9680) (0.2892,0.9480)
\rput(0.2892,0.0840){ 0.5} \PST@Border(0.3714,0.1260) (0.3714,0.1460)
\PST@Border(0.3714,0.9680) (0.3714,0.9480) \rput(0.3714,0.0840){ 1}
\PST@Border(0.4537,0.1260) (0.4537,0.1460) \PST@Border(0.4537,0.9680)
(0.4537,0.9480) \rput(0.4537,0.0840){ 1.5} \PST@Border(0.5359,0.1260)
(0.5359,0.1460) \PST@Border(0.5359,0.9680) (0.5359,0.9480)
\rput(0.5359,0.0840){ 2} \PST@Border(0.6181,0.1260) (0.6181,0.1460)
\PST@Border(0.6181,0.9680) (0.6181,0.9480) \rput(0.6181,0.0840){ 2.5}
\PST@Border(0.7003,0.1260) (0.7003,0.1460) \PST@Border(0.7003,0.9680)
(0.7003,0.9480) \rput(0.7003,0.0840){ 3} \PST@Border(0.7826,0.1260)
(0.7826,0.1460) \PST@Border(0.7826,0.9680) (0.7826,0.9480)
\rput(0.7826,0.0840){ 3.5} \PST@Border(0.8648,0.1260) (0.8648,0.1460)
\PST@Border(0.8648,0.9680) (0.8648,0.9480) \rput(0.8648,0.0840){ 4}
\PST@Border(0.9470,0.1260) (0.9470,0.1460) \PST@Border(0.9470,0.9680)
(0.9470,0.9480) \rput(0.9470,0.0840){ 4.5} \PST@Border(0.2070,0.9680)
(0.2070,0.1260) (0.9470,0.1260) (0.9470,0.9680) (0.2070,0.9680)

\rput{L}(0.0820,0.5470){$E\frac{dN}{d^3p}$ [GeV$^{-2}$]}
\rput(0.5770,0.0410){$p_\bot$~[GeV]}

\rput[r](0.8200,0.9270){sum}
\PST@inklusiv(0.8360,0.9270)(0.9150,0.9270)

\PST@inklusiv(0.2510,0.9680) (0.2646,0.9081) (0.2810,0.8736) (0.2974,0.8488)
(0.3139,0.8253) (0.3303,0.8115) (0.3468,0.7651) (0.3632,0.7635) (0.3797,0.7374)
(0.3961,0.7029) (0.4126,0.7071) (0.4290,0.6662) (0.4454,0.6634) (0.4619,0.6255)
(0.4783,0.6190) (0.4948,0.5943) (0.5112,0.5648) (0.5277,0.5691) (0.5441,0.5333)
(0.5606,0.5239) (0.5770,0.4922) (0.5934,0.4726) (0.6099,0.4732) (0.6263,0.4294)
(0.6428,0.4282) (0.6592,0.3944) (0.6757,0.3783) (0.6921,0.3636) (0.7086,0.3428)
(0.7250,0.3342) (0.7414,0.3101) (0.7579,0.3072) (0.7743,0.2999) (0.7908,0.2855)
(0.8072,0.2842) (0.8237,0.2645) (0.8401,0.2607) (0.8566,0.2478) (0.8730,0.2434)
(0.8894,0.2308) (0.9059,0.2200) (0.9223,0.2188) (0.9388,0.2091) (0.9470,0.2073)

\rput[r](0.8200,0.8850){$\pi\rho\rightarrow\gamma\pi$}
\PST@pirhogpi(0.8360,0.8850)(0.9150,0.8850)

\PST@pirhogpi(0.2152,0.5773) (0.2152,0.5773) (0.2317,0.7439) (0.2481,0.7706)
(0.2646,0.7639) (0.2810,0.7569) (0.2974,0.7488) (0.3139,0.7361) (0.3303,0.7285)
(0.3468,0.6922) (0.3632,0.6904) (0.3797,0.6683) (0.3961,0.6375) (0.4126,0.6410)
(0.4290,0.6035) (0.4454,0.6002) (0.4619,0.5648) (0.4783,0.5568) (0.4948,0.5334)
(0.5112,0.5056) (0.5277,0.5073) (0.5441,0.4738) (0.5606,0.4620) (0.5770,0.4325)
(0.5934,0.4130) (0.6099,0.4100) (0.6263,0.3727) (0.6428,0.3618) (0.6592,0.3329)
(0.6757,0.3163) (0.6921,0.2975) (0.7086,0.2836) (0.7250,0.2714) (0.7414,0.2626)
(0.7579,0.2612) (0.7743,0.2474) (0.7908,0.2381) (0.8072,0.2320) (0.8237,0.2135)
(0.8401,0.2095) (0.8566,0.2030) (0.8730,0.1924) (0.8894,0.1874) (0.9059,0.1752)
(0.9223,0.1782) (0.9388,0.1804) (0.9470,0.1775)

\rput[r](0.8200,0.8430){$\pi\pi\rightarrow\gamma\rho$}
\PST@pipigrho(0.8360,0.8430)(0.9150,0.8430)

\PST@pipigrho(0.2384,0.9680) (0.2481,0.9437) (0.2646,0.8428) (0.2810,0.7906)
(0.2974,0.7489) (0.3139,0.7132) (0.3303,0.6876) (0.3468,0.6291) (0.3632,0.6217)
(0.3797,0.5869) (0.3961,0.5488) (0.4126,0.5448) (0.4290,0.5030) (0.4454,0.4928)
(0.4619,0.4581) (0.4783,0.4461) (0.4948,0.4238) (0.5112,0.4018) (0.5277,0.3942)
(0.5441,0.3710) (0.5606,0.3643) (0.5770,0.3422) (0.5934,0.3273) (0.6099,0.3196)
(0.6263,0.3044) (0.6428,0.2944) (0.6592,0.2870) (0.6757,0.2767) (0.6921,0.2542)
(0.7086,0.2499) (0.7250,0.2498) (0.7414,0.2346) (0.7579,0.2185) (0.7743,0.2218)
(0.7908,0.2157) (0.8072,0.2117) (0.8237,0.2044) (0.8401,0.1928) (0.8566,0.1746)
(0.8730,0.1699) (0.8894,0.1512) (0.9059,0.1513) (0.9223,0.1338) (0.9288,0.1260)

\rput[r](0.8200,0.8010){Processes inv. $\eta$}
\PST@etachann(0.8360,0.8010)(0.9150,0.8010)

\PST@etachann(0.2152,0.5920) (0.2152,0.5920) (0.2317,0.5941) (0.2481,0.5633)
(0.2646,0.5299) (0.2810,0.5015) (0.2974,0.4785) (0.3139,0.4637) (0.3303,0.4512)
(0.3468,0.4410) (0.3632,0.4313) (0.3797,0.4227) (0.3961,0.4182) (0.4126,0.4115)
(0.4290,0.4021) (0.4454,0.3969) (0.4619,0.3894) (0.4783,0.3819) (0.4948,0.3687)
(0.5112,0.3618) (0.5277,0.3554) (0.5441,0.3511) (0.5606,0.3418) (0.5770,0.3334)
(0.5934,0.3208) (0.6099,0.3159) (0.6263,0.3046) (0.6428,0.2943) (0.6592,0.2873)
(0.6757,0.2818) (0.6921,0.2703) (0.7086,0.2699) (0.7250,0.2571) (0.7414,0.2463)
(0.7579,0.2457) (0.7743,0.2500) (0.7908,0.2377) (0.8072,0.2412) (0.8237,0.2193)
(0.8401,0.2204) (0.8566,0.2066) (0.8730,0.2082) (0.8894,0.1929) (0.9059,0.1784)
(0.9223,0.1805) (0.9388,0.1598) (0.9470,0.1555)

\rput[r](0.8200,0.7590){Processes inv. K}
\PST@kaonchan(0.8360,0.7590)(0.9150,0.7590)

\PST@kaonchan(0.2317,0.8987) (0.2481,0.8612) (0.2646,0.7698) (0.2810,0.7242)
(0.2974,0.6919) (0.3139,0.6645) (0.3303,0.6476) (0.3468,0.5964) (0.3632,0.5936)
(0.3797,0.5636) (0.3961,0.5252) (0.4126,0.5281) (0.4290,0.4828) (0.4454,0.4770)
(0.4619,0.4348) (0.4783,0.4260) (0.4948,0.3974) (0.5112,0.3644) (0.5277,0.3664)
(0.5441,0.3259) (0.5606,0.3124) (0.5770,0.2772) (0.5934,0.2535) (0.6099,0.2501)
(0.6263,0.1996) (0.6428,0.1929) (0.6592,0.1519) (0.6747,0.1260)

\rput[r](0.8200,0.7170){$\pi\pi\rightarrow\gamma\gamma$}
\PST@gammagam(0.8360,0.7170)(0.9150,0.7170)

\PST@gammagam(0.2152,0.6274) (0.2152,0.6274) (0.2317,0.6169) (0.2481,0.5700)
(0.2646,0.5291) (0.2810,0.4947) (0.2974,0.4647) (0.3139,0.4320) (0.3303,0.4033)
(0.3468,0.3790) (0.3632,0.3520) (0.3797,0.3281) (0.3961,0.2997) (0.4126,0.2875)
(0.4290,0.2694) (0.4454,0.2628) (0.4619,0.2528) (0.4783,0.2391) (0.4948,0.2044)
(0.5112,0.1690) (0.5277,0.1627) (0.5441,0.1516) (0.5606,0.1290) (0.5770,0.1270)
(0.5773,0.1260)

\rput[r](0.8200,0.6750){QGP}
\PST@qgp(0.8360,0.6750)(0.9150,0.6750)

\PST@qgp(0.2340,0.9680) (0.2481,0.9469) (0.2646,0.8897) (0.2810,0.8586)
(0.2974,0.8351) (0.3139,0.8116) (0.3303,0.7979) (0.3468,0.7501) (0.3632,0.7490)
(0.3797,0.7222) (0.3961,0.6868) (0.4126,0.6917) (0.4290,0.6498) (0.4454,0.6474)
(0.4619,0.6084) (0.4783,0.6027) (0.4948,0.5773) (0.5112,0.5467) (0.5277,0.5527)
(0.5441,0.5150) (0.5606,0.5065) (0.5770,0.4725) (0.5934,0.4524) (0.6099,0.4552)
(0.6263,0.4040) (0.6428,0.4088) (0.6592,0.3665) (0.6757,0.3475) (0.6921,0.3354)
(0.7086,0.2997) (0.7250,0.2926) (0.7414,0.2378) (0.7579,0.2400) (0.7743,0.2178)
(0.7908,0.1676) (0.8072,0.1724) (0.8189,0.1260)

\rput[r](0.9150,0.6010){Hybrid, Bag Model EoS}
\rput[r](0.9150,0.5590){Pb+Pb 158 AGeV}
\rput[r](0.9150,0.5170){$b < 4.5$~fm}
\rput[r](0.9150,0.4750){$|y_{\rm c.m.}| < 0.5$}

\PST@Border(0.2070,0.9680) (0.2070,0.1260) (0.9470,0.1260) (0.9470,0.9680)
(0.2070,0.9680)

\catcode`@=12
\fi
\endpspicture

%% file: directcomp.tex
\ifx\PSTloaded\undefined
\def\PSTloaded{t}
\psset{arrowsize=.01 3.2 1.4 .3}
\psset{dotsize=.01}
\catcode`@=11

\newpsobject{PST@Border}{psline}{linewidth=.0015,linestyle=solid}
\newpsobject{PST@cascade}{psdots}{linewidth=.02,linestyle=solid,dotsize=.05pt 1,dotstyle=x,linecolor=red}
\newpsobject{PST@hadrong}{psline}{linecolor=blue,     linewidth=.0025,linestyle=solid}
\catcode`@=12

\fi
\psset{unit=5.0in,xunit=1.123\columnwidth,yunit=.75\columnwidth}
\pspicture(0.100000,0.022000)(.985,.89)
\ifx\nofigs\undefined
\catcode`@=11

\PST@Border(0.2390,0.1344)
(0.2540,0.1344)

\PST@Border(0.9630,0.1344)
(0.9480,0.1344)

\rput[r](0.2230,0.1344){$10^{-7}$}
\PST@Border(0.2390,0.1623)
(0.2465,0.1623)

\PST@Border(0.9630,0.1623)
(0.9555,0.1623)

\PST@Border(0.2390,0.1991)
(0.2465,0.1991)

\PST@Border(0.9630,0.1991)
(0.9555,0.1991)

\PST@Border(0.2390,0.2180)
(0.2465,0.2180)

\PST@Border(0.9630,0.2180)
(0.9555,0.2180)

\PST@Border(0.2390,0.2270)
(0.2540,0.2270)

\PST@Border(0.9630,0.2270)
(0.9480,0.2270)

\rput[r](0.2230,0.2270){$10^{-6}$}
\PST@Border(0.2390,0.2549)
(0.2465,0.2549)

\PST@Border(0.9630,0.2549)
(0.9555,0.2549)

\PST@Border(0.2390,0.2918)
(0.2465,0.2918)

\PST@Border(0.9630,0.2918)
(0.9555,0.2918)

\PST@Border(0.2390,0.3107)
(0.2465,0.3107)

\PST@Border(0.9630,0.3107)
(0.9555,0.3107)

\PST@Border(0.2390,0.3196)
(0.2540,0.3196)

\PST@Border(0.9630,0.3196)
(0.9480,0.3196)

\rput[r](0.2230,0.3196){$10^{-5}$}
\PST@Border(0.2390,0.3475)
(0.2465,0.3475)

\PST@Border(0.9630,0.3475)
(0.9555,0.3475)

\PST@Border(0.2390,0.3844)
(0.2465,0.3844)

\PST@Border(0.9630,0.3844)
(0.9555,0.3844)

\PST@Border(0.2390,0.4033)
(0.2465,0.4033)

\PST@Border(0.9630,0.4033)
(0.9555,0.4033)

\PST@Border(0.2390,0.4123)
(0.2540,0.4123)

\PST@Border(0.9630,0.4123)
(0.9480,0.4123)

\rput[r](0.2230,0.4123){$10^{-4}$}
\PST@Border(0.2390,0.4401)
(0.2465,0.4401)

\PST@Border(0.9630,0.4401)
(0.9555,0.4401)

\PST@Border(0.2390,0.4770)
(0.2465,0.4770)

\PST@Border(0.9630,0.4770)
(0.9555,0.4770)

\PST@Border(0.2390,0.4959)
(0.2465,0.4959)

\PST@Border(0.9630,0.4959)
(0.9555,0.4959)

\PST@Border(0.2390,0.5049)
(0.2540,0.5049)

\PST@Border(0.9630,0.5049)
(0.9480,0.5049)

\rput[r](0.2230,0.5049){$10^{-3}$}
\PST@Border(0.2390,0.5328)
(0.2465,0.5328)

\PST@Border(0.9630,0.5328)
(0.9555,0.5328)

\PST@Border(0.2390,0.5696)
(0.2465,0.5696)

\PST@Border(0.9630,0.5696)
(0.9555,0.5696)

\PST@Border(0.2390,0.5885)
(0.2465,0.5885)

\PST@Border(0.9630,0.5885)
(0.9555,0.5885)

\PST@Border(0.2390,0.5975)
(0.2540,0.5975)

\PST@Border(0.9630,0.5975)
(0.9480,0.5975)

\rput[r](0.2230,0.5975){ 0.01}
\PST@Border(0.2390,0.6254)
(0.2465,0.6254)

\PST@Border(0.9630,0.6254)
(0.9555,0.6254)

\PST@Border(0.2390,0.6623)
(0.2465,0.6623)

\PST@Border(0.9630,0.6623)
(0.9555,0.6623)

\PST@Border(0.2390,0.6812)
(0.2465,0.6812)

\PST@Border(0.9630,0.6812)
(0.9555,0.6812)

\PST@Border(0.2390,0.6901)
(0.2540,0.6901)

\PST@Border(0.9630,0.6901)
(0.9480,0.6901)

\rput[r](0.2230,0.6901){ 0.1}
\PST@Border(0.2390,0.7180)
(0.2465,0.7180)

\PST@Border(0.9630,0.7180)
(0.9555,0.7180)

\PST@Border(0.2390,0.7549)
(0.2465,0.7549)

\PST@Border(0.9630,0.7549)
(0.9555,0.7549)

\PST@Border(0.2390,0.7738)
(0.2465,0.7738)

\PST@Border(0.9630,0.7738)
(0.9555,0.7738)

\PST@Border(0.2390,0.7828)
(0.2540,0.7828)

\PST@Border(0.9630,0.7828)
(0.9480,0.7828)

\rput[r](0.2230,0.7828){ 1}
\PST@Border(0.2390,0.8106)
(0.2465,0.8106)

\PST@Border(0.9630,0.8106)
(0.9555,0.8106)

\PST@Border(0.2390,0.8475)
(0.2465,0.8475)

\PST@Border(0.9630,0.8475)
(0.9555,0.8475)

\PST@Border(0.2390,0.8664)
(0.2465,0.8664)

\PST@Border(0.9630,0.8664)
(0.9555,0.8664)

\PST@Border(0.2390,0.8754)
(0.2540,0.8754)

\PST@Border(0.9630,0.8754)
(0.9480,0.8754)

\rput[r](0.2230,0.8754){ 10}

\PST@Border(0.2390,0.1344)
(0.2390,0.1544)

\PST@Border(0.2390,0.8754)
(0.2390,0.8554)

\rput(0.2390,0.0924){ 0}
\PST@Border(0.3194,0.1344)
(0.3194,0.1544)

\PST@Border(0.3194,0.8754)
(0.3194,0.8554)

\rput(0.3194,0.0924){ 0.5}
\PST@Border(0.3999,0.1344)
(0.3999,0.1544)

\PST@Border(0.3999,0.8754)
(0.3999,0.8554)

\rput(0.3999,0.0924){ 1}
\PST@Border(0.4803,0.1344)
(0.4803,0.1544)

\PST@Border(0.4803,0.8754)
(0.4803,0.8554)

\rput(0.4803,0.0924){ 1.5}
\PST@Border(0.5608,0.1344)
(0.5608,0.1544)

\PST@Border(0.5608,0.8754)
(0.5608,0.8554)

\rput(0.5608,0.0924){ 2}
\PST@Border(0.6412,0.1344)
(0.6412,0.1544)

\PST@Border(0.6412,0.8754)
(0.6412,0.8554)

\rput(0.6412,0.0924){ 2.5}
\PST@Border(0.7217,0.1344)
(0.7217,0.1544)

\PST@Border(0.7217,0.8754)
(0.7217,0.8554)

\rput(0.7217,0.0924){ 3}
\PST@Border(0.8021,0.1344)
(0.8021,0.1544)

\PST@Border(0.8021,0.8754)
(0.8021,0.8554)

\rput(0.8021,0.0924){ 3.5}
\PST@Border(0.8826,0.1344)
(0.8826,0.1544)

\PST@Border(0.8826,0.8754)
(0.8826,0.8554)

\rput(0.8826,0.0924){ 4}
\PST@Border(0.9630,0.1344)
(0.9630,0.1544)

\PST@Border(0.9630,0.8754)
(0.9630,0.8554)

\rput(0.9630,0.0924){ 4.5}

\PST@Border(0.2390,0.8754) (0.2390,0.1344)
(0.9630,0.1344) (0.9630,0.8754) (0.2390,0.8754)

\rput{L}(0.1220,0.5470){$E\frac{dN}{d^3p}$ [GeV$^{-2}$]}
\rput(0.6010,0.0494){$p_\bot$ [GeV]}

\rput[r](0.8360,0.8144){Pure UrQMD}
\rput[r](0.8360,0.7524){Hybrid, hadron gas EoS}
\PST@cascade(0.8915,0.8144)
\PST@hadrong(0.8520,0.7524) (0.9310,0.7524)

\PST@cascade(0.2631,0.8276) (0.2792,0.7787) (0.2953,0.7567) (0.3114,0.7422)
(0.3275,0.7275) (0.3436,0.7110) (0.3597,0.6932) (0.3758,0.6743) (0.3918,0.6551)
(0.4079,0.6354) (0.4240,0.6154) (0.4401,0.5955) (0.4562,0.5759) (0.4723,0.5561)
(0.4884,0.5370) (0.5045,0.5171) (0.5206,0.4988) (0.5366,0.4804) (0.5527,0.4634)
(0.5688,0.4450) (0.5849,0.4279) (0.6010,0.4120) (0.6171,0.3971) (0.6332,0.3808)
(0.6493,0.3675) (0.6654,0.3525) (0.6814,0.3416) (0.6975,0.3301) (0.7136,0.3177)
(0.7297,0.3067) (0.7458,0.2980) (0.7619,0.2883) (0.7780,0.2795) (0.7941,0.2676)
(0.8102,0.2586) (0.8262,0.2493) (0.8423,0.2423) (0.8584,0.2365) (0.8745,0.2276)
(0.8906,0.2220) (0.9067,0.2177) (0.9228,0.2119) (0.9389,0.2026) (0.9550,0.1961)

\PST@hadrong(0.2631,0.8636) (0.2792,0.8277) (0.2953,0.7626) (0.3114,0.7340)
(0.3275,0.7141) (0.3436,0.6969) (0.3597,0.6841) (0.3758,0.6574) (0.3918,0.6489)
(0.4079,0.6303) (0.4240,0.6079) (0.4401,0.6022) (0.4562,0.5761) (0.4723,0.5676)
(0.4884,0.5428) (0.5045,0.5321) (0.5206,0.5139) (0.5366,0.4935) (0.5527,0.4889)
(0.5688,0.4648) (0.5849,0.4530) (0.6010,0.4315) (0.6171,0.4163) (0.6332,0.4110)
(0.6493,0.3759) (0.6654,0.3701) (0.6814,0.3543) (0.6975,0.3406) (0.7136,0.3337)
(0.7297,0.3157) (0.7458,0.3025) (0.7619,0.2874) (0.7780,0.2770) (0.7941,0.2658)
(0.8102,0.2551) (0.8262,0.2460) (0.8423,0.2394) (0.8584,0.2196) (0.8745,0.2217)
(0.8906,0.2198) (0.9067,0.2106) (0.9228,0.2052) (0.9389,0.2081) (0.9550,0.2031)

\rput[r](0.9150,0.6330){Only common channels}
\rput[r](0.9150,0.5910){Pb+Pb 158~AGeV}
\rput[r](0.9150,0.5490){$b<4.5$~fm}
\rput[r](0.9150,0.5070){$|y_{\rm c.m.}| < 0.5$}

\catcode`@=12
\fi
\endpspicture

%% file: comparison.tex
\ifx\PSTloaded\undefined
\def\PSTloaded{t}
\psset{arrowsize=.01 3.2 1.4 .3}
\psset{dotsize=.01}
\catcode`@=11

\definecolor{darkgreen}{rgb}{0,.5,0}
\definecolor{violett}{rgb}{.5,0,.5}
\definecolor{orange}{rgb}{.8,.4,0}
\newpsobject{PST@Border}{psline}{linewidth=.0015,linestyle=solid}
\newpsobject{PST@Axes}{psline}{linewidth=.0015,linestyle=dotted,dotsep=.004}
\newpsobject{PST@cascade}{psdots}{linewidth=.02,linestyle=solid,dotsize=.05pt 1,dotstyle=x,linecolor=red}
\newpsobject{PST@hadrong}{psline}{linecolor=blue,     linewidth=.0025,linestyle=solid}
\newpsobject{PST@bagmode}{psline}{linecolor=darkgreen,linewidth=.0040,linestyle=dotted,dotsep=.004}
\newpsobject{PST@pqcd}   {psline}{linecolor=violett,  linewidth=.0025,linestyle=dashed,dash=.01 .004 .004 .004}

\newpsobject{PST@square}{psdots}{linewidth=.005,linestyle=solid,dotsize=.05pt 2,dotstyle=square,linecolor=black}
\newpsobject{PST@Limit}{psline}{linecolor=black,linewidth=.001,arrowsize=4pt,arrowlength=1.7,arrowinset=.7,arrows=->}
\newpsobject{PST@limit}{psline}{linecolor=black,linewidth=.001,arrowsize=4pt,arrowlength=1.7,arrowinset=.7,arrows=|->}
\newpsobject{PST@error}{psline}{linecolor=black,linewidth=.001,linestyle=solid,arrows=|-|}
\catcode`@=12

\fi
\psset{unit=5.0in,xunit=1.075\columnwidth,yunit=.8\columnwidth}
\pspicture(0.0500,0.0220)(0.980000,0.9880)
\ifx\nofigs\undefined
\catcode`@=11

\PST@Border(0.2070,0.1260) (0.2220,0.1260) \PST@Border(0.9470,0.1260)
(0.9320,0.1260) \rput[r](0.1910,0.1260){$10^{-7}$} \PST@Border(0.2070,0.1577)
(0.2145,0.1577) \PST@Border(0.9470,0.1577) (0.9395,0.1577)
\PST@Border(0.2070,0.1996) (0.2145,0.1996) \PST@Border(0.9470,0.1996)
(0.9395,0.1996) \PST@Border(0.2070,0.2211) (0.2145,0.2211)
\PST@Border(0.9470,0.2211) (0.9395,0.2211) \PST@Border(0.2070,0.2313)
(0.2220,0.2313) \PST@Border(0.9470,0.2313) (0.9320,0.2313)
\rput[r](0.1910,0.2313){$10^{-6}$} \PST@Border(0.2070,0.2629) (0.2145,0.2629)
\PST@Border(0.9470,0.2629) (0.9395,0.2629) \PST@Border(0.2070,0.3048)
(0.2145,0.3048) \PST@Border(0.9470,0.3048) (0.9395,0.3048)
\PST@Border(0.2070,0.3263) (0.2145,0.3263) \PST@Border(0.9470,0.3263)
(0.9395,0.3263) \PST@Border(0.2070,0.3365) (0.2220,0.3365)
\PST@Border(0.9470,0.3365) (0.9320,0.3365) \rput[r](0.1910,0.3365){$10^{-5}$}
\PST@Border(0.2070,0.3682) (0.2145,0.3682) \PST@Border(0.9470,0.3682)
(0.9395,0.3682) \PST@Border(0.2070,0.4101) (0.2145,0.4101)
\PST@Border(0.9470,0.4101) (0.9395,0.4101) \PST@Border(0.2070,0.4316)
(0.2145,0.4316) \PST@Border(0.9470,0.4316) (0.9395,0.4316)
\PST@Border(0.2070,0.4418) (0.2220,0.4418) \PST@Border(0.9470,0.4418)
(0.9320,0.4418) \rput[r](0.1910,0.4418){$10^{-4}$} \PST@Border(0.2070,0.4734)
(0.2145,0.4734) \PST@Border(0.9470,0.4734) (0.9395,0.4734)
\PST@Border(0.2070,0.5153) (0.2145,0.5153) \PST@Border(0.9470,0.5153)
(0.9395,0.5153) \PST@Border(0.2070,0.5368) (0.2145,0.5368)
\PST@Border(0.9470,0.5368) (0.9395,0.5368) \PST@Border(0.2070,0.5470)
(0.2220,0.5470) \PST@Border(0.9470,0.5470) (0.9320,0.5470)
\rput[r](0.1910,0.5470){ 0.001} \PST@Border(0.2070,0.5787) (0.2145,0.5787)
\PST@Border(0.9470,0.5787) (0.9395,0.5787) \PST@Border(0.2070,0.6206)
(0.2145,0.6206) \PST@Border(0.9470,0.6206) (0.9395,0.6206)
\PST@Border(0.2070,0.6421) (0.2145,0.6421) \PST@Border(0.9470,0.6421)
(0.9395,0.6421) \PST@Border(0.2070,0.6523) (0.2220,0.6523)
\PST@Border(0.9470,0.6523) (0.9320,0.6523) \rput[r](0.1910,0.6523){ 0.01}
\PST@Border(0.2070,0.6839) (0.2145,0.6839) \PST@Border(0.9470,0.6839)
(0.9395,0.6839) \PST@Border(0.2070,0.7258) (0.2145,0.7258)
\PST@Border(0.9470,0.7258) (0.9395,0.7258) \PST@Border(0.2070,0.7473)
(0.2145,0.7473) \PST@Border(0.9470,0.7473) (0.9395,0.7473)
\PST@Border(0.2070,0.7575) (0.2220,0.7575) \PST@Border(0.9470,0.7575)
(0.9320,0.7575) \rput[r](0.1910,0.7575){ 0.1} \PST@Border(0.2070,0.7892)
(0.2145,0.7892) \PST@Border(0.9470,0.7892) (0.9395,0.7892)
\PST@Border(0.2070,0.8311) (0.2145,0.8311) \PST@Border(0.9470,0.8311)
(0.9395,0.8311) \PST@Border(0.2070,0.8526) (0.2145,0.8526)
\PST@Border(0.9470,0.8526) (0.9395,0.8526) \PST@Border(0.2070,0.8628)
(0.2220,0.8628) \PST@Border(0.9470,0.8628) (0.9320,0.8628)
\rput[r](0.1910,0.8628){ 1} \PST@Border(0.2070,0.8944) (0.2145,0.8944)
\PST@Border(0.9470,0.8944) (0.9395,0.8944) \PST@Border(0.2070,0.9363)
(0.2145,0.9363) \PST@Border(0.9470,0.9363) (0.9395,0.9363)
\PST@Border(0.2070,0.9578) (0.2145,0.9578) \PST@Border(0.9470,0.9578)
(0.9395,0.9578) \PST@Border(0.2070,0.9680) (0.2220,0.9680)
\PST@Border(0.9470,0.9680) (0.9320,0.9680) \rput[r](0.1910,0.9680){ 10}
\PST@Border(0.2070,0.1260) (0.2070,0.1460) \PST@Border(0.2070,0.9680)
(0.2070,0.9480) \rput(0.2070,0.0840){ 0} \PST@Border(0.2892,0.1260)
(0.2892,0.1460) \PST@Border(0.2892,0.9680) (0.2892,0.9480)
\rput(0.2892,0.0840){ 0.5} \PST@Border(0.3714,0.1260) (0.3714,0.1460)
\PST@Border(0.3714,0.9680) (0.3714,0.9480) \rput(0.3714,0.0840){ 1}
\PST@Border(0.4537,0.1260) (0.4537,0.1460) \PST@Border(0.4537,0.9680)
(0.4537,0.9480) \rput(0.4537,0.0840){ 1.5} \PST@Border(0.5359,0.1260)
(0.5359,0.1460) \PST@Border(0.5359,0.9680) (0.5359,0.9480)
\rput(0.5359,0.0840){ 2} \PST@Border(0.6181,0.1260) (0.6181,0.1460)
\PST@Border(0.6181,0.9680) (0.6181,0.9480) \rput(0.6181,0.0840){ 2.5}
\PST@Border(0.7003,0.1260) (0.7003,0.1460) \PST@Border(0.7003,0.9680)
(0.7003,0.9480) \rput(0.7003,0.0840){ 3} \PST@Border(0.7826,0.1260)
(0.7826,0.1460) \PST@Border(0.7826,0.9680) (0.7826,0.9480)
\rput(0.7826,0.0840){ 3.5} \PST@Border(0.8648,0.1260) (0.8648,0.1460)
\PST@Border(0.8648,0.9680) (0.8648,0.9480) \rput(0.8648,0.0840){ 4}
\PST@Border(0.9470,0.1260) (0.9470,0.1460) \PST@Border(0.9470,0.9680)
(0.9470,0.9480) \rput(0.9470,0.0840){ 4.5}

\PST@Border(0.2070,0.9680)(0.2070,0.1260) (0.9470,0.1260) (0.9470,0.9680) (0.2070,0.9680)

\rput{L}(0.0820,0.5470){$E\frac{dN}{d^3p}$ [GeV$^{-2}$]}
\rput(0.5770,0.0410){$p_\bot$~[GeV]}

\PST@Limit(0.2481,0.9680)(0.2481,0.8882)
\PST@Limit(0.2646,0.9680)(0.2646,0.7475)
\PST@Limit(0.2810,0.9680)(0.2810,0.6732)
\PST@limit(0.2974,0.9337)(0.2974,0.6180)
\PST@limit(0.3139,0.8905)(0.3139,0.5747)
\PST@limit(0.3303,0.8534)(0.3303,0.5377)
\PST@limit(0.3468,0.8263)(0.3468,0.5105)
\PST@limit(0.3632,0.7971)(0.3632,0.4813)
\PST@limit(0.3797,0.7779)(0.3797,0.4621)
\PST@limit(0.3961,0.7554)(0.3961,0.4397)
\PST@limit(0.4126,0.7355)(0.4126,0.4197)
\PST@limit(0.4290,0.7167)(0.4290,0.4009)
\PST@limit(0.4454,0.6956)(0.4454,0.3799)
\PST@limit(0.7579,0.3433)(0.7579,0.1260)
\PST@limit(0.8072,0.2982)(0.8072,0.1260)
\PST@limit(0.8237,0.2598)(0.8237,0.1260)
\PST@limit(0.8401,0.2741)(0.8401,0.1260)
\PST@error(0.4619,0.4854)(0.4619,0.6751)
\PST@error(0.4783,0.5042)(0.4783,0.6537)
\PST@error(0.4948,0.5643)(0.4948,0.6390)
\PST@error(0.5112,0.5732)(0.5112,0.6244)
\PST@error(0.5277,0.5613)(0.5277,0.6060)
\PST@error(0.5441,0.5462)(0.5441,0.5873)
\PST@error(0.5606,0.5385)(0.5606,0.5721)
\PST@error(0.5770,0.5150)(0.5770,0.5505)
\PST@error(0.5934,0.4880)(0.5934,0.5282)
\PST@error(0.6099,0.4806)(0.6099,0.5135)
\PST@error(0.6263,0.4495)(0.6263,0.4898)
\PST@error(0.6428,0.4454)(0.6428,0.4773)
\PST@error(0.6592,0.3806)(0.6592,0.4438)
\PST@error(0.6757,0.3325)(0.6757,0.4208)
\PST@error(0.6921,0.3618)(0.6921,0.4134)
\PST@error(0.7086,0.3750)(0.7086,0.4080)
\PST@error(0.7250,0.3643)(0.7250,0.3947)
\PST@error(0.7414,0.3313)(0.7414,0.3726)
\PST@error(0.7743,0.3270)(0.7743,0.3554)
\PST@error(0.7908,0.2734)(0.7908,0.3278)
\PST@error(0.8566,0.2837)(0.8566,0.3051)

\rput[r](0.8200,0.9270){WA98 Pb+Pb 158 AGeV}
\PST@square(0.8755,0.9270)

\PST@square(0.3961,0.6159) \PST@square(0.4126,0.6548)
\PST@square(0.4290,0.6613) \PST@square(0.4454,0.6476)
\PST@square(0.4619,0.6466) \PST@square(0.4783,0.6262)
\PST@square(0.4948,0.6173) \PST@square(0.5112,0.6070)
\PST@square(0.5277,0.5902) \PST@square(0.5441,0.5724)
\PST@square(0.5606,0.5593) \PST@square(0.5770,0.5372)
\PST@square(0.5934,0.5135) \PST@square(0.6099,0.5009)
\PST@square(0.6263,0.4751) \PST@square(0.6428,0.4649)
\PST@square(0.6592,0.4238) \PST@square(0.6757,0.3971)
\PST@square(0.6921,0.3956) \PST@square(0.7086,0.3951)
\PST@square(0.7250,0.3825) \PST@square(0.7414,0.3571)
\PST@square(0.7579,0.3057) \PST@square(0.7743,0.3436)
\PST@square(0.7908,0.3086) \PST@square(0.8072,0.2410)
\PST@square(0.8401,0.2056) \PST@square(0.8566,0.2957)

\rput[r](0.8200,0.8850){Pure UrQMD}
\PST@cascade(0.8755,0.8850)

\PST@cascade(0.2257,0.9680) (0.2317,0.9146) (0.2481,0.8595) (0.2646,0.8342)
(0.2810,0.8175) (0.2974,0.8006) (0.3139,0.7817) (0.3303,0.7614) (0.3468,0.7399)
(0.3632,0.7180) (0.3797,0.6957) (0.3961,0.6731) (0.4126,0.6505) (0.4290,0.6283)
(0.4454,0.6059) (0.4619,0.5845) (0.4783,0.5621) (0.4948,0.5416) (0.5112,0.5212)
(0.5277,0.5022) (0.5441,0.4820) (0.5606,0.4634) (0.5770,0.4462) (0.5934,0.4300)
(0.6099,0.4127) (0.6263,0.3989) (0.6428,0.3835) (0.6592,0.3721) (0.6757,0.3603)
(0.6921,0.3474) (0.7086,0.3359) (0.7250,0.3260) (0.7414,0.3158) (0.7579,0.3064)
(0.7743,0.2953) (0.7908,0.2853) (0.8072,0.2747) (0.8237,0.2666) (0.8401,0.2611)
(0.8566,0.2520) (0.8730,0.2488) (0.8894,0.2425) (0.9059,0.2351) (0.9223,0.2254)
(0.9388,0.2162)

\rput[r](0.8200,0.8430){Hybrid, Hadron gas EoS}
\PST@hadrong(0.8360,0.8430)(0.9150,0.8430)

\PST@hadrong(0.2299,0.9680) (0.2317,0.9604) (0.2481,0.9205) (0.2646,0.8458)
(0.2810,0.8124) (0.2974,0.7891) (0.3139,0.7691) (0.3303,0.7544) (0.3468,0.7230)
(0.3632,0.7136) (0.3797,0.6921) (0.3961,0.6662) (0.4126,0.6601) (0.4290,0.6300)
(0.4454,0.6205) (0.4619,0.5920) (0.4783,0.5801) (0.4948,0.5594) (0.5112,0.5362)
(0.5277,0.5311) (0.5441,0.5040) (0.5606,0.4909) (0.5770,0.4669) (0.5934,0.4502)
(0.6099,0.4433) (0.6263,0.4052) (0.6428,0.3985) (0.6592,0.3813) (0.6757,0.3665)
(0.6921,0.3597) (0.7086,0.3414) (0.7250,0.3264) (0.7414,0.3113) (0.7579,0.3026)
(0.7743,0.2884) (0.7908,0.2773) (0.8072,0.2677) (0.8237,0.2639) (0.8401,0.2406)
(0.8566,0.2464) (0.8730,0.2391) (0.8894,0.2299) (0.9059,0.2245) (0.9223,0.2231)
(0.9388,0.2167)

\rput[r](0.8200,0.8010){Hybrid, Bag Model EoS}
\PST@bagmode(0.8360,0.8010)(0.9150,0.8010)

\PST@bagmode(0.2510,0.9680) (0.2646,0.9081) (0.2810,0.8736) (0.2974,0.8488)
(0.3139,0.8253) (0.3303,0.8115) (0.3468,0.7651) (0.3632,0.7635) (0.3797,0.7374)
(0.3961,0.7029) (0.4126,0.7071) (0.4290,0.6662) (0.4454,0.6634) (0.4619,0.6255)
(0.4783,0.6190) (0.4948,0.5943) (0.5112,0.5648) (0.5277,0.5691) (0.5441,0.5333)
(0.5606,0.5239) (0.5770,0.4922) (0.5934,0.4726) (0.6099,0.4732) (0.6263,0.4294)
(0.6428,0.4282) (0.6592,0.3944) (0.6757,0.3783) (0.6921,0.3636) (0.7086,0.3428)
(0.7250,0.3342) (0.7414,0.3101) (0.7579,0.3072) (0.7743,0.2999) (0.7908,0.2855)
(0.8072,0.2842) (0.8237,0.2645) (0.8401,0.2607) (0.8566,0.2478) (0.8730,0.2434)
(0.8894,0.2308) (0.9059,0.2200) (0.9223,0.2188) (0.9388,0.2091) (0.9470,0.2073)


\rput[r](0.9150,0.5910){Pb+Pb 158~AGeV}
\rput[r](0.9150,0.5490){$b<4.5$~fm}
\rput[r](0.9150,0.5070){$|y_\textnormal{c.m.}| < 0.5$}

\PST@Border(0.2070,0.9680) (0.2070,0.1260) (0.9470,0.1260) (0.9470,0.9680)
(0.2070,0.9680)

\catcode`@=12
\fi
\endpspicture

%% file: comparison_pqcd.tex
\ifx\PSTloaded\undefined
\def\PSTloaded{t}
\psset{arrowsize=.01 3.2 1.4 .3}
\psset{dotsize=.01}
\catcode`@=11

\definecolor{darkgreen}{rgb}{0,.5,0}
\definecolor{violett}{rgb}{.5,0,.5}
\definecolor{orange}{rgb}{.8,.4,0}
\newpsobject{PST@Border}{psline}{linewidth=.0015,linestyle=solid}
\newpsobject{PST@Axes}{psline}{linewidth=.0015,linestyle=dotted,dotsep=.004}
\newpsobject{PST@cascade}{psdots}{linewidth=.02,linestyle=solid,dotsize=.05pt 1,dotstyle=x,linecolor=red}
\newpsobject{PST@hadrong}{psline}{linecolor=blue,     linewidth=.0025,linestyle=solid}
\newpsobject{PST@bagmode}{psline}{linecolor=darkgreen,linewidth=.0040,linestyle=dotted,dotsep=.004}
\newpsobject{PST@pqcd}   {psline}{linecolor=violett,  linewidth=.0025,linestyle=dashed,dash=.01 .004 .004 .004}

\newpsobject{PST@square}{psdots}{linewidth=.005,linestyle=solid,dotsize=.05pt 2,dotstyle=square,linecolor=black}
\newpsobject{PST@Limit}{psline}{linecolor=black,linewidth=.001,arrowsize=4pt,arrowlength=1.7,arrowinset=.7,arrows=->}
\newpsobject{PST@limit}{psline}{linecolor=black,linewidth=.001,arrowsize=4pt,arrowlength=1.7,arrowinset=.7,arrows=|->}
\newpsobject{PST@error}{psline}{linecolor=black,linewidth=.001,linestyle=solid,arrows=|-|}
\catcode`@=12

\fi
\psset{unit=5.0in,xunit=1.075\columnwidth,yunit=.8\columnwidth}
\pspicture(0.0500,0.0220)(0.980000,0.9880)
\ifx\nofigs\undefined
\catcode`@=11

\PST@Border(0.2390,0.1344) (0.2540,0.1344) \PST@Border(0.9630,0.1344)
(0.9480,0.1344) \rput[r](0.2230,0.1344){$10^{-7}$} \PST@Border(0.2390,0.1817)
(0.2465,0.1817) \PST@Border(0.9630,0.1817) (0.9555,0.1817)
\PST@Border(0.2390,0.2443) (0.2465,0.2443) \PST@Border(0.9630,0.2443)
(0.9555,0.2443) \PST@Border(0.2390,0.2764) (0.2465,0.2764)
\PST@Border(0.9630,0.2764) (0.9555,0.2764) \PST@Border(0.2390,0.2917)
(0.2540,0.2917) \PST@Border(0.9630,0.2917) (0.9480,0.2917)
\rput[r](0.2230,0.2917){$10^{-6}$} \PST@Border(0.2390,0.3390) (0.2465,0.3390)
\PST@Border(0.9630,0.3390) (0.9555,0.3390) \PST@Border(0.2390,0.4016)
(0.2465,0.4016) \PST@Border(0.9630,0.4016) (0.9555,0.4016)
\PST@Border(0.2390,0.4337) (0.2465,0.4337) \PST@Border(0.9630,0.4337)
(0.9555,0.4337) \PST@Border(0.2390,0.4489) (0.2540,0.4489)
\PST@Border(0.9630,0.4489) (0.9480,0.4489) \rput[r](0.2230,0.4489){$10^{-5}$}
\PST@Border(0.2390,0.4962) (0.2465,0.4962) \PST@Border(0.9630,0.4962)
(0.9555,0.4962) \PST@Border(0.2390,0.5588) (0.2465,0.5588)
\PST@Border(0.9630,0.5588) (0.9555,0.5588) \PST@Border(0.2390,0.5909)
(0.2465,0.5909) \PST@Border(0.9630,0.5909) (0.9555,0.5909)
\PST@Border(0.2390,0.6062) (0.2540,0.6062) \PST@Border(0.9630,0.6062)
(0.9480,0.6062) \rput[r](0.2230,0.6062){$10^{-4}$} \PST@Border(0.2390,0.6535)
(0.2465,0.6535) \PST@Border(0.9630,0.6535) (0.9555,0.6535)
\PST@Border(0.2390,0.7161) (0.2465,0.7161) \PST@Border(0.9630,0.7161)
(0.9555,0.7161) \PST@Border(0.2390,0.7482) (0.2465,0.7482)
\PST@Border(0.9630,0.7482) (0.9555,0.7482) \PST@Border(0.2390,0.7634)
(0.2540,0.7634) \PST@Border(0.9630,0.7634) (0.9480,0.7634)
\rput[r](0.2230,0.7634){ 0.001} \PST@Border(0.2390,0.8107) (0.2465,0.8107)
\PST@Border(0.9630,0.8107) (0.9555,0.8107) \PST@Border(0.2390,0.8733)
(0.2465,0.8733) \PST@Border(0.9630,0.8733) (0.9555,0.8733)
\PST@Border(0.2390,0.9054) (0.2465,0.9054) \PST@Border(0.9630,0.9054)
(0.9555,0.9054) \PST@Border(0.2390,0.9207) (0.2540,0.9207)
\PST@Border(0.9630,0.9207) (0.9480,0.9207) \rput[r](0.2230,0.9207){ 0.01}
\PST@Border(0.2390,0.9680) (0.2465,0.9680) \PST@Border(0.9630,0.9680)
(0.9555,0.9680) \PST@Border(0.2390,0.1344) (0.2390,0.1544)
\PST@Border(0.2390,0.9680) (0.2390,0.9480) \rput(0.2390,0.0924){ 1.5}
\PST@Border(0.3597,0.1344) (0.3597,0.1544) \PST@Border(0.3597,0.9680)
(0.3597,0.9480) \rput(0.3597,0.0924){ 2} \PST@Border(0.4803,0.1344)
(0.4803,0.1544) \PST@Border(0.4803,0.9680) (0.4803,0.9480)
\rput(0.4803,0.0924){ 2.5} \PST@Border(0.6010,0.1344) (0.6010,0.1544)
\PST@Border(0.6010,0.9680) (0.6010,0.9480) \rput(0.6010,0.0924){ 3}
\PST@Border(0.7217,0.1344) (0.7217,0.1544) \PST@Border(0.7217,0.9680)
(0.7217,0.9480) \rput(0.7217,0.0924){ 3.5} \PST@Border(0.8423,0.1344)
(0.8423,0.1544) \PST@Border(0.8423,0.9680) (0.8423,0.9480)
\rput(0.8423,0.0924){ 4} \PST@Border(0.9630,0.1344) (0.9630,0.1544)
\PST@Border(0.9630,0.9680) (0.9630,0.9480) \rput(0.9630,0.0924){ 4.5}

\PST@Border(0.2390,0.9680) (0.2390,0.1344) (0.9630,0.1344) (0.9630,0.9680) (0.2390,0.9680)

\rput{L}(0.0820,0.5470){$E\frac{dN}{d^3p}$ [GeV$^{-2}$]}
\rput(0.5770,0.0410){$p_\bot$~[GeV]}

\PST@limit(0.6855,0.4591)(0.6855,0.1344)
\PST@limit(0.7579,0.3917)(0.7579,0.1344)
\PST@limit(0.7820,0.3342)(0.7820,0.1344)
\PST@error(0.8061,0.3556)(0.8061,0.1344)
\PST@error(0.2511,0.6713)(0.2511,0.9548)
\PST@error(0.2752,0.6995)(0.2752,0.9228)
\PST@error(0.2993,0.7893)(0.2993,0.9008)
\PST@error(0.3235,0.8025)(0.3235,0.8790)
\PST@error(0.3476,0.7847)(0.3476,0.8515)
\PST@error(0.3717,0.7622)(0.3717,0.8236)
\PST@error(0.3959,0.7507)(0.3959,0.8010)
\PST@error(0.4200,0.7156)(0.4200,0.7687)
\PST@error(0.4441,0.6752)(0.4441,0.7352)
\PST@error(0.4683,0.6643)(0.4683,0.7134)
\PST@error(0.4924,0.6177)(0.4924,0.6779)
\PST@error(0.5165,0.6117)(0.5165,0.6592)
\PST@error(0.5407,0.5148)(0.5407,0.6092)
\PST@error(0.5648,0.4429)(0.5648,0.5749)
\PST@error(0.5889,0.4866)(0.5889,0.5637)
\PST@error(0.6131,0.5065)(0.6131,0.5558)
\PST@error(0.6372,0.4905)(0.6372,0.5358)
\PST@error(0.6613,0.4412)(0.6613,0.5029)
\PST@error(0.7096,0.4347)(0.7096,0.4772)
\PST@error(0.7337,0.3546)(0.7337,0.4359)
\PST@error(0.8303,0.3700)(0.8303,0.4020)

\rput[r](0.8200,0.9270){WA98 Pb+Pb 158 AGeV}
\PST@square(0.8915,0.9270)

\PST@square(0.2511,0.9123) \PST@square(0.2752,0.8818)
\PST@square(0.2993,0.8685) \PST@square(0.3235,0.8531)
\PST@square(0.3476,0.8279) \PST@square(0.3717,0.8013)
\PST@square(0.3959,0.7819) \PST@square(0.4200,0.7487)
\PST@square(0.4441,0.7134) \PST@square(0.4683,0.6946)
\PST@square(0.4924,0.6559) \PST@square(0.5165,0.6408)
\PST@square(0.5407,0.5794) \PST@square(0.5648,0.5394)
\PST@square(0.5889,0.5372) \PST@square(0.6131,0.5364)
\PST@square(0.6372,0.5176) \PST@square(0.6613,0.4796)
\PST@square(0.6855,0.4028) \PST@square(0.7096,0.4596)
\PST@square(0.7337,0.4072) \PST@square(0.7579,0.3062)
\PST@square(0.8061,0.2533) \PST@square(0.8303,0.3879)

\rput[r](0.8360,0.8850){UrQMD + pQCD}
\PST@cascade(0.8915,0.8850)

\PST@cascade(0.2511,0.9035) (0.2752,0.8724) (0.2993,0.8454)
(0.3235,0.8191) (0.3476,0.7921) (0.3717,0.7660) (0.3959,0.7392) (0.4200,0.7135)
(0.4441,0.6889) (0.4683,0.6625) (0.4924,0.6387) (0.5165,0.6123) (0.5407,0.5871)
(0.5648,0.5632) (0.5889,0.5396) (0.6131,0.5186) (0.6372,0.4965) (0.6613,0.4767)
(0.6855,0.4571) (0.7096,0.4375) (0.7337,0.4198) (0.7579,0.4018) (0.7820,0.3862)
(0.8061,0.3731) (0.8303,0.3574) (0.8544,0.3471) (0.8785,0.3352) (0.9027,0.3224)
(0.9268,0.3070) (0.9509,0.2925)

\rput[r](0.8360,0.8430){Hybrid + pQCD, HG EoS}
\PST@hadrong(0.8520,0.8430)
(0.9310,0.8430)

\PST@hadrong(0.2390,0.8901) (0.2511,0.9070) (0.2752,0.8811) (0.2993,0.8536)
(0.3235,0.8255) (0.3476,0.8056) (0.3717,0.7752) (0.3959,0.7510) (0.4200,0.7219)
(0.4441,0.6971) (0.4683,0.6761) (0.4924,0.6411) (0.5165,0.6188) (0.5407,0.5913)
(0.5648,0.5663) (0.5889,0.5463) (0.6131,0.5217) (0.6372,0.4967) (0.6613,0.4739)
(0.6855,0.4545) (0.7096,0.4328) (0.7337,0.4141) (0.7579,0.3966) (0.7820,0.3840)
(0.8061,0.3569) (0.8303,0.3525) (0.8544,0.3381) (0.8785,0.3231) (0.9027,0.3118)
(0.9268,0.3045) (0.9509,0.2931)

\rput[r](0.8360,0.8010){Hybrid + pQCD, BM EoS}
\PST@bagmode(0.8520,0.8010)
(0.9310,0.8010)

\PST@bagmode(0.2390,0.9325) (0.2511,0.9277) (0.2752,0.9086) (0.2993,0.8766)
(0.3235,0.8422) (0.3476,0.8329) (0.3717,0.7929) (0.3959,0.7726) (0.4200,0.7363)
(0.4441,0.7096) (0.4683,0.6962) (0.4924,0.6528) (0.5165,0.6365) (0.5407,0.5984)
(0.5648,0.5729) (0.5889,0.5487) (0.6131,0.5224) (0.6372,0.5016) (0.6613,0.4731)
(0.6855,0.4576) (0.7096,0.4409) (0.7337,0.4200) (0.7579,0.4095) (0.7820,0.3845)
(0.8061,0.3727) (0.8303,0.3537) (0.8544,0.3420) (0.8785,0.3239) (0.9027,0.3076)
(0.9268,0.3001) (0.9509,0.2851) (0.9630,0.2692)

\rput[r](0.9350,0.6750){pQCD: Turbide {\it et al.}}
\rput[r](0.9350,0.6330){Pb+Pb 158~AGeV}
\rput[r](0.9350,0.5910){$b<4.5$~fm}
\rput[r](0.9350,0.5490){$|y_\textnormal{c.m.}| < 0.5$}

\PST@Border(0.2390,0.9680) (0.2390,0.1344) (0.9630,0.1344) (0.9630,0.9680)
(0.2390,0.9680)

\catcode`@=12
\fi
\endpspicture

%% file: filterp.tex
\ifx\PSTloaded\undefined
\def\PSTloaded{t}
\psset{arrowsize=.01 3.2 1.4 .3}
\psset{dotsize=.01}
\catcode`@=11

\newpsobject{PST@Border}{psline}{linewidth=.0015,linestyle=solid}
\newpsobject{PST@Axes}{psline}{linewidth=.0015,linestyle=dotted,dotsep=.004}
\newpsobject{PST@lineone}{psline}{linecolor=black,    linewidth=.0025,linestyle=solid}
\newpsobject{PST@linetwo}{psline}{linecolor=red,      linewidth=.0025,linestyle=dashed,dash=.01 .01}
\newpsobject{PST@linethr}{psline}{linecolor=blue,     linewidth=.0035,linestyle=dotted,dotsep=.004}
\definecolor{hellgrau}{rgb}{.8,.8,.8}
\definecolor{hellblau}{rgb}{.8,.8,1}
\definecolor{hellrot}{rgb} {1,.8,.8}
\catcode`@=12

\fi
\psset{unit=5.0in,xunit=1.078\columnwidth,yunit=.8\columnwidth}
\pspicture(0.040000,0.8420000)(0.968000,1.8100)
\ifx\nofigs\undefined
\catcode`@=11

\psframe[fillstyle=solid,fillcolor=hellgrau,linecolor=hellgrau](0.5241,0.9680)(0.9470,1.8100)
\psframe[fillstyle=solid,fillcolor=white,linecolor=white](0.5700,1.3940)(0.9470,1.8100)

\PST@Border(0.2070,0.9680) (0.2220,0.9680) \PST@Border(0.9470,0.9680)
(0.9320,0.9680) \PST@Border(0.2070,0.9962)
(0.2145,0.9962) \PST@Border(0.9470,0.9962) (0.9395,0.9962)
\rput[r](0.1910,0.9680){$10^{-9}$}
\PST@Border(0.2070,1.0334) (0.2145,1.0334) \PST@Border(0.9470,1.0334)
(0.9395,1.0334) \PST@Border(0.2070,1.0525) (0.2145,1.0525)
\PST@Border(0.9470,1.0525) (0.9395,1.0525) \PST@Border(0.2070,1.0616)
(0.2220,1.0616) \PST@Border(0.9470,1.0616) (0.9320,1.0616)
\rput[r](0.1910,1.0616){$10^{-8}$} \PST@Border(0.2070,1.0897) (0.2145,1.0897)
\PST@Border(0.9470,1.0897) (0.9395,1.0897) \PST@Border(0.2070,1.1269)
(0.2145,1.1269) \PST@Border(0.9470,1.1269) (0.9395,1.1269)
\PST@Border(0.2070,1.1460) (0.2145,1.1460) \PST@Border(0.9470,1.1460)
(0.9395,1.1460) \PST@Border(0.2070,1.1551) (0.2220,1.1551)
\PST@Border(0.9470,1.1551) (0.9320,1.1551) \rput[r](0.1910,1.1551){$10^{-7}$}
\PST@Border(0.2070,1.1833) (0.2145,1.1833) \PST@Border(0.9470,1.1833)
(0.9395,1.1833) \PST@Border(0.2070,1.2205) (0.2145,1.2205)
\PST@Border(0.9470,1.2205) (0.9395,1.2205) \PST@Border(0.2070,1.2396)
(0.2145,1.2396) \PST@Border(0.9470,1.2396) (0.9395,1.2396)
\PST@Border(0.2070,1.2487) (0.2220,1.2487) \PST@Border(0.9470,1.2487)
(0.9320,1.2487) \rput[r](0.1910,1.2487){$10^{-6}$} \PST@Border(0.2070,1.2768)
(0.2145,1.2768) \PST@Border(0.9470,1.2768) (0.9395,1.2768)
\PST@Border(0.2070,1.3141) (0.2145,1.3141) \PST@Border(0.9470,1.3141)
(0.9395,1.3141) \PST@Border(0.2070,1.3332) (0.2145,1.3332)
\PST@Border(0.9470,1.3332) (0.9395,1.3332) \PST@Border(0.2070,1.3422)
(0.2220,1.3422) \PST@Border(0.9470,1.3422) (0.9320,1.3422)
\rput[r](0.1910,1.3422){$10^{-5}$} \PST@Border(0.2070,1.3704) (0.2145,1.3704)
\PST@Border(0.9470,1.3704) (0.9395,1.3704) \PST@Border(0.2070,1.4076)
(0.2145,1.4076) \PST@Border(0.9470,1.4076) (0.9395,1.4076)
\PST@Border(0.2070,1.4267) (0.2145,1.4267) \PST@Border(0.9470,1.4267)
(0.9395,1.4267) \PST@Border(0.2070,1.4358) (0.2220,1.4358)
\PST@Border(0.9470,1.4358) (0.9320,1.4358) \rput[r](0.1910,1.4358){$10^{-4}$}
\PST@Border(0.2070,1.4639) (0.2145,1.4639) \PST@Border(0.9470,1.4639)
(0.9395,1.4639) \PST@Border(0.2070,1.5012) (0.2145,1.5012)
\PST@Border(0.9470,1.5012) (0.9395,1.5012) \PST@Border(0.2070,1.5203)
(0.2145,1.5203) \PST@Border(0.9470,1.5203) (0.9395,1.5203)
\PST@Border(0.2070,1.5293) (0.2220,1.5293) \PST@Border(0.9470,1.5293)
(0.9320,1.5293) \rput[r](0.1910,1.5293){$10^{-3}$} \PST@Border(0.2070,1.5575)
(0.2145,1.5575) \PST@Border(0.9470,1.5575) (0.9395,1.5575)
\PST@Border(0.2070,1.5947) (0.2145,1.5947) \PST@Border(0.9470,1.5947)
(0.9395,1.5947) \PST@Border(0.2070,1.6138) (0.2145,1.6138)
\PST@Border(0.9470,1.6138) (0.9395,1.6138) \PST@Border(0.2070,1.6229)
(0.2220,1.6229) \PST@Border(0.9470,1.6229) (0.9320,1.6229)
\rput[r](0.1910,1.6229){ 0.01} \PST@Border(0.2070,1.6511) (0.2145,1.6511)
\PST@Border(0.9470,1.6511) (0.9395,1.6511) \PST@Border(0.2070,1.6883)
(0.2145,1.6883) \PST@Border(0.9470,1.6883) (0.9395,1.6883)
\PST@Border(0.2070,1.7074) (0.2145,1.7074) \PST@Border(0.9470,1.7074)
(0.9395,1.7074) \PST@Border(0.2070,1.7164) (0.2220,1.7164)
\PST@Border(0.9470,1.7164) (0.9320,1.7164) \rput[r](0.1910,1.7164){ 0.1}
\PST@Border(0.2070,1.7446) (0.2145,1.7446) \PST@Border(0.9470,1.7446)
(0.9395,1.7446) \PST@Border(0.2070,1.7818) (0.2145,1.7818)
\PST@Border(0.9470,1.7818) (0.9395,1.7818) \PST@Border(0.2070,1.8009)
(0.2145,1.8009) \PST@Border(0.9470,1.8009) (0.9395,1.8009)
\PST@Border(0.2070,1.8100) (0.2220,1.8100) \PST@Border(0.9470,1.8100)
(0.9320,1.8100) \PST@Border(0.2070,0.9680)
(0.2070,0.9880) \PST@Border(0.2070,1.8100) (0.2070,1.7900)

\rput(0.2070,0.9260){ 0} \PST@Border(0.3127,0.9680) (0.3127,0.9880)
\PST@Border(0.3127,1.8100) (0.3127,1.7900) \rput(0.3127,0.9260){ 1}
\PST@Border(0.4184,0.9680) (0.4184,0.9880) \PST@Border(0.4184,1.8100)
(0.4184,1.7900) \rput(0.4184,0.9260){ 2} \PST@Border(0.5241,0.9680)
(0.5241,0.9880) \PST@Border(0.5241,1.8100) (0.5241,1.7900)
\rput(0.5241,0.9260){ 3} \PST@Border(0.6299,0.9680) (0.6299,0.9880)
\PST@Border(0.6299,1.8100) (0.6299,1.7900) \rput(0.6299,0.9260){ 4}
\PST@Border(0.7356,0.9680) (0.7356,0.9880) \PST@Border(0.7356,1.8100)
(0.7356,1.7900) \rput(0.7356,0.9260){ 5} \PST@Border(0.8413,0.9680)
(0.8413,0.9880) \PST@Border(0.8413,1.8100) (0.8413,1.7900)
\rput(0.8413,0.9260){ 6} \PST@Border(0.9470,0.9680) (0.9470,0.9880)
\PST@Border(0.9470,1.8100) (0.9470,1.7900) \rput(0.9470,0.9260){ 7}
\PST@Border(0.2070,1.8100) (0.2070,0.9680) (0.9470,0.9680) (0.9470,1.8100)
(0.2070,1.8100)

\rput{L}(0.0820,1.3890){$E\frac{dN}{d^3p}$~[GeV$^{-2}$]}
\rput(0.5770,0.8630){$p_\bot^\gamma$~[GeV]}

\rput[r](0.8200,1.7490){inclusive}
\rput[r](0.8200,1.6890){$q_\bot^{\rm coll.} > 3$~GeV}
\rput[r](0.8200,1.6290){$\sqrt{s_{\rm coll.}} > 4$~GeV}
\rput[r](0.9150,1.5790){UrQMD}
\rput[r](0.9150,1.5290){Pb+Pb @ 158 AGeV}
\rput[r](0.9150,1.4790){$b < 4.5$~fm}
\rput[r](0.9150,1.4290){$|y_{\rm c.m.}| < 0.5$}

\PST@lineone(0.8360,1.7490)(0.9150,1.7490)
\PST@linethr(0.8360,1.6890)(0.9150,1.6890)
\PST@linetwo(0.8360,1.6290) (0.9150,1.6290)

\PST@lineone(0.2328,1.8100) (0.2334,1.8071) (0.2440,1.7847) (0.2546,1.7698)
(0.2651,1.7546) (0.2757,1.7379) (0.2863,1.7199) (0.2969,1.7008) (0.3074,1.6812)
(0.3180,1.6614) (0.3286,1.6415) (0.3391,1.6215) (0.3497,1.6016) (0.3603,1.5820)
(0.3709,1.5627) (0.3814,1.5437) (0.3920,1.5248) (0.4026,1.5066) (0.4131,1.4889)
(0.4237,1.4715) (0.4343,1.4551) (0.4449,1.4392) (0.4554,1.4241) (0.4660,1.4098)
(0.4766,1.3970) (0.4871,1.3843) (0.4977,1.3724) (0.5083,1.3619) (0.5189,1.3511)
(0.5294,1.3415) (0.5400,1.3318) (0.5506,1.3232) (0.5611,1.3143) (0.5717,1.3056)
(0.5823,1.2981) (0.5929,1.2900) (0.6034,1.2819) (0.6140,1.2753) (0.6246,1.2677)
(0.6351,1.2597) (0.6457,1.2524) (0.6563,1.2450) (0.6669,1.2379) (0.6774,1.2309)
(0.6880,1.2231) (0.6986,1.2164) (0.7091,1.2092) (0.7197,1.2020) (0.7303,1.1953)
(0.7409,1.1880) (0.7514,1.1802) (0.7620,1.1735) (0.7726,1.1651) (0.7831,1.1582)
(0.7937,1.1488) (0.8043,1.1434) (0.8149,1.1360) (0.8254,1.1273) (0.8360,1.1208)
(0.8466,1.1138) (0.8571,1.1059) (0.8677,1.0949) (0.8783,1.0883) (0.8889,1.0787)
(0.8994,1.0756) (0.9100,1.0663) (0.9206,1.0577) (0.9311,1.0468) (0.9417,1.0402)
(0.9470,1.0392)

\PST@linethr(0.2123,1.6073) (0.2123,1.6073) (0.2229,1.5125) (0.2334,1.4730)
(0.2440,1.4447) (0.2546,1.4261) (0.2651,1.4086) (0.2757,1.3963) (0.2863,1.3849)
(0.2969,1.3760) (0.3074,1.3651) (0.3180,1.3592) (0.3286,1.3514) (0.3391,1.3453)
(0.3497,1.3396) (0.3603,1.3339) (0.3709,1.3295) (0.3814,1.3252) (0.3920,1.3200)
(0.4026,1.3161) (0.4131,1.3123) (0.4237,1.3078) (0.4343,1.3032) (0.4449,1.3006)
(0.4554,1.2970) (0.4660,1.2944) (0.4766,1.2903) (0.4871,1.2884) (0.4977,1.2855)
(0.5083,1.2834) (0.5189,1.2814) (0.5294,1.2788) (0.5400,1.2707) (0.5506,1.2582)
(0.5611,1.2464) (0.5717,1.2357) (0.5823,1.2245) (0.5929,1.2151) (0.6034,1.2020)
(0.6140,1.1949) (0.6246,1.1861) (0.6351,1.1787) (0.6457,1.1709) (0.6563,1.1604)
(0.6669,1.1520) (0.6774,1.1459) (0.6880,1.1357) (0.6986,1.1380) (0.7091,1.1249)
(0.7197,1.1172) (0.7303,1.1101) (0.7409,1.1119) (0.7514,1.1015) (0.7620,1.0950)
(0.7726,1.0944) (0.7831,1.0842) (0.7937,1.0812) (0.8043,1.0696) (0.8149,1.0588)
(0.8254,1.0500) (0.8360,1.0565) (0.8466,1.0443) (0.8571,1.0392) (0.8677,1.0405)
(0.8783,1.0411) (0.8889,1.0307) (0.8994,1.0282) (0.9100,1.0167) (0.9206,1.0022)
(0.9311,0.9996) (0.9417,0.9885) (0.9470,0.9940)

\PST@linetwo(0.2521,0.9680) (0.2546,0.9953) (0.2651,1.0815) (0.2757,1.1792)
(0.2863,1.2182) (0.2969,1.2470) (0.3074,1.2733) (0.3180,1.2935) (0.3286,1.3080)
(0.3391,1.3222) (0.3497,1.3334) (0.3603,1.3443) (0.3709,1.3541) (0.3814,1.3625)
(0.3920,1.3696) (0.4026,1.3747) (0.4131,1.3771) (0.4237,1.3782) (0.4343,1.3771)
(0.4449,1.3739) (0.4554,1.3697) (0.4660,1.3644) (0.4766,1.3585) (0.4871,1.3524)
(0.4977,1.3459) (0.5083,1.3391) (0.5189,1.3322) (0.5294,1.3263) (0.5400,1.3187)
(0.5506,1.3115) (0.5611,1.3046) (0.5717,1.2974) (0.5823,1.2908) (0.5929,1.2837)
(0.6034,1.2766) (0.6140,1.2691) (0.6246,1.2635) (0.6351,1.2559) (0.6457,1.2475)
(0.6563,1.2406) (0.6669,1.2332) (0.6774,1.2264) (0.6880,1.2196) (0.6986,1.2123)
(0.7091,1.2061) (0.7197,1.1984) (0.7303,1.1914) (0.7409,1.1836) (0.7514,1.1754)
(0.7620,1.1682) (0.7726,1.1601) (0.7831,1.1524) (0.7937,1.1462) (0.8043,1.1387)
(0.8149,1.1315) (0.8254,1.1231) (0.8360,1.1154) (0.8466,1.1090) (0.8571,1.0974)
(0.8677,1.0892) (0.8783,1.0805) (0.8889,1.0733) (0.8994,1.0674) (0.9100,1.0585)
(0.9206,1.0535) (0.9311,1.0447) (0.9417,1.0352) (0.9470,1.0316)

\PST@Border(0.2070,1.8100)
(0.2070,0.9680)
(0.9470,0.9680)
(0.9470,1.8100)
(0.2070,1.8100)

\catcode`@=12
\fi
\endpspicture

%% file: filterq.tex
\ifx\PSTloaded\undefined
\def\PSTloaded{t}
\psset{arrowsize=.01 3.2 1.4 .3}
\psset{dotsize=.01}
\catcode`@=11

\newpsobject{PST@Border}{psline}{linewidth=.0015,linestyle=solid}
\newpsobject{PST@Axes}{psline}{linewidth=.0015,linestyle=dotted,dotsep=.004}
\newpsobject{PST@lineone}{psline}{linecolor=black,    linewidth=.0025,linestyle=solid}
\newpsobject{PST@linetwo}{psline}{linecolor=red,      linewidth=.0025,linestyle=dashed,dash=.01 .01}
\newpsobject{PST@linethr}{psline}{linecolor=blue,     linewidth=.0035,linestyle=dotted,dotsep=.004}
\definecolor{hellgrau}{rgb}{.8,.8,.8}
\definecolor{hellblau}{rgb}{.8,.8,1}
\definecolor{hellrot}{rgb} {1,.8,.8}
\catcode`@=12

\fi
\psset{unit=5.0in,xunit=1.078\columnwidth,yunit=.8\columnwidth}
\pspicture(0.040000,0.0000000)(0.968000,0.9680)
\ifx\nofigs\undefined
\catcode`@=11

\psframe[fillstyle=solid,fillcolor=hellblau,linecolor=hellblau](.5241,.126)(.9470,.968)
\psframe[fillstyle=solid,fillcolor=white,linecolor=white](.5700,.5620)(.9470,.9680)

\PST@Border(0.2070,0.1260) (0.2220,0.1260) \PST@Border(0.9470,0.1260)
(0.9320,0.1260) \rput[r](0.1910,0.1260){$10^{-8}$} \PST@Border(0.2070,0.1577)
(0.2145,0.1577) \PST@Border(0.9470,0.1577) (0.9395,0.1577)
\PST@Border(0.2070,0.1996) (0.2145,0.1996) \PST@Border(0.9470,0.1996)
(0.9395,0.1996) \PST@Border(0.2070,0.2211) (0.2145,0.2211)
\PST@Border(0.9470,0.2211) (0.9395,0.2211) \PST@Border(0.2070,0.2313)
(0.2220,0.2313) \PST@Border(0.9470,0.2313) (0.9320,0.2313)
\rput[r](0.1910,0.2313){$10^{-7}$} \PST@Border(0.2070,0.2629) (0.2145,0.2629)
\PST@Border(0.9470,0.2629) (0.9395,0.2629) \PST@Border(0.2070,0.3048)
(0.2145,0.3048) \PST@Border(0.9470,0.3048) (0.9395,0.3048)
\PST@Border(0.2070,0.3263) (0.2145,0.3263) \PST@Border(0.9470,0.3263)
(0.9395,0.3263) \PST@Border(0.2070,0.3365) (0.2220,0.3365)
\PST@Border(0.9470,0.3365) (0.9320,0.3365) \rput[r](0.1910,0.3365){$10^{-6}$}
\PST@Border(0.2070,0.3682) (0.2145,0.3682) \PST@Border(0.9470,0.3682)
(0.9395,0.3682) \PST@Border(0.2070,0.4101) (0.2145,0.4101)
\PST@Border(0.9470,0.4101) (0.9395,0.4101) \PST@Border(0.2070,0.4316)
(0.2145,0.4316) \PST@Border(0.9470,0.4316) (0.9395,0.4316)
\PST@Border(0.2070,0.4418) (0.2220,0.4418) \PST@Border(0.9470,0.4418)
(0.9320,0.4418) \rput[r](0.1910,0.4418){$10^{-5}$} \PST@Border(0.2070,0.4734)
(0.2145,0.4734) \PST@Border(0.9470,0.4734) (0.9395,0.4734)
\PST@Border(0.2070,0.5153) (0.2145,0.5153) \PST@Border(0.9470,0.5153)
(0.9395,0.5153) \PST@Border(0.2070,0.5368) (0.2145,0.5368)
\PST@Border(0.9470,0.5368) (0.9395,0.5368) \PST@Border(0.2070,0.5470)
(0.2220,0.5470) \PST@Border(0.9470,0.5470) (0.9320,0.5470)
\rput[r](0.1910,0.5470){$10^{-4}$} \PST@Border(0.2070,0.5787) (0.2145,0.5787)
\PST@Border(0.9470,0.5787) (0.9395,0.5787) \PST@Border(0.2070,0.6206)
(0.2145,0.6206) \PST@Border(0.9470,0.6206) (0.9395,0.6206)
\PST@Border(0.2070,0.6421) (0.2145,0.6421) \PST@Border(0.9470,0.6421)
(0.9395,0.6421) \PST@Border(0.2070,0.6523) (0.2220,0.6523)
\PST@Border(0.9470,0.6523) (0.9320,0.6523) \rput[r](0.1910,0.6523){$10^{-3}$}
\PST@Border(0.2070,0.6839) (0.2145,0.6839) \PST@Border(0.9470,0.6839)
(0.9395,0.6839) \PST@Border(0.2070,0.7258) (0.2145,0.7258)
\PST@Border(0.9470,0.7258) (0.9395,0.7258) \PST@Border(0.2070,0.7473)
(0.2145,0.7473) \PST@Border(0.9470,0.7473) (0.9395,0.7473)
\PST@Border(0.2070,0.7575) (0.2220,0.7575) \PST@Border(0.9470,0.7575)
(0.9320,0.7575) \rput[r](0.1910,0.7575){ 0.01} \PST@Border(0.2070,0.7892)
(0.2145,0.7892) \PST@Border(0.9470,0.7892) (0.9395,0.7892)
\PST@Border(0.2070,0.8311) (0.2145,0.8311) \PST@Border(0.9470,0.8311)
(0.9395,0.8311) \PST@Border(0.2070,0.8526) (0.2145,0.8526)
\PST@Border(0.9470,0.8526) (0.9395,0.8526) \PST@Border(0.2070,0.8628)
(0.2220,0.8628) \PST@Border(0.9470,0.8628) (0.9320,0.8628)
\rput[r](0.1910,0.8628){ 0.1} \PST@Border(0.2070,0.8944) (0.2145,0.8944)
\PST@Border(0.9470,0.8944) (0.9395,0.8944) \PST@Border(0.2070,0.9363)
(0.2145,0.9363) \PST@Border(0.9470,0.9363) (0.9395,0.9363)
\PST@Border(0.2070,0.9578) (0.2145,0.9578) \PST@Border(0.9470,0.9578)
(0.9395,0.9578) \PST@Border(0.2070,0.9680) (0.2220,0.9680)
\PST@Border(0.9470,0.9680) (0.9320,0.9680)
\PST@Border(0.2070,0.1260) (0.2070,0.1460) \PST@Border(0.2070,0.9680)
(0.2070,0.9480) \rput(0.2070,0.0840){ 0} \PST@Border(0.3127,0.1260)
(0.3127,0.1460) \PST@Border(0.3127,0.9680) (0.3127,0.9480)
\rput(0.3127,0.0840){ 1} \PST@Border(0.4184,0.1260) (0.4184,0.1460)
\PST@Border(0.4184,0.9680) (0.4184,0.9480) \rput(0.4184,0.0840){ 2}
\PST@Border(0.5241,0.1260) (0.5241,0.1460) \PST@Border(0.5241,0.9680)
(0.5241,0.9480) \rput(0.5241,0.0840){ 3} \PST@Border(0.6299,0.1260)
(0.6299,0.1460) \PST@Border(0.6299,0.9680) (0.6299,0.9480)
\rput(0.6299,0.0840){ 4} \PST@Border(0.7356,0.1260) (0.7356,0.1460)
\PST@Border(0.7356,0.9680) (0.7356,0.9480) \rput(0.7356,0.0840){ 5}
\PST@Border(0.8413,0.1260) (0.8413,0.1460) \PST@Border(0.8413,0.9680)
(0.8413,0.9480) \rput(0.8413,0.0840){ 6} \PST@Border(0.9470,0.1260)
(0.9470,0.1460) \PST@Border(0.9470,0.9680) (0.9470,0.9480)
\rput(0.9470,0.0840){ 7} \PST@Border(0.2070,0.9680) (0.2070,0.1260)
(0.9470,0.1260) (0.9470,0.9680) (0.2070,0.9680)

\rput{L}(0.0820,0.5470){$q_0^{\rm coll.}\frac{dN}{d^3q^{\rm coll.}}$~[GeV$^{-2}$]}
\rput(0.5770,0.0210){$q_\bot^{\rm coll.}$~[GeV]}

\rput[r](0.8200,0.9070){inclusive}
\rput[r](0.8200,0.8570){$p_\bot^\gamma > 3$~GeV}
\rput[r](0.8200,0.8070){$\sqrt{s_{\rm coll.}} > 4$~GeV}
\rput[r](0.9150,0.7590){UrQMD}
\rput[r](0.9150,0.7090){Pb+Pb @ 158 AGeV}
\rput[r](0.9150,0.6590){$b < 4.5$~fm}
\rput[r](0.9150,0.6090){$|y_{\rm c.m.}| < 0.5$}

\PST@linethr(0.8360,0.9070)(0.9150,0.9070)
\PST@lineone(0.8360,0.8570)(0.9150,0.8570)
\PST@linetwo(0.8360,0.8070)(0.9150,0.8070)

\PST@linethr(0.2687,0.9680) (0.2757,0.9582) (0.2863,0.9425) (0.2969,0.9259)
(0.3074,0.9091) (0.3180,0.8916) (0.3286,0.8739) (0.3391,0.8563) (0.3497,0.8378)
(0.3603,0.8195) (0.3709,0.8016) (0.3814,0.7831) (0.3920,0.7652) (0.4026,0.7466)
(0.4131,0.7291) (0.4237,0.7096) (0.4343,0.6913) (0.4449,0.6734) (0.4554,0.6555)
(0.4660,0.6374) (0.4766,0.6215) (0.4871,0.6022) (0.4977,0.5857) (0.5083,0.5658)
(0.5189,0.5501) (0.5294,0.5329) (0.5400,0.5253) (0.5506,0.4974) (0.5611,0.4835)
(0.5717,0.4611) (0.5823,0.4473) (0.5929,0.4360) (0.6034,0.4186) (0.6140,0.4279)
(0.6246,0.3826) (0.6351,0.4043) (0.6457,0.3899) (0.6563,0.3756) (0.6669,0.3453)
(0.6774,0.3168) (0.6880,0.3260) (0.6986,0.3288) (0.7091,0.3104) (0.7197,0.3209)
(0.7303,0.2807) (0.7409,0.2757) (0.7514,0.2639) (0.7620,0.2454) (0.7726,0.2761)
(0.7831,0.2771) (0.7937,0.2157) (0.8043,0.2426) (0.8149,0.2676) (0.8254,0.2757)
(0.8360,0.2388) (0.8466,0.2588) (0.8571,0.2882) (0.8677,0.2747) (0.8783,0.1923)
(0.8889,0.2406) (0.8994,0.1936) (0.9100,0.1958) (0.9206,0.1396) (0.9311,0.2039)
(0.9410,0.1260)

\PST@lineone(0.2123,0.4970) (0.2123,0.4970) (0.2229,0.4948) (0.2334,0.4924)
(0.2440,0.4876) (0.2546,0.4818) (0.2651,0.4746) (0.2757,0.4691) (0.2863,0.4597)
(0.2969,0.4498) (0.3074,0.4452) (0.3180,0.4318) (0.3286,0.4210) (0.3391,0.4091)
(0.3497,0.4031) (0.3603,0.3891) (0.3709,0.3811) (0.3814,0.3662) (0.3920,0.3588)
(0.4026,0.3513) (0.4131,0.3391) (0.4237,0.3353) (0.4343,0.3287) (0.4449,0.3214)
(0.4554,0.3142) (0.4660,0.3092) (0.4766,0.3096) (0.4871,0.3001) (0.4977,0.3003)
(0.5083,0.3176) (0.5189,0.3322) (0.5294,0.3406) (0.5400,0.3383) (0.5506,0.3356)
(0.5611,0.3312) (0.5717,0.3276) (0.5823,0.3150) (0.5929,0.3049) (0.6034,0.3015)
(0.6140,0.2928) (0.6246,0.2731) (0.6351,0.2808) (0.6457,0.2614) (0.6563,0.2579)
(0.6669,0.2387) (0.6774,0.2288) (0.6880,0.2214) (0.6986,0.2142) (0.7091,0.2464)
(0.7197,0.2245) (0.7303,0.2195) (0.7409,0.1982) (0.7514,0.1980) (0.7620,0.1988)
(0.7726,0.2060) (0.7831,0.2010) (0.7937,0.1692) (0.8043,0.1563) (0.8149,0.1655)
(0.8254,0.1960) (0.8360,0.1772) (0.8466,0.1743) (0.8571,0.1488) (0.8677,0.1354)
(0.8783,0.1674) (0.8860,0.1260) \PST@lineone(0.8923,0.1260) (0.8994,0.1575)
(0.9100,0.1497) (0.9171,0.1260) \PST@lineone(0.9224,0.1260) (0.9311,0.1810)
(0.9370,0.1260)

\PST@linetwo(0.2123,0.5849) (0.2123,0.5849) (0.2229,0.5816) (0.2334,0.5783)
(0.2440,0.5746) (0.2546,0.5673) (0.2651,0.5578) (0.2757,0.5503) (0.2863,0.5381)
(0.2969,0.5265) (0.3074,0.5157) (0.3180,0.5029) (0.3286,0.4909) (0.3391,0.4773)
(0.3497,0.4652) (0.3603,0.4513) (0.3709,0.4420) (0.3814,0.4245) (0.3920,0.4154)
(0.4026,0.4027) (0.4131,0.3927) (0.4237,0.3786) (0.4343,0.3696) (0.4449,0.3648)
(0.4554,0.3475) (0.4660,0.3293) (0.4766,0.3256) (0.4871,0.3093) (0.4977,0.3046)
(0.5083,0.3109) (0.5189,0.3074) (0.5294,0.2715) (0.5400,0.2495) (0.5506,0.2397)
(0.5611,0.2369) (0.5717,0.2211) (0.5823,0.1933) (0.5929,0.2462) (0.6034,0.2208)
(0.6140,0.1802) (0.6246,0.1341) (0.6351,0.1403) (0.6457,0.1470) (0.6563,0.1724)
(0.6669,0.1639) (0.6752,0.1260) \PST@linetwo(0.6878,0.1260) (0.6880,0.1262)
(0.6884,0.1260) \PST@linetwo(0.7144,0.1260) (0.7197,0.1610) (0.7303,0.1419)
(0.7365,0.1260) \PST@linetwo(0.7717,0.1260) (0.7726,0.1329) (0.7733,0.1260)
\PST@linetwo(0.8014,0.1260) (0.8043,0.1491) (0.8113,0.1260)

\PST@Border(0.2070,0.9680) (0.2070,0.1260) (0.9470,0.1260) (0.9470,0.9680)
(0.2070,0.9680)

\catcode`@=12
\fi
\endpspicture

%% file: filters.tex
\ifx\PSTloaded\undefined
\def\PSTloaded{t}
\psset{arrowsize=.01 3.2 1.4 .3}
\psset{dotsize=.01}
\catcode`@=11

\newpsobject{PST@Border}{psline}{linewidth=.0015,linestyle=solid}
\newpsobject{PST@Axes}{psline}{linewidth=.0015,linestyle=dotted,dotsep=.004}
\newpsobject{PST@lineone}{psline}{linecolor=black,    linewidth=.0025,linestyle=solid}
\newpsobject{PST@linetwo}{psline}{linecolor=red,      linewidth=.0025,linestyle=dashed,dash=.01 .01}
\newpsobject{PST@linethr}{psline}{linecolor=blue,     linewidth=.0035,linestyle=dotted,dotsep=.004}
\definecolor{hellgrau}{rgb}{.8,.8,.8}
\definecolor{hellblau}{rgb}{.8,.8,1}
\definecolor{hellrot}{rgb} {1,.8,.8}
\catcode`@=12

\fi
\psset{unit=5.0in,xunit=1.078\columnwidth,yunit=.8\columnwidth}
\pspicture(0.040000,-1.0000000)(0.968000,0.0000)
\ifx\nofigs\undefined
\catcode`@=11

\psframe[fillstyle=solid,fillcolor=hellrot, linecolor=hellrot] (0.4184,-0.8740)(0.9470,-0.0320)
\psframe[fillstyle=solid,fillcolor=white,linecolor=white](0.5700,-0.4380)(0.9470,-0.0320)

\PST@Border(0.2070,-0.8740) (0.2220,-0.8740) \PST@Border(0.9470,-0.8740)
(0.9320,-0.8740) \rput[r](0.1910,-0.8740){$10^{-8}$} \PST@Border(0.2070,-0.8423)
(0.2145,-0.8423) \PST@Border(0.9470,-0.8423) (0.9395,-0.8423)
\PST@Border(0.2070,-0.8004) (0.2145,-0.8004) \PST@Border(0.9470,-0.8004)
(0.9395,-0.8004) \PST@Border(0.2070,-0.7789) (0.2145,-0.7789)
\PST@Border(0.9470,-0.7789) (0.9395,-0.7789) \PST@Border(0.2070,-0.7687)
(0.2220,-0.7687) \PST@Border(0.9470,-0.7687) (0.9320,-0.7687)
\rput[r](0.1910,-0.7687){$10^{-7}$} \PST@Border(0.2070,-0.7371) (0.2145,-0.7371)
\PST@Border(0.9470,-0.7371) (0.9395,-0.7371) \PST@Border(0.2070,-0.6952)
(0.2145,-0.6952) \PST@Border(0.9470,-0.6952) (0.9395,-0.6952)
\PST@Border(0.2070,-0.6737) (0.2145,-0.6737) \PST@Border(0.9470,-0.6737)
(0.9395,-0.6737) \PST@Border(0.2070,-0.6635) (0.2220,-0.6635)
\PST@Border(0.9470,-0.6635) (0.9320,-0.6635) \rput[r](0.1910,-0.6635){$10^{-6}$}
\PST@Border(0.2070,-0.6318) (0.2145,-0.6318) \PST@Border(0.9470,-0.6318)
(0.9395,-0.6318) \PST@Border(0.2070,-0.5899) (0.2145,-0.5899)
\PST@Border(0.9470,-0.5899) (0.9395,-0.5899) \PST@Border(0.2070,-0.5684)
(0.2145,-0.5684) \PST@Border(0.9470,-0.5684) (0.9395,-0.5684)
\PST@Border(0.2070,-0.5582) (0.2220,-0.5582) \PST@Border(0.9470,-0.5582)
(0.9320,-0.5582) \rput[r](0.1910,-0.5582){$10^{-5}$} \PST@Border(0.2070,-0.5266)
(0.2145,-0.5266) \PST@Border(0.9470,-0.5266) (0.9395,-0.5266)
\PST@Border(0.2070,-0.4847) (0.2145,-0.4847) \PST@Border(0.9470,-0.4847)
(0.9395,-0.4847) \PST@Border(0.2070,-0.4632) (0.2145,-0.4632)
\PST@Border(0.9470,-0.4632) (0.9395,-0.4632) \PST@Border(0.2070,-0.4530)
(0.2220,-0.4530) \PST@Border(0.9470,-0.4530) (0.9320,-0.4530)
\rput[r](0.1910,-0.4530){$10^{-4}$} \PST@Border(0.2070,-0.4213) (0.2145,-0.4213)
\PST@Border(0.9470,-0.4213) (0.9395,-0.4213) \PST@Border(0.2070,-0.3794)
(0.2145,-0.3794) \PST@Border(0.9470,-0.3794) (0.9395,-0.3794)
\PST@Border(0.2070,-0.3579) (0.2145,-0.3579) \PST@Border(0.9470,-0.3579)
(0.9395,-0.3579) \PST@Border(0.2070,-0.3477) (0.2220,-0.3477)
\PST@Border(0.9470,-0.3477) (0.9320,-0.3477) \rput[r](0.1910,-0.3477){$10^{-3}$}
\PST@Border(0.2070,-0.3161) (0.2145,-0.3161) \PST@Border(0.9470,-0.3161)
(0.9395,-0.3161) \PST@Border(0.2070,-0.2742) (0.2145,-0.2742)
\PST@Border(0.9470,-0.2742) (0.9395,-0.2742) \PST@Border(0.2070,-0.2527)
(0.2145,-0.2527) \PST@Border(0.9470,-0.2527) (0.9395,-0.2527)
\PST@Border(0.2070,-0.2425) (0.2220,-0.2425) \PST@Border(0.9470,-0.2425)
(0.9320,-0.2425) \rput[r](0.1910,-0.2425){ 0.01} \PST@Border(0.2070,-0.2108)
(0.2145,-0.2108) \PST@Border(0.9470,-0.2108) (0.9395,-0.2108)
\PST@Border(0.2070,-0.1689) (0.2145,-0.1689) \PST@Border(0.9470,-0.1689)
(0.9395,-0.1689) \PST@Border(0.2070,-0.1474) (0.2145,-0.1474)
\PST@Border(0.9470,-0.1474) (0.9395,-0.1474) \PST@Border(0.2070,-0.1372)
(0.2220,-0.1372) \PST@Border(0.9470,-0.1372) (0.9320,-0.1372)
\rput[r](0.1910,-0.1372){ 0.1} \PST@Border(0.2070,-0.1056) (0.2145,-0.1056)
\PST@Border(0.9470,-0.1056) (0.9395,-0.1056) \PST@Border(0.2070,-0.0637)
(0.2145,-0.0637) \PST@Border(0.9470,-0.0637) (0.9395,-0.0637)
\PST@Border(0.2070,-0.0422) (0.2145,-0.0422) \PST@Border(0.9470,-0.0422)
(0.9395,-0.0422) \PST@Border(0.2070,-0.0320) (0.2220,-0.0320)
\PST@Border(0.9470,-0.0320) (0.9320,-0.0320) \rput[r](0.1910,-0.0320){ 1}
\PST@Border(0.2070,-0.8740) (0.2070,-0.8540) \PST@Border(0.2070,-0.0320)
(0.2070,-0.0520) \rput(0.2070,-0.9160){ 0} \PST@Border(0.3127,-0.8740)
(0.3127,-0.8540) \PST@Border(0.3127,-0.0320) (0.3127,-0.0520)
\rput(0.3127,-0.9160){ 2} \PST@Border(0.4184,-0.8740) (0.4184,-0.8540)
\PST@Border(0.4184,-0.0320) (0.4184,-0.0520) \rput(0.4184,-0.9160){ 4}
\PST@Border(0.5241,-0.8740) (0.5241,-0.8540) \PST@Border(0.5241,-0.0320)
(0.5241,-0.0520) \rput(0.5241,-0.9160){ 6} \PST@Border(0.6299,-0.8740)
(0.6299,-0.8540) \PST@Border(0.6299,-0.0320) (0.6299,-0.0520)
\rput(0.6299,-0.9160){ 8} \PST@Border(0.7356,-0.8740) (0.7356,-0.8540)
\PST@Border(0.7356,-0.0320) (0.7356,-0.0520) \rput(0.7356,-0.9160){ 10}
\PST@Border(0.8413,-0.8740) (0.8413,-0.8540) \PST@Border(0.8413,-0.0320)
(0.8413,-0.0520) \rput(0.8413,-0.9160){ 12} \PST@Border(0.9470,-0.8740)
(0.9470,-0.8540) \PST@Border(0.9470,-0.0320) (0.9470,-0.0520)
\rput(0.9470,-0.9160){ 14} \PST@Border(0.2070,-0.0320) (0.2070,-0.8740)
(0.9470,-0.8740) (0.9470,-0.0320) (0.2070,-0.0320)

\rput{L}(0.0820,-0.4530){$\frac{dN}{d\sqrt{s_{\rm coll.}}}$~[GeV$^{-1}$]}
\rput(0.5770,-0.9790){$\sqrt{s_{\rm coll.}}$~[GeV]}

\rput[r](0.8200,-.0930){inclusive}
\rput[r](0.8200,-.1430){$p_\bot^\gamma > 3$~GeV}
\rput[r](0.8200,-.1930){$q_\bot^{\rm coll.} > 3$~GeV}
\rput[r](0.9150,-.2430){UrQMD}
\rput[r](0.9150,-.2930){Pb+Pb @ 158 AGeV}
\rput[r](0.9150,-.3430){$b < 4.5$~fm}
\rput[r](0.9150,-.3930){$|y_{\rm c.m.}| < 0.5$}

\PST@linetwo(0.8360,-.0930)(0.9150,-.0930)
\PST@lineone(0.8360,-.1430)(0.9150,-.1430)
\PST@linethr(0.8360,-.1930)(0.9150,-.1930)

\PST@linetwo(0.2202,-0.1986) (0.2202,-0.1986) (0.2255,-0.1587) (0.2308,-0.1135)
(0.2361,-0.0498) (0.2374,-0.0320) \PST@linetwo(0.2754,-0.0320) (0.2784,-0.0620)
(0.2836,-0.0966) (0.2889,-0.1238) (0.2942,-0.1486) (0.2995,-0.1736) (0.3048,-0.1950)
(0.3101,-0.2151) (0.3154,-0.2347) (0.3206,-0.2519) (0.3259,-0.2671) (0.3312,-0.2808)
(0.3365,-0.2937) (0.3418,-0.3046) (0.3471,-0.3154) (0.3524,-0.3244) (0.3576,-0.3346)
(0.3629,-0.3435) (0.3682,-0.3500) (0.3735,-0.3562) (0.3788,-0.3604) (0.3841,-0.3686)
(0.3894,-0.3719) (0.3946,-0.3761) (0.3999,-0.3800) (0.4052,-0.3841) (0.4105,-0.3893)
(0.4158,-0.3940) (0.4211,-0.3960) (0.4264,-0.4014) (0.4316,-0.4034) (0.4369,-0.4055)
(0.4422,-0.4118) (0.4475,-0.4153) (0.4528,-0.4190) (0.4581,-0.4202) (0.4634,-0.4254)
(0.4686,-0.4289) (0.4739,-0.4297) (0.4792,-0.4340) (0.4845,-0.4384) (0.4898,-0.4415)
(0.4951,-0.4451) (0.5004,-0.4469) (0.5056,-0.4507) (0.5109,-0.4569) (0.5162,-0.4582)
(0.5215,-0.4643) (0.5268,-0.4655) (0.5321,-0.4704) (0.5374,-0.4731) (0.5426,-0.4758)
(0.5479,-0.4774) (0.5532,-0.4820) (0.5585,-0.4844) (0.5638,-0.4917) (0.5691,-0.4893)
(0.5744,-0.4963) (0.5796,-0.4983) (0.5849,-0.4920) (0.5902,-0.5090) (0.5955,-0.5080)
(0.6008,-0.5138) (0.6061,-0.5160) (0.6114,-0.5184) (0.6166,-0.5226) (0.6219,-0.5268)
(0.6272,-0.5225) (0.6325,-0.5309) (0.6378,-0.5376) (0.6431,-0.5402) (0.6484,-0.5420)
(0.6536,-0.5428) (0.6589,-0.5481) (0.6642,-0.5533) (0.6695,-0.5566) (0.6748,-0.5575)
(0.6801,-0.5652) (0.6854,-0.5651) (0.6906,-0.5726) (0.6959,-0.5717) (0.7012,-0.5734)
(0.7065,-0.5801) (0.7118,-0.5802) (0.7171,-0.5912) (0.7224,-0.5948) (0.7276,-0.5924)
(0.7329,-0.5942) (0.7382,-0.6013) (0.7435,-0.6006) (0.7488,-0.6104) (0.7541,-0.6155)
(0.7594,-0.6178) (0.7646,-0.6131) (0.7699,-0.6241) (0.7752,-0.6327) (0.7805,-0.6319)
(0.7858,-0.6428) (0.7911,-0.6434) (0.7964,-0.6495) (0.8016,-0.6618)
\PST@linetwo(0.8016,-0.6618) (0.8069,-0.6524) (0.8122,-0.6583) (0.8175,-0.6516)
(0.8228,-0.6641) (0.8281,-0.6776) (0.8334,-0.6719) (0.8386,-0.6763) (0.8439,-0.6644)
(0.8492,-0.6853) (0.8545,-0.6935) (0.8598,-0.6909) (0.8651,-0.6984) (0.8704,-0.7359)
(0.8756,-0.7183) (0.8809,-0.7031) (0.8862,-0.6987) (0.8915,-0.7356) (0.8968,-0.7308)
(0.9021,-0.7379) (0.9074,-0.7462) (0.9126,-0.7209) (0.9179,-0.7385) (0.9232,-0.7407)
(0.9285,-0.7323) (0.9338,-0.7948) (0.9391,-0.7667) (0.9444,-0.7293) (0.9470,-0.7585)

\PST@lineone(0.2255,-0.8098) (0.2255,-0.8098) (0.2308,-0.8637) (0.2361,-0.8693)
(0.2414,-0.7207) (0.2466,-0.6867) (0.2519,-0.7777) (0.2572,-0.6053) (0.2625,-0.5686)
(0.2678,-0.5170) (0.2731,-0.4937) (0.2784,-0.5307) (0.2836,-0.5460) (0.2889,-0.5547)
(0.2942,-0.5539) (0.2995,-0.5637) (0.3048,-0.5697) (0.3101,-0.5814) (0.3154,-0.5756)
(0.3206,-0.5843) (0.3259,-0.5892) (0.3312,-0.5919) (0.3365,-0.5895) (0.3418,-0.6066)
(0.3471,-0.5974) (0.3524,-0.5990) (0.3576,-0.6080) (0.3629,-0.6000) (0.3682,-0.6072)
(0.3735,-0.6079) (0.3788,-0.6026) (0.3841,-0.6130) (0.3894,-0.6171) (0.3946,-0.5943)
(0.3999,-0.6102) (0.4052,-0.6054) (0.4105,-0.6098) (0.4158,-0.6073) (0.4211,-0.6032)
(0.4264,-0.5814) (0.4316,-0.5923) (0.4369,-0.6015) (0.4422,-0.6008) (0.4475,-0.5839)
(0.4528,-0.5939) (0.4581,-0.5701) (0.4634,-0.5772) (0.4686,-0.5768) (0.4739,-0.5651)
(0.4792,-0.5604) (0.4845,-0.5612) (0.4898,-0.5508) (0.4951,-0.5465) (0.5004,-0.5421)
(0.5056,-0.5356) (0.5109,-0.5326) (0.5162,-0.5278) (0.5215,-0.5242) (0.5268,-0.5194)
(0.5321,-0.5174) (0.5374,-0.5134) (0.5426,-0.5114) (0.5479,-0.5091) (0.5532,-0.5087)
(0.5585,-0.5087) (0.5638,-0.5115) (0.5691,-0.5071) (0.5744,-0.5129) (0.5796,-0.5113)
(0.5849,-0.5021) (0.5902,-0.5178) (0.5955,-0.5162) (0.6008,-0.5214) (0.6061,-0.5226)
(0.6114,-0.5245) (0.6166,-0.5280) (0.6219,-0.5309) (0.6272,-0.5264) (0.6325,-0.5343)
(0.6378,-0.5409) (0.6431,-0.5420) (0.6484,-0.5433) (0.6536,-0.5457) (0.6589,-0.5503)
(0.6642,-0.5542) (0.6695,-0.5579) (0.6748,-0.5585) (0.6801,-0.5654) (0.6854,-0.5658)
(0.6906,-0.5731) (0.6959,-0.5728) (0.7012,-0.5744) (0.7065,-0.5806) (0.7118,-0.5825)
(0.7171,-0.5918) (0.7224,-0.5943) (0.7276,-0.5922) (0.7329,-0.5954) (0.7382,-0.6002)
(0.7435,-0.6002) (0.7488,-0.6111) \PST@lineone(0.7488,-0.6111) (0.7541,-0.6154)
(0.7594,-0.6165) (0.7646,-0.6123) (0.7699,-0.6248) (0.7752,-0.6361) (0.7805,-0.6322)
(0.7858,-0.6454) (0.7911,-0.6434) (0.7964,-0.6485) (0.8016,-0.6617) (0.8069,-0.6532)
(0.8122,-0.6553) (0.8175,-0.6512) (0.8228,-0.6645) (0.8281,-0.6775) (0.8334,-0.6707)
(0.8386,-0.6745) (0.8439,-0.6659) (0.8492,-0.6871) (0.8545,-0.6942) (0.8598,-0.6897)
(0.8651,-0.6953) (0.8704,-0.7415) (0.8756,-0.7111) (0.8809,-0.7028) (0.8862,-0.6959)
(0.8915,-0.7287) (0.8968,-0.7351) (0.9021,-0.7378) (0.9074,-0.7497) (0.9126,-0.7221)
(0.9179,-0.7420) (0.9232,-0.7433) (0.9285,-0.7315) (0.9338,-0.7940) (0.9391,-0.7671)
(0.9444,-0.7352) (0.9470,-0.7556)

\PST@linethr(0.2223,-0.8740) (0.2255,-0.7421) (0.2308,-0.5503) (0.2361,-0.6163)
(0.2414,-0.4130) (0.2466,-0.3471) (0.2519,-0.3436) (0.2572,-0.3683) (0.2625,-0.3881)
(0.2678,-0.3775) (0.2731,-0.3824) (0.2784,-0.4117) (0.2836,-0.4358) (0.2889,-0.4452)
(0.2942,-0.4447) (0.2995,-0.4733) (0.3048,-0.4840) (0.3101,-0.4989) (0.3154,-0.5044)
(0.3206,-0.5119) (0.3259,-0.5365) (0.3312,-0.5396) (0.3365,-0.5422) (0.3418,-0.5520)
(0.3471,-0.5622) (0.3524,-0.5603) (0.3576,-0.5752) (0.3629,-0.5739) (0.3682,-0.5769)
(0.3735,-0.5846) (0.3788,-0.5856) (0.3841,-0.6080) (0.3894,-0.6127) (0.3946,-0.5836)
(0.3999,-0.6256) (0.4052,-0.6116) (0.4105,-0.6228) (0.4158,-0.6297) (0.4211,-0.6378)
(0.4264,-0.6339) (0.4316,-0.6152) (0.4369,-0.6543) (0.4422,-0.6677) (0.4475,-0.6236)
(0.4528,-0.6383) (0.4581,-0.6578) (0.4634,-0.6853) (0.4686,-0.6580) (0.4739,-0.6705)
(0.4792,-0.6750) (0.4845,-0.6968) (0.4898,-0.6837) (0.4951,-0.6844) (0.5004,-0.6916)
(0.5056,-0.6718) (0.5109,-0.6808) (0.5162,-0.6918) (0.5215,-0.7124) (0.5268,-0.7569)
(0.5321,-0.6951) (0.5374,-0.6889) (0.5426,-0.7025) (0.5479,-0.7003) (0.5532,-0.7467)
(0.5585,-0.7252) (0.5638,-0.7456) (0.5691,-0.7384) (0.5744,-0.7612) (0.5796,-0.7414)
(0.5849,-0.7332) (0.5902,-0.8291) (0.5955,-0.7352) (0.6008,-0.7718) (0.6061,-0.7349)
(0.6114,-0.7333) (0.6166,-0.8532) (0.6219,-0.8367) (0.6272,-0.6991) (0.6325,-0.7372)
(0.6378,-0.8162) (0.6431,-0.7665) (0.6477,-0.8740) \PST@linethr(0.6490,-0.8740)
(0.6536,-0.7623) (0.6589,-0.8628) (0.6642,-0.7856) (0.6695,-0.7675) (0.6746,-0.8740)
\PST@linethr(0.6752,-0.8740) (0.6801,-0.8362) (0.6906,-0.7757) (0.6959,-0.7878)
(0.7012,-0.7535) (0.7118,-0.8274) (0.7276,-0.7937) (0.7301,-0.8740)
\PST@linethr(0.7381,-0.8740) (0.7435,-0.7762) (0.7462,-0.8740)
\PST@linethr(0.7544,-0.8740) (0.7594,-0.7843) (0.7646,-0.8463) (0.7699,-0.8029)
(0.7788,-0.8740) \PST@linethr(0.8091,-0.8740) (0.8175,-0.8286) (0.8363,-0.8740)
\PST@linethr(0.8457,-0.8740) (0.8492,-0.8373) (0.8598,-0.8232) (0.9470,-0.8338)

\catcode`@=12
\fi
\endpspicture

%% file: timeavg.tex
\ifx\PSTloaded\undefined
\def\PSTloaded{t}
\psset{arrowsize=.01 3.2 1.4 .3}
\psset{dotsize=.01}
\catcode`@=11

\definecolor{darkgreen}{rgb}{0,.5,0}
\newpsobject{PST@Border}{psline}{linewidth=.0015,linestyle=solid}
\newpsobject{PST@Axes}{psline}{linewidth=.0015,linestyle=dotted,dotsep=.004}
\newpsobject{PST@lineone}{psline}{linecolor=black,    linewidth=.0025,linestyle=solid}
\newpsobject{PST@linetwo}{psline}{linecolor=red,      linewidth=.0025,linestyle=dashed,dash=.01 .01}
\newpsobject{PST@linethr}{psline}{linecolor=blue,     linewidth=.0035,linestyle=dotted,dotsep=.004}
\newpsobject{PST@linefou}{psline}{linecolor=darkgreen,linewidth=.0025,linestyle=dashed,dash=.01 .004 .004 .004}
\catcode`@=12

\fi
\psset{unit=5.0in,xunit=\columnwidth,yunit=.6\columnwidth}
\pspicture(0.000000,0.000000)(1.000000,1.000000)
\ifx\nofigs\undefined
\catcode`@=11

\definecolor{hellgrau}{rgb}{.8,.8,.8}
\definecolor{orange}{rgb}{.8,.4,0}
\definecolor{violett}{rgb}{.5,0,.5}
\definecolor{hellorange}{rgb}{1,.75,.5}
\definecolor{hellviolett}{rgb}{1,.75,1}
\psframe[fillstyle=solid,fillcolor=hellgrau,   linecolor=black,  linewidth=.002](.3216,.126)(.4110,.968)
\psframe[fillstyle=solid,fillcolor=hellorange, linecolor=orange, linestyle=dashed,dash=.01 .01,linewidth=.002](.5003,.126)(.5897,.968)
\psframe[fillstyle=solid,fillcolor=hellviolett,linecolor=violett,linestyle=dotted,dotsep=.004, linewidth=.002](.6790,.126)(.7683,.968)

\PST@Border(0.1430,0.1260)(0.1580,0.1260)
\PST@Border(0.9470,0.1260)(0.9320,0.1260) \rput[r](0.1270,0.1260){ 0}
\PST@Border(0.1430,0.2102)(0.1580,0.2102)
\PST@Border(0.9470,0.2102)(0.9320,0.2102) \rput[r](0.1270,0.2102){ 2}
\PST@Border(0.1430,0.2944)(0.1580,0.2944)
\PST@Border(0.9470,0.2944)(0.9320,0.2944) \rput[r](0.1270,0.2944){ 4}
\PST@Border(0.1430,0.3786)(0.1580,0.3786)
\PST@Border(0.9470,0.3786)(0.9320,0.3786) \rput[r](0.1270,0.3786){ 6}
\PST@Border(0.1430,0.4628)(0.1580,0.4628)
\PST@Border(0.9470,0.4628)(0.9320,0.4628) \rput[r](0.1270,0.4628){ 8}
\PST@Border(0.1430,0.5470)(0.1580,0.5470)
\PST@Border(0.9470,0.5470)(0.9320,0.5470) \rput[r](0.1270,0.5470){ 10}
\PST@Border(0.1430,0.6312)(0.1580,0.6312)
\PST@Border(0.9470,0.6312)(0.9320,0.6312) \rput[r](0.1270,0.6312){ 12}
\PST@Border(0.1430,0.7154)(0.1580,0.7154)
\PST@Border(0.9470,0.7154)(0.9320,0.7154) \rput[r](0.1270,0.7154){ 14}
\PST@Border(0.1430,0.7996)(0.1580,0.7996)
\PST@Border(0.9470,0.7996)(0.9320,0.7996) \rput[r](0.1270,0.7996){ 16}
\PST@Border(0.1430,0.8838)(0.1580,0.8838)
\PST@Border(0.9470,0.8838)(0.9320,0.8838) \rput[r](0.1270,0.8838){ 18}
\PST@Border(0.1430,0.9680)(0.1580,0.9680)
\PST@Border(0.9470,0.9680)(0.9320,0.9680) \rput[r](0.1270,0.9680){ 20}
\PST@Border(0.1430,0.1260)(0.1430,0.1460)
\PST@Border(0.1430,0.9680)(0.1430,0.9480) \rput(0.1430,0.0840){ 0}
\PST@Border(0.2323,0.1260)(0.2323,0.1460)
\PST@Border(0.2323,0.9680)(0.2323,0.9480) \rput(0.2323,0.0840){ 0.5}
\PST@Border(0.3217,0.1260)(0.3217,0.1460)
\PST@Border(0.3217,0.9680)(0.3217,0.9480) \rput(0.3217,0.0840){ 1}
\PST@Border(0.4110,0.1260)(0.4110,0.1460)
\PST@Border(0.4110,0.9680)(0.4110,0.9480) \rput(0.4110,0.0840){ 1.5}
\PST@Border(0.5003,0.1260)(0.5003,0.1460)
\PST@Border(0.5003,0.9680)(0.5003,0.9480) \rput(0.5003,0.0840){ 2}
\PST@Border(0.5897,0.1260)(0.5897,0.1460)
\PST@Border(0.5897,0.9680)(0.5897,0.9480) \rput(0.5897,0.0840){ 2.5}
\PST@Border(0.6790,0.1260)(0.6790,0.1460)
\PST@Border(0.6790,0.9680)(0.6790,0.9480) \rput(0.6790,0.0840){ 3}
\PST@Border(0.7683,0.1260)(0.7683,0.1460)
\PST@Border(0.7683,0.9680)(0.7683,0.9480) \rput(0.7683,0.0840){ 3.5}
\PST@Border(0.8577,0.1260)(0.8577,0.1460)
\PST@Border(0.8577,0.9680)(0.8577,0.9480) \rput(0.8577,0.0840){ 4}
\PST@Border(0.9470,0.1260)(0.9470,0.1460)
\PST@Border(0.9470,0.9680)(0.9470,0.9480) \rput(0.9470,0.0840){ 4.5}
\PST@Border(0.1430,0.9680)(0.1430,0.1260)(0.9470,0.1260)(0.9470,0.9680)(0.1430,0.9680)

\rput{L}(0.0420,0.5470){$\langle t_{\rm emission}\rangle$~[fm]}
\rput(0.5450,0.0210){$p_\bot$~[GeV]}

\rput[r](0.9150,0.58){Pb+Pb @ 158 AGeV}
\rput[r](0.9150,0.52){$b < 4.5$~fm}
\rput[r](0.9150,0.46){$|y_{\rm c.m.}| < 0.5$}
\rput[r](0.9150,0.40){UrQMD}

\rput[r](0.8200,0.9270){all}
\PST@lineone(0.8360,0.9270)(0.9150,0.9270)

\PST@lineone(0.1519,0.5676) (0.1519,0.5676) (0.1698,0.5464) (0.1877,0.5340)
(0.2055,0.5088) (0.2234,0.4888) (0.2413,0.4783) (0.2591,0.4734) (0.2770,0.4721)
(0.2949,0.4725) (0.3127,0.4750) (0.3306,0.4774) (0.3485,0.4786) (0.3663,0.4801)
(0.3842,0.4806) (0.4021,0.4787) (0.4199,0.4770) (0.4378,0.4728) (0.4557,0.4656)
(0.4735,0.4564) (0.4914,0.4505) (0.5093,0.4425) (0.5271,0.4205) (0.5450,0.4124)
(0.5629,0.3966) (0.5807,0.3811) (0.5986,0.3569) (0.6165,0.3492) (0.6343,0.3026)
(0.6522,0.3039) (0.6701,0.2665) (0.6879,0.2530) (0.7058,0.2315) (0.7237,0.2275)
(0.7415,0.2172) (0.7594,0.2185) (0.7773,0.1930) (0.7951,0.1840) (0.8130,0.1954)
(0.8309,0.1903) (0.8487,0.1919) (0.8666,0.1780) (0.8845,0.1837) (0.9023,0.1873)
(0.9202,0.1872) (0.9381,0.1767) (0.9470,0.1738)

\rput[r](0.8200,0.8750){$\pi\rho\rightarrow\gamma\pi$}
\PST@linetwo(0.8360,0.8750)(0.9150,0.8750)

\PST@linetwo(0.1519,0.8116) (0.1519,0.8116) (0.1698,0.6239) (0.1877,0.5379)
(0.2055,0.4993) (0.2234,0.4813) (0.2413,0.4736) (0.2591,0.4711) (0.2770,0.4711)
(0.2949,0.4727) (0.3127,0.4763) (0.3306,0.4800) (0.3485,0.4823) (0.3663,0.4854)
(0.3842,0.4877) (0.4021,0.4881) (0.4199,0.4890) (0.4378,0.4891) (0.4557,0.4859)
(0.4735,0.4823) (0.4914,0.4815) (0.5093,0.4819) (0.5271,0.4686) (0.5450,0.4675)
(0.5629,0.4588) (0.5807,0.4585) (0.5986,0.4432) (0.6165,0.4420) (0.6343,0.3890)
(0.6522,0.3948) (0.6701,0.3543) (0.6879,0.3267) (0.7058,0.3006) (0.7237,0.2952)
(0.7415,0.2830) (0.7594,0.2817) (0.7773,0.2350) (0.7951,0.2081) (0.8130,0.2285)
(0.8309,0.2227) (0.8487,0.2287) (0.8666,0.1985) (0.8845,0.2070) (0.9023,0.2265)
(0.9202,0.2175) (0.9381,0.1928) (0.9470,0.1866)

\rput[r](0.8200,0.8230){$\pi\pi\rightarrow\gamma\rho$}
\PST@linethr(0.8360,0.8230)(0.9150,0.8230)

\PST@linethr(0.1519,0.5673) (0.1519,0.5673) (0.1698,0.5384) (0.1877,0.5252)
(0.2055,0.5144) (0.2234,0.5043) (0.2413,0.4930) (0.2591,0.4775) (0.2770,0.4638)
(0.2949,0.4495) (0.3127,0.4313) (0.3306,0.4078) (0.3485,0.3869) (0.3663,0.3623)
(0.3842,0.3357) (0.4021,0.3156) (0.4199,0.2985) (0.4378,0.2721) (0.4557,0.2584)
(0.4735,0.2382) (0.4914,0.2297) (0.5093,0.2171) (0.5271,0.2081) (0.5450,0.2045)
(0.5629,0.1985) (0.5807,0.1936) (0.5986,0.1919) (0.6165,0.1921) (0.6343,0.1874)
(0.6522,0.1865) (0.6701,0.1761) (0.6879,0.1818) (0.7058,0.1706) (0.7237,0.1744)
(0.7415,0.1724) (0.7594,0.1725) (0.7773,0.1694) (0.7951,0.1735) (0.8130,0.1792)
(0.8309,0.1757) (0.8487,0.1763) (0.8666,0.1701) (0.8845,0.1759) (0.9023,0.1700)
(0.9202,0.1741) (0.9381,0.1758) (0.9470,0.1725)

\rput[r](0.8200,0.7710){Processes with $\eta$ and $\pi\pi\rightarrow\gamma\gamma$}
\PST@linefou(0.8360,0.7710)(0.9150,0.7710)

\PST@linefou(0.1519,0.9197) (0.1519,0.9197) (0.1698,0.7118) (0.1877,0.6877)
(0.2055,0.6843) (0.2234,0.6757) (0.2413,0.6676) (0.2591,0.6477) (0.2770,0.6256)
(0.2949,0.5920) (0.3127,0.5532) (0.3306,0.5210) (0.3485,0.4830) (0.3663,0.4393)
(0.3842,0.4074) (0.4021,0.3534) (0.4199,0.3264) (0.4378,0.2903) (0.4557,0.2695)
(0.4735,0.2463) (0.4914,0.2366) (0.5093,0.2165) (0.5271,0.2066) (0.5450,0.1964)
(0.5629,0.1982) (0.5807,0.1852) (0.5986,0.1827) (0.6165,0.1797) (0.6343,0.1758)
(0.6522,0.1742) (0.6701,0.1711) (0.6879,0.1697) (0.7058,0.1681) (0.7237,0.1664)
(0.7415,0.1655) (0.7594,0.1665) (0.7773,0.1653) (0.7951,0.1660) (0.8130,0.1651)
(0.8309,0.1663) (0.8487,0.1649) (0.8666,0.1657) (0.8845,0.1651) (0.9023,0.1644)
(0.9202,0.1649) (0.9381,0.1631) (0.9470,0.1627)

\PST@Border(0.1430,0.9680)(0.1430,0.1260)(0.9470,0.1260)(0.9470,0.9680)(0.1430,0.9680)

\catcode`@=12
\fi
\endpspicture

%% file: timefilter.tex
\ifx\PSTloaded\undefined
\def\PSTloaded{t}
\psset{arrowsize=.01 3.2 1.4 .3}
\psset{dotsize=.01}
\catcode`@=11

\definecolor{orange}{rgb}{.8,.4,0}
\definecolor{violett}{rgb}{.5,0,.5}
\newpsobject{PST@Border}{psline}{linewidth=.0015,linestyle=solid}
\newpsobject{PST@Axes}{psline}{linewidth=.0015,linestyle=dotted,dotsep=.004}
\newpsobject{PST@lineone}{psline}{linecolor=black,  linewidth=.0025,linestyle=solid}
\newpsobject{PST@linetwo}{psline}{linecolor=orange, linewidth=.0025,linestyle=dashed,dash=.01 .01}
\newpsobject{PST@linethr}{psline}{linecolor=violett,linewidth=.0035,linestyle=dotted,dotsep=.004}
\catcode`@=12

\fi
\psset{unit=5.0in,xunit=1.17\columnwidth,yunit=.8\columnwidth}
\pspicture(0.115,0)(.97,1.04)
\ifx\nofigs\undefined
\catcode`@=11

\rput[c](.5187,.9980){$\langle t_1 \rangle$, \color{orange}{$\langle t_2 \rangle$}}
\rput[c](.3835,.9980){\color{violett}{$\langle t_3 \rangle$}}
\PST@lineone(.5178,.1260)(.5178,.9680)
\PST@linetwo(.5196,.1260)(.5196,.9680)
\PST@linethr(.3835,.1260)(.3835,.9680)

\PST@Border(0.2070,0.1260) (0.2220,0.1260) \PST@Border(0.9470,0.1260)
(0.9320,0.1260) \rput[r](0.1910,0.1260){ 0} \PST@Border(0.2070,0.2196)
(0.2220,0.2196) \PST@Border(0.9470,0.2196) (0.9320,0.2196)
\PST@Border(0.2070,0.3131) (0.2220,0.3131)
\PST@Border(0.9470,0.3131) (0.9320,0.3131) \rput[r](0.1910,0.3131){ 1}
\PST@Border(0.2070,0.4067) (0.2220,0.4067) \PST@Border(0.9470,0.4067)
(0.9320,0.4067) \PST@Border(0.2070,0.5002)
(0.2220,0.5002) \PST@Border(0.9470,0.5002) (0.9320,0.5002)
\rput[r](0.1910,0.5002){ 2} \PST@Border(0.2070,0.5938) (0.2220,0.5938)
\PST@Border(0.9470,0.5938) (0.9320,0.5938) 
\PST@Border(0.2070,0.6873) (0.2220,0.6873) \PST@Border(0.9470,0.6873)
(0.9320,0.6873) \rput[r](0.1910,0.6873){ 3} \PST@Border(0.2070,0.7809)
(0.2220,0.7809) \PST@Border(0.9470,0.7809) (0.9320,0.7809)
\PST@Border(0.2070,0.8744) (0.2220,0.8744)
\PST@Border(0.9470,0.8744) (0.9320,0.8744) \rput[r](0.1910,0.8744){ 4}
\PST@Border(0.2070,0.9680) (0.2220,0.9680) \PST@Border(0.9470,0.9680)
(0.9320,0.9680) \PST@Border(0.2070,0.1260)
(0.2070,0.1460) \PST@Border(0.2070,0.9680) (0.2070,0.9480)
\rput(0.2070,0.0840){ 0} \PST@Border(0.3920,0.1260) (0.3920,0.1460)
\PST@Border(0.3920,0.9680) (0.3920,0.9480) \rput(0.3920,0.0840){ 5}
\PST@Border(0.5770,0.1260) (0.5770,0.1460) \PST@Border(0.5770,0.9680)
(0.5770,0.9480) \rput(0.5770,0.0840){ 10} \PST@Border(0.7620,0.1260)
(0.7620,0.1460) \PST@Border(0.7620,0.9680) (0.7620,0.9480)
\rput(0.7620,0.0840){ 15} \PST@Border(0.9470,0.1260) (0.9470,0.1460)
\PST@Border(0.9470,0.9680) (0.9470,0.9480) \rput(0.9470,0.0840){ 20}
\PST@Border(0.2070,0.9680) (0.2070,0.1260) (0.9470,0.1260) (0.9470,0.9680)
(0.2070,0.9680)

\rput{L}(0.1420,0.5470){$\frac{dN}{dt}~[10^{-3}$~fm$^{-1}]$}
\rput(0.5770,0.0210){$t$~[fm]}

\rput[r](0.8200,0.9270){$p_\bot = 1 - 1.5~{\rm GeV}$}
\rput[r](0.8200,0.8800){$p_\bot = 2 - 2.5~{\rm GeV}$}
\rput[r](0.8200,0.8330){$p_\bot = 3 - 3.5~{\rm GeV}$}
\rput[r](0.9150,0.7860){Only $\pi\rho\rightarrow\gamma\pi$}

\PST@lineone(0.8360,0.9270)(0.9150,0.9270)
\PST@linetwo(0.8360,0.8800)(0.9150,0.8800)
\PST@linethr(0.8360,0.8330)(0.9150,0.8330)

\PST@lineone(0.2089,0.1260) (0.2089,0.1260) (0.2125,0.1293) (0.2163,0.1328)
(0.2199,0.1549) (0.2237,0.2113) (0.2274,0.2822) (0.2311,0.3775) (0.2348,0.4677)
(0.2384,0.5382) (0.2421,0.5495) (0.2459,0.5522) (0.2495,0.5378) (0.2533,0.4825)
(0.2570,0.3938) (0.2606,0.3322) (0.2644,0.2998) (0.2680,0.2774) (0.2718,0.2829)
(0.2755,0.2675) (0.2791,0.2996) (0.2828,0.3177) (0.2865,0.3204) (0.2903,0.3290)
(0.2940,0.3396) (0.2977,0.3668) (0.3013,0.3889) (0.3050,0.4157) (0.3088,0.4226)
(0.3125,0.4694) (0.3162,0.4849) (0.3198,0.5148) (0.3235,0.5211) (0.3273,0.6108)
(0.3310,0.5347) (0.3347,0.5864) (0.3383,0.6264) (0.3420,0.6940) (0.3458,0.6804)
(0.3495,0.7027) (0.3532,0.6980) (0.3568,0.7450) (0.3606,0.7192) (0.3643,0.7715)
(0.3679,0.7589) (0.3717,0.7437) (0.3753,0.7908) (0.3791,0.7615) (0.3828,0.8278)
(0.3864,0.8145) (0.3902,0.7809) (0.3938,0.8268) (0.3976,0.8292) (0.4013,0.8323)
(0.4049,0.8236) (0.4087,0.8409) (0.4123,0.8829) (0.4161,0.8639) (0.4198,0.8679)
(0.4234,0.8515) (0.4272,0.8490) (0.4308,0.8621) (0.4346,0.8795) (0.4383,0.8681)
(0.4419,0.8508) (0.4457,0.8434) (0.4493,0.8253) (0.4531,0.7840) (0.4568,0.8689)
(0.4604,0.8174) (0.4642,0.7929) (0.4678,0.8238) (0.4716,0.7846) (0.4753,0.8205)
(0.4789,0.8295) (0.4827,0.7739) (0.4863,0.8580) (0.4901,0.7461) (0.4938,0.7845)
(0.4974,0.7507) (0.5012,0.7893) (0.5049,0.7743) (0.5086,0.7516) (0.5123,0.7776)
(0.5159,0.7947) (0.5196,0.7395) (0.5234,0.7645) (0.5271,0.7475) (0.5308,0.7114)
(0.5344,0.7186) (0.5381,0.7468) (0.5419,0.7492) (0.5456,0.7387) (0.5493,0.7652)
(0.5529,0.7360) (0.5566,0.7320) (0.5604,0.7561) (0.5641,0.6999) (0.5678,0.7115)
(0.5714,0.7343) (0.5751,0.7265) (0.5789,0.7197) (0.5826,0.6939) (0.5863,0.6986)
(0.5899,0.7087) (0.5936,0.6770) (0.5974,0.7061) (0.6011,0.7350) (0.6048,0.6806)
(0.6084,0.6605) (0.6121,0.7117) (0.6159,0.6718) (0.6196,0.6523) (0.6233,0.6458)
(0.6269,0.6603) (0.6306,0.6892) (0.6344,0.6565) (0.6381,0.6235) (0.6418,0.6610)
(0.6454,0.5921) (0.6491,0.6046) (0.6529,0.6006) (0.6566,0.6261) (0.6603,0.6014)
(0.6639,0.6133) (0.6676,0.5906) (0.6714,0.5215) (0.6751,0.5613) (0.6788,0.5284)
(0.6824,0.5226) (0.6861,0.5180) (0.6899,0.4870) (0.6936,0.5273) (0.6973,0.4838)
(0.7009,0.4700) (0.7046,0.4666) (0.7084,0.4438) (0.7121,0.4837) (0.7158,0.4245)
(0.7194,0.4503) (0.7231,0.4366) (0.7269,0.4029) (0.7306,0.4074) (0.7343,0.3801)
(0.7379,0.4137) (0.7416,0.3890) (0.7454,0.3909) (0.7491,0.3562) (0.7528,0.3556)
(0.7564,0.3400) (0.7601,0.3703) (0.7639,0.3434) (0.7676,0.3256) (0.7713,0.3005)
(0.7749,0.3352) (0.7786,0.3145) (0.7824,0.3058) (0.7861,0.2795) (0.7898,0.2944)
(0.7934,0.2887) (0.7971,0.2885) (0.8009,0.2727) (0.8045,0.2742) (0.8083,0.2933)
(0.8120,0.2621) (0.8156,0.2649) (0.8194,0.2731) (0.8230,0.2611) (0.8268,0.2641)
(0.8305,0.2335) (0.8341,0.2219) (0.8379,0.2256) (0.8415,0.2234) (0.8453,0.2246)
(0.8490,0.2146) (0.8526,0.2156) (0.8564,0.2096) (0.8600,0.1950) (0.8638,0.2002)
(0.8675,0.1906) (0.8711,0.2070) (0.8749,0.1962) (0.8785,0.2001) (0.8823,0.1944)
(0.8860,0.1988) (0.8896,0.1796) (0.8934,0.1868) (0.8970,0.1946) (0.9008,0.1718)
(0.9045,0.1723) (0.9081,0.1646) (0.9119,0.1730) (0.9155,0.1759) (0.9193,0.1775)
(0.9230,0.1672) (0.9266,0.1700) (0.9304,0.1678) (0.9340,0.1613) (0.9378,0.1639)
(0.9415,0.1592) (0.9451,0.1674) (0.9470,0.1649)

\rput(0.45,.35){\rnode{B}{$\times 10$}}
\pnode(0.4112,.2106){b}
\ncarc{<-}{b}{B}

\PST@linetwo(0.2125,0.1262) (0.2125,0.1262) (0.2163,0.1330) (0.2199,0.1526)
(0.2237,0.1957) (0.2274,0.2494) (0.2311,0.3067) (0.2348,0.4653) (0.2384,0.4118)
(0.2421,0.4421) (0.2459,0.4167) (0.2495,0.3672) (0.2533,0.2871) (0.2570,0.2205)
(0.2606,0.1990) (0.2644,0.1758) (0.2680,0.1631) (0.2718,0.1527) (0.2755,0.1609)
(0.2791,0.1658) (0.2828,0.1689) (0.2865,0.1578) (0.2903,0.1682) (0.2940,0.1587)
(0.2977,0.1569) (0.3013,0.1884) (0.3050,0.1524) (0.3088,0.1832) (0.3125,0.2173)
(0.3162,0.2168) (0.3198,0.2061) (0.3235,0.1762) (0.3273,0.1667) (0.3310,0.1795)
(0.3347,0.2233) (0.3383,0.2013) (0.3420,0.1938) (0.3458,0.2002) (0.3495,0.1774)
(0.3532,0.1924) (0.3568,0.1984) (0.3606,0.1903) (0.3643,0.2106) (0.3679,0.1930)
(0.3717,0.2271) (0.3753,0.1975) (0.3791,0.2128) (0.3828,0.2037) (0.3864,0.2239)
(0.3902,0.1741) (0.3938,0.1904) (0.3976,0.1881) (0.4013,0.1830) (0.4049,0.2073)
(0.4087,0.2571) (0.4123,0.1895) (0.4161,0.2039) (0.4198,0.2110) (0.4234,0.1955)
(0.4272,0.2162) (0.4308,0.2134) (0.4346,0.2439) (0.4383,0.1786) (0.4419,0.1798)
(0.4457,0.2038) (0.4493,0.2385) (0.4531,0.2218) (0.4568,0.1954) (0.4604,0.2056)
(0.4642,0.1900) (0.4678,0.2039) (0.4716,0.1870) (0.4753,0.2099) (0.4789,0.2528)
(0.4827,0.1894) (0.4863,0.2134) (0.4901,0.2445) (0.4938,0.2384) (0.4974,0.2227)
(0.5012,0.2041) (0.5049,0.2251) (0.5086,0.2047) (0.5123,0.2274) (0.5159,0.1888)
(0.5196,0.1946) (0.5234,0.2124) (0.5271,0.1824) (0.5308,0.2187) (0.5344,0.2464)
(0.5381,0.2261) (0.5419,0.2140) (0.5456,0.2073) (0.5493,0.2030) (0.5529,0.2659)
(0.5566,0.2012) (0.5604,0.2308) (0.5641,0.1809) (0.5678,0.2062) (0.5714,0.1902)
(0.5751,0.2140) (0.5789,0.2002) (0.5826,0.1974) (0.5863,0.1981) (0.5899,0.1865)
(0.5936,0.1903) (0.5974,0.2207) (0.6011,0.2003) (0.6048,0.2196) (0.6084,0.1839)
(0.6121,0.1992) (0.6159,0.2467) (0.6196,0.2257) (0.6233,0.1939) (0.6269,0.2376)
(0.6306,0.1807) (0.6344,0.2073) (0.6381,0.2036) (0.6418,0.2109) (0.6454,0.2387)
(0.6491,0.2099) (0.6529,0.2221) (0.6566,0.1924) (0.6603,0.1773) (0.6639,0.2376)
(0.6676,0.2001) (0.6714,0.2048) (0.6751,0.1963) (0.6788,0.2611) (0.6824,0.1778)
(0.6861,0.1717) (0.6899,0.1670) (0.6936,0.1687) (0.6973,0.2320) (0.7009,0.1863)
(0.7046,0.1939) (0.7084,0.1677) (0.7121,0.1676) (0.7158,0.1743) (0.7194,0.2111)
(0.7231,0.1979) (0.7269,0.1857) (0.7306,0.2184) (0.7343,0.1592) (0.7379,0.1728)
(0.7416,0.1663) (0.7454,0.1924) (0.7491,0.1808) (0.7528,0.1547) (0.7564,0.1691)
(0.7601,0.1764) (0.7639,0.1701) (0.7676,0.1668) (0.7713,0.1549) (0.7749,0.1435)
(0.7786,0.1662) (0.7824,0.1395) (0.7861,0.1635) (0.7898,0.1675) (0.7934,0.1566)
(0.7971,0.1653) (0.8009,0.1548) (0.8045,0.1549) (0.8083,0.1578) (0.8120,0.1430)
(0.8156,0.1415) (0.8194,0.1629) (0.8230,0.1476) (0.8268,0.1403) (0.8305,0.1754)
(0.8341,0.1340) (0.8379,0.1475) (0.8415,0.1601) (0.8453,0.1439) (0.8490,0.1569)
(0.8526,0.1554) (0.8564,0.1494) (0.8600,0.1445) (0.8638,0.1366) (0.8675,0.1453)
(0.8711,0.1481) (0.8749,0.1651) (0.8785,0.1510) (0.8823,0.1644) (0.8860,0.1357)
(0.8896,0.1403) (0.8934,0.1367) (0.8970,0.1309) (0.9008,0.1336) (0.9045,0.1441)
(0.9081,0.1370) (0.9119,0.1297) (0.9155,0.1421) (0.9193,0.1573) (0.9230,0.1465)
(0.9266,0.1285) (0.9304,0.1315) (0.9340,0.1326) (0.9378,0.1311) (0.9415,0.1747)
(0.9451,0.1260) (0.9470,0.1267)

\rput(0.60,.35){\rnode{C}{$\times 100$}}
\pnode(0.6612,.1437){c}
\ncarc{->}{C}{c}

\PST@linethr(0.2163,0.1266) (0.2163,0.1266) (0.2199,0.1470) (0.2237,0.1880)
(0.2274,0.2271) (0.2311,0.2617) (0.2348,0.3784) (0.2384,0.3823) (0.2421,0.3352)
(0.2459,0.3312) (0.2495,0.3075) (0.2533,0.2185) (0.2570,0.1842) (0.2606,0.1613)
(0.2644,0.1692) (0.2680,0.1624) (0.2718,0.1320) (0.2755,0.1327) (0.2791,0.1273)
(0.2828,0.1489) (0.2865,0.1439) (0.2903,0.1348) (0.2940,0.1385) (0.2977,0.1260)
(0.3013,0.1368) (0.3050,0.1370) (0.3088,0.1383) (0.3125,0.1302) (0.3162,0.1445)
(0.3198,0.1317) (0.3235,0.1267) (0.3273,0.1339) (0.3310,0.1299) (0.3347,0.1285)
(0.3383,0.1437) (0.3420,0.1495) (0.3458,0.1402) (0.3495,0.1543) (0.3532,0.1291)
(0.3568,0.1336) (0.3606,0.1311) (0.3643,0.1498) (0.3679,0.1262) (0.3717,0.1302)
(0.3753,0.1362) (0.3791,0.1333) (0.3828,0.1411) (0.3864,0.1269) (0.3902,0.1265)
(0.3938,0.1309) (0.4013,0.1358) (0.4049,0.1313) (0.4087,0.1382) (0.4123,0.1272)
(0.4161,0.1281) (0.4198,0.1418) (0.4234,0.1346) (0.4272,0.1298) (0.4308,0.1338)
(0.4346,0.1422) (0.4383,0.1313) (0.4419,0.1396) (0.4457,0.1407) (0.4493,0.1867)
(0.4531,0.1268) (0.4568,0.1396) (0.4604,0.1361) (0.4642,0.1284) (0.4716,0.1272)
(0.4753,0.1260) (0.4789,0.1260) (0.4827,0.1276) (0.4863,0.1332) (0.4901,0.1292)
(0.4938,0.1394) (0.4974,0.1276) (0.5012,0.1264) (0.5049,0.1292) (0.5086,0.1479)
(0.5123,0.1340) (0.5159,0.1269) (0.5196,0.1260) (0.5234,0.1260) (0.5271,0.1354)
(0.5308,0.1279) (0.5344,0.1274) (0.5381,0.1260) (0.5456,0.1355) (0.5529,0.1264)
(0.5566,0.1303) (0.5604,0.1274) (0.5641,0.1260) (0.5678,0.1500) (0.5714,0.1408)
(0.5751,0.1260) (0.5789,0.1359) (0.5826,0.1314) (0.5863,0.1388) (0.5899,0.1359)
(0.5936,0.1260) (0.5974,0.1306) (0.6011,0.1315) (0.6048,0.1260) (0.6084,0.1313)
(0.6121,0.1273) (0.6196,0.1280) (0.6233,0.1324) (0.6269,0.1261) (0.6306,0.1285)
(0.6344,0.1281) (0.6381,0.1734) (0.6454,0.1535) (0.6491,0.1271) (0.6529,0.1388)
(0.6566,0.1298) (0.6603,0.1437) (0.6639,0.1529) (0.6676,0.1308) (0.6714,0.1260)
(0.6751,0.1278) (0.6788,0.1260) (0.6824,0.1288) (0.6861,0.1281) (0.6899,0.1268)
(0.6973,0.1260) (0.7009,0.1260) (0.7046,0.1580) (0.7084,0.1305) (0.7158,0.1263)
(0.7343,0.1347) (0.7379,0.1868) (0.7416,0.1273) (0.7491,0.1288) (0.7528,0.1296)
(0.7564,0.1263) (0.7601,0.1262) (0.7676,0.1893) (0.7786,0.1410) (0.7898,0.1480)
(0.7934,0.1263) (0.8009,0.1286) (0.8083,0.1533) (0.8120,0.1262) (0.8156,0.1266)
(0.8415,0.1260) (0.8490,0.1276) (0.8600,0.1360) (0.8638,0.1260) (0.8675,0.1265)
(0.8749,0.1260) (0.9155,0.1293) (0.9340,0.1260) (0.9470,0.1267)

\PST@Border(0.2070,0.9680)
(0.2070,0.1260)
(0.9470,0.1260)
(0.9470,0.9680)
(0.2070,0.9680)

\catcode`@=12
\fi
\endpspicture

%% file: strings.tex
\ifx\PSTloaded\undefined
\def\PSTloaded{t}
\psset{arrowsize=.01 3.2 1.4 .3}
\psset{dotsize=.01}
\catcode`@=11

\definecolor{darkgreen}{rgb}{0,.5,0}
\newpsobject{PST@Border}{psline}{linewidth=.0015,linestyle=solid}
\newpsobject{PST@lineone}{psline}{linewidth=.0025,linestyle=solid}
\newpsobject{PST@linetwo}{psline}{linewidth=.0025,linecolor=red,linestyle=dashed,dash=.01 .01}
\newpsobject{PST@linethr}{psline}{linewidth=.0035,linecolor=blue,linestyle=dotted,dotsep=.004}
\newpsobject{PST@linefou}{psline}{linewidth=.0025,linecolor=darkgreen,linestyle=dashed,dash=.01 .004 .004 .004}
\catcode`@=12

\fi
\psset{unit=5.0in,xunit=1.099\columnwidth,yunit=.8\columnwidth}
\pspicture(0.070000,0.000000)(0.980000,1.000000)
\ifx\nofigs\undefined
\catcode`@=11

\PST@Border(0.2390,0.1344)
(0.2540,0.1344)

\PST@Border(0.9630,0.1344)
(0.9480,0.1344)

\rput[r](0.2230,0.1344){$10^{-7}$}
\PST@Border(0.2390,0.1658)
(0.2465,0.1658)

\PST@Border(0.9630,0.1658)
(0.9555,0.1658)

\PST@Border(0.2390,0.2072)
(0.2465,0.2072)

\PST@Border(0.9630,0.2072)
(0.9555,0.2072)

\PST@Border(0.2390,0.2285)
(0.2465,0.2285)

\PST@Border(0.9630,0.2285)
(0.9555,0.2285)

\PST@Border(0.2390,0.2386)
(0.2540,0.2386)

\PST@Border(0.9630,0.2386)
(0.9480,0.2386)

\rput[r](0.2230,0.2386){$10^{-6}$}
\PST@Border(0.2390,0.2700)
(0.2465,0.2700)

\PST@Border(0.9630,0.2700)
(0.9555,0.2700)

\PST@Border(0.2390,0.3114)
(0.2465,0.3114)

\PST@Border(0.9630,0.3114)
(0.9555,0.3114)

\PST@Border(0.2390,0.3327)
(0.2465,0.3327)

\PST@Border(0.9630,0.3327)
(0.9555,0.3327)

\PST@Border(0.2390,0.3428)
(0.2540,0.3428)

\PST@Border(0.9630,0.3428)
(0.9480,0.3428)

\rput[r](0.2230,0.3428){$10^{-5}$}
\PST@Border(0.2390,0.3742)
(0.2465,0.3742)

\PST@Border(0.9630,0.3742)
(0.9555,0.3742)

\PST@Border(0.2390,0.4156)
(0.2465,0.4156)

\PST@Border(0.9630,0.4156)
(0.9555,0.4156)

\PST@Border(0.2390,0.4369)
(0.2465,0.4369)

\PST@Border(0.9630,0.4369)
(0.9555,0.4369)

\PST@Border(0.2390,0.4470)
(0.2540,0.4470)

\PST@Border(0.9630,0.4470)
(0.9480,0.4470)

\rput[r](0.2230,0.4470){$10^{-4}$}
\PST@Border(0.2390,0.4784)
(0.2465,0.4784)

\PST@Border(0.9630,0.4784)
(0.9555,0.4784)

\PST@Border(0.2390,0.5198)
(0.2465,0.5198)

\PST@Border(0.9630,0.5198)
(0.9555,0.5198)

\PST@Border(0.2390,0.5411)
(0.2465,0.5411)

\PST@Border(0.9630,0.5411)
(0.9555,0.5411)

\PST@Border(0.2390,0.5512)
(0.2540,0.5512)

\PST@Border(0.9630,0.5512)
(0.9480,0.5512)

\rput[r](0.2230,0.5512){ 0.001}
\PST@Border(0.2390,0.5826)
(0.2465,0.5826)

\PST@Border(0.9630,0.5826)
(0.9555,0.5826)

\PST@Border(0.2390,0.6240)
(0.2465,0.6240)

\PST@Border(0.9630,0.6240)
(0.9555,0.6240)

\PST@Border(0.2390,0.6453)
(0.2465,0.6453)

\PST@Border(0.9630,0.6453)
(0.9555,0.6453)

\PST@Border(0.2390,0.6554)
(0.2540,0.6554)

\PST@Border(0.9630,0.6554)
(0.9480,0.6554)

\rput[r](0.2230,0.6554){ 0.01}
\PST@Border(0.2390,0.6868)
(0.2465,0.6868)

\PST@Border(0.9630,0.6868)
(0.9555,0.6868)

\PST@Border(0.2390,0.7282)
(0.2465,0.7282)

\PST@Border(0.9630,0.7282)
(0.9555,0.7282)

\PST@Border(0.2390,0.7495)
(0.2465,0.7495)

\PST@Border(0.9630,0.7495)
(0.9555,0.7495)

\PST@Border(0.2390,0.7596)
(0.2540,0.7596)

\PST@Border(0.9630,0.7596)
(0.9480,0.7596)

\rput[r](0.2230,0.7596){ 0.1}
\PST@Border(0.2390,0.7910)
(0.2465,0.7910)

\PST@Border(0.9630,0.7910)
(0.9555,0.7910)

\PST@Border(0.2390,0.8324)
(0.2465,0.8324)

\PST@Border(0.9630,0.8324)
(0.9555,0.8324)

\PST@Border(0.2390,0.8537)
(0.2465,0.8537)

\PST@Border(0.9630,0.8537)
(0.9555,0.8537)

\PST@Border(0.2390,0.8638)
(0.2540,0.8638)

\PST@Border(0.9630,0.8638)
(0.9480,0.8638)

\rput[r](0.2230,0.8638){ 1}
\PST@Border(0.2390,0.8952)
(0.2465,0.8952)

\PST@Border(0.9630,0.8952)
(0.9555,0.8952)

\PST@Border(0.2390,0.9366)
(0.2465,0.9366)

\PST@Border(0.9630,0.9366)
(0.9555,0.9366)

\PST@Border(0.2390,0.9579)
(0.2465,0.9579)

\PST@Border(0.9630,0.9579)
(0.9555,0.9579)

\PST@Border(0.2390,0.9680)
(0.2540,0.9680)

\PST@Border(0.9630,0.9680)
(0.9480,0.9680)

\rput[r](0.2230,0.9680){ 10}
\PST@Border(0.2390,0.1344)
(0.2390,0.1544)

\PST@Border(0.2390,0.9680)
(0.2390,0.9480)

\rput(0.2390,0.0924){ 0}
\PST@Border(0.3194,0.1344)
(0.3194,0.1544)

\PST@Border(0.3194,0.9680)
(0.3194,0.9480)

\rput(0.3194,0.0924){ 0.5}
\PST@Border(0.3999,0.1344)
(0.3999,0.1544)

\PST@Border(0.3999,0.9680)
(0.3999,0.9480)

\rput(0.3999,0.0924){ 1}
\PST@Border(0.4803,0.1344)
(0.4803,0.1544)

\PST@Border(0.4803,0.9680)
(0.4803,0.9480)

\rput(0.4803,0.0924){ 1.5}
\PST@Border(0.5608,0.1344)
(0.5608,0.1544)

\PST@Border(0.5608,0.9680)
(0.5608,0.9480)

\rput(0.5608,0.0924){ 2}
\PST@Border(0.6412,0.1344)
(0.6412,0.1544)

\PST@Border(0.6412,0.9680)
(0.6412,0.9480)

\rput(0.6412,0.0924){ 2.5}
\PST@Border(0.7217,0.1344)
(0.7217,0.1544)

\PST@Border(0.7217,0.9680)
(0.7217,0.9480)

\rput(0.7217,0.0924){ 3}
\PST@Border(0.8021,0.1344)
(0.8021,0.1544)

\PST@Border(0.8021,0.9680)
(0.8021,0.9480)

\rput(0.8021,0.0924){ 3.5}
\PST@Border(0.8826,0.1344)
(0.8826,0.1544)

\PST@Border(0.8826,0.9680)
(0.8826,0.9480)

\rput(0.8826,0.0924){ 4}
\PST@Border(0.9630,0.1344)
(0.9630,0.1544)

\PST@Border(0.9630,0.9680)
(0.9630,0.9480)

\rput(0.9630,0.0924){ 4.5}
\PST@Border(0.2390,0.9680)
(0.2390,0.1344)
(0.9630,0.1344)
(0.9630,0.9680)
(0.2390,0.9680)

\rput[l](0.27,.24){Pb+Pb 158 AGeV}
\rput[l](0.27,.19){$b < 4.5$~fm, $|y_{\rm c.m.}| < 0.5$}

\rput{L}(0.1120,0.5470){$E\frac{dN}{d^3p}$ [GeV$^{-2}$]}
\rput(0.6010,0.0294){$p_\bot$~[GeV]}

\rput[r](0.8360,0.9270){UrQMD w/o string ends}
\PST@lineone(0.8520,0.9270)(0.9310,0.9270)

\PST@lineone(0.2566,0.9680) (0.2631,0.9090) (0.2792,0.8527) (0.2953,0.8261)
(0.3114,0.8089) (0.3275,0.7919) (0.3436,0.7731) (0.3597,0.7532) (0.3758,0.7320)
(0.3918,0.7103) (0.4079,0.6879) (0.4240,0.6654) (0.4401,0.6424) (0.4562,0.6202)
(0.4723,0.5968) (0.4884,0.5750) (0.5045,0.5522) (0.5206,0.5286) (0.5366,0.5066)
(0.5527,0.4885) (0.5688,0.4647) (0.5849,0.4426) (0.6010,0.4172) (0.6171,0.3982)
(0.6332,0.3789) (0.6493,0.3574) (0.6654,0.3342) (0.6814,0.3153) (0.6975,0.3169)
(0.7136,0.2976) (0.7297,0.2614) (0.7458,0.2491) (0.7619,0.2417) (0.7780,0.2425)
(0.7941,0.1850) (0.8102,0.1880) (0.8262,0.1849) (0.8423,0.1826) (0.8584,0.1355)
(0.8745,0.1791) (0.8884,0.1344)

\rput[r](0.8360,0.8850){UrQMD with string ends}
\PST@linetwo(0.8520,0.8850)(0.9310,0.8850)

\PST@linetwo(0.2573,0.9680) (0.2631,0.9151) (0.2792,0.8606) (0.2953,0.8355)
(0.3114,0.8190) (0.3275,0.8023) (0.3436,0.7836) (0.3597,0.7634) (0.3758,0.7422)
(0.3918,0.7205) (0.4079,0.6984) (0.4240,0.6760) (0.4401,0.6537) (0.4562,0.6317)
(0.4723,0.6095) (0.4884,0.5883) (0.5045,0.5662) (0.5206,0.5459) (0.5366,0.5256)
(0.5527,0.5069) (0.5688,0.4869) (0.5849,0.4684) (0.6010,0.4514) (0.6171,0.4354)
(0.6332,0.4182) (0.6493,0.4046) (0.6654,0.3894) (0.6814,0.3781) (0.6975,0.3664)
(0.7136,0.3536) (0.7297,0.3422) (0.7458,0.3324) (0.7619,0.3224) (0.7780,0.3130)
(0.7941,0.3020) (0.8102,0.2921) (0.8262,0.2816) (0.8423,0.2736) (0.8584,0.2682)
(0.8745,0.2592) (0.8906,0.2559) (0.9067,0.2497) (0.9228,0.2424) (0.9389,0.2328)
(0.9550,0.2237)

\rput[r](0.8360,0.8430){pQCD-photons (Gale)}
\PST@linethr(0.8520,0.8430)(0.9310,0.8430)

\PST@linethr(0.4884,0.6143) (0.4884,0.6143) (0.5043,0.5961) (0.5202,0.5768)
(0.5368,0.5562) (0.5527,0.5369) (0.5688,0.5181) (0.5848,0.4999) (0.6013,0.4812)
(0.6172,0.4635) (0.6332,0.4459) (0.6493,0.4289) (0.6652,0.4113) (0.6811,0.3961)
(0.6977,0.3779) (0.7136,0.3626) (0.7297,0.3462) (0.7456,0.3304) (0.7616,0.3151)
(0.7781,0.2999) (0.7941,0.2852) (0.8102,0.2699) (0.8261,0.2553) (0.8427,0.2394)
(0.8586,0.2248) (0.8745,0.2101) (0.8906,0.1954) (0.9065,0.1813) (0.9231,0.1661)
(0.9390,0.1526) (0.9550,0.1408)

\rput[r](0.8360,0.8010){pQCD-photons (Turbide et al.)}
\PST@linefou(0.8520,0.8010)(0.9310,0.8010)

\PST@linefou(0.4884,0.6284) (0.4884,0.6284) (0.5043,0.6085) (0.5209,0.5915)
(0.5368,0.5750) (0.5527,0.5574) (0.5688,0.5410) (0.5848,0.5234) (0.6013,0.5064)
(0.6172,0.4899) (0.6332,0.4724) (0.6493,0.4559) (0.6652,0.4377) (0.6818,0.4190)
(0.6977,0.4014) (0.7136,0.3843) (0.7297,0.3691) (0.7456,0.3515) (0.7616,0.3362)
(0.7781,0.3204) (0.7941,0.3057) (0.8102,0.2922) (0.8261,0.2787) (0.8420,0.2658)
(0.8586,0.2529) (0.8745,0.2406) (0.8906,0.2277) (0.9065,0.2165) (0.9231,0.2054)
(0.9390,0.1937) (0.9550,0.1831)

\PST@Border(0.2390,0.9680) (0.2390,0.1344) (0.9630,0.1344) (0.9630,0.9680)
(0.2390,0.9680)

\catcode`@=12
\fi
\endpspicture

%% file: onoffpole.tex
%

\ifx\PSTloaded\undefined
\def\PSTloaded{t}
\psset{arrowsize=.01 3.2 1.4 .3}
\psset{dotsize=.01}
\catcode`@=11

\definecolor{orange}{rgb}{.8,.4,0}
\definecolor{violett}{rgb}{.5,0,.5}
\newpsobject{PST@Border}{psline}{linewidth=.0015,linestyle=solid}
\newpsobject{PST@Axes}{psline}{linewidth=.0015,linestyle=dotted,dotsep=.004}
\newpsobject{PST@lineone}{psline}{linecolor=red,    linewidth=.0015,linestyle=solid}
\newpsobject{PST@linetwo}{psline}{linecolor=orange, linewidth=.0015,linestyle=dashed,dash=.01 .01}
\newpsobject{PST@linethr}{psline}{linecolor=blue,   linewidth=.0025,linestyle=dotted,dotsep=.004}
\newpsobject{PST@linefou}{psline}{linecolor=violett,linewidth=.0015,linestyle=dashed,dash=.01 .004 .004 .004}
\catcode`@=12

\fi
\psset{unit=5.0in,xunit=1.087\columnwidth,yunit=.8\columnwidth}
\noindent
\pspicture(0.050000,0.126000)(0.970000,1.08100000)
\ifx\nofigs\undefined
\catcode`@=11

\PST@Border(0.1910,0.1260) (0.2060,0.1260) \PST@Border(0.9470,0.1260)
(0.9320,0.1260) \PST@Border(0.1910,0.1682)
(0.1985,0.1682) \PST@Border(0.9470,0.1682) (0.9395,0.1682)
\PST@Border(0.1910,0.2241) (0.1985,0.2241) \PST@Border(0.9470,0.2241)
(0.9395,0.2241) \PST@Border(0.1910,0.2527) (0.1985,0.2527)
\PST@Border(0.9470,0.2527) (0.9395,0.2527) \PST@Border(0.1910,0.2663)
(0.2060,0.2663) \PST@Border(0.9470,0.2663) (0.9320,0.2663)
\rput[r](0.1750,0.2663){ 0.01} \PST@Border(0.1910,0.3086) (0.1985,0.3086)
\PST@Border(0.9470,0.3086) (0.9395,0.3086) \PST@Border(0.1910,0.3644)
(0.1985,0.3644) \PST@Border(0.9470,0.3644) (0.9395,0.3644)
\PST@Border(0.1910,0.3931) (0.1985,0.3931) \PST@Border(0.9470,0.3931)
(0.9395,0.3931) \PST@Border(0.1910,0.4067) (0.2060,0.4067)
\PST@Border(0.9470,0.4067) (0.9320,0.4067) \rput[r](0.1750,0.4067){ 0.1}
\PST@Border(0.1910,0.4489) (0.1985,0.4489) \PST@Border(0.9470,0.4489)
(0.9395,0.4489) \PST@Border(0.1910,0.5048) (0.1985,0.5048)
\PST@Border(0.9470,0.5048) (0.9395,0.5048) \PST@Border(0.1910,0.5334)
(0.1985,0.5334) \PST@Border(0.9470,0.5334) (0.9395,0.5334)
\PST@Border(0.1910,0.5470) (0.2060,0.5470) \PST@Border(0.9470,0.5470)
(0.9320,0.5470) \rput[r](0.1750,0.5470){ 1} \PST@Border(0.1910,0.5892)
(0.1985,0.5892) \PST@Border(0.9470,0.5892) (0.9395,0.5892)
\PST@Border(0.1910,0.6451) (0.1985,0.6451) \PST@Border(0.9470,0.6451)
(0.9395,0.6451) \PST@Border(0.1910,0.6737) (0.1985,0.6737)
\PST@Border(0.9470,0.6737) (0.9395,0.6737) \PST@Border(0.1910,0.6873)
(0.2060,0.6873) \PST@Border(0.9470,0.6873) (0.9320,0.6873)
\rput[r](0.1750,0.6873){ 10} \PST@Border(0.1910,0.7296) (0.1985,0.7296)
\PST@Border(0.9470,0.7296) (0.9395,0.7296) \PST@Border(0.1910,0.7854)
(0.1985,0.7854) \PST@Border(0.9470,0.7854) (0.9395,0.7854)
\PST@Border(0.1910,0.8141) (0.1985,0.8141) \PST@Border(0.9470,0.8141)
(0.9395,0.8141) \PST@Border(0.1910,0.8277) (0.2060,0.8277)
\PST@Border(0.9470,0.8277) (0.9320,0.8277) \rput[r](0.1750,0.8277){ 100}
\PST@Border(0.1910,0.8699) (0.1985,0.8699) \PST@Border(0.9470,0.8699)
(0.9395,0.8699) \PST@Border(0.1910,0.9258) (0.1985,0.9258)
\PST@Border(0.9470,0.9258) (0.9395,0.9258) \PST@Border(0.1910,0.9544)
(0.1985,0.9544) \PST@Border(0.9470,0.9544) (0.9395,0.9544)
\PST@Border(0.1910,0.9680) (0.2060,0.9680) \PST@Border(0.9470,0.9680)
(0.9320,0.9680) \rput[r](0.1750,0.9680){$10^3$} \PST@Border(0.1910,0.1260)
(0.1910,0.1460) \PST@Border(0.1910,0.9680) (0.1910,0.9480)
\rput(0.1910,0.9940){ 0} \PST@Border(0.3422,0.1260) (0.3422,0.1460)
\PST@Border(0.3422,0.9680) (0.3422,0.9480) \rput(0.3422,0.9940){ 0.2}
\PST@Border(0.4934,0.1260) (0.4934,0.1460) \PST@Border(0.4934,0.9680)
(0.4934,0.9480) \rput(0.4934,0.9940){ 0.4} \PST@Border(0.6446,0.1260)
(0.6446,0.1460) \PST@Border(0.6446,0.9680) (0.6446,0.9480)
\rput(0.6446,0.9940){ 0.6} \PST@Border(0.7958,0.1260) (0.7958,0.1460)
\PST@Border(0.7958,0.9680) (0.7958,0.9480) \rput(0.7958,0.9940){ 0.8}
\PST@Border(0.9470,0.1260) (0.9470,0.1460) \PST@Border(0.9470,0.9680)
(0.9470,0.9480) \rput(0.9470,0.9940){ 1} \PST@Border(0.1910,0.9680)
(0.1910,0.1260) (0.9470,0.1260) (0.9470,0.9680) (0.1910,0.9680)

\rput{L}(0.0820,0.5470){$E\frac{dN}{d^3p}$ [GeV$^{-2}$]}
\rput(0.5690,1.0610){$p_\bot$~[GeV]}

\rput[r](0.9150,0.7590){UrQMD}
\rput[r](0.9150,0.7170){Pb+Pb 158~AGeV}
\rput[r](0.9150,0.6750){$b<4.5$~fm}
\rput[r](0.9150,0.6330){$|y_{\rm c.m.}| < 0.5$}

\rput[r](0.8200,0.9270){$m_\rho$~fixed}
\rput[r](0.8200,0.8850){$m_\rho$~Breit-Wigner}
\PST@lineone(0.8360,0.9320)(0.9150,0.9320)
\PST@linetwo(0.8360,0.9220)(0.9150,0.9220)
\PST@linethr(0.8360,0.8900)(0.9150,0.8900)
\PST@linefou(0.8360,0.8800)(0.9150,0.8800)

\rput[r](0.6000,0.2500){$\pi^\pm\pi^0\rightarrow\gamma\rho^\pm$}
\rput[l](0.5500,0.5000){$\pi^\pm\pi^\mp\rightarrow\gamma\rho^0\ \times 10$}

\PST@lineone(0.2288,0.7982) (0.2288,0.7982) (0.3044,0.5881) (0.3800,0.4915)
(0.4556,0.4206) (0.5312,0.3620) (0.6068,0.3101) (0.6824,0.2644) (0.7580,0.2220)
(0.8336,0.1825) (0.9092,0.1417) (0.9433,0.1260)

\PST@linetwo(0.2288,0.9191) (0.2288,0.9191) (0.3044,0.7016) (0.3800,0.6039)
(0.4556,0.5329) (0.5312,0.4713) (0.6068,0.4187) (0.6824,0.3693) (0.7580,0.3254)
(0.8336,0.2869) (0.9092,0.2498) (0.9470,0.2300)

\PST@linethr(0.2288,0.7786) (0.2288,0.7786) (0.3044,0.5806) (0.3800,0.4856)
(0.4556,0.4123) (0.5312,0.3557) (0.6068,0.3054) (0.6824,0.2573) (0.7580,0.2146)
(0.8336,0.1768) (0.9092,0.1363) (0.9310,0.1260)

\PST@linefou(0.2288,0.8990) (0.2288,0.8990) (0.3044,0.6945) (0.3800,0.5974)
(0.4556,0.5259) (0.5312,0.4663) (0.6068,0.4126) (0.6824,0.3650) (0.7580,0.3207)
(0.8336,0.2798) (0.9092,0.2459) (0.9470,0.2271)

\PST@Border(0.1910,0.9680)(0.1910,0.1260)(0.9470,0.1260)(0.9470,0.9680)(0.1910,0.9680)

\catcode`@=12
\fi
\endpspicture\\
\psset{yunit=.2\columnwidth}%
\pspicture(0.050000,-.200000)(0.970000,.9680000)
\ifx\nofigs\undefined
\catcode`@=11

\PST@Border(0.1910,0.1260) (0.2057,0.1260)
\PST@Border(0.9470,0.1260) (0.9323,0.1260)
\rput[r](0.1753,0.1260){ 0}
\PST@Border(0.1910,0.3365) (0.1985,0.3365)
\PST@Border(0.9470,0.3365) (0.9395,0.3365)
\PST@Border(0.1910,0.5470) (0.2057,0.5470)
\PST@Border(0.9470,0.5470) (0.9323,0.5470)
\rput[r](0.1753,0.5470){ 0.2}
\PST@Border(0.1910,0.7575) (0.1985,0.7575)
\PST@Border(0.9470,0.7575) (0.9395,0.7575)
\PST@Border(0.1910,0.1260) (0.1910,0.1460)
\PST@Border(0.1910,0.9680) (0.1910,0.9480)
\rput(0.1910,0.0240){ 0} \PST@Border(0.2666,0.1260) (0.2666,0.1460)
\PST@Border(0.2666,0.9680) (0.2666,0.9480)
\PST@Border(0.3422,0.1260) (0.3422,0.1460)
\PST@Border(0.3422,0.9680) (0.3422,0.9480)
\rput(0.3422,0.0240){ 0.2} \PST@Border(0.4178,0.1260) (0.4178,0.1460)
\PST@Border(0.4178,0.9680) (0.4178,0.9480)
\PST@Border(0.4934,0.1260) (0.4934,0.1460)
\PST@Border(0.4934,0.9680) (0.4934,0.9480)
\rput(0.4934,0.0240){ 0.4} \PST@Border(0.5690,0.1260) (0.5690,0.1460)
\PST@Border(0.5690,0.9680) (0.5690,0.9480)
\PST@Border(0.6446,0.1260) (0.6446,0.1460)
\PST@Border(0.6446,0.9680) (0.6446,0.9480)
\rput(0.6446,0.0240){ 0.6} \PST@Border(0.7202,0.1260) (0.7202,0.1460)
\PST@Border(0.7202,0.9680) (0.7202,0.9480)
\PST@Border(0.7958,0.1260) (0.7958,0.1460)
\PST@Border(0.7958,0.9680) (0.7958,0.9480)
\rput(0.7958,0.0240){ 0.8} \PST@Border(0.8714,0.1260) (0.8714,0.1460)
\PST@Border(0.8714,0.9680) (0.8714,0.9480)
\PST@Border(0.9470,0.1260) (0.9470,0.1460)
\PST@Border(0.9470,0.9680) (0.9470,0.9480)
\rput(0.9470,0.0240){ 1} \PST@Border(0.1910,0.9680) (0.1910,0.1260)
(0.9470,0.1260) (0.9470,0.9680) (0.1910,0.9680)

\rput{L}(0.0999,0.4500){(on-off)/off}
\rput(0.5690,-.1240){$p_\bot$~[GeV]}

\rput[r](0.8200,0.8070){$\pi^\pm\pi^0\rightarrow\gamma\rho^\pm$}
\rput[r](0.8200,0.6550){$\pi^\pm\pi^\mp\rightarrow\gamma\rho^0$}
\PST@lineone(0.8360,0.8070)(0.9150,0.8070)
\PST@linefou(0.8360,0.6550)(0.9150,0.6550)

\PST@lineone(0.2288,0.9221) (0.2288,0.9221) (0.3044,0.4001) (0.3800,0.3386)
(0.4556,0.4367) (0.5312,0.3550) (0.6068,0.2927) (0.6824,0.3888) (0.7580,0.3976)
(0.8336,0.3347) (0.9092,0.3189)

\PST@linefou(0.2288,0.9512) (0.2288,0.9512) (0.3044,0.3860) (0.3800,0.3666)
(0.4556,0.3822) (0.5312,0.3077) (0.6068,0.3474) (0.6824,0.2807) (0.7580,0.2964)
(0.8336,0.3866) (0.9092,0.2667)

\PST@Border(0.1910,0.9680) (0.1910,0.1260) (0.9470,0.1260) (0.9470,0.9680)
(0.1910,0.9680)

\catcode`@=12
\fi
\endpspicture

%% file: dumitru.tex
\ifx\PSTloaded\undefined
\def\PSTloaded{t}
\psset{arrowsize=.01 3.2 1.4 .3}
\psset{dotsize=.01}
\catcode`@=11

\newpsobject{PST@Border}{psline}{linewidth=.0015,linestyle=solid}
\newpsobject{PST@adrian}{psline}{linewidth=.0015,linestyle=solid}
\newpsobject{PST@oldver}{psline}{linewidth=.0025,linecolor=blue,linestyle=dotted,dotsep=.008}
\newpsobject{PST@thiswo}{psline}{linewidth=.0015,linecolor=red,linestyle=dashed,dash=.01 .01}
\catcode`@=12

\fi
\psset{unit=5.0in,xunit=1.149\columnwidth,yunit=.5\columnwidth}
\pspicture(0.11,0.0)(0.98,0.988)
\ifx\nofigs\undefined
\catcode`@=11

\rput[r](0.2230,0.1344){$10^{-6}$}
\rput[r](0.2230,0.3011){$10^{-5}$}
\rput[r](0.2230,0.4678){$10^{-4}$}
\rput[r](0.2230,0.6346){$10^{-3}$}
\rput[r](0.2230,0.8013){ 0.01}
\rput[r](0.2230,0.9680){ 0.1}
\rput(0.2390,0.0924){ 1}
\rput(0.4200,0.0924){ 1.5}
\rput(0.6010,0.0924){ 2}
\rput(0.7820,0.0924){ 2.5}
\rput(0.9630,0.0924){ 3}

\PST@Border(0.2390,0.1344)(0.2540,0.1344)
\PST@Border(0.9630,0.1344)(0.9480,0.1344)
\PST@Border(0.2390,0.1846)(0.2465,0.1846)
\PST@Border(0.9630,0.1846)(0.9555,0.1846)
\PST@Border(0.2390,0.2509)(0.2465,0.2509)
\PST@Border(0.9630,0.2509)(0.9555,0.2509)
\PST@Border(0.2390,0.2850)(0.2465,0.2850)
\PST@Border(0.9630,0.2850)(0.9555,0.2850)
\PST@Border(0.2390,0.3011)(0.2540,0.3011)
\PST@Border(0.9630,0.3011)(0.9480,0.3011)
\PST@Border(0.2390,0.3513)(0.2465,0.3513)
\PST@Border(0.9630,0.3513)(0.9555,0.3513)
\PST@Border(0.2390,0.4177)(0.2465,0.4177)
\PST@Border(0.9630,0.4177)(0.9555,0.4177)
\PST@Border(0.2390,0.4517)(0.2465,0.4517)
\PST@Border(0.9630,0.4517)(0.9555,0.4517)
\PST@Border(0.2390,0.4678)(0.2540,0.4678)
\PST@Border(0.9630,0.4678)(0.9480,0.4678)
\PST@Border(0.2390,0.5180)(0.2465,0.5180)
\PST@Border(0.9630,0.5180)(0.9555,0.5180)
\PST@Border(0.2390,0.5844)(0.2465,0.5844)
\PST@Border(0.9630,0.5844)(0.9555,0.5844)
\PST@Border(0.2390,0.6184)(0.2465,0.6184)
\PST@Border(0.9630,0.6184)(0.9555,0.6184)
\PST@Border(0.2390,0.6346)(0.2540,0.6346)
\PST@Border(0.9630,0.6346)(0.9480,0.6346)
\PST@Border(0.2390,0.6847)(0.2465,0.6847)
\PST@Border(0.9630,0.6847)(0.9555,0.6847)
\PST@Border(0.2390,0.7511)(0.2465,0.7511)
\PST@Border(0.9630,0.7511)(0.9555,0.7511)
\PST@Border(0.2390,0.7851)(0.2465,0.7851)
\PST@Border(0.9630,0.7851)(0.9555,0.7851)
\PST@Border(0.2390,0.8013)(0.2540,0.8013)
\PST@Border(0.9630,0.8013)(0.9480,0.8013)
\PST@Border(0.2390,0.8515)(0.2465,0.8515)
\PST@Border(0.9630,0.8515)(0.9555,0.8515)
\PST@Border(0.2390,0.9178)(0.2465,0.9178)
\PST@Border(0.9630,0.9178)(0.9555,0.9178)
\PST@Border(0.2390,0.9518)(0.2465,0.9518)
\PST@Border(0.9630,0.9518)(0.9555,0.9518)
\PST@Border(0.2390,0.9680)(0.2540,0.9680)
\PST@Border(0.9630,0.9680)(0.9480,0.9680)
\PST@Border(0.2390,0.1344)(0.2390,0.1544)
\PST@Border(0.2390,0.9680)(0.2390,0.9480)
\PST@Border(0.4200,0.1344)(0.4200,0.1544)
\PST@Border(0.4200,0.9680)(0.4200,0.9480)
\PST@Border(0.6010,0.1344)(0.6010,0.1544)
\PST@Border(0.6010,0.9680)(0.6010,0.9480)
\PST@Border(0.7820,0.1344)(0.7820,0.1544)
\PST@Border(0.7820,0.9680)(0.7820,0.9480)
\PST@Border(0.9630,0.1344)(0.9630,0.1544)
\PST@Border(0.9630,0.9680)(0.9630,0.9480)
\PST@Border(0.2390,0.9680)(0.2390,0.1344) (0.9630,0.1344)(0.9630,0.9680)
(0.2390,0.9680)

\rput{L}(0.1360,0.5512){$E\frac{dN}{d^3p}$ [GeV$^{-2}$]}
\rput(0.6010,0.0294){$p_\bot$~[GeV]}

\rput[r](0.8360,0.9070){Dumitru {\it et al.} (UrQMD v1.0)}
\rput[r](0.8360,0.8450){UrQMD v1.3, this work}
\rput[r](0.8360,0.7830){UrQMD v2.3, this work}

\PST@adrian(0.8520,0.9070)(0.9310,0.9070)
\PST@oldver(0.8520,0.8450)(0.9310,0.8450)
\PST@thiswo(0.8520,0.7830)(0.9310,0.7830)

\PST@adrian(0.2390,0.8874) (0.3299,0.7949) (0.4377,0.6893) (0.5456,0.6376)
(0.6549,0.5859) (0.7643,0.5084) (0.8721,0.4394) (0.9630,0.4341)

\PST@oldver(0.2390,0.8698) (0.2571,0.8503) (0.2933,0.8106) (0.3295,0.7713)
(0.3657,0.7334) (0.4019,0.7057) (0.4381,0.6654) (0.4743,0.6315) (0.5105,0.6055)
(0.5467,0.5911) (0.5829,0.5669) (0.6191,0.5109) (0.6553,0.4941) (0.6915,0.4822)
(0.7277,0.4657) (0.7639,0.4422) (0.8001,0.4477) (0.8363,0.4426) (0.8725,0.4119)
(0.9087,0.3626) (0.9449,0.3667) (0.9630,0.3411)

\PST@thiswo(0.2390,0.8250) (0.2571,0.8067) (0.2933,0.7683) (0.3295,0.7357)
(0.3657,0.7018) (0.4019,0.6661) (0.4381,0.6222) (0.4743,0.5861) (0.5105,0.5553)
(0.5467,0.5205) (0.5829,0.4948) (0.6191,0.4681) (0.6553,0.4443) (0.6915,0.4180)
(0.7277,0.4154) (0.7639,0.3828) (0.8001,0.3373) (0.8363,0.3394) (0.8725,0.2876)
(0.9087,0.2841) (0.9449,0.2311) (0.9630,0.2350)

\PST@Border(0.2390,0.9680) (0.2390,0.1344) (0.9630,0.1344) (0.9630,0.9680)
(0.2390,0.9680)

\catcode`@=12
\fi
\endpspicture

%% file: paper.bbl
\begin{thebibliography}{}

\bibitem{Harris:1996zx}
  J.~W.~Harris and B.~Muller,
  Ann.\ Rev.\ Nucl.\ Part.\ Sci.\  {\bf 46} (1996) 71

\bibitem{Bass:1998vz}
  S.~A.~Bass, M.~Gyulassy, H.~Stoecker and W.~Greiner,
  J.\ Phys.\ G {\bf 25} (1999) R1

\bibitem{Adams:2005dq}
  J.~Adams {\it et al.}  [STAR Collaboration],
  Nucl.\ Phys.\  A {\bf 757} (2005) 102

\bibitem{Back:2004je}
  B.~B.~Back {\it et al.},
  Nucl.\ Phys.\  A {\bf 757} (2005) 28

\bibitem{Arsene:2004fa}
  I.~Arsene {\it et al.}  [BRAHMS Collaboration],
  Nucl.\ Phys.\  A {\bf 757} (2005) 1

\bibitem{Adcox:2004mh}
  K.~Adcox {\it et al.}  [PHENIX Collaboration],
  Nucl.\ Phys.\  A {\bf 757} (2005) 184

\bibitem{:2007fe}
  C.~Alt {\it et al.}  [NA49 Collaboration],
  Phys.\ Rev.\  C {\bf 77} (2008) 024903

\bibitem{Aggarwal:2000ps}
  M.~M.~Aggarwal {\it et al.}  [WA98 Collaboration],

\bibitem{Adler:2005ig}
  S.~S.~Adler {\it et al.}  [PHENIX Collaboration],
  Phys.\ Rev.\ Lett.\  {\bf 94} (2005) 232301

\bibitem{:2008fqa}
  A.~Adare {\it et al.}  [PHENIX Collaboration],

\bibitem{Kapusta:1991qp}
  J.~I.~Kapusta, P.~Lichard and D.~Seibert,
  Phys.\ Rev.\  D {\bf 44} (1991) 2774
  [Erratum-ibid.\  D {\bf 47} (1993) 4171].

\bibitem{Xiong:1992ui}
  L.~Xiong, E.~V.~Shuryak and G.~E.~Brown,
  Phys.\ Rev.\  D {\bf 46} (1992) 3798

\bibitem{Geiger:1997pf}
  K.~Geiger,
  Comput.\ Phys.\ Commun.\  {\bf 104} (1997) 70

\bibitem{Bass:1998ca}
  S.~A.~Bass {\it et al.},
  Prog.\ Part.\ Nucl.\ Phys.\  {\bf 41} (1998) 255
  [Prog.\ Part.\ Nucl.\ Phys.\  {\bf 41} (1998) 225]

\bibitem{Bleicher:1999xi}
  M.~Bleicher {\it et al.},
  J.\ Phys.\ G {\bf 25} (1999) 1859

\bibitem{Ehehalt:1995is}
  W.~Ehehalt and W.~Cassing,

\bibitem{Molnar:2004yh}
  D.~Molnar and P.~Huovinen,
  Phys.\ Rev.\ Lett.\  {\bf 94} (2005) 012302

\bibitem{Xu:2004mz}
  Z.~Xu and C.~Greiner,
  Phys.\ Rev.\  C {\bf 71} (2005) 064901

\bibitem{Lin:2004en}
  Z.~W.~Lin, C.~M.~Ko, B.~A.~Li, B.~Zhang and S.~Pal,
  Phys.\ Rev.\  C {\bf 72} (2005) 064901

\bibitem{Burau:2004ev}
  G.~Burau, J.~Bleibel, C.~Fuchs, A.~Faessler, L.~V.~Bravina and E.~E.~Zabrodin,
  Phys.\ Rev.\  C {\bf 71} (2005) 054905

\bibitem{Bass:2007hy}
  S.~A.~Bass, T.~Renk and D.~K.~Srivastava,
  Nucl.\ Phys.\  A {\bf 783} (2007) 367.

\bibitem{Barz:2000zz}
  H.~W.~Barz and B.~Kampfer,
  Nucl.\ Phys.\  A {\bf 683} (2001) 594

\bibitem{Cassing:2001ds}
  W.~Cassing,
  Nucl.\ Phys.\  A {\bf 700} (2002) 618

\bibitem{Larionov:2007hy}
  A.~B.~Larionov, O.~Buss, K.~Gallmeister and U.~Mosel,
  Phys.\ Rev.\  C {\bf 76} (2007) 044909

\bibitem{Bleibel:2006xx}
  J.~Bleibel, G.~Burau, A.~Faessler and C.~Fuchs,
  Phys.\ Rev.\  C {\bf 76} (2007) 024912

\bibitem{Bleibel:2007se}
  J.~Bleibel, G.~Burau and C.~Fuchs,
  Phys.\ Lett.\  B {\bf 659} (2008) 520

\bibitem{Cleymans:1985wp}
  J.~Cleymans and K.~Redlich,

\bibitem{McLerran:1986nc}
  L.~D.~McLerran, M.~Kataja, P.~V.~Ruuskanen and H.~von Gersdorff,
  Phys.\ Rev.\  D {\bf 34} (1986) 2755.

\bibitem{PHRVA.D34.794}
  H.~Von Gersdorff, L.~D.~McLerran, M.~Kataja and P.~V.~Ruuskanen,
  Phys.\ Rev.\  D {\bf 34} (1986) 794.

\bibitem{Kataja:1988iq}
  M.~Kataja,
  Z.\ Phys.\  C {\bf 38} (1988) 419.

\bibitem{Srivastava:1991ju}
  D.~K.~Srivastava and B.~Sinha,
  Phys.\ Lett.\  B {\bf 261} (1991) 1.

\bibitem{Srivastava:1991nc}
  D.~K.~Srivastava, B.~Sinha and T.~C.~Awes,
  Phys.\ Lett.\  B {\bf 387} (1996) 21.

\bibitem{Srivastava:1991dm}
  D.~K.~Srivastava, B.~Sinha, M.~Gyulassy and X.~N.~Wang,
  Phys.\ Lett.\  B {\bf 276} (1992) 285.

\bibitem{Srivastava:1992gh}
  D.~K.~Srivastava, J.~Alam, S.~Chakrabarty, S.~Raha and B.~Sinha,
  Phys.\ Lett.\  B {\bf 278} (1992) 225.

\bibitem{Cleymans:1992zc}
  J.~Cleymans and H.~Satz,
  Z.\ Phys.\  C {\bf 57} (1993) 135

\bibitem{Rischke:1995mt}
  D.~H.~Rischke, Y.~Pursun and J.~A.~Maruhn,
  Nucl.\ Phys.\  A {\bf 595} (1995) 383
  [Erratum-ibid.\  A {\bf 596} (1996) 717]

\bibitem{Hirano:2001eu}
  T.~Hirano,
  Phys.\ Rev.\  C {\bf 65} (2002) 011901

\bibitem{Huovinen:2001wx}
  P.~Huovinen, P.~V.~Ruuskanen and S.~S.~Rasanen,
  Phys.\ Lett.\  B {\bf 535} (2002) 109

\bibitem{Huovinen:2002im}
  P.~Huovinen, M.~Belkacem, P.~J.~Ellis and J.~I.~Kapusta,
  Phys.\ Rev.\  C {\bf 66} (2002) 014903

\bibitem{Kolb:2003dz}
  P.~F.~Kolb and U.~W.~Heinz,
  arXiv:nucl-th/0305084.

\bibitem{Nonaka:2006yn}
  C.~Nonaka and S.~A.~Bass,
  Phys.\ Rev.\  C {\bf 75} (2007) 014902

\bibitem{Frodermann:2007ab}
  E.~Frodermann, R.~Chatterjee and U.~Heinz,
  J.\ Phys.\ G {\bf 34} (2007) 2249

\bibitem{Dusling:2007gi}
  K.~Dusling and D.~Teaney,
  Phys.\ Rev.\  C {\bf 77} (2008) 034905

\bibitem{Baier:2006um}
  R.~Baier, P.~Romatschke and U.~A.~Wiedemann,
  Phys.\ Rev.\  C {\bf 73} (2006) 064903

\bibitem{Song:2008si}
  H.~Song and U.~W.~Heinz,
  Phys.\ Rev.\  C {\bf 78} (2008) 024902

\bibitem{Dumitru:1998sd}
  A.~Dumitru, M.~Bleicher, S.~A.~Bass, C.~Spieles, L.~Neise, H.~Stoecker and W.~Greiner,
  Phys.\ Rev.\  C {\bf 57} (1998) 3271

\bibitem{Srivastava:1998tf}
  D.~K.~Srivastava and K.~Geiger,
  Phys.\ Rev.\  C {\bf 58} (1998) 1734

\bibitem{Bratkovskaya:2008iq}
  E.~L.~Bratkovskaya, S.~M.~Kiselev and G.~B.~Sharkov,
  Phys.\ Rev.\  C {\bf 78} (2008) 034905

\bibitem{Petersen:2008kb}
  H.~Petersen, M.~Bleicher, S.~A.~Bass and H.~Stocker,
  arXiv:0805.0567 [hep-ph].

\bibitem{Sjostrand:2006za}
  T.~Sjostrand, S.~Mrenna and P.~Z.~Skands,
  JHEP {\bf 0605} (2006) 026

\bibitem{Petersen:2008dd}
  H.~Petersen, J.~Steinheimer, G.~Burau, M.~Bleicher and H.~Stocker,
  Phys.\ Rev.\  C {\bf 78} (2008) 044901

\bibitem{Steinheimer:2008hr}
  J.~Steinheimer, M.~Mitrovski, T.~Schuster, H.~Petersen, M.~Bleicher and H.~Stoecker,
  Phys.\ Lett.\  B {\bf 676} (2009) 126

\bibitem{Petersen:2009zi}
  H.~Petersen, M.~Mitrovski, T.~Schuster and M.~Bleicher,
  Phys.\ Rev.\  C {\bf 80} (2009) 054910

\bibitem{Li:2008qm}
  Q.~f.~Li, J.~Steinheimer, H.~Petersen, M.~Bleicher and H.~Stocker,
  Phys.\ Lett.\  B {\bf 674} (2009) 111

\bibitem{Petersen:2009mz}
  H.~Petersen, J.~Steinheimer, M.~Bleicher and H.~Stocker,
  J.\ Phys.\ G {\bf 36} (2009) 055104

\bibitem{Petersen:1900zz}
  H.~Petersen, J.~Steinheimer, G.~Burau and M.~Bleicher,
  Eur.\ Phys.\ J.\  C {\bf 62} (2009) 31.

\bibitem{Turbide:2003si}
  S.~Turbide, R.~Rapp and C.~Gale,
  Phys.\ Rev.\  C {\bf 69} (2004) 014903

\bibitem{Arnold:2001ms}
  P.~B.~Arnold, G.~D.~Moore and L.~G.~Yaffe,
  JHEP {\bf 0112} (2001) 009

\bibitem{Liu:2007tw}
  F.~M.~Liu and K.~Werner,
  J.\ Phys.\ G {\bf 36} (2009) 035101

\bibitem{Gale:2001yh}
  C.~Gale,
  Nucl.\ Phys.\  A {\bf 698} (2002) 143

\bibitem{Wong:1998pq}
  C.~Y.~Wong and H.~Wang,
  Phys.\ Rev.\  C {\bf 58} (1998) 376

\bibitem{nucl-th/9804058}
  M.~Belkacem {\it et al.},
  Phys.\ Rev.\  C {\bf 58} (1998) 1727

\bibitem{Steinheimer:2009nn}
  J.~Steinheimer, V.~Dexheimer, H.~Petersen, M.~Bleicher, S.~Schramm and H.~Stoecker,
  arXiv:0905.3099 [hep-ph].



\end{thebibliography}
